\documentclass[12pt]{article}
\usepackage{cite}
\usepackage{epsfig}
\usepackage{graphicx}
\usepackage{amsmath}
\usepackage{amssymb}
\usepackage{ulem}
\usepackage{mdwlist} 
\usepackage{color}  
\usepackage[OT2, T1]{fontenc}
\usepackage[russian,english]{babel}
\usepackage{pifont}
\usepackage{hyperref}

\usepackage{a41}
\usepackage{color}
\usepackage[rflt]{floatflt}
\usepackage{float}
\usepackage{slashed}
\usepackage{multirow}

\newcommand{\GA}{G}

\def\q{\slashed{q}} 

\usepackage{array}

\usepackage[english]{babel}

\usepackage{url}

\usepackage{amsmath, amsthm, amssymb}
\newtheorem{thm}{Theorem}[section]

\def\b0{\beta_0}

\newtheorem{definition}[thm]{Definition}

\renewcommand{\P}{\mathbb P}

\newcommand{\Li}{{\rm Li}}
\newcommand{\HA}{{\rm H}}

\newcommand{\KK}{\mathbb{K}}
\newcommand{\QQ}{\mathbb{Q}}

\newcommand{\tc}{\tilde{c}}
\newcommand{\hs}{\hat{s}}
\newcommand{\tk}{\tilde{\kappa}}

\newcommand{\tz}{\tilde{z}}

\newcommand{\cm}{\scriptsize \checkmark}

\newcommand{\ep}{\varepsilon}

\def\b0{\beta_0}

\usepackage{rotating}

\usepackage{graphicx}

\newcounter{mmacnt}
\def\restartmma{\setcounter{mmacnt}{0}}
\restartmma \catcode`|=\active
\def|#1|{\mathrm{#1}}
\catcode`|=12
\newenvironment{mma}{
 \par\smallskip
 \catcode`|=\active
 \parskip=0pt\parindent=0pt 
 \small
 \def\In##1\\{%
\def\linebreak{\hfill\break\null\qquad}%
\refstepcounter{mmacnt}
\hangindent=2.5em\hangafter=0
\leavevmode
\llap{\tiny\sffamily n[\arabic{mmacnt}]:=\kern.5em}%
\mathversion{bold}\footnotesize$\displaystyle##1$\normalsize
\mathversion{normal}\par
 }%
 \def\Print##1\\{%
\def\linebreak{\hfill\break}%
\hangindent=2.5em\hangafter=0
\leavevmode ##1\par}%
 \def\Out##1\\{%
\def\linebreak{$\hfill\break\null\hfill$}%
\kern\abovedisplayskip\par
\hangindent=2.5em\hangafter=0
\leavevmode
\llap{\tiny\sffamily Out[\arabic{mmacnt}]=\kern.5em}
\footnotesize$\displaystyle##1$\normalsize\hfill\null\par
\kern\belowdisplayskip
 }%
 \def\Warning##1##2\\{%
\def\linebreak{\hfill\break}%
\hangindent=2.5em\hangafter=0
\leavevmode
{\scriptsize##1 : ##2}\par}%
}{%
 \par\smallskip
}

\usepackage{color}

\newenvironment{fshaded}{%
\MakeFramed {\FrameRestore}
}%
{\endMakeFramed}

\catcode`,\active

\catcode`\,12

\allowdisplaybreaks[4]

\begin{document}
\setlength{\baselineskip}{0.515cm}
\sloppy
\thispagestyle{empty}
\begin{flushleft}
DESY 23--012
\\
DO--TH 23/02\\
RISC Report Series 23-08\\
July 2023\\
\end{flushleft}

\mbox{}
\vspace*{\fill}
\begin{center}

{\LARGE\bf  }

\vspace*{2mm}
{\LARGE\bf  Analytic results on the massive three-loop} 

\vspace*{3mm}
{\LARGE\bf form factors: quarkonic contributions}

\vspace{3cm}
\large 
J.~Bl\"umlein$^a$,
A.~De~Freitas$^b$,
P.~Marquard$^a$, 
N.~Rana$^{a,c}$, 
 and
C.~Schneider$^b$

\vspace{1.cm}
\normalsize
{\it  $^a$ Deutsches Elektronen--Synchrotron DESY,
Platanenallee 6, 15738 Zeuthen, Germany}

\vspace*{3mm}
{\it $^b$~  Johannes Kepler University,
Research Institute for Symbolic Computation (RISC), Altenbergerstra{\ss}e 69,
A--4040, Linz, Austria}

\vspace*{3mm}
{\it $^c$~  School of Physical Sciences, National Institute of Science Education and Research,\\ An OCC of Homi Bhabha National Institute, 752050 Jatni, India}\\


\end{center}
\normalsize
\vspace{\fill}
\begin{abstract}
\noindent
The quarkonic contributions to the three--loop heavy-quark form factors for vector, axial-vector, scalar and 
pseudoscalar currents are described by closed form difference equations for the expansion coefficients in the 
limit of small virtualities $q^2/m^2$. A part of the contributions can
be solved analytically and expressed in terms 
of harmonic and cyclotomic harmonic polylogarithms and square-root valued iterated integrals. Other contributions obey equations which 
are not first--order factorizable. For them still infinite series expansions around the singularities of the form 
factors can be obtained by matching the expansions at intermediate points and using differential equations which 
are obeyed directly by the form factors and are derived by guessing algorithms. One may determine all expansion 
coefficients for $q^2/m^2 \rightarrow \infty$ analytically in terms of multiple zeta values. By expanding around 
the threshold and pseudo--threshold, the corresponding constants are multiple zeta values supplemented by a finite 
amount of new constants, which can be computed at high precision. For a part of these coefficients, the infinite 
series in front of these constants may be even resummed into harmonic polylogarithms. In this way, one obtains a 
deeper analytic description of the massive form factors, beyond their pure numerical evaluation. The calculations 
of these analytic results are based on sophisticated computer algebra techniques. We also compare our results with 
numerical results in the literature.
\end{abstract}

\vspace*{\fill}
\noindent
\numberwithin{equation}{section}

\newpage

\section{Introduction} 
\label{sec:1}

\vspace*{1mm}
\noindent
The massive form factors for the vector, axial--vector, scalar and pseudoscalar currents
describe the decay of virtual bosons into two massive quarks in Quantum Chromodynamics (QCD). They are important 
building blocks to the virtual corrections to a given loop order for processes like bottom- and top--quark pair 
production at $e^+e^-$ and $pp$ colliders. In a similar way, they also contribute to the corresponding corrections
in Quantum Electrodynamics (QED). These processes can be induced by virtual photons, $Z$--bosons and Higgs bosons in
the Standard Model and its extensions. The form factors depend on the ratio of the current virtuality $q^2$ and the 
square of the heavy quark mass $m$. 

The massive form factors were computed at two-loop order in Refs. \cite{Bernreuther:2004ih, Bernreuther:2004th, 
Bernreuther:2005rw, Bernreuther:2005gw} for the first time. Later in Ref. \cite{Gluza:2009yy}, the two-loop calculation 
was extended to include $O(\ep)$ terms in the dimensional parameter $\ep = (4-D)/2$. In Ref.~\cite{Ablinger:2017hst}, 
the $O(\ep^2)$ terms were obtained. The further corrections in $\ep$ contribute to the respective higher order 
corrections. A first part of the analytic three--loop corrections has been obtained in Refs.~\cite{Henn:2016kjz,Henn:2016tyf,
Ahmed:2017gyt,Ablinger:2018yae,Ablinger:2018zwz,Lee:2018nxa,Lee:2018rgs,Blumlein:2018tmz,Blumlein:2019oas}. 
Numerical results on the three--loop form factors have been presented in Refs.~\cite{Fael:2022miw,Fael:2022rgm},
where the first order differential equations for the master integrals were solved numerically in the whole region 
of the kinematic parameters expanding up to a certain order using formal Taylor series around a series of 
points to obtain a numerical representation with overlapping convergence radii.
Very recently, also the anomaly contribution has been calculated in the same way \cite{Fael:2023zqr}.

The purpose of the present paper is to derive analytic results for the expansion coefficients of the quarkonic contributions
to the form factors around the four characteristic kinematic points, namely, the low and high energy limits, as well 
as the two-- and four--particle thresholds, by using the difference equations derived in \cite{Blumlein:2019oas} to very 
high order. The difference equations were obtained by using the method of arbitrary high moments \cite{Blumlein:2017dxp} 
and its implementation within the Mathematica package \texttt{SolveCoupledSystem}~\cite{Blumlein:2019oas,SolveCoupledSystem}. 
Our representations consist of (logarithmically modulated) Taylor series around the four critical points, which 
can be  computed analytically to arbitrary high order. The coefficients in these expansions are given in terms of 
multiple zeta values \cite{Blumlein:2009cf} and a series of new special constants, which we present at very high accuracy. 
For a part of these terms, we can even resum the complete series into special functions like classical polylogarithms 
\cite{LEWIN1}. 
The different expansions are found by using large scale computer algebra calculations. In particular, new difference and differential equation techniques based on the holonomic framework~\cite{HOLONOMIC,SageOre} and symbolic summation~\cite{DRAlgorithms} are non-trivially used for this challenge. 
Our analytic representations allow one to perform detailed numerical comparisons with the work 
in Refs.~\cite{Fael:2022miw,Fael:2022rgm}.

The paper is organized as follows. In Section~\ref{sec:2}, we describe the form factors in general and give a brief summary 
of the calculation done in \cite{Blumlein:2019oas}, which is the basis of the calculations performed in the subsequent 
sections of the present paper. In Section \ref{sec:3}, we introduce our method for the calculation of power--log 
expansions at the singular points of the form factors and apply it to the non-solvable parts described in 
\cite{Blumlein:2019oas} in the high energy limit. The method is based on the matching of expansions at different points 
(Section \ref{sec:31}) and differential equations obeyed directly by the form factors (Section \ref{sec:32}).
We explain how to derive these differential equations, relying on guessing methods, cf.~Refs.~\cite{SageOre} and 
obtain numerical expansions. 
We show that the precision achieved for these expansions can be very high, allowing us to determine the expansion 
coefficients in terms of known constants (Section \ref{sec:33}). For some of the sequences of rational numbers 
associated to these constants, recursion relations can be found and in some cases solved, leading to closed form solutions.
In Section~\ref{sec:4}, we apply this method to the two-- and four--particle thresholds, which required a transformation 
of the expansions used in \cite{Blumlein:2019oas}
(Section \ref{Sec:ComputeSMoments}). In the case of the two--particle threshold (Section \ref{Sec:s4-expansions}), 
we found again that many coefficients in the expansions could be written in terms of known constants. However, this was 
not the case for all of them, which required the introduction of new constants that were computed with very high precision. 
Again, for some of the sequences of rational numbers associated to these constants, we were able to find recursions, 
and in some cases, closed form solutions. Section \ref{sec:5} contains graphs and numerical results, which we compare 
with results in the literature. Section \ref{sec:6} contains the conclusions. In the appendices A--E, we present a series 
of analytic results. Ancillary files are provided in computer--readable form for analytic and numerical studies.
\section{The heavy quark form factors} 
\label{sec:2}

\vspace*{1mm}
\noindent
The decay of a virtual neutral boson into a pair of heavy quarks can be 
described by four different types of vertex amplitudes depending on the spin 
of the boson under consideration and the way it couples to the quarks. For a 
spin-1 boson, we may have vector and/or axial-vector couplings, while in the 
case of a spin-0 particle, we may deal with a scalar or a pseudo-scalar. The 
spin-1 particles typically studied are the photon, which couples as a vector 
to the fermions, and the $Z$-boson, which has both types of couplings. In the 
latter case, the vertex $\Gamma^{\mu}_{cd}$ can be written in terms of form 
factors $F_{V,i}$ and $F_{A,i}$ (with $i=1,2$) as follows
\begin{align}
 \Gamma_{cd}^{\mu} = \Gamma_{V,cd}^{\mu} + \Gamma_{A,cd}^{\mu}
 = -i \delta_{cd} & \Biggl[ v_Q \Biggl( \gamma^{\mu}~F_{V,1}\left(\frac{q^2}{m^2}\right) + \frac{i}{2 m} \sigma^{\mu\nu} q_{\nu} ~F_{V,2}\left(\frac{q^2}{m^2}\right) \Biggr)
\nonumber\\& 
+ a_Q \Biggl( \gamma^{\mu} \gamma_5~F_{A,1}\left(\frac{q^2}{m^2}\right) 
         + \frac{1}{2 m} q^{\mu} \gamma_5 ~ F_{A,2}\left(\frac{q^2}{m^2}\right)  
\Biggr) \Biggr] \,,
\end{align}
where $\sigma^{\mu\nu} = \frac{i}{2} [\gamma^{\mu},\gamma^{\nu}]$, $q$ is the 
momentum of the $Z$-boson, $m$  the mass of the heavy quarks, $c$ and $d$ are 
their color indices, and $v_Q$ and $a_Q$ are the Standard Model vector and 
axial--vector couplings, respectively. 

For scalar and pseudo-scalar currents  the generic vertex structure is given by the following, where $F_S$ and $F_P$ are the corresponding form factors
\begin{align}
 \Gamma_{cd} &= \Gamma_{S,cd} +  \Gamma_{P,cd}
 = - \frac{m}{v} \delta_{cd} ~ \Biggl[ s_Q \, F_{S}\left(\frac{q^2}{m^2}\right) 
+ i p_Q \gamma_5 \, F_{P}\left(\frac{q^2}{m^2}\right) \Biggr] \,,
\end{align}
where $v$ is the vacuum expectation value of the Higgs field of the Standard 
Model, and $s_Q$ and $p_Q$ are the scalar and pseudo-scalar couplings, 
respectively.

\begin{center}
\begin{figure}[H]
\begin{center}
     \includegraphics[width=0.7\textwidth]{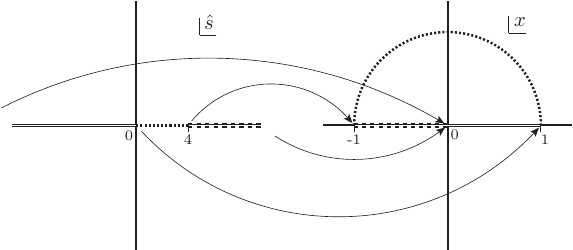}
\end{center}
\caption{\small\sf Mapping of regions from $\hat{s} \in \mathbb{R}$ to the complex plane in $x$. 
The region $\hat{s} \leq 0$ is mapped to the interval $x \in (0,1]$, while the region $0 \leq \hat{s} \leq 
4$ is mapped to the arc of unit radius, and the region $\hat{s}>4$ is mapped to $x \in (-1,0)$.} 
\label{fig:sxregions}
\end{figure}
\end{center}

We introduce the variable $x$ defined by
\begin{equation}
\label{eq:1}
\hs = \frac{q^2}{m^2} := - \frac{(1-x)^2}{x},
\end{equation}
from which one obtains
\begin{equation}
x = \frac{\sqrt{4-\hs}-\sqrt{-\hs}}{\sqrt{4-\hs}+\sqrt{-\hs}}.
\label{eq:x2s}
\end{equation}
Our notation follows widely that of Ref.~\cite{Blumlein:2019oas}.
In Figure~\ref{fig:sxregions}, we can see how different regions of $\hs \in 
\mathbb{R}$ are mapped to the 
complex plane in $x$. We will later present results both in the 
variables $\hs$ and $x$.

To obtain the form factors $F_{I,i}, ~ I=V,A$, we multiply the following projectors on 
$\Gamma^{\mu}_{cd}$ and perform a trace over the spinor and color indices
\begin{align}
 P_{V,i} &= \frac{i}{v_Q} \frac{\delta_{cd}}{N_c} \frac{\q_2 - m}{m} \Big( \gamma_{\mu} g_{V,i}^{(1)} 
+ \frac{1}{2 m} (q_{2 \mu} - q_{1 \mu}) g_{V,i}^{(2)} \Big) \frac{\q_1 + m}{m} \,,
 \nonumber\\
 P_{A,i} &= \frac{i}{a_Q} \frac{\delta_{cd}}{N_c} \frac{\q_2 - m}{m} \Big( \gamma_{\mu} \gamma_5 g_{A,i}^{(1)} 
          + \frac{1}{2 m} (q_{1 \mu} + q_{2 \mu}) \gamma_5 g_{A,i}^{(2)} \Big) \frac{\q_1 + m}{m} \,,
\end{align}
where $q_1$ and $q_2$ are the momenta of the heavy quarks ($q_1+q_2 = q$) and $N_c$ denotes the number of colors. 
In the present paper we relate all results to $N_c$ as an explicit variable, i.e. the color factors of $\mathrm{SU}(N_C)$ are given by $C_A = N_c, C_F = (N_c^2 -  1)/(2N_c), T_F = 1/2$. In the following we set 
$N_c = 3$. The factors $g_{I,i}^{(k)}$ are given by
\begin{align}
\label{eq:G1}
g_{V,1}^{(1)} &= \frac{x}{4 (1-\ep) (1+x)^2} \,, & g_{V,1}^{(2)} &= \frac{(3-2 \ep) x^2}{(1-\ep) (1+x)^4} \,, \\
g_{V,2}^{(1)} &= \frac{x^2}{(1-\ep) (1-x^2)^2} \,,
& g_{V,2}^{(2)} &= \frac{2 x^2 [-1+\ep (1-x)^2+(4-x) x]}{(1-\ep) (1-x)^2 (1+x)^4} \,, \\
g_{A,1}^{(1)} &= \frac{x}{4 (1 - \ep) (1 + x)^2} \,, & g_{A,1}^{(2)} &= \frac{x^2}{(1 - \ep) (1 - x^2)^2} \,, \\
\label{eq:G4}
g_{A,2}^{(1)} &= \frac{x^2}{(1-\ep) (1-x^2)^2} \,,  & g_{A,2}^{(2)} &= \frac{2 x^2 [1-\ep (1+x)^2+x (4+x)]}{(1-\ep) 
(1-x)^4 (1+x)^2} \,.
\end{align}
We will later also refer to each separate term attached to the functions $g_{I,i}^{(k)}$'s after the 
projection is performed. We define $F_{v,i}(x)$ and $F_{a,i}(x)$ such that 
\begin{eqnarray}
F_{V,i}(x) &=& g_{V,i}^{(1)} F_{v,1}(x) + g_{V,i}^{(2)} F_{v,2}(x) \,, \quad i=1,2. \label{gidecomp1} \\
F_{A,i}(x) &=& g_{A,i}^{(1)} F_{a,1}(x) + g_{A,i}^{(2)} F_{a,2}(x) \,, \quad i=1,2. \label{gidecomp2}
\end{eqnarray}

The form factors $F_S$ and $F_P$ can be obtained from $\Gamma_{cd}$ using the projectors given below
and performing the trace over the bi-spinor and color indices 
\begin{align}
 P_{S} &= - \frac{v}{2 m s_Q} \frac{\delta_{cd}}{N_c} \frac{x}{(1+x)^2} \frac{\q_2 - m}{m} \frac{\q_1 + m}{m} \,, \\
 P_{P} &= \frac{v}{2 m p_Q} \frac{\delta_{cd}}{N_c} \frac{x}{(1-x)^2} \frac{\q_2 - m}{m} i \gamma_5  \frac{\q_1 + m}{m} \,.
\end{align}
For convenience, and in a similar way as in Eqs. (\ref{gidecomp1}--\ref{gidecomp2}), we extract the $x$-dependent part of the
projectors by defining $F_s(x)$ and $F_p(x)$ as follows
\begin{eqnarray}
F_S(x) &=& -\frac{x}{2 (1+x)^2} F_s(x), \label{Sdecomp} \\
F_P(x) &=& \frac{x}{2 (1-x)^2} F_p(x). \label{Pdecomp}
\end{eqnarray}

In the following, we will identify a generic form factor as $F_I(x)$, where the index $I$ 
will denote either $I \in \left\{S,P\right\}$ or one of the pair of indices $I \in \left\{(V,1), \, (V,2), \, 
(A,1), \, (A,2)\right\}$. It may also denote any of the lower case versions $I \in \left\{(v,1), \, (v,2), \, 
(a,1), \, (a,2), \, s, \, p\right\}$.

We first summarize the results of our previous paper \cite{Blumlein:2019oas} on the quarkonic 
contributions to the three--loop form factor, i.e. for the contributions characterized by an overall factor 
of $n_h$ or $n_h^2$, where $n_h$ is the number of heavy quarks.
The form factors $F_{I}(x)$ can be expanded perturbatively in QCD as follows
\begin{equation}
 F_{I}(x) = \sum_{l=0}^{\infty} a_s^l F_{I}^{(l)}(x) \,,
\end{equation}
where $a_s = g_s^2/(4\pi)^2$ is the strong coupling constant. In Ref. 
\cite{Blumlein:2019oas}, four of the 
present authors computed all of the solvable parts of the $O(a_s^3)$ heavy fermion contributions. 
By `solvable parts' we mean all of the parts for which the corresponding differential equations or 
difference equations were factorizable to first order, and the corresponding solutions could be expressed 
in terms of harmonic polylogarithms (or any of their first--order factorizing generalizations) 
depending on the variable $x$, cf.~Ref.~\cite{Blumlein:2018cms} for a detailed discussion. 
The calculations were performed in $D = 4 - 2 \varepsilon$ dimensions, which 
allows one to get the expansion in the dimensional parameter $\varepsilon$ at each loop order as follows
\begin{equation}
F_{I}^{(l)}(x) = \sum_{k=-l}^0 \varepsilon^k F_{I}^{(l,k)}(x) + O(\varepsilon) \,.
\label{epsilonexpansion1}
\end{equation}
In general, if we want to obtain the $O(a_s^l)$ corrections, we need to compute the $l$-loop form factors up 
to $O(\varepsilon^0)$, while the $j$-loop form factors, with $j<l$, will be needed up to $O(\varepsilon^{l-j})$, 
in order to properly perform the corresponding renormalization procedure. This is the reason why 
one--loop and 
two--loop form factors were computed to higher orders in $\ep$ in \cite{Gluza:2009yy, 
Ablinger:2017hst}. 
In this paper, we are interested in three--loop corrections ($l=3$), so in what follows we will drop 
the label $l$ 
from the expansion terms in Eq. (\ref{epsilonexpansion1}). In other words, from now on, $F_{I}^{(k)}(x)$ 
will denote the $O(\varepsilon^k)$ term in the $\varepsilon$ expansion of the corresponding 
three--loop form 
factor, instead of the $O(a_s^k)$ term in the perturbative expansion.

\begin{center}
\begin{figure}[H]
\begin{center}
\begin{minipage}[c]{0.16\linewidth}
     \includegraphics[width=1\textwidth]{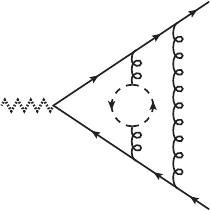}
\vspace*{-10mm}
\begin{center}
{\footnotesize (a)}
\end{center}
\end{minipage}
\hspace*{0.01\linewidth}
\begin{minipage}[c]{0.16\linewidth}
     \includegraphics[width=1\textwidth]{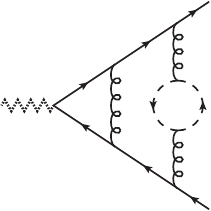}
\vspace*{-10mm}
\begin{center}
{\footnotesize (b)}
\end{center}
\end{minipage}
\hspace*{0.01\linewidth}
\begin{minipage}[c]{0.16\linewidth}
     \includegraphics[width=1\textwidth]{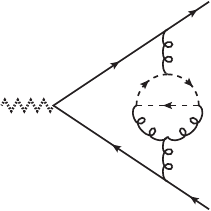}
\vspace*{-10mm}
\begin{center}
{\footnotesize (b)}
\end{center}
\end{minipage}
\hspace*{0.01\linewidth}
\begin{minipage}[c]{0.16\linewidth}
     \includegraphics[width=1\textwidth]{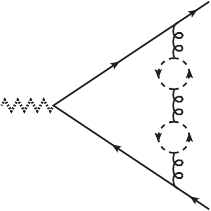}
\vspace*{-10mm}
\begin{center}
{\footnotesize (c)}
\end{center}
\end{minipage}
\hspace*{0.01\linewidth}
\begin{minipage}[c]{0.16\linewidth}
     \includegraphics[width=1\textwidth]{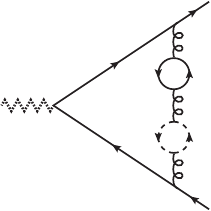}
\vspace*{-10mm}
\begin{center}
{\footnotesize (d)}
\end{center}
\end{minipage}

\vspace*{3mm}
\begin{minipage}[c]{0.16\linewidth}
     \includegraphics[width=1\textwidth]{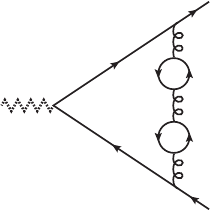}
\vspace*{-10mm}
\begin{center}
{\footnotesize (e)}
\end{center}
\end{minipage}
\hspace*{0.01\linewidth}
\begin{minipage}[c]{0.16\linewidth}
     \includegraphics[width=1\textwidth]{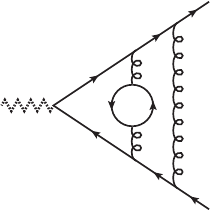}
\vspace*{-10mm}
\begin{center}
{\footnotesize (f)}
\end{center}
\end{minipage}
\hspace*{0.01\linewidth}
\begin{minipage}[c]{0.16\linewidth}
     \includegraphics[width=1\textwidth]{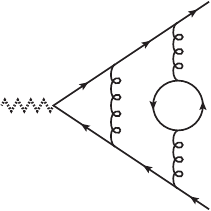}
\vspace*{-10mm}
\begin{center}
{\footnotesize (g)}
\end{center}
\end{minipage}
\hspace*{0.01\linewidth}
\begin{minipage}[c]{0.16\linewidth}
     \includegraphics[width=1\textwidth]{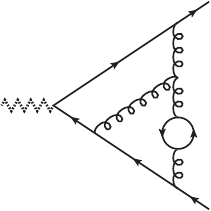}
\vspace*{-10mm}
\begin{center}
{\footnotesize (g)}
\end{center}
\end{minipage}
\hspace*{0.01\linewidth}
\begin{minipage}[c]{0.16\linewidth}
     \includegraphics[width=1\textwidth]{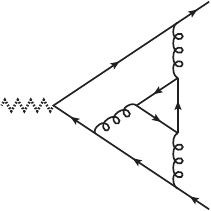}
\vspace*{-10mm}
\begin{center}
{\footnotesize (h)}
\end{center}
\end{minipage}

\vspace*{3mm}
\begin{minipage}[c]{0.16\linewidth}
     \includegraphics[width=1\textwidth]{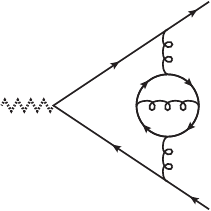}
\vspace*{-10mm}
\begin{center}
{\footnotesize (i)}
\end{center}
\end{minipage}
\hspace*{0.01\linewidth}
\begin{minipage}[c]{0.16\linewidth}
     \includegraphics[width=1\textwidth]{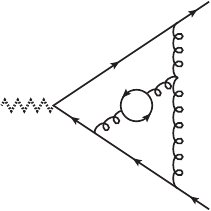}
\vspace*{-10mm}
\begin{center}
{\footnotesize (i)}
\end{center}
\end{minipage}
\hspace*{0.01\linewidth}
\begin{minipage}[c]{0.16\linewidth}
     \includegraphics[width=1\textwidth]{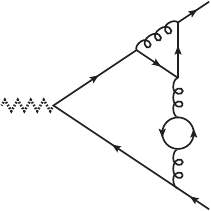}
\vspace*{-10mm}
\begin{center}
{\footnotesize (j)}
\end{center}
\end{minipage}
\hspace*{0.01\linewidth}
\begin{minipage}[c]{0.16\linewidth}
     \includegraphics[width=1\textwidth]{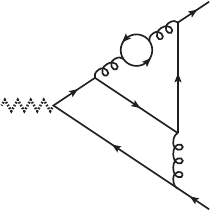}
\vspace*{-10mm}
\begin{center}
{\footnotesize (k)}
\end{center}
\end{minipage}
\hspace*{0.01\linewidth}
\begin{minipage}[c]{0.16\linewidth}
     \includegraphics[width=1\textwidth]{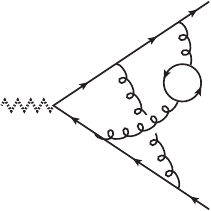}
\vspace*{-10mm}
\begin{center}
{\footnotesize (l)}
\end{center}
\end{minipage}
\end{center}
\caption{\small \sf Sample of diagrams required for the calculation of quarkonic contributions to 
three--loop massive form factors. 
Dashed arrow lines represent massless quarks, while solid arrow lines represent massive quarks. 
The dotted zigzag line represents any of the possible currents. 
Diagrams with closed massless quark loops, such as diagrams (a)-(d) were computed in Ref. \cite{Lee:2018nxa}.
Diagrams with closed heavy quark loops, such as (e)-(l), were first considered in Ref. \cite{Blumlein:2019oas}.
Numerical expansions spanning all regions in $\hat{s}$ were calculated in Ref. 
\cite{Fael:2022miw,Fael:2022rgm}. 
In this paper, we use a different method to get very deep expansions in all regions of $\hat{s}$
with exact numerical representations around $x=0$, and even analytic results in some cases.
}
\label{fig:samplediagrams}
\end{figure}
\end{center}

In Ref. \cite{Blumlein:2019oas}, the required Feynman diagrams with closed quark loops were generated using the program 
{\tt QGRAF} \cite{Nogueira:1991ex}, a sample of topologies is shown in 
Figure~\ref{fig:samplediagrams}.
After introducing the projectors described in the previous section, the Lorentz algebra was performed using 
{\tt Form} \cite{Vermaseren:2000nd, Tentyukov:2007mu} and the color algebra by {\tt Color}
\cite{vanRitbergen:1998pn}. After this, the results were expressed in terms of a linear combination of 
Feynman integrals, which were reduced by applying the integration--by--parts relations 
\cite{IBP} using the package {\tt Crusher} \cite{CRUSHER} to master integrals. A system of 
differential equations was obtained for the corresponding master integrals $I_i(x)$, with $i=1,2 \ldots$ 
\begin{eqnarray}
\label{eq:DEQEp}
\hspace*{2cm}\frac{d}{dx} 
\left(\begin{matrix} I_1(x,\ep) \\ I_2(x,\ep) \\ \vdots \end{matrix}\right)
= A(x,\ep)
\left(\begin{matrix} I_1(x,\ep) \\ I_2(x,\ep) \\ \vdots \end{matrix}\right) \,,
\label{diffeqsys0}
\end{eqnarray}
where $A(x,\ep)$ is an invertible matrix with entries from the polynomial ring\footnote{We assume that $\KK$ 
is a computable field containing the rational numbers $\QQ$ as a sub--field.} $\KK[x,\ep]$. Taking advantage 
of the fact that at $q^2=0$ (i.e. $x=1$), the non-singlet massive 3-point Feynman integrals reduce to 
self-energies, it was possible to obtain initial conditions for this system of differential equations. 
In fact, the integrals turn out to be regular at $x=1$, and a formal power series expansion around $y=0$ 
(where $y=1-x$) can be performed
\begin{equation}
I_i(x,\ep) = \sum_{n=0}^{\infty} c_i(n,\ep) y^n \,.
\label{MIexpansion1}
\end{equation}
Since the form factors are then linear combinations of the master integrals, they too can be expanded in the 
same way as in Eq. (\ref{MIexpansion1})
\begin{equation}
F_{I}(x,\ep) = \sum_{n=0}^{\infty} c_{I}(n,\ep) y^n \,,
\label{FFexpansion1}
\end{equation}
or
\begin{equation}
F_{I}^{(k)}(x,\ep) = \sum_{k=-3}^0 \ep^k \sum_{n=0}^{\infty} c_{I}^{(k)}(n) y^n \,.
\label{FFexpansion2}
\end{equation}
Introducing these expansions in (\ref{diffeqsys0}), and using the initial conditions, it was possible to 
generate thousands of expansion coefficients for all the different form factors. This is a highly non-trivial 
task, and it required the use of two different strategies described in Section 5 of Ref. \cite{Blumlein:2019oas} that have been implemented
within the package \texttt{SolveCoupledSystem}~\cite{Blumlein:2019oas,SolveCoupledSystem}.
For each color factor, the coefficients thus obtained are given in terms of linear combinations of the 
following constants
\begin{equation}
\left\{\zeta_2, \, \zeta_3, \, \zeta_4, \, \zeta_5, \, \zeta_2 \zeta_3, \, 
       \zeta_2 l_2, \, \zeta_2 l_2^2, \, l_2^4, \, a_4 \right\}
\label{listofconsts1}
\end{equation}
where 
\begin{equation}
l_2 = \ln(2),~~~~~~a_4 = {\rm Li}_4\left(\frac{1}{2}\right), 
\end{equation}
and $\zeta_k = \zeta(k)$ is the Riemann 
$\zeta$ function evaluated at integer values $k \geq 2$. The numbers 
in front of these constants in the linear 
combinations are just rational numbers and 
\begin{equation}
\Li_n(x) = \sum_{k=1}^\infty \frac{x^k}{k^n},~~~|x| \leq 1
\end{equation}
denote the classical polylogarithms \cite{LEWIN1}.

Therefore, for each form factor and for each constant in (\ref{listofconsts1}), we get a sequence of rational 
numbers, which we can feed to the {\tt Sage} package {\tt ore\_algebra} presented in \cite{SageOre} in order to 
guess a recursion relation satisfied by each one of the sequences. 
In several cases, the obtained recursion 
relations factorize at first order and can be solved even 
in terms of harmonic sums using the algorithms from~\cite{LinearSolver
,DRAlgorithms} implemented in the \texttt{Mathematica} 
package \texttt{Sigma}~\cite{SIG1,SIG2,Schneider:2013zna}, which where then 
introduced in (\ref{FFexpansion1}). The corresponding infinite sums can then be performed
with the help of 
{\tt Sigma}  and the \texttt{Mathematica} package {\tt HarmonicSums} 
\cite{HARMSU,Blumlein:2009ta,Vermaseren:1998uu,Blumlein:1998if,Remiddi:1999ew,Ablinger:2011te,
Ablinger:2013cf,Ablinger:2014bra,ALG,Ablinger:2021fnc} and one obtains
solutions in terms of harmonic polylogarithms. However, for a few of the 
color-factors/constants, the recursion relations turned out to be non--first order factorizable. 
In these cases no general techniques are available to obtain closed form solutions.
Therefore, in Ref. \cite{Blumlein:2019oas} we had to limit ourselves to just 
giving the corresponding expansions around $y=0$ in these cases. 

In this paper, we go a step further and obtain thousands of expansion coefficients for the power--log expansion around 
$x=0$ and other singular points, thus obtaining complete coverage of the whole kinematic range of $\hs$.
We decompose the unrenormalized form factors in terms of their first-order factorizable  and 
non first--order factorizable parts as follows,
\begin{equation}
F_I^{(0)}(x) = F_I^{(0), \rm sol}(x) + n_h \left(F_{I,1}^{(0)}(x) + \zeta_2 F_{I,2}^{(0)}(x) + \zeta_3 F_{I,3}^{(0)}(x)\right). \label{FFdecomp}
\end{equation}
The term $F_{I}^{(0), \rm sol}(x)$ corresponds to the solvable parts. The results for this term were given in Ref. \cite{Blumlein:2019oas}, 
together with the expressions for all the pole terms $F_I^{(k)}(x)$, $k=-3,-2,-1$, which were always solvable.
The remaining terms, $F_{I,i}^{(0)}(x)$ with $i=1,2,3$, are the non-solvable ones.
This decomposition indicates that all of the recursions associated with the constants 
in (\ref{listofconsts1}) were solvable except for the recursions coming from the factor
$n_h$
at $O(\ep^0)$ associated with $\zeta_2$, $\zeta_3$ and the rational term\footnote{Since all the poles in $\ep$ were solvable, and there are recursions coming
from these poles that are also associated with theses constants, there are also terms proportional to $\zeta_2$ and $\zeta_3$ popping up in $F_I^{(0), \rm sol}(x)$ 
as a consequence of the $\ep$ dependence of the factors in Eqs. (\ref{eq:G1})-(\ref{eq:G4}).}
(i.e., the term arising from just rational numbers without any of the constants in (\ref{listofconsts1})). 
Following the decomposition in Eqs. (\ref{gidecomp1}--\ref{gidecomp2}) and the explicit form of the factors $g_{I,i}^{(k)}$ given in (\ref{eq:G1}), we will also have
\begin{eqnarray}
F_{V,1,i}^{(0)}(x) &=& \frac{x}{4 (1+x)^2} F_{v,1,i}^{(0)}(x) + \frac{3 x^2}{(1+x)^4} F_{v,2,i}^{(0)}(x) \label{gidecomp2a} \\
F_{V,2,i}^{(0)}(x) &=& \frac{x^2}{(1-x^2)^2} F_{v,1,i}^{(0)}(x) + \frac{2 x^2 \left(4 x-1-x^2\right)}{(1-x)^2 (1+x)^4} F_{v,2,i}^{(0)}(x) \label{gidecomp2b} \\
F_{A,1,i}^{(0)}(x) &=& \frac{x}{4 (1+x)^2} F_{a,1,i}^{(0)}(x) + \frac{x^2}{(1-x^2)^2} F_{a,2,i}^{(0)}(x) \label{gidecomp2c} \\
F_{A,2,i}^{(0)}(x) &=& \frac{x^2}{(1-x^2)^2} F_{a,1,i}^{(0)}(x) 
+ \frac{2 x^2 \left(x^2+4 x+1\right)}{(1-x)^4 (1+x)^2} F_{a,2,i}^{(0)}(x). \label{gidecomp2d}
\end{eqnarray} 

Before describing the method we used to obtain power-log expansions
around $x=0$, i.e., in the high-energy limit $q^2 \rightarrow
-\infty$, for all of the $F_{I,i}^{(0)}(x)$'s, we summarize here the ultraviolet renormalization procedure. 
We have followed a mixed scheme to renormalize all the fields and parameters appearing in the form factors.
The mass and the wave function of the heavy quark have been renormalized using the on-shell scheme,
while the strong coupling constant renormalization has been performed in the $\overline{\text{MS}}$ scheme.
For the Yukawa coupling in the scalar and pseudoscalar form factors,
we also used the $\overline{\text{MS}}$ renormalization
scheme.\footnote{Note, that this is different then the procedure we used in \cite{Blumlein:2019oas}.}
All the necessary renormalization constants are available in \cite{Tarasov:1980au,Larin:1993tp,vanRitbergen:1997va,Czakon:2004bu,Chetyrkin:2004mf,Baikov:2016tgj,Herzog:2017ohr,Chetyrkin:2017bjc,Luthe:2017ttg,Luthe:2017ttc,Chetyrkin:1999ys,Chetyrkin:1999qi,Melnikov:2000qh,Broadhurst:1991fy,Marquard:2018rwx,Marquard:2016dcn,Marquard:2015qpa}.

The ultraviolet renormalized form factors maintain a universal infrared (IR) structure. 
The IR singularities of the form factors can be factorized as a
multiplicative renormalization factor
\cite{Becher:2009kw,Mitov:2006xs}, the structure of which is
controlled by the renormalization group equation through the massive cusp anomalous dimension \cite{Korchemsky:1987wg,Kidonakis:2009ev,Grozin:2014hna,Grozin:2015kna}.
We note that one needs to consider the decoupling relation
\cite{Schroder:2005hy,Chetyrkin:2005ia} of the
strong coupling constant to obtain the complete (full-QCD) IR
structure of the form factors. All final results for the form factors
will thus be given in terms of $\alpha_s^{(n_l)}$.

\section{The high energy limit}
\label{sec:3}

\vspace*{1mm}
\noindent
Unlike the expansions around $x=1$, the expansions around singular points will involve logarithms 
$\ln^k(x)$. In particular, the expansions around $x=0$ will be given in terms of power series 
multiplying various powers of $\ln(x)$. In principle, the coefficients of each power series could 
be obtained numerically by directly matching a truncated version of the expansion we want to determine 
with a truncated version of the known expansion (in this case, the one at $x=1$) at intermediate points 
$0<x<1$. However, this naive method will fail to produce precise enough coefficients in most cases, 
unless exact relations among the coefficients can be obtained, thus reducing the number of free 
coefficients to be determined. This can be achieved using differential equations, which can be derived 
from the large number of coefficients generated in Ref.~\cite{Blumlein:2019oas} for the expansions 
around $x=1$. In the next subsections, we will describe the matching of expansions at intermediate points, 
initially without relying on differential equations, and then we will explain how we derive those 
differential equations and how we use them in combination with the matching conditions to obtain the 
coefficients numerically. As we will see, the precision of the coefficients in the power--log expansions 
can be increased almost indefinitely by increasing the number of terms in both of the expansions to be 
matched. 

A closely related method based also on differential equations was presented in 
Ref.~\cite{Fael:2022miw,Fael:2022rgm}
for the master integrals. There, however, numerical solutions were aimed at. In our case, we use 
guessing methods to obtain differential equations obeyed directly by the non--solvable parts of the form 
factors themselves, determining the expansion coefficients analytically. This allows 
us to work with very deep expansions and also avoid many spurious singularities, and therefore, unlike 
the case of Ref. \cite{Fael:2022miw,Fael:2022rgm}, in our case obtaining expansions at intermediate points 
is not 
strictly necessary, although it may still be useful for practical applications.

In this section we focus on the high energy limit $q^2 \rightarrow \pm \infty$. The same methods
are applicable to all other singular points.
\subsection{Matching at intermediate points} 
\label{sec:31}

\vspace*{1mm}
\noindent
The form factors are regular at $x=1$ and have an expansion of the form
\begin{equation}
f(x) = \sum_{j=0}^{\infty} a(j) (1-x)^j, 
\label{x=1expansion}
\end{equation}
where $f(x)$ can be any of the non-solvable contributions in (\ref{FFdecomp}), which were split into three 
terms so that the corresponding coefficients $a(j)$ contain only rational numbers. 

We assume now that the function $f(x)$ in (\ref{x=1expansion}) has also a logarithmic expansion around $x=0$
\begin{equation}
f(x) =  \sum_{i=0}^{r} \sum_{j=\lambda_i}^\infty c(i,j) x^j \ln^i(x), 
\label{x=0exactexpansion}
\end{equation}
with $\lambda_i \leq 0$.
We want to determine as many coefficients as possible of the expansion (\ref{x=0exactexpansion}).
The two expansions (\ref{x=1expansion}) and (\ref{x=0exactexpansion}) should be convergent and equal to each other in a region within $0<x<1$.
Note that the expansion about $x=1$ is convergent in $x \in (0,1]$, but the expansion about $x=0$ converges only in $x \in (3-2\sqrt{2},-3+2\sqrt{2})$ in the case of $F^{(0)}_{I,2}(x)$, due to a spurious singularity at $\hat{s}=-4$, while in the case of $F^{(0)}_{I,1}(x)$ and $F^{(0)}_{I,3}(x)$, the expansions about $x=0$ converge in $x \in (4\sqrt{3}-7,-4\sqrt{3}+7)$, due to a four particle cut at $\hat{s}=16$.
Of course, since we are considering non-solvable cases, we cannot perform the infinite sum in (\ref{x=1expansion}). 
We therefore consider the truncated expansion 
\begin{equation}
\tilde{f}_1(x) = \sum_{j=0}^l a(j) (1-x)^j, 
\label{x=1trunc}
\end{equation}
where $l$ is large but finite. As we have already said, in all cases we have recursion relations for the 
coefficients $a(j)$ and enough initial values to be able to compute hundreds of thousands of expansion 
coefficients. 

We want to use the truncated expression (\ref{x=1trunc}) in order to obtain approximate values for the
coefficients $c(i,j)$ of the expansion (\ref{x=0exactexpansion}). For this purpose, we also define 
a truncated expansion around $x=0$,
\begin{equation}
\tilde{f}_0(x) =  \sum_{i=0}^r \sum_{j=\lambda_i}^{\nu_i} \tilde{c}(i,j) x^j \ln^i(x). 
\label{x=0trunc}
\end{equation}
The truncated expansions have to be matched at values $x_i \in \chi$ 
\begin{equation}
\chi = \left\{ x_k = x_0+k \delta \right. \, \left| \,\, k=1,2,\ldots,n_p \right\}\,,
\label{setofpoints1}
\end{equation}
where $x_0>0$, $\delta$ is a real number, which in general should be small,
and $n_p$ is a positive integer such that $x_0+n_p\delta<1$. 
This leads to a system of equations
\begin{equation}
 \frac{d^k}{dx^k} \tilde{f}_1(x_i) = \frac{d^k}{dx^k} \tilde{f}_0(x_i), \quad x_i \in \chi, \quad k=0,\ldots,
n_d \,,
\label{syseqs}
\end{equation}
which consists of $n_p (n_d+1)$ equations for the unknown coefficients $\tilde{c}(i,j)$.

To find a solution for all $\tilde{c}(i,j)$ the parameters $\nu_j$, $n_p$, and $n_d$ have to be chosen such that
\begin{equation}
\sum_{j=0}^r (\nu_j-\lambda_j+1) = n_p (n_d+1) \,.
\end{equation}

In general, we try to distribute the coefficients evenly among 
the different powers of the logarithms in (\ref{x=0trunc}) by choosing
\begin{eqnarray}
\nu_j &=& \left\lfloor \frac{n_p (n_d+1)}{r+1}+\frac{1}{2} \right\rfloor + \lambda_j - 1, \quad {\rm for} \quad j=0,\ldots,r-1, \\
\nu_r &=& n_p (n_d+1) + \lambda_r -1 - \sum_{j=0}^{r-1} \left(\nu_j - \lambda_j + 1\right).
\end{eqnarray}
Our goal is to minimize the difference between the coefficients obtained by solving (\ref{syseqs}) and the 
actual yet unknown coefficients of the expansion around $x=0$ given in (\ref{x=0exactexpansion}), 
i.e., we want to minimize the quantities
\begin{equation}
\Delta_{i,j} = \left|c(i,j) - \tilde{c}(i,j)\right|. \label{delta2minimize}
\end{equation}
These differences between the actual coefficients and the ones obtained from the system of equations (\ref{syseqs}) will depend 
on the set of points in $\chi$, the number of derivatives $n_d$ and the number of coefficients $l$ in the truncated series (\ref{x=1trunc}), i.e.,
\begin{equation}
\Delta_{i,j} = \Delta_{i,j}(x_0,\delta,n_p,n_d,l). \label{deltaparameters}
\end{equation}
Here the coefficients $\tilde{c}(i,j)$ will be numerical approximations of the coefficients $c(i,j)$, 
which we will obtain by evaluating $\tilde{f}_1(x)$ and $\tilde{f}_0(x)$ as well as their derivatives 
at different points in a region within $0<x<1$ where both series converge, and equating the corresponding 
results. This will then lead to a linear system of equations for the coefficients $\tilde{c}(i,j)$.

The task is then to find values for $x_0$, $\delta$, $n_p$, $n_d$ and $l$ such that $\Delta_{i,j} \sim 0$. This may seem
impossible at first sight, given the fact that we do not know the coefficients $c(i,j)$ in (\ref{delta2minimize})
to begin with. Moreover, the values of $r$ and $\lambda_j$ in (\ref{x=0trunc}) are also not known a priori. 
As will be described in the next section, we know, however, that $r \leq 6$ and $\lambda_j \geq -2$, 
so if we choose $r$ high enough or $\lambda_j < -2$, we should expect the corresponding coefficients 
$\tilde{c}_r(\lambda_j)$ to be close to zero, as long as the parameters in (\ref{deltaparameters}) are in 
the right region. We must therefore search for values of $x_0$, $\delta$, $n_p$, $n_d$ and $l$ such that these 
coefficients are as close to zero as possible. In general, higher values of $l$ and $n_d$ lead to better results, but determining good values for $x_0$, $\delta$ and $n_p$ is not as simple.  In fact, we have observed that the optimal value for 
$x_0$, with $\delta$ and $n_p$ fixed, shifts to the left as we increase $l$. 

The number of points used ranged from $n_p=400$ to $n_p=600$. Such a large number of points makes it necessary 
to choose them very close to each other, in order to avoid moving too
much to the right of the starting point 
$x_0$, where the evaluation of the expansion at $x=0$ will be too imprecise. We used values of $\delta$ that 
ranged between 
\begin{equation}
\delta = \left[\frac{1}{120000},~\frac{1}{20000}\right].
\end{equation}

We proceeded in this way for all of the $F_{I,2}^{(0)}(x)$ terms, and
found very high numerical precision for a relatively large number of
the coefficients in (\ref{x=0trunc}) and also applied this procedure
to solvable cases and verified that it reproduces the coefficients
obtained by directly expanding the corresponding exact solutions
around $x=0$.

Unfortunately, it is difficult to apply this method in the case of
$F_{I,1}^{(0)}(x)$ and $F_{I,3}^{(0)}(x)$. This is due to the fact
that, as we mentioned above, in the case of $F_{I,2}^{(0)}(x)$ the
radius of convergence is $3-2\sqrt{2} \sim 0.1716$, while in the case
of $F_{I,1}^{(0)}(x)$ and $F_{I,3}^{(0)}(x)$ it is
$4\sqrt{3}-7 \sim 0.0718$, which is closer to zero, making it
difficult to find a good point to match the expansions without using a
prohibitively large amount of coefficients for the expansions around
$x=1$ and $x=0$. We will now see how the use of differential equations
solves this problem.

\subsection{Differential equations obeyed directly by the form factors} 
\label{sec:32}

\vspace*{1mm}
\noindent
As we mentioned before, in Ref. \cite{Blumlein:2019oas} a large number of expansion coefficients around $x=1$ ($y=0$) were computed in order to obtain recursion 
relations satisfied by the sequence of rational numbers that multiply each constant in (\ref{listofconsts1}) in these coefficients.
We can also use these sequences to derive the  differential equations by using the method of guessing,
cf.~\cite{SageOre}, such that the coefficients of the power expansion of the respective solutions match the 
corresponding sequences.
This package has been used by us in many applications since 
2008, and in calculating the three--loop massive form factor in Ref.~\cite{Blumlein:2019oas} to guess the recursions. 
We obtained differential equations for all of the cases where the recursions did not factorize to first order. These differential equations have the following form
\begin{equation}
\sum_{k=0}^{n_{I,i}-1} \hat{p}_{I,i,k}(y) \frac{d^k}{dy^k} F_{I,i}^{(0)}(y) = 0, \quad I \in \left\{(v,1), \, (v,2), \, 
(a,1), \, (a,2), \, s, \, p\right\}, \quad i=1,2,3,
\label{diffeqiny1}
\end{equation}
where $n_{I,i}-1$ is the order of the differential equation and $\hat{p}_{I,i,k}(y)$ is a polynomial in $y$. We can then 
perform the change of variables $y \rightarrow 1-x$,
which allows us to obtain differential equations in $x$,
\begin{equation}
\sum_{k=0}^{n_{I,i}} p_{I,i,k}(x) \frac{d^k}{dx^k} F_{I,i}^{(0)}(x) = 0.
\label{diffeqinx1}
\end{equation}
If we look for a power expansion solution of the form $x^{\alpha} \sum_{i=0}^{\infty} c_i x^i$, along the 
lines of what is done in the case of second order differential equations,
we get an indicial equation where all the solutions in $\alpha$ are integers. For example, in the case of $F_{v,1,1}^{(0)}(x)$, the indicial equation
turns out to be
\begin{equation}
(\alpha+2) (\alpha+1)^6 \alpha^5 (\alpha-1)^5 (\alpha-2)^4  (\alpha-7)^2 \Pi_{i=3}^6 (\alpha-i)^3 \Pi_{j=8}^{18} (\alpha-j) = 0 \,.
\label{indicialeq1}
\end{equation}
Since $\alpha=-2$ is one of the solutions of this equation, we can expect our expansions to start from $1/x^2$ onward,\footnote{However, the final physical result will be free of negative
powers of $x$.}
and since all roots differ by integer values, the solutions should involve powers of $\ln(x)$.
The solution of this differential equation can then be expanded in a power-log series like the one in Eq. (\ref{x=0exactexpansion}), i.e.,
\begin{equation}
F_{I,i}^{(0)}(x) = \sum_{j=0}^r \sum_{k=-2}^{\infty} c_{I,i}(j,k) x^k \ln^j(x) \,.
\label{Fpowlog}
\end{equation}
The ansatz (\ref{x=0exactexpansion}) is thus justified for some integer $r$. The value of $r$ depends on the differential equation under consideration; 
if we use an ansatz with a value of $r$ that is too large, and then insert the ansatz in the differential equation, 
we will find that all of the coefficients for $j$ larger than a certain value will vanish.

It is important to remark that the first few coefficients of the expansions obtained in \cite{Blumlein:2019oas} usually did not satisfy the guessed recursion relations,
which were valid only after the second or third coefficient. This means that we cannot include these coefficients when guessing the differential equations, which
leads to solutions where the sequence of expansion coefficients is shifted. We correct this by subtracting these first few terms from the expansion of the
solution of the differential equation and performing a shift in the powers of $y$ before doing the change to $x$. The resulting differential equation ends up with an inhomogeneous part,
which can be removed at the expense of increasing the order of the differential equation by one. This is the reason why in Eq. (\ref{diffeqiny1}) the order of 
the differential equation is $n_{I,i}-1$, while in (\ref{diffeqinx1}) it is $n_{I,i}$. 

In Table~\ref{orderstable} we summarize the orders of the differential equations obtained for each 
$F_{I,i}^{(0)}(x)$, 
while in Table~\ref{degreetable}, the degree of the polynomial of highest degree in each differential 
equation is given. 
From the numbers in these tables, we can see that the guessed differential equations are quite formidable. 
Not only are the orders and especially the degrees of the polynomials in these equations quite large, but 
so are the integer coefficients in these polynomials. For example, the coefficient in front of $x^{2734}$ 
in $p_{v,2,1,48}(x)$ is an integer with 358 digits. Nevertheless, the orders of the differential 
equations, albeit high, are still manageable. The highest of them is 48 (for $F_{v,2,1}^{(0)}(x)$), 
which means that only 48 initial conditions are needed to determine the solution. This is in sharp contrast 
with the method discussed in the previous section, where the expansions had to 
be evaluated at hundreds of 
different points in order to obtain precise enough coefficients. When we insert the power-log expansion 
(\ref{Fpowlog}) in (\ref{diffeqinx1}), we therefore find that all of the coefficients can be written in 
terms of $n_{I,i}$ of them.  We will refer to this subset of $n_{I,i}$ coefficients as the base 
coefficients.\footnote{We have some freedom in the choice of coefficients $c_{I,k}(i,j)$ in the basis, 
of course, but in general it is a good idea to choose them with the lowest possible values of $j$.}
The relations between the base coefficients and the rest can be obtained by observing that equation 
(\ref{diffeqinx1}) will be satisfied after inserting (\ref{Fpowlog}) only if the collected coefficients for 
each power of the log and each power in $x$ vanish.
\begin{table}[H]
\begin{center}
\begin{tabular}{|c|c|c|c|c|c|c|}
\hline    
$i$ & $F_{v,1,i}^{(0)}$ & $F_{v,2,i}^{(0)}$ & $F_{a,1,i}^{(0)}$ & $F_{a,2,i}^{(0)}$ &   $F_{s,i}^{(0)}$   &   $F_{p,i}^{(0)}$   \\ 
\hline
  1 &        46         &        48         &         46        &        43         &          43         &          43         \\
  2 &        20         &        22         &         20        &        18         &          18         &          18         \\ 
  3 &        25         &        26         &         25        &        23         &          23         &          23         \\
\hline
\end{tabular}
\caption{\sf \small Orders of the differential equations for the $F_{I,i}^{(0)}(x)$'s.}
\label{orderstable}
\end{center}
\end{table}

\vspace*{-1cm}
\begin{table}[H]
\begin{center}
\begin{tabular}{|c|c|c|c|c|c|c|}
\hline    
$i$ & $F_{v,1,i}^{(0)}$ & $F_{v,2,i}^{(0)}$ & $F_{a,1,i}^{(0)}$ & $F_{a,2,i}^{(0)}$ &   $F_{s,i}^{(0)}$   &   $F_{p,i}^{(0)}$   \\ 
\hline
  1 &       2668        &       2734        &       2720        &       2347        &         2315        &         2347        \\
  2 &       593         &       637         &       605         &       511         &         505         &         511         \\ 
  3 &       875         &       920         &       895         &       759         &         753         &         759         \\ 
\hline
\end{tabular}
\caption{\sf \small Maximum degree of the polynomials in the differential equations for the $F_{I,i}^{(0)}(x)$'s.}
\label{degreetable}
\end{center}
\end{table}


\vspace*{-1cm}
Notice the difference of the present method with the one used in Ref. \cite{Fael:2022miw,Fael:2022rgm}, 
where the 
systems of differential equations obeyed by the master integrals  were solved numerically 
in terms of power-log expansions. This required one initial condition per integral since the differential 
equations are of first order for each integral in the system. In our case, we solve differential equations 
which are obeyed directly by the form factors themselves. This allows us to get higher numerical 
precision and express our results in terms of rational numbers and a few known constants, as we will see in the next 
section.

In Table~\ref{freecoeffs1} we can see the list of base coefficients that are left to be determined by 
initial conditions in each case. In that Table, we use the following sets of pairs of indices $(i,j)$ to
identify the corresponding base coefficients $c_{I,k}(i,j)$,
\begin{eqnarray}
W_1 &=& \left\{(5,-1), (4,-1),\ldots,(4,1), \, (3,-1),\ldots(3,2), \, (2,-1),\ldots,(2,5), \, 
(1,-1),\ldots,(1,6), \, 
\right.
\nonumber\\ && \left.
(0,-2), \ldots,(0,17)\right\}, \nonumber\\
W_2 &=& \left\{(3,-1), \, (2,-1), \, (2,1), \, (1,-1),\ldots,(1,4), \, 
(0,-2),\ldots,(0,6)\right\}, 
\nonumber\\
W_3 &=& \left\{(4,-1), \, (3,-1), \, (2,-1), \, (2,0), \, (2,1), \, (1,-1),\ldots,(1,5), \, 
(0,-2),\ldots,(0,8)\right\}.
\end{eqnarray}
The following are some examples of the relations between the remaining
coefficients of $F_{v,2,2}^{(0)}(x)$ and the ones shown in Table~\ref{freecoeffs1},
\begin{eqnarray}
 c_{v,2,2}(4,2) &=& -\frac{1}{4} c_{v,2,2}(3,-1),
\\
c_{v,2,2}(3,-2) &=& 0, 
\\
c_{v,2,2}(3,1)  &=& \frac{27}{31} c_{v,2,2}(0,-2) + \frac{1499}{54} c_{v,2,2}(3,-1),
\\
c_{v,2,2}(3,2)  &=& -\frac{14}{31} c_{v,2,2}(0,-2)-\frac{1}{3} c_{v,2,2}(2,-1)
-\frac{521}{486} c_{v,2,2}(3,-1),
\\
c_{v,2,2}(2,-2) &=& 0,
\\
c_{v,2,2}(2,3) &=& \frac{655}{186} c_{v,2,2}(0,-2)
+\frac{329}{18} c_{v,2,2}(2,-1)
+c_{v,2,2}(2,1)
+\frac{25307 c_{v,2,2}(3,-1)}{5832},
\\
c_{v,2,2}(1,-2) &=& 0,
\\
c_{v,2,2}(1,5) &=& -\frac{28281678020281 c_{v,2,2}(0,-2)}{158513905800}
+\frac{994883 c_{v,2,2}(1,-1)}{1578195} \nonumber\\ &&
+\frac{6235153 c_{v,2,2}(1,0)}{526065}
+\frac{994883 c_{v,2,2}(1,1)}{1578195}
-\frac{11628593 c_{v,2,2}(1,2)}{1052130} \nonumber\\ &&
+\frac{583312 c_{v,2,2}(1,3)}{1578195}
+\frac{539344 c_{v,2,2}(1,4)}{105213}
+\frac{576002152169 c_{v,2,2}(2,-1)}{7670027700} \nonumber\\ &&
-\frac{881799097 c_{v,2,2}(2,0)}{14203755}
+\frac{16510826 c_{v,2,2}(2,1)}{14203755}
-\frac{11311570189148899 c_{v,2,2}(3,-1)}{6212722437000},
\nonumber\\
\label{crels15}
\\
c_{v,2,2}(1,6) &=& -\frac{21830660599313 c_{v,2,2}(0,-2)}{36986578020}
+\frac{2765144 c_{v,2,2}(1,-1)}{526065}
+\frac{23030548 c_{v,2,2}(1,0)}{526065} \nonumber\\ && 
+\frac{2765144 c_{v,2,2}(1,1)}{526065}
-\frac{7082437 c_{v,2,2}(1,2)}{175355}
-\frac{2765144 c_{v,2,2}(1,3)}{526065} \nonumber\\ && 
+\frac{10258102 c_{v,2,2}(1,4)}{526065}
+\frac{1439537964281 c_{v,2,2}(2,-1)}{8948365650}
-\frac{181544612 c_{v,2,2}(2,0)}{946917} \nonumber\\ &&
+\frac{51133604 c_{v,2,2}(2,1)}{4734585}
-\frac{272690648223213781 c_{v,2,2}(3,-1)}{50737233235500},
\label{crels16}
\end{eqnarray}
etc. In general, for a given value of $i$ and $k$ in $c_{I,k}(i,j)$, the numbers in front of the coefficients on the right-hand side of these relations keep increasing as
we increase $j$. We can see this clearly in the case of Eqs. (\ref{crels15}) and (\ref{crels16}) by writing the numbers in front of the coefficients in floating point.
\begin{eqnarray}
c_{v,2,2}(1,5) &=& 
-178.418 c_{v,2,2}(0,-2) +0.630 c_{v,2,2}(1,-1)+11.852 c_{v,2,2}(1,0) \nonumber \\ &&
+0.630 c_{v,2,2}(1,1)-11.052 c_{v,2,2}(1,2)+0.369 c_{v,2,2}(1,3) \nonumber \\ &&
+5.126 c_{v,2,2}(1,4)+75.098 c_{v,2,2}(2,-1)-62.082 c_{v,2,2}(2,0) \nonumber \\ &&
+1.162 c_{v,2,2}(2,1)-1820.710 c_{v,2,2}(3,-1) \, , \label{eq:bcfp1} \\
c_{v,2,2}(1,6) &=& 
-590.232 c_{v,2,2}(0,-2)+5.256 c_{v,2,2}(1,-1)+43.779 c_{v,2,2}(1,0) \nonumber \\ &&
+5.256 c_{v,2,2}(1,1)-40.389 c_{v,2,2}(1,2)-5.256 c_{v,2,2}(1,3) \nonumber \\ &&
+19.499 c_{v,2,2}(1,4)+160.872 c_{v,2,2}(2,-1)-191.722 c_{v,2,2}(2,0) \nonumber \\ &&
+10.800 c_{v,2,2}(2,1)-5374.567 c_{v,2,2}(3,-1) \, .\label{eq:bcfp2}
\end{eqnarray}
Given that we will determine the base coefficients in Table~\ref{freecoeffs1} numerically with a certain precision, in principle we might lose precision on the coefficients on the left-hand side of relations like the ones given above, due to possible large cancellations. However, in general, we have observed that this does not happen. The coefficients on the left-hand side maintain the precision and are of a similar size to the largest rational number on the right-hand side of these relations.
\bgroup
\def\arraystretch{1.5}
\begin{table}[H]
\begin{center}
\begin{tabular}{|c|l|}
\hline
Coefficients & $\phantom{aaaaaaaaaaaa}$  Values of $i$ and $j$ \\
\hline \hline 
$c_{v,1,1}(i,j), \, c_{a,1,1}(i,j)$         & $(i,j) \in W_1 \cup \left\{(2,6), \, (1,7), \, (0,18)\right\}$ \\
\hline
$c_{v,2,1}(i,j)$                                   & $(i,j) \in W_1 \cup \left\{(5,0), \, (2,6), \, (1,7), \, (1,8), \, (0,18)\right\}$ \\
\hline          
$c_{a,2,1}(i,j), \, c_{s,1}(i,j), \, c_{p,1}(i,j)$ & $(i,j) \in W_1$ \\
\hline  
$c_{v,1,2}(i,j), \, c_{a,1,2}(i,j)$         & $(i,j) \in W_2 \cup \left\{(0,7), \, (0,8)\right\}$ \\
\hline
$c_{v,2,2}(i,j)$                                   & $(i,j) \in W_2 \cup \left\{(2,0), \, (0,7), \, (0,8), \, (0,9)\right\}$ \\
\hline          
$c_{a,2,2}(i,j), \, c_{s,2}(i,j), \, c_{p,2}(i,j)$ & $(i,j) \in W_2$ \\
\hline  
$c_{v,1,3}(i,j), \, c_{a,1,3}(i,j)$         & $(i,j) \in W_3 \cup \left\{(1,6), \, (0,9)\right\}$ \\
\hline
$c_{v,2,3}(i,j)$                                   & $(i,j) \in W_3 \cup \left\{(3,0), \, (1,6), \, (0,9)\right\}$ \\
\hline          
$c_{a,2,3}(i,j), \, c_{s,3}(i,j), \, c_{p,3}(i,j)$ & $(i,j) \in W_3$ \\
\hline  
\end{tabular}
\caption{\small \sf Base coefficients to be determined by initial conditions.
} 
\label{freecoeffs1}
\end{center}
\end{table}
\egroup

\vspace*{-1cm}
Similar relations can be obtained for all other form factors using the differential equations. 
In principle, we can obtain relations as deep in the expansions around $x=0$ as we want, only limited by computational power. 
We have obtained relations for the coefficients $c_{I,k}(i,j)$ as deep as $j=3000$ in the most complicated cases.

Once we have these relations, all that remains to be done is to determine the base coefficients numerically using initial conditions. 
To do so, first we insert these relations in Eq. (\ref{Fpowlog}) in order to obtain expansions involving only these coefficients. 
As we will emphasize below, the more relations we insert, and therefore, the more coefficients of (\ref{Fpowlog}) we include, the more precision 
we gain for the base coefficients.

After the relations are inserted, we can then proceed in a similar way as we did in the previous section, i.e., 
truncated versions of the expansions around $x=0$ and around $x=1$ (and their derivatives with respect to $x$) can be 
evaluated at different points in $x$, and the results can be equated as in Eq. (\ref{syseqs}). This produces a linear system 
of equations for the base coefficients that we can solve numerically, as long as the number of points and derivatives
is chosen in such a way that the number of equations equals the order of the corresponding differential 
equation,\footnote{If we produce more equations, in principle the extra ones will be linearly dependent on the others and 
should not cause any harm, but we are looking for numerical solutions, and doing this may cause numerical instabilities.}
or equivalently, the number of base coefficients. In this way, we only need to determine $n_{I,i}$ coefficients as supposed 
to the hundreds of them
that we needed to determine with the method of Section \ref{sec:31}. 
This is a more efficient way of dealing with the problem that makes it easier to achieve very high precision.
We remark that the entries of the corresponding matrix can be obtained with arbitrary precision, since they are given in terms
of linear combinations of powers of $x$ and of the log over rational numbers. The precision of the right-hand side of the equations
can be increased dramatically by increasing the number of terms of the expansion around $x=1$, which can be determined by 
the available recurrences.

As we will see later, we were able to obtain the base coefficients with a precision of more than a thousand decimal places. 
For physical applications, the precision needed is far below this. However, in the next subsection we will see that having such a high precision
allows us to express our results in ways that go beyond what can be achieved with the method of Section 
\ref{sec:31} or the method of Ref. \cite{Fael:2022miw,Fael:2022rgm}. 
In fact, the results we will present may in a sense be considered as exact solutions.

In Table~\ref{precision1}, we show the numerical precision that is achieved for the base coefficients using 
different configurations of the
number of coefficients in the truncated expansions around $x=1$ and $x=0$. We took 3000, 5000 and 10000 coefficients for the expansion around $x=1$
in combination with 500 and 1000 coefficients per power of the log for the expansion around $x=0$. The precision
of the base coefficients is quite similar for all of the base coefficients of any given $F_{I,k}^{(0)}(x)$, and also for different values of $I$ 
with $k$ fixed, so the numbers shown in Table~\ref{precision1} are an
average of those precisions and represent the number of correct decimal places that can be obtained in each case. We can see that the least precise
base coefficients are those of $F_{I,1}^{(0)}(x)$, with a precision of only a few digits when 3000 and 500 (per power of the log) expansions coefficients
are used for the expansions around $x=1$ and $x=0$, respectively. The precision increases as expected when we use deeper expansions.
In all cases the base coefficients were obtained by evaluating directly at a few 
points in $x$ without taking derivatives, choosing a set of points that optimizes the precision.\footnote{This is done by scanning different sets of points
like those in Eq. (\ref{setofpoints1}) and choosing the ones that give the highest precision.} The precision could presumably be increased if derivatives were also used. 
However, even this straightforward evaluation at the points gives pretty good precision using a reasonably small number of coefficients for the expansions. We
can see in Table~\ref{precision1} that taking 5000 coefficients for the expansion around $x=1$ and 1000 
coefficients per power of the log around $x=0$ leads to
a precision of at least 38 decimal places for the base coefficients. This number of coefficients is much lower than the number of coefficients required to guess
the differential equations themselves, and the precision achieved is more than enough for physical applications.
\begin{table}[t]
\begin{center}
\begin{tabular}{|cc|ccc|ccc|ccc|}
\hline
 & & \multicolumn{9}{c|}{$x=1$:} \\
\cline{3-11}
 & & \multicolumn{3}{c|}{3000} & \multicolumn{3}{c|}{5000} &\multicolumn{3}{c|}{10000}   \\
\cline{3-11}
                        &       &  $F_{I,1}^{(0)}$  &  $F_{I,2}^{(0)}$  &  $F_{I,3}^{(0)}$&  $F_{I,1}^{(0)}$  &  $F_{I,2}^{(0)}$  &  $F_{I,3}^{(0)}$&  $F_{I,1}^{(0)}$  &  $F_{I,2}^{(0)}$  &  $F_{I,3}^{(0)}$   \\
\hline 
\multirow{2}{*}{$x=0$:} &  500  &  2  & 90  & 28  & 22  & 134 & 57 & 70  & 205 & 110 \\
                        &  1000 &  5  & 120 & 36  & 38  & 185 & 74 & 110 & 300 & 150 \\
\hline
\end{tabular}
\caption{\sf \small Average number of correct decimal places for the base coefficients obtained using 3000, 5000 and 
10000 coefficients for the truncated expansion around $x=1$, and 500 and 1000 coefficients per power of the 
log for the expansion around $x=0$.
}.
\label{precision1}
\end{center}
\end{table}
After we combine the vector form factors $F_{v,1,i}^{(0)}(x)$ and $F_{v,2,i}^{(0)}(x)$ using Eqs. (\ref{gidecomp2a}--\ref{gidecomp2b}) 
to get  $F_{V,1,i}^{(0)}(x)$, followed by inserting the results in Eq. (\ref{FFdecomp}), replacing $x$ by $\hs$, 
Eq.~(\ref{eq:1}) subtracting the infrared divergences, and
expanding everything, including the solvable part $F_{V,1}^{(0), \rm sol}(x)$, in $\hs$, we reproduce the 
result given in Ref. \cite{Fael:2022miw,Fael:2022rgm}, and give more digits
\begin{eqnarray}
F_{V,1}^{(0)}(x) &=&
-32.897722507313999578 n_h^2
-87.77373705582837517 n_h n_l \nonumber \\ &&
+1147.4240120750982840 n_h
+(
-27.400233959497856255 n_h^2 \nonumber \\ &&
-60.87855567202544303 n_h n_l
+820.7777180906884097 n_h
) l_s \nonumber \\ &&
+(
-8.632679058651395941 n_h^2
-17.265358117302791882 n_h n_l \nonumber \\ &&
+186.38969938287483682 n_h
) l_s^2
+ (
-1.2510288065843621399 n_h^2 \nonumber \\ &&
-2.5020576131687242798 n_h n_l
+10.547828331577832952 n_h
) l_s^3 \nonumber \\ &&
+ (
-0.09876543209876543210 n_h^2
-0.19753086419753086420 n_h n_l \nonumber \\ &&
-1.6790123456790123457 n_h
) l_s^4
-0.39506172839506172840 n_h l_s^5 \nonumber \\ &&
+ [
-253.82786033439203372 n_h^2
-404.93106546060673261 n_h n_l \nonumber \\ &&
+6970.035736369961772 n_h
+ (
-154.28901373485052021 n_h^2 \nonumber \\ &&
-185.93693993861074771 n_h n_l
+3125.4534231017452606 n_h
) l_s \nonumber \\ &&
+(
-38.518518518518518519 n_h^2
-53.33333333333333333 n_h n_l \nonumber \\ &&
+526.7543200860128681 n_h
) l_s^2
+ (
-5.925925925925925926 n_h^2 \nonumber \\ &&
-7.111111111111111111 n_h n_l
+49.976937394787551232 n_h
) l_s^3 \nonumber \\ &&
-19.193415637860082305 n_h l_s^4
+0.049382716049382716049 n_h l_s^5
] \frac{1}{\hs} 
+O\left(\frac{1}{\hs^2}\right) \,,
\label{eq:numKIT13}
\nonumber\\ 
\end{eqnarray}
where
\begin{eqnarray}
l_s = \ln\left(-\frac{1}{\hs}\right)
\end{eqnarray}
and $n_l$ is the number of light quarks. In order to compare these numbers with the ones in 
\cite{Fael:2022miw,Fael:2022rgm}, we need to replace the color factors for $SU(3)_c$ in that reference. 
We will perform a thorough numerical comparison of our results to the ones of  \cite{Fael:2022miw,Fael:2022rgm}
in Section~\ref{sec:5} below. We can gain confidence in the precision of our results by increasing as much as 
possible the precision of the base coefficients, which can be done to almost arbitrary levels, and will be the 
topic of the next section.
\subsection{Coefficients in terms of known constants}
\label{sec:33}

\vspace*{1mm}
\noindent
There is a total of 530 base coefficients, shown in Table~\ref{freecoeffs1}. This number corresponds to the 
sum of the orders of all the differential equations given
in Table~\ref{orderstable}. It turns out that all of these coefficients can be expressed in terms of linear 
combinations of known constants over rational numbers. 
To achieve this, we needed to substantially increase the precision of the base coefficients. Using over 3000 coefficients per power of the log for the expansion
at $x=0$, written in terms of the base coefficients through the corresponding relations, and using 500.000 coefficients for the expansion around $x=1$, 
we were able to obtain the base coefficients with over a thousand correct decimal places. Such a high 
precision allowed us to fully exploit the {\tt PSLQ} algorithm
\cite{PSLQ} implemented in {\tt Mathematica}.
We found that all of the base coefficients in the case of $F_{I,2}^{(0)}(x)$ (for all values of $I$) can be expressed in terms of the following set of constants,
\begin{equation}
\left\{\zeta_2, \, \zeta_3, \, \zeta_4, \, l_2, \, l_2^2, \, \zeta_2 l_2\right\}\, . \label{zeta2consts1}
\end{equation}
Similiarly, the base coefficients in the case of $F_{I,3}^{(0)}(x)$ can be written as linear combinations 
over rational numbers of the constants,
\begin{equation}
\left\{\zeta_2, \, \zeta_3, \, l_2,  \, \zeta_2 l_2, \, \frac{1}{\pi^2}, \,  \frac{\zeta_3}{\pi^2}, \, \frac{\zeta_5}{\pi^2}\right\}\, . \label{zeta3consts1} 
\end{equation}
In the case of $F_{I,1}^{(0)}(x)$, the following nineteen constants were needed in order to express all 
base coefficients,
\begin{eqnarray}
&& \biggl\{\zeta_2, \, \zeta_3, \, \zeta_4, \, \zeta_5, \, \zeta_6, \, \zeta_2 \zeta_3, \, \zeta_3^2, \, l_2^4, \, \zeta_2 l_2^4, \, \zeta_2 l_2^2, \, \zeta_4 l_2^2, \, 
\zeta_3 l_2, \, \zeta_4 l_2, \, \zeta_2 \zeta_3 l_2, \, a_4, \, \zeta_2 a_4, \,
\frac{\zeta_3}{\pi^2}, \, \frac{\zeta_3^2}{\pi^2}, \,  \frac{\zeta_3 \zeta_5}{\pi^2} \biggr\}\,. \nonumber \\ &&  \label{zeta0consts1}
\end{eqnarray}
Since every coefficient in the expansions can be written in terms of the base coefficients, all coefficients can be written in terms of these constants.
Here are some examples of the corresponding expressions,
\begin{eqnarray}
c_{a,2,1}(5,-1) &=& \frac{352}{81} \,, \label{z0eq1}
\\
c_{p,1}(5,-1)   &=& - \frac{352}{81} \,, \label{z0eq2}
\\
c_{s,3}(4,-1)   &=& \frac{952}{81 \pi^2} \,, \label{z0eq3}
\\
c_{p,3}(3,1)    &=& \frac{160}{9} - \frac{193424}{81 \pi^2} \,, \label{z0eq4}
\\
c_{a,1,2}(0,-2) &=& - \frac{177152}{81} l_2 \,,  \label{z0eq5} 
\\
c_{v,2,2}(3,0)  &=& \frac{6592}{27} + 128 l_2 \,, \label{z0eq6}
\\  
c_{a,1,1}(3,0)  &=& \frac{10048}{27}
+\frac{47872 \zeta_2}{81}
+\frac{98560 \zeta_3}{81 \pi ^2} \,,
\label{z0eq7} 
\\
c_{a,2,2}(2,1) &=& \frac{58816}{27} - 192 \zeta_2 - 256 l_2 \,, \label{z0eq8} 
\\
c_{s,1}(3,1) &=& \frac{888016}{729} - \frac{10880}{81} \zeta_2 - \frac{992}{9} \zeta_3 + \frac{10192}{27 
\pi^2} \zeta_3 \,, \label{z0eq9} 
\\
c_{p,1}(2,1) &=& \frac{2709004}{729} 
+ \frac{15536}{27} \zeta_2 
- \frac{96}{5} \zeta_2^2 
- \frac{24272}{27} \zeta_3 
+ \frac{3925264}{243 \pi ^2} \zeta_3 
\nonumber\\ &&
- 896 \frac{\zeta_3^2}{\pi ^2}
\,, \label{z0eq10} \\
%
c_{p,1}(1,0) &=& 
- \frac{1551536}{243} 
-\frac{6656}{9} a_4
- \frac{832}{27} l_2^4
+ \frac{399584}{243} \zeta_2 
+ \frac{1664}{9} \zeta_2 l_2^2 
\nonumber \\ &&
+ \frac{11776}{135} \zeta_2^2 
+ \frac{481120}{243} \zeta_3 
- \frac{5824}{9} \zeta_3 l_2 
- \frac{101696}{27 \pi^2} \zeta_3 
+ \frac{17920}{3 \pi ^2} \zeta_3^2     
                \,, \label{z0eq11}
  \\
c_{a,1,1}(0,2) &=& 
-\frac{182732572}{2187}
-\frac{544768  a_4 }{81}
-\frac{68096 l_2^4}{243}
-\frac{5075968 \zeta_2}{243} 
\nonumber\\ &&
+\frac{136192 l_2^2 \zeta_2}{81}
-\frac{942976 \zeta_2^2}{27}
+\frac{2048 l_2 \zeta_2^2}{9}
+\frac{32441728 \zeta_3}{729} 
\nonumber\\ &&
-\frac{318976 l_2 \zeta_3}{81}
+\frac{174528536 \zeta_3}{2187 \pi ^2}
+\frac{626432 \zeta_2 \zeta_3}{405}
-\frac{28672 \zeta_3^2}{3 \pi ^2} 
\nonumber\\ &&
+\frac{45440 \zeta_5}{27} \,,
\\
c_{s,1}(0,3) &=& \frac{3841800953}{6561000}
-\frac{40960  a_4 }{81}
-\frac{5120 l_2^4}{243}
+\frac{2843353 \zeta_2}{729} \nonumber\\ &&
+\frac{10240 l_2^2 \zeta_2}{81}
-\frac{1903352 \zeta_2^2}{405}
-\frac{512}{9} l_2 \zeta_2^2
+\frac{1664 \zeta_2^3}{7} \nonumber\\ &&
+\frac{47204788 \zeta_3}{729}
-\frac{17920 l_2 \zeta_3}{81}
+\frac{55173839 \zeta_3}{4374 \pi ^2}
+\frac{2799872 \zeta_2 \zeta_3}{405} \nonumber\\ &&
-128 \zeta_3^2
+\frac{1132768 \zeta_3^2}{27 \pi ^2}
-\frac{175232 \zeta_5}{9}
+\frac{15008 \zeta_3 \zeta_5}{\pi ^2}.
\label{z0eq14}
\end{eqnarray}
More examples can be found in Appendix~\ref{app:coefficientresults}, where we give the explicit expressions 
for the first few coefficients $c_{I,k}(i,j)$, 
with $j=-2,\ldots,3$ in the vector cases $I \in \left\{(v,1), \, (v,2)\right\}$.

The presence of constants with  $\pi^2$ in the denominator, such as $\frac{\zeta_3}{\pi^2}$, $\frac{\zeta_5}{\pi^2}$, etc., is rather surprising,
since such constants do not typically appear in these types of calculations. However, when we combine all the non-solvable parts in Eq. (\ref{FFdecomp}), 
it turns out that these constants disappear. We illustrate this later in an explicit example. 

Notice that in some cases the coefficients turned out to be quite simple, such as in the case of Eqs. (\ref{z0eq1}) and (\ref{z0eq2}), or even Eq. (\ref{z0eq3}).
In fact, it is precisely using this type of relations that we were able to estimate and improve the precision of the coefficients in the previous section. We can start
by choosing some arbitrary intermediate set of points for the matching conditions, and then the simplicity of the
coefficients under consideration will most likely allow us to determine them using the {\tt PSLQ} 
algorithm, even if the choice of points is sub-optimal. 
Once this is done, we can change the parameters of the matching conditions, as described in Section 
\ref{sec:31}, to increase the precision systematically. 
In particular, we may scan different sets of intermediate points in the region $0<x<1$ until we find one that maximizes the precision.

The precision required to obtain relations such as the ones shown in Eqs. (\ref{z0eq1})--(\ref{z0eq14}) depends on the coefficient under consideration. 
In general, the lower the values of $i$, and the higher the values of
$j$ in $c_{I,k}(i,j)$, the higher the precision is needed to obtain the coefficient. 
For example, Eqs. (\ref{z0eq1})--(\ref{z0eq3}) require less than 13 digits in the {\tt PSLQ} algorithm. Eq. 
(\ref{z0eq9}) can be derived using just 20 digits, 
and Eq. (\ref{z0eq14}) required 192 digits. Now consider the relation for the base coefficient $c_{v,2,1}(0,18)$, namely,
\begin{eqnarray}
c_{v,2,1}(0,18) &=&
4096 \zeta_2 a_4
+\frac{10923574219252432}{1527701175} a_4
-\frac{112576}{105} \zeta_2^3    \nonumber \\ &&
-\frac{417009593483425474341751033}{167122870039125} \zeta_2^2
-1024 \zeta_2^2 l_2^2
+\frac{512}{9} \zeta_2^2 l_2    \nonumber \\ &&
+\frac{25177667228688306512}{11486475} \zeta_2 \zeta_3
+3584 \zeta_2 \zeta_3 l_2
+\frac{512}{3} \zeta_2 l_2^4    \nonumber \\ &&
-\frac{132151944249327423407357442311}{2661099315365336289444000} \zeta_2
-\frac{2730893554813108}{1527701175} \zeta_2 l_2^2    \nonumber \\ &&
+\frac{39715135923817931568319382}{682134163425 \pi^2} \zeta_3^2
+\frac{4096}{3} \zeta_3^2
-21504 \frac{\zeta_3 \zeta_5}{\pi ^2}    \nonumber \\ &&
-\frac{165654702709673585159148398370738061879104893331245527}{2113636351824179636864371971609600000 \pi^2} \zeta_3    \nonumber \\ &&
+\frac{33016255719498077965765879287030858299}{4668595290114625069200} \zeta_3    \nonumber \\ &&
+\frac{2047720334163982}{654729075} \zeta_3 l_2
-\frac{36745784642748584464}{2297295} \zeta_5    \nonumber \\ &&
+\frac{119197070568465928467850610893335774543954251223039329947645441}{12270218005177147973665342631327199732864000000}   \nonumber \\ &&
+\frac{1365446777406554}{4583103525} l_2^4
\,. 
\end{eqnarray}
To obtain such a relation, we need to compute this coefficient with at least 1400 digits. A few hundred additional decimal places allowed us to make sure
that the relation was stable. Another way we made sure that relations like these are not spurious was by introducing extra constants while applying
the {\tt PSLQ} algorithm and noticing that the corresponding coefficient vanishes (this is extremely 
unlikely unless the relation is correct).

Having the expansions in terms of the constants in (\ref{zeta2consts1}--\ref{zeta0consts1}), 
we can now insert them in Eqs. (\ref{gidecomp2a}--\ref{gidecomp2d}) in the case of the vector and axial-vector form factors, 
and in Eqs. (\ref{Sdecomp}) and (\ref{Pdecomp}) in the case of the scalar and pseudo-scalar form factors, respectively,
after which we can expand in $x$ the terms coming from the projectors 
to obtain a full expansion of the non-solvable parts in Eq. (\ref{FFdecomp}). In the case of $F_{V,1}^{(0)}(x)$, up to $O(x^3)$ we obtain

\begin{eqnarray}
F^{(0)}_{V,1}(x) &=& F^{(0), \rm sol}_{V,1}(x)+n_h \biggl\{
-\frac{44288}{81 x} l_2 \zeta_2
-1024 l_2 \zeta_2^2 - \frac{1120}{81} l_2^2 \zeta_2 
- \frac{1168}{243} l_2^4 
\nonumber \\ &&
- \frac{1640}{9} \zeta_2 \zeta_3 
+ \frac{3964}{9} \zeta_5 + \frac{5068}{27} \zeta_2^2 
- \frac{7226}{9} - \frac{9344}{81} a_4 + \frac{46324}{81} \zeta_3 
\nonumber \\ &&
+ \frac{303056}{81} l_2 \zeta_2 
+ \frac{1862569}{243} \zeta_2 
+ \biggl(
\frac{64}{27} l_2^4 + \frac{512}{9} a_4 - \frac{512}{9} l_2^2 \zeta_2 
\nonumber \\ &&
+ \frac{15520}{27} l_2 \zeta_2 + \frac{21728}{135} \zeta_2^2 
+ \frac{34660}{243} \zeta_2 
- \frac{55676}{81} \zeta_3 + \frac{670460}{243} 
\biggr) \ln(x)
\nonumber \\ &&
+ \biggl(
\frac{128}{9} l_2 \zeta_2 - \frac{512}{9} \zeta_3 
+ \frac{2264}{81} \zeta_2 + \frac{18274}{27} 
\biggr) \ln^2(x)
\nonumber \\ &&
+ \biggl(
60 - \frac{32}{3} \zeta_2
\biggr) \ln^3(x)  
-\frac{172}{243} \ln^4(x) 
-\frac{176}{81} \ln^5(x)
\nonumber \\ &&
+\biggl[
544 \zeta_2 \zeta_3 - 4688 \zeta_5 - \frac{5056}{243} l_2^4 
+ \frac{36032}{81} l_2^2 \zeta_2 
- \frac{40448}{81} a_4 
\nonumber \\ &&
+ \frac{41728}{9} l_2 \zeta_2^2 
- \frac{92176}{27} \zeta_3 + \frac{235288}{243} + \frac{355376}{135} \zeta_2^2 
- \frac{722464}{81} l_2 \zeta_2 
\nonumber \\ &&
- \frac{5251684}{243} \zeta_2
+ \biggl(
-800 \zeta_3 
+ \frac{512}{27} l_2^4 - \frac{640}{9} l_2^2 \zeta_2 
+ \frac{4096}{9} a_4 
\nonumber \\ &&
+ \frac{9376}{15} \zeta_2^2 
- \frac{24272}{9} \zeta_2 - \frac{55168}{27} l_2 \zeta_2 + \frac{105760}{243} 
\biggr) \ln(x)
\nonumber \\ &&
+ \biggl(\frac{1744}{9} \zeta_3 - \frac{3392}{9} l_2 \zeta_2 
- \frac{27968}{81} \zeta_2 + \frac{87068}{243} 
\biggr) \ln^2(x) 
\nonumber \\ &&
+ \biggl(
-\frac{7312}{81} \zeta_2 
+ \frac{10496}{243}
 \biggr) \ln^3(x) 
+ \frac{8480}{243} \ln^4(x)
-\frac{44}{405} \ln^5(x) 
\biggr] x
\nonumber \\ &&
+\biggl[
5120 a_4 \zeta_2 - 11520 l_2 \zeta_2^2 - 1280 l_2^2 \zeta_2^2 - 304 \zeta_3^2 
+ \frac{640}{3} l_2^4 \zeta_2 + \frac{2768}{35} \zeta_2^3 
\nonumber \\ &&
- \frac{50704}{9} \zeta_2 \zeta_3 
+ \frac{72704}{243} l_2^4 - \frac{249088}{81} l_2^2 \zeta_2 + \frac{328544}{9} \zeta_3 
+ \frac{581632}{81} a_4 
\nonumber \\ &&
+ \frac{1230400}{27} l_2 \zeta_2 + \frac{1581448}{27} \zeta_5 
- \frac{2711600}{81} \zeta_2^2 + \frac{9890900}{81} \zeta_2 
\nonumber \\ &&
+ \frac{116957693}{13122} 
+ \biggl(-128 \zeta_2 \zeta_3 + 1088 \zeta_5 
+ \frac{640}{27} l_2^4 + \frac{5120}{9} a_4 
\nonumber \\ &&
- \frac{5120}{9} l_2^2 \zeta_2 
+ \frac{21560}{3} \zeta_3 - \frac{68720}{9} \zeta_2^2 + \frac{241664}{27} l_2 \zeta_2 
+ \frac{391184}{27} \zeta_2 
\nonumber \\ &&
+ \frac{47925197}{2187} 
\biggr) \ln(x)
+ \biggl(
1280 a_4 - 320 l_2^2 \zeta_2 + \frac{160}{3} l_2^4 - \frac{1512}{5} \zeta_2^2 
\nonumber \\ &&
+ \frac{42368}{9} l_2 \zeta_2 - \frac{79840}{27} \zeta_3 + \frac{121480}{81} \zeta_2 
+ \frac{4100885}{729} 
\biggr) \ln^2(x)  
\nonumber \\ &&
+ \biggl(
448 l_2 \zeta_2 + 32 \zeta_3 + \frac{87088}{81} \zeta_2 - \frac{761942}{729} 
\biggr) \ln^3(x)
+ \biggl(
4 \zeta_2 
\nonumber \\ &&
- \frac{56711}{243} 
\biggr) \ln^4(x)  
+ \frac{154}{81} \ln^5(x)
\biggr] x^2
+\biggl[
-22528 a_4 \zeta_2 + 5632 l_2^2 \zeta_2^2 
\nonumber \\ &&
- 448 \zeta_3^2 
- \frac{2816}{3} l_2^4 \zeta_2 - \frac{147104}{35} \zeta_2^3 
+ \frac{222976}{9} l_2 \zeta_2^2 - \frac{393184}{3} l_2 \zeta_2 
\nonumber \\ &&
- \frac{557504}{243} l_2^4 + \frac{739936}{27} \zeta_2 \zeta_3 
+ \frac{1400128}{81} l_2^2 \zeta_2 - \frac{4460032}{81} a_4 
\nonumber \\ &&
- \frac{7842800}{27} \zeta_5 + \frac{7863616}{45} \zeta_2^2 
- \frac{10205680}{81} \zeta_3 - \frac{84004240}{2187} 
\nonumber \\ &&
- \frac{300606748}{729} \zeta_2 
+ \biggl(
640 \zeta_2 \zeta_3 - 12352 \zeta_5 + \frac{4480}{9} l_2^2 \zeta_2 
+ \frac{6784}{27} l_2^4 
\nonumber \\ &&
+ \frac{21248}{27} l_2 \zeta_2 
+ \frac{54272}{9} a_4 + \frac{672448}{15} \zeta_2^2 
- \frac{1037608}{2187} 
- \frac{6858880}{81} \zeta_3 
\nonumber \\ &&
- \frac{11355152}{243} \zeta_2
\biggr) \ln(x) 
+ \biggl(
-5632 a_4 + 1408 l_2^2 \zeta_2 - \frac{704}{3} l_2^4 
+ \frac{6672}{5} \zeta_2^2 
\nonumber \\ &&
- \frac{170816}{9} l_2 \zeta_2 + \frac{184880}{27} \zeta_2 + \frac{390416}{27} \zeta_3 
- \frac{1958972}{81} 
\biggr) \ln^2(x) 
\nonumber \\ &&
+ \biggl(
-3200 l_2 \zeta_2 - 256 \zeta_3 - \frac{511504}{81} \zeta_2 
+ \frac{6621976}{729} 
\biggr) \ln^3(x) 
\nonumber \\ &&
+ \biggl(
-8 \zeta_2 + \frac{399020}{243} 
\biggr) \ln^4(x) 
-\frac{40264}{405} \ln^5(x)
-\frac{34}{15} \ln^6(x) 
\biggr] x^3
+O(x^4) \biggr\}.
\label{eq:FV1x-expansion}
\end{eqnarray}

Replacing $x$ in terms of $\hs$ using Eq. (\ref{eq:x2s}), and expanding everything 
(including the solvable part, as well as the renormalization and infrared subtraction terms) in $\hs \rightarrow -\infty$ up to $O(1/\hs^3)$, 
we obtain the following result for the term proportional to $n_h$ (which we call $F_{V,1}^{n_h}(\hs)$),
\begin{eqnarray}
F^{n_h}_{V,1}(\hs) &=&
\frac{64}{3} l_2 \zeta_2^2 - \frac{464}{3} \zeta_2 \zeta_3 - \frac{3808}{243} l_2^4 - \frac{4856}{9} \zeta_5 + \frac{5828}{27} \zeta_2^2 + \frac{7616}{81} l_2^2 \zeta_2 
\nonumber \\ &&
- \frac{30464}{81} a_4 - \frac{52708}{81} \zeta_3 - \frac{107008}{81} l_2 \zeta_2 + \frac{281176}{243} \zeta_2 + \frac{461144}{243} 
\nonumber \\ &&
+ \biggl(
-\frac{64}{81} l_2^4 + \frac{128}{27} l_2^2 \zeta_2 + \frac{368}{27} \zeta_2^2 - \frac{512}{27} a_4 - \frac{1804}{9} \zeta_3 + \frac{47936}{243} \zeta_2 
\nonumber \\ &&
+ \frac{171704}{243} 
\biggr) l_s 
+ \biggl(
-\frac{160}{9} \zeta_3 + \frac{896}{81} \zeta_2 + \frac{46064}{243} 
\biggr) l_s^2 
\nonumber \\ &&
+ \biggl(
-\frac{416}{81} \zeta_2 
+ \frac{4616}{243} 
\biggr) l_s^3 
-\frac{136}{81} l_s^4
-\frac{32}{81} l_s^5 
\nonumber \\ &&
+\biggl[
-\frac{1664}{27} \zeta_2 \zeta_3 + \frac{2560}{243} l_2^4 + \frac{4672}{9} \zeta_2^2 
- \frac{5120}{81} l_2^2 \zeta_2 - \frac{5632}{9} \zeta_5 
\nonumber \\ &&
+ \frac{20480}{81} a_4 - \frac{54968}{27} \zeta_3 - \frac{70400}{3} l_2 \zeta_2 
+ \frac{329392}{243} + \frac{5037584}{243} \zeta_2 
\nonumber \\ &&
+ \biggl(
-1632 \zeta_3 + \frac{928}{27} \zeta_2^2 + \frac{96304}{81} \zeta_2 + \frac{738352}{243} 
\biggr) l_s 
+ \biggl(
-\frac{16}{27} \zeta_3 
\nonumber \\ &&
+ \frac{6896}{81} \zeta_2 + \frac{94144}{243} 
\biggr) l_s^2
+ \biggl(
\frac{64}{243} + \frac{272}{9} \zeta_2
\biggr) l_s^3 
-\frac{4664}{243} l_s^4
+ \frac{4}{81} l_s^5 
\biggr] \frac{1}{\hs}
\nonumber \\ &&
+\biggl[
-304 \zeta_3^2 - \frac{128}{3} l_2 \zeta_2^2 - \frac{14984}{27} \zeta_5 + \frac{22032}{35} \zeta_2^3 - \frac{45136}{27} \zeta_2 \zeta_3 
\nonumber \\ &&
+ \frac{85312}{243} l_2^4 
- \frac{170624}{81} l_2^2 \zeta_2 + \frac{682496}{81} a_4 - \frac{765568}{405} \zeta_2^2 - \frac{1017860}{81} \zeta_3 
\nonumber \\ &&
+ \frac{3636101}{13122} 
- \frac{11654656}{81} l_2 \zeta_2 + \frac{34490284}{243} \zeta_2 
+ \biggl(
1168 \zeta_5 - 288 \zeta_2 \zeta_3
\nonumber \\ &&
+ \frac{128}{81} l_2^4 - \frac{256}{27} l_2^2 \zeta_2 + \frac{1024}{27} a_4 - \frac{19456}{27} l_2 \zeta_2 
- \frac{24976}{135} \zeta_2^2 
\nonumber \\ &&
- \frac{1149688}{81} \zeta_3 + \frac{2440840}{243} \zeta_2 + \frac{57528233}{2187} 
\biggr) l_s
- \biggl(
\frac{824}{5} \zeta_2^2 + \frac{1856}{3} \zeta_3 
\nonumber \\ &&
- \frac{126104}{81} \zeta_2 - \frac{2518829}{729} 
\biggr) l_s^2  
+ \biggl(
32 \zeta_3 + \frac{39152}{81} \zeta_2 - \frac{446750}{729} 
\biggr) l_s^3 
\nonumber \\ &&
+ \biggl(
- \frac{28079}{243} + 4 \zeta_2 
\biggr) l_s^4 
+ \frac{1306}{405} l_s^5
\biggr] \frac{1}{\hs^2}
\nonumber \\ &&
+\biggl[
-768 \zeta_3^2 - \frac{4096}{3} l_2 \zeta_2^2 + \frac{14176}{9} \zeta_5 + \frac{268384}{35} \zeta_2^3 - \frac{306368}{27} \zeta_2 \zeta_3 
\nonumber \\ &&
+ \frac{879616}{243} l_2^4 - \frac{1759232}{81} l_2^2 \zeta_2 - \frac{4436752}{81} \zeta_2^2 + \frac{7036928}{81} a_4 
\nonumber \\ &&
- \frac{13444000}{243} \zeta_3 
- \frac{62817280}{81} l_2 \zeta_2 - \frac{145774160}{6561} + \frac{615879368}{729} \zeta_2 
\nonumber \\ &&
+ \biggl(
-3456 \zeta_2 \zeta_3 + 18624 \zeta_5 + \frac{4096}{81} l_2^4 - \frac{8192}{27} l_2^2 \zeta_2 + \frac{32768}{27} a_4 
\nonumber \\ &&
- \frac{44288}{45} \zeta_2^2 
- \frac{269312}{27} l_2 \zeta_2 - \frac{8223584}{81} \zeta_3 + \frac{13513568}{243} \zeta_2 
\nonumber \\ &&
+ \frac{354234328}{2187} 
\biggr) l_s 
+ \biggl(
-\frac{8784}{5} \zeta_2^2 - \frac{46528}{9} \zeta_3 + \frac{788512}{81} \zeta_2 
\nonumber \\ &&
+ \frac{10524520}{729} 
\biggr) l_s^2 
+ \biggl(
384 \zeta_3 + \frac{361600}{81} \zeta_2 - \frac{5575000}{729} 
\biggr) l_s^3 
\nonumber \\ &&
+ \biggl(
24 \zeta_2 - \frac{223328}{243} 
\biggr) l_s^4 
+ \frac{1652}{27} l_s^5
+ \frac{34}{15} l_s^6 
\biggr] \frac{1}{\hs^3}
                +O\left(\frac{1}{\hs^4}\right)
                ,
\label{eq:FV1sinf}
\end{eqnarray}
which, of course, agrees numerically with Eq. (\ref{eq:numKIT13}). 
Similar expansions for the other form factors can be found in Appendix~\ref{app:x=0expansions}.
Notice that the constants with $\pi^2$ in the denominator
are indeed not present in the final results (\ref{eq:FV1x-expansion}) or (\ref{eq:FV1sinf}). In fact, only the following constants appear
in the final result of any of the form factors
\begin{equation}
\bigl\{1, \zeta_2, \, \zeta_3, \, \zeta_4, \, \zeta_5, \, \zeta_6, \, \zeta_2 \zeta_3, \, \zeta_3^2, \, l_2^4, \, \zeta_2 l_2, \, \zeta_2 l_2^2, \, \zeta_2 l_2^4, \,
\zeta_4 l_2, \, \zeta_4 l_2^2, \, a_4, \, \zeta_2 a_4 \bigr\}\,,  
\label{finalconsts1}
\end{equation}
so even the constants $\zeta_3 l_2$ and $\zeta_2 \zeta_3 l_2$, which
are present in (\ref{zeta0consts1}), disappear from the final
results. These cancellations give another strong indication that the guessed 
components of the individual terms using {\tt PSLQ} are correct.

In general, the form factors can be written as follows,
\begin{eqnarray}
F^{(0)}_{I}(x) &=& F^{(0), \rm sol}_{I}(x) + n_h \left(F^{(0)}_{I,1}(x) + \zeta_2 F^{(0)}_{I,2}(x) + \zeta_3 F^{(0)}_{I,3}(x)\right) \nonumber \\
               &=& F^{(0), \rm sol}_{I}(x) + n_h \sum_{n=-1}^{\infty} \sum_{j=0}^6 \sum_{i=1}^{16} r_{I;j,n}^{(C_i)} C_i x^n \ln^j(x),
\end{eqnarray}
where $C_i$ is any of the constants in (\ref{finalconsts1}) and the $r_{I;j,n}^{(C_i)}$'s are rational numbers.

One may consider now the possibility of generating several thousand coefficients for the expansions around $x=0$. 
These can be used to generate recursions by applying guessing methods on each of the series of rational numbers $r_{I;j,n}^{(C_i)}$ with $i$ and $j$ fixed.
We did this and found that some of the recursions were solvable, allowing us to find closed form solutions in these cases. 
This is similar to what we did in the case of the expansions at $x=1$ in Ref. \cite{Blumlein:2019oas}. The main difference now is the presence of powers of $\ln(x)$,
but this does not represent a major difficulty. It merely means that the procedure applied in \cite{Blumlein:2019oas} needs to be repeated on each power series
multiplying each power of the log.

Let us consider again the case of the non-solvable parts of $F_{V,1}^{(0)}(x)$, 
and take as an example the sequence of rational numbers multiplying the term $\zeta_3 \ln^3(x)$, which is given by
\begin{equation}
\left\{ r_{V,1;3,n}^{(\zeta_3)} \, | \, n \geq 1 \right\} =
\bigl\{0, 32, -256, 896, -2304, 4896, -9216, 15872, -25600, \ldots \bigr\}.
\label{eq:z2ln3_list}
\end{equation}
In other words, the rational number multiplying the term $\zeta_3 x^n \ln^3(x)$ in the expansion around $x=0$ of the non-solvable part of $F_{V,1}^{(0)}(x)$ with
$n \geq 1$ is given by the $n$th term in the list (\ref{eq:z2ln3_list}). 
We can now use this list and the {\tt Sage} package {\tt ore\_algebra} to
obtain a recursion satisfied by this sequence. We get
\begin{equation}
(n+2)^2 f_1(n) - 4 (n+1) f_1(n+1) - n^2 f_1(n+2) = 0,
\label{eq:rec1}
\end{equation}
where $f_1(n)$ is the $n$th term in the list (\ref{eq:z2ln3_list}). 
This recursion can be obtained using just the first 15 terms in that list.
As another example, the list associated with the term $l_2 \zeta_2$ ($=l_2 \zeta_2 \ln^0(x)$) is given by
\begin{equation}
\left\{ r_{V,1;0,n}^{(l_2 \zeta_2)} \, | \, n \geq 1 \right\} =
\left\{-\frac{722464}{81},\frac{1230400}{27},-\frac{393184}{3},\frac{2295712}{9},-\frac{260063392}{675},\ldots\right\}.
\label{eq:z2l2_list}
\end{equation}
We can verify that the $n$th term in the lists (\ref{eq:z2ln3_list}) and (\ref{eq:z2l2_list}) agree with the corresponding rational numbers multiplying the
corresponding factor in Eq. (\ref{eq:FV1x-expansion}).
A recursion of order 7 can also be found in the case of the list (\ref{eq:z2l2_list}), 
\begin{equation}
\sum_{i=0}^7 p_i(n) f_2(n+i) = 0,
\label{eq:rec2}
\end{equation}
where $f_2(n)$ is the $n$th term in (\ref{eq:z2l2_list}), and the $p_i(n)$'s are polynomials in $n$ of degree 35 with very large integer coefficients.
In order to obtain the recursion (\ref{eq:rec2}), we must apply guessing algorithms using at least the first 142 terms in (\ref{eq:z2l2_list}).

In the case of (\ref{eq:rec1}), a very simple solution can be found with \texttt{Sigma}, namely
\begin{equation}
f_1(n) = -2+2 \big(1-4 n^2+2 n^4\big) (-1)^n,
\end{equation}
while in the case of the $l_2 \zeta_2$ term, the solution found by \texttt{Sigma} is somewhat 
more complicated
\begin{eqnarray}
f_2(n) &=& 
-\frac{16 T_1}{81 n^2}
-\frac{32 (-1)^n T_2}{405 n^2}
+\frac{128}{9} n \big(
        112+13 n^2\big) (-1)^n S_1
+\Biggl[
        -\frac{1168}{3}
\nonumber\\ &&       
 -\frac{32}{9} (-1)^n \big(
                413+264 n^2+34 n^4\big)
\Biggr] S_2
+\frac{128}{9} n \big(
        -14+13 n^2\big) (-1)^n S_{-1}
\nonumber\\ &&
+\Biggl[
        -\frac{992}{9}
        -\frac{16}{9} (-1)^n \big(
                237-16 n^2+14 n^4\big)
\Biggr] S_{-2} \,,
\end{eqnarray}
with
\begin{eqnarray}
T_1 &=& 32 n^5+405 n^4-67 n^3-432 n^2-216 n-2052,
\\
T_2 &=& 1032 n^7-71100 n^5-1080 n^4-33847 n^3+1080 n^2+1620 n-9990.
\end{eqnarray}
Here the $S_k(n)$ are the single harmonics sums 
\cite{Vermaseren:1998uu,Blumlein:1998if}, which are defined as
\begin{equation}
S_k(n) = \sum_{i=1}^n \frac{({\rm sign}(k))^i}{i^{|k|}} \,;
\end{equation}
for convenience we sometimes drop the $n$-dependence 
of the harmonic sums.
We can now perform the sums. In the case of $f_1(n)$ we get
\begin{equation}
\sum_{n=1}^{\infty} f_1(n) x^n = \frac{32 x^2 \left(1-4 
x+x^2\right)}{(1-x) (1+x)^5},
\label{eq:f1summed}
\end{equation}
while in the case of $f_2(n)$ we obtain
\begin{eqnarray}
\sum_{n=1}^{\infty} f_2(n) x^n &=&
\frac{128 \HA_1(x) R_1}{9 (1+x)^4}
+\frac{128 \HA_{-1}(x) R_2}{9 (1+x)^4}
-\frac{64 \HA_{0,1}(x) R_3}{9 (1-x) (1+x)^5}
+\frac{64 \HA_{0,-1}(x) R_4}{9 (1-x) (1+x)^5}
\nonumber\\ &&
-\frac{32 x R_5}{81 (1-x)^4 (1+x)^6} \,,
\label{eq:f2summed}
\end{eqnarray}
with
\begin{eqnarray}
R_1 &=& 3 x^4+16 x^3+62 x^2+16 x+3,
\\
R_2 &=& 9 x^4-5 x^3+98 x^2-5 x+9,
\\
R_3 &=& 57 x^6+224 x^5+423 x^4+344 x^3+423 x^2+224 x+57,
\\
R_4 &=& 111 x^6+104 x^5+217 x^4-368 x^3+217 x^2+104 x+111,
\\
R_5 &=& 22597 x^8-78400 x^7+43500 x^6+179008 x^5-327266 x^4+179008 
x^3 +43500 x^2
\nonumber\\ &&
-78400 x+22597
\end{eqnarray}
and $\HA_{\vec{a}}(z)$ denotes a harmonic polylogarithm \cite{Remiddi:1999ew}. The term multiplying $\ln(2) \zeta_2$ in $F^{(0)}_{V, 1}$ is thus given by

\begin{eqnarray}
F_{\ln(2) \zeta_2,V,1}^{(0)} =  - \frac{44288}{81 x} + \frac{303056}{81}  
- \frac{44288}{81} x + \sum_{n=1}^{\infty} f_2(n) x^n.
\end{eqnarray}

We found recursions for all but one, cf. Table~\ref{t:difficulty1}, of the 
sequences of rational numbers multiplying all of the terms in the expansions around $x=0$ of each form factor. 
In all cases where a solution could be found, we performed the corresponding sum as we did in (\ref{eq:f1summed}) and (\ref{eq:f2summed}),
and added all of the results, leaving a remainder of non-solvable parts.
In the case of $F^{(0)}_{V,1}(x)$, we found that the formerly unsolvable parts at $x=1$ on the right-hand 
side of $F_{V,1}(x)$ in Eq. (\ref{FFdecomp}) can now be resummed as follows

\begin{eqnarray}
\lefteqn{F^{(0)}_{V,1,1}(x) + \zeta_2 F^{(0)}_{V,1,2}(x) + \zeta_3 F^{(0)}_{V,1,3}(x) =}  
\nonumber\\ &&
\ln(2) \Biggl\{
        \frac{256 \zeta_2^2 P_{50}}{9 (x-1) (1+x)^5}
        +\Biggl[
                \frac{128 \ln(1+x) P_{17}}{9 (1+x)^4}
                +\frac{64 \Li_2(x) P_{53}}{9 (x-1) (1+x)^5}
                +\frac{64 \Li_2(-x) P_{59}}{9 (x-1) (1+x)^5}
\nonumber\\ &&             
    -\frac{64 \ln^2(x) P_{65}}{9 (x-1)^2 (1+x)^6}
                -\frac{16 P_{82}}{81 (x-1)^4 x (1+x)^6}
                -\ln(1-x) \Biggl(
                        \frac{128 P_9}{9 (1+x)^4}
\nonumber\\ &&                
         -\frac{64 \ln(x) P_{54}}{9 (x-1) (1+x)^5}
                \Biggr)
                +\ln(x) \Biggl(
                        \frac{64 \ln(1+x) P_{58}}{9 (x-1) (1+x)^5}
                        -\frac{32 P_{78}}{27 (x-1)^5 (1+x)^5}
                \Biggr)
\nonumber\\ &&                 
-\frac{64 x^2 \big(
                        7-22 x+7 x^2\big) \ln^3(x)}{(x-1) (1+x)^5}
        \Biggr] \zeta_2
\Biggr\}
+\ln(2)^2 \Biggl\{
        \Biggl[
                \frac{128 \ln(x) P_{45}}{9 (x-1) (1+x)^5}
                -\frac{32 P_{73}}{81 (x-1)^5 (1+x)^5}
\nonumber\\ &&                
 -\ln(1-x) \Biggl(
                        -\frac{128 \big(
                                1+x^2\big)
\big(11-14 x+11 x^2\big)}{3 (x-1) (1+x)^3}
                        +\frac{256 x^2 \big(
                                5-2 x+5 x^2\big) \ln(x)}{(x-1) 
(1+x)^5}
                \Biggr)
\nonumber\\ &&                
 +\frac{64 x^2 \big(
                        5-2 x+5 x^2\big) \ln^2(x)}{(x-1) (1+x)^5}
                -\frac{128 \big(
                        1+x^2
                \big)
\big(11-14 x+11 x^2\big) \ln(1+x)}{3 (x-1) (1+x)^3}
\nonumber\\ &&                
 -\frac{256 x^2 \big(
                        5-2 x+5 x^2\big) \Li
                _2(x)}{(x-1) (1+x)^5}
        \Biggr] \zeta_2
        +
        \frac{256 x^2 \big(
                5-2 x+5 x^2\big) \zeta_2^2}{(x-1) (1+x)^5}
\Biggr\}
\nonumber\\ && 
+\ln^4(2) \Biggl\{
        -\frac{64 \ln(x) P_{43}}{27 (x-1) (1+x)^5}
        -\frac{16 P_{68}}{243 (x-1)^5 (1+x)^5}
        -\ln(1-x) 
\nonumber\\ && \times 
\Biggl[
                \frac{64 \big(
                        1+x^2
                \big)
\big(11-14 x+11 x^2\big)}{9 (x-1) (1+x)^3}
                -\frac{128 x^2 \big(
                        5-2 x+5 x^2\big) \ln(x)}{3 (x-1) (1+x)^5}
        \Biggr]
\nonumber\\ && 
        -\frac{128 x^2 \big(
                5-2 x+5 x^2\big) \zeta_2}{3 (x-1) (1+x)^5}
        -\frac{32 x^2 \big(
                5-2 x+5 x^2\big) \ln^2(x)}{3 (x-1) (1+x)^5}
\nonumber\\ &&       
+\frac{64 \big(
                1+x^2
        \big)
\big(11-14 x+11 x^2\big) \ln(1+x)}{9 (x-1) (1+x)^3}
        +\frac{128 x^2 \big(
                5-2 x+5 x^2\big) \Li_2(x)}{3 (x-1) (1+x)^5}
\Biggr\}
\nonumber\\ && 
+\Li_4\left(\frac{1}{2}\right) \Biggl\{
        -\frac{512 \ln(x) P_{43}}{9 (x-1) (1+x)^5}
        -\frac{128 P_{68}}{81 (x-1)^5 (1+x)^5}
        -\ln(1-x) 
\nonumber\\ &&  \times
\Biggl[
                \frac{512 \big(
                        1+x^2
                \big)
\big(11-14 x+11 x^2\big)}{3 (x-1) (1+x)^3}
                -\frac{1024 x^2 \big(
                        5-2 x+5 x^2\big) \ln(x)}{(x-1) (1+x)^5}
        \Biggr]
\nonumber\\ && 
        -\frac{256 x^2 \big(
                5-2 x+5 x^2\big) \ln^2(x)}{(x-1) (1+x)^5}
        +\frac{512 \big(
                1+x^2
        \big)
\big(11-14 x+11 x^2\big) \ln(1+x)}{3 (x-1) (1+x)^3}
\nonumber\\ &&       
  +
        \frac{1024 x^2 \big(
                5-2 x+5 x^2\big) \Li_2(x)}{(x-1) (1+x)^5}
\Biggr\}
+\Biggl\{
        -\frac{1024 \Li_4\left(\frac{1}{2}\right) x^2 \big(
                5-2 x+5 x^2\big)}{(x-1) (1+x)^5}
\nonumber\\ &&   
      +\frac{128 x^2 \big(
                1-x+x^2\big) \zeta_3 \ln(x)}{(x-1) (1+x)^5}
        -\frac{4 x^2 \ln^4(x)}{(x-1) (1+x)^3}
\Biggr\} \zeta_2
+\frac{16 x^2 \big(
        19+104 x+19 x^2\big) \zeta_3^2}{(x-1) (1+x)^5}
\nonumber\\ && 
-\frac{16 x^2 \big(
        173-8502 x+173 x^2\big) \zeta_2^3}{35 (x-1) (1+x)^5}
-\frac{64 x^2 \big(
        17-125 x+17 x^2\big) \zeta_5 \ln(x)}{(x-1) (1+x)^5}
\nonumber\\ && 
+\frac{24 x^2 \big(
        63-26 x+63 x^2\big) \zeta_2^2 \ln^2(x)}{5 (x-1) (1+x)^5}
-\frac{32 x^2 \big(
        1-4 x+x^2\big) \zeta_3 \ln^3(x)}{(x-1) (1+x)^5}
\nonumber\\ &&
+\frac{34 x^3 \ln^6(x)}{15 (x-1) (1+x)^5} + \tilde{F}_{V,1}^{(0), \rm rest}(x),
\label{eq:FV1resummed2}
\end{eqnarray}

\noindent
and the polynomials $P_i$ are given in Appendix~\ref{app:resum}.
The function  $\tilde{F}^{(0), \rm rest}_{V,1}(x)$ includes all the 
terms in the 
expansion at $x=0$ for which no closed form solutions could be found. Up to $O(x^3)$, we have
\begin{eqnarray}
\tilde{F}_{V,1}^{(0), \rm rest}(x) &=&
\frac{5068}{27} \zeta_2^2
-\frac{1640}{9} \zeta_2 \zeta_3
+\frac{1862569}{243} \zeta_2
+\frac{46324}{81} \zeta_3
+\frac{3964}{9} \zeta_5
-\frac{7226}{9}
\nonumber \\ &&
+\biggl(
\frac{21728}{135} \zeta_2^2
+\frac{34660}{243} \zeta_2
-\frac{55676}{81} \zeta_3
+\frac{670460}{243}
\biggr)\ln(x) 
\nonumber \\ &&
+\biggl(
\frac{2264}{81} \zeta_2
-\frac{512}{9} \zeta_3
+\frac{18274}{27}
\biggr) \ln^2(x)
+\biggl(
60
-\frac{32}{3} \zeta_2
\biggr) \ln^3(x)
\nonumber \\ &&
-\frac{172}{243} \ln^4(x)
-\frac{176}{81} \ln^5(x)
+\biggl[
\frac{355376}{135} \zeta_2^2
+544 \zeta_2 \zeta_3
-\frac{5251684}{243} \zeta_2
\nonumber \\ &&
-\frac{92176}{27} \zeta_3
-4688 \zeta_5
+\frac{235288}{243}
+\biggl(
\frac{9376}{15} \zeta_2^2
-\frac{24272}{9} \zeta_2
-800 \zeta_3
\nonumber \\ &&
+\frac{105760}{243}
\biggr) \ln(x)
+\biggl(
-\frac{27968}{81} \zeta_2
+\frac{1744}{9} \zeta_3
+\frac{87068}{243}
\biggr) \ln^2(x)
\nonumber \\ &&
+\biggl(
\frac{10496}{243}
-\frac{7312}{81} \zeta_2
\biggr) \ln^3(x)
+\frac{8480}{243} \ln^4(x)
-\frac{44}{405} \ln^5(x)
\biggr] x
\nonumber \\ &&
+\biggl[
-\frac{2711600}{81} \zeta_2^2
-\frac{50704}{9} \zeta_2 \zeta_3
+\frac{9890900}{81} \zeta_2
+\frac{328544}{9} \zeta_3
\nonumber \\ &&
+\frac{1581448}{27} \zeta_5
+\frac{116957693}{13122}
+\biggl(
\frac{391184}{27} \zeta_2
-\frac{68720}{9} \zeta_2^2
+\frac{21560}{3} \zeta_3
\nonumber \\ &&
+\frac{47925197}{2187}
\biggr) \ln(x)
+\biggl(
\frac{121480}{81} \zeta_2
-\frac{79840}{27} \zeta_3
+\frac{4100885}{729}
\biggr) \ln^2(x) 
\nonumber \\ &&
+\biggl(
\frac{87088}{81} \zeta_2
-\frac{761942}{729}
\biggr) \ln^3(x)
-\frac{56711}{243} \ln^4(x)
+\frac{154}{81} \ln^5(x)
\biggr] x^2
\nonumber \\ &&
+\biggl[
\frac{7863616}{45} \zeta_2^2
+\frac{739936}{27} \zeta_2 \zeta_3
-\frac{300606748}{729} \zeta_2
-\frac{10205680}{81} \zeta_3
\nonumber \\ &&
-\frac{7842800}{27} \zeta_5
-\frac{84004240}{2187}
+\biggl(
\frac{672448}{15} \zeta_2^2
-\frac{11355152}{243} \zeta_2
\nonumber \\ &&
-\frac{6858880}{81} \zeta_3
-\frac{1037608}{2187}
\biggr) \ln(x)
+\biggl(
\frac{184880}{27} \zeta_2
+\frac{390416}{27} \zeta_3
\nonumber \\ &&
-\frac{1958972}{81}
\biggr) \ln^2(x)
+\biggl(
\frac{6621976}{729}
-\frac{511504}{81} \zeta_2
\biggr) \ln^3(x)
\nonumber \\ &&
+\frac{399020}{243} \ln^4(x)
-\frac{40264}{405} \ln^5(x)
\biggr] x^3
+O(x^4).
\end{eqnarray}
We can see that the result obtained in (\ref{eq:f1summed}) appears in the second to last line of Eq. (\ref{eq:FV1resummed2}) multiplying the corresponding factor $\zeta_3 \ln^3(x)$, while the result obtained in (3.69) appears in the first and second lines of (\ref{eq:FV1resummed2}) multiplying the factor $\ln(2) \zeta_2$. 
The corresponding expressions for the other form factors can be found in Appendix~\ref{app:resum}. 
To get the full solvable parts of the form factors at $x=0$, we need to add the resummed expressions, like the one in Eq. (\ref{eq:FV1resummed2}),
to the solvable parts arising from the expansion at $x=1$, as shown in Eq. (\ref{FFdecomp}).
\begin{table}[H]
\begin{center}
\begin{tabular}{|c|c|c|c|c|c|c|c|}
\hline
 \phantom{\Large I}    & $\ln^6(x)$ & $\ln^5(x)$ & $\ln^4(x)$ & $\ln^3(x)$ & $\ln^2(x)$ & $\ln^1(x)$ & $\ln^0(x)$ \\
\hline
         1             &   9\,\cm   &    136     &    512     &    3680    &   12100    &    23000   &   ?        \\
\hline
       $\zeta_2$       &            &            &   9\,\cm   &    136     &    512     &    1440    &    2970    \\
\hline
       $\zeta_3$       &            &            &            &   12\,\cm  &    136     &    510     &    3560    \\
\hline
    $l_2 \zeta_2$      &            &            &            &   15\,\cm  &   18\,\cm  &   70\,\cm  &  140\,\cm  \\
\hline
       $l_2^4$         &            &            &            &            &   15\,\cm  &   28\,\cm  &   88\,\cm  \\
\hline
    $l_2^2 \zeta_2$    &            &            &            &            &   15\,\cm  &   28\,\cm  &   88\,\cm  \\
\hline
         $a_4$         &            &            &            &            &   15\,\cm  &   28\,\cm  &   88\,\cm  \\
\hline
       $\zeta_4$       &            &            &            &            &   15\,\cm  &    136     &    510     \\
\hline
       $\zeta_5$       &            &            &            &            &            &   15\,\cm  &    136     \\
\hline
   $\zeta_2 \zeta_3$   &            &            &            &            &            &   15\,\cm  &    136     \\
\hline
       $\zeta_6$       &            &            &            &            &            &            &   15\,\cm  \\
\hline
      $\zeta_3^2$      &            &            &            &            &            &            &   15\,\cm  \\
\hline
    $l_2 \zeta_4$      &            &            &            &            &            &            &   15\,\cm  \\
\hline
     $l_2^2 \zeta_4$   &            &            &            &            &            &            &   15\,\cm  \\
\hline
     $l_2^4 \zeta_2$   &            &            &            &            &            &            &   15\,\cm  \\
\hline
      $\zeta_2 a_4$    &            &            &            &            &            &            &   15\,\cm  \\
\hline
\end{tabular}
\caption{\sf\small{Number of terms needed to derive recursions for the sequence of rational numbers associated to
a given constant in the first column times a given power of $\ln(x)$ in the case of $F_{V,1}^{(0)}(x)$'s.
A check mark indicates that the corresponding recursion was solvable. An empty cell means that the corresponding product
of constant and power of the log is not present in this form factor. The interrogation sign in one of the entries indicates 
that no recursion could be found through guessing algorithms in this case.}}
\label{t:difficulty1}
\end{center}
\end{table}

\vspace*{-1cm}
\noindent
Depending on the value of $i$ and $j$ in $r_{I;j,n}^{(C_i)}$, finding the corresponding recursion can be quite difficult, 
requiring sometimes the use of thousands of coefficients for the guessing algorithms. 
In Table~\ref{t:difficulty1}, we show the minimum number of terms required to guess the recursions for each 
of the constants listed in (\ref{finalconsts1})
that may appear in the power series multiplying each power of $\ln(x)$ in the case of $F^{(0)}_{V,1}(x)$. 
A check mark near a number indicates that a solution for the corresponding recursion was found (25 out of 42 cases).
We can see that, for example, we needed 512 terms to get a recursion for the series of rational numbers defined by $r_{V,1;2,n}^{(\zeta_2)}$.
The most complicated recursion that we found was the one associated to $r_{V,1;1,n}^{(1)}$, which required a sequence of 23000 terms.
This was also the recursion of highest order that we derived in the vector case, with an order of 90. Finding this recursion, as well as similar ones in other form factors,
pushed the limits of our computational resources, exhausting all 2 Tb of RAM memory in one of our machines. 
It was therefore not possible to find recursions for the sequence of numbers defined by $r_{I;0,n}^{(1)}$, which
are even more demanding. This was the only case where no recursion was derived in this way.

The advantage of having all of these recursions, even in the cases where no solutions can be found, is that they can be used to calculate a very large
number of expansion coefficients in a simple way compared with their direct computation from the differential equations as we did initially. 
We were able to compute up to 30.000 coefficients for all of the constants in Table \ref{t:difficulty1}, except the one where no recursion was found,
in which case 20.000 coefficients were derived by inserting the coefficients associated to the other constants to the recursion coming
from the differential equations associated to this term.

\section{Threshold expansions}
\label{sec:4}

\vspace*{1mm}
\noindent
Power--log expansions around the two- and four--particle (pseudo)thresholds can also be 
found using the method presented in the previous section, provided 
an appropriate expansion parameter is used. The expansions around $\hs=4$ (two-particle threshold) 
and around $\hs=16$ (four-particle threshold), respectively, 
can be written in terms of the variables 
\begin{eqnarray}
z       &=& \sqrt{4-\hs} \,,
 \\
\tilde{z} &=& \sqrt{16-\hs}.
\end{eqnarray}
In both cases the expansions will have the form
\begin{equation}
F_{I,i}^{(0)}(z) = \sum_{j=0}^3 \sum_{k=k_{\rm min}}^{\infty} \tc_{I,i}(j,k) z^k \ln^j(z).
\label{eq:s4ansatz}
\end{equation}
Notice that the variable $z=\sqrt{4-\hs}$ is similar to the velocity of the produced quarks $\beta=\sqrt{1-4/\hs}$, but unlike the velocity, $z$ is
defined to be real for $\hs<4$, which is convenient since $F_{I,i}^{(0)}(z)$ is also real in this region, and therefore the corresponding
expansion coefficients in (\ref{eq:s4ansatz}) will be real.

The point $z=0$ ($\hs=4$) corresponds to $x=-1$ in $x$-space. One might also think about finding expansions around $x=-1$ using the variable $\hat{y}=1+x$ instead of $z$.
In this case, the expansions should involve powers of $\ln(\hat{y})=\ln(1+x)$. Deriving differential equations in $\hat{y}$ from the ones we already have would be simple, 
but after that we would also need to match the expansion at $\hat{y}=0$ with the expansion at another point. 
So far, we have obtained the expansions at $x=1$ and then at $x=0$. As we can see in 
Figure~\ref{fig:sxregions}, the point $x=1$ is directly connected to the point $x=-1$ 
through the unit arc. 
However, trying to match {\it directly} the expansions at $x=1$ and $x=-1$ through this arc is not possible. The radius of convergence of the expansion around $x=1$ is equal to one, which
means that for $x=e^{i \phi}$ the expansion diverges for $\phi>\pi/3$. The radius of convergence of the expansion around $x=-1$ is even smaller, 
which means it diverges for $\phi<2\pi/3$, and therefore there is no common region along the unit arc where both series can be matched.\footnote{One may still try to go through
the unit arc by matching first at some intermediate points before getting all the way to $x=-1$.} 

On the other hand, to go from the expansion at $x=0$ to the expansion at $x=-1$, we first need to go through the four-particle threshold at $x=4\sqrt{3}-7$ ($\hs=16$). 
This is certainly an option, and we indeed found expansions at this (pseudo-)threshold by matching with the expansions at $x=0$, as we will see later in Section \ref{Sec:s16-expansions}. 
However, we did not use the expansions at $\hs=16$ to obtain the expansions at $x=-1$ ($\hs=4$).
Instead, we found expansions at $\hs=0$, derived the corresponding differential equations, 
and matched with the expansions at $\hs=4$ given by Eq. (\ref{eq:s4ansatz}). 
We found this more convenient for several reasons.
First of all, as we already mentioned, in the region $0<\hs<4$, we deal with real numbers, 
while in the region $-1<x<4\sqrt{3}-7$, the form factors will also have imaginary parts.\footnote{Working with complex coefficients is not really problematic, 
but we will later use the {\tt PSLQ} algorithm, as we did in the case of the expansions at $x=0$, and having complex numbers may double the amount of searches.}
Second, with the expansions around $\hs=0$ and $\hs=4$ ($z=0$), we cover the entire
unit arc, which is something we would still have to do if we stay in $x$, as discussed above.
Third, we can find much deeper expansions at $\hs=0$ than at $\hs=16$, and all the coefficients of the expansions at $\hs=0$
are given in terms of known constants, which allows us to evaluate these expansions at different points with very high precision. 
Finally, although it is possible to derive differential equations in $z=\sqrt{4-\hs}$ from the differential equations in $x$, 
the change of variables from $x$ to $z$ would lead to very clumsy expressions. 
From $\hs$ to $z$, on the other hand, the replacement of the derivatives is a simple one, namely, $\frac{d}{d\hs} \rightarrow -\frac{1}{2 z} \frac{d}{dz}$, 
and the differential equations look similar to the ones in Eq. (\ref{diffeqinx1}),
\begin{equation}
\sum_{k=0}^{o_{I,i}} \tilde{p}_{I,i,k}(z) \frac{d^k}{dz^k} F_{I,i}^{(0)}(z) = 0,
\label{diffeqinz1}
\end{equation}
where $o_{I,i}$ is the order of the differential equation and the $\tilde{p}_{I,i,k}(z)$'s are polynomials in $z$.

The expansions in $\hs$ around $\hs=0$ can be obtained from the corresponding expansions at the level of the master integrals 
in the same way as we did in the case of the expansions in $x$ around $x=1$ in Ref. \cite{Blumlein:2019oas}. 
However, since this is very laborious, and since the expansions around $x=1$ were already available from previous works, it was easier to use them to derive the ones at $\hs=0$. 
This is, nevertheless,  harder than it sounds, since many thousands of coefficients are needed to find matching conditions that lead to high precision, 
and also to guess differential equations in $\hs$.
Simply replacing $x$ by $\hs$ using Eq. (\ref{eq:x2s}) in the expansions at $x=1$ 
and expanding in $\hs$ may take a lot of time if done naively, for example, using directly the {\tt Mathematica} function {\tt Series}. 
In Section~\ref{Sec:ComputeSMoments}, we will describe the  computer algebra methods we used to derive arbitrary many coefficients of the expansions at $\hs=0$.
These methods may find applications in other areas of physics or computer algebra where large expansions are needed.
In Sections~\ref{Sec:s4-expansions}, we will describe the calculation of the threshold expansions at $\hs=4$ exploiting these ingredients. In Section~\ref{Sec:s16-expansions}, we will
discuss the four-particle expansions at $\hs=16$. 
\subsection{\boldmath Computing the recurrences and differential equations in $\hat{s}$}
\label{Sec:ComputeSMoments}

\vspace*{1mm}
\noindent
In the following we suppose that $f(x)$ stands for any of the 18 possible non-solvable parts of the 
form factors, namely, $F_{I,k}^{(0)}(x)$ with $I \in \left\{(v,1), \, (v,2), \, (a,1), \, (a,2), \, s, \, p\right\}$ and $k=1,2,3$.
Moreover, we assume that we are given the first $\nu+1$ coefficients $f_n$ in its power series expansion 
\begin{equation}\label{Equ:XExpansion}
	f(x)=\sum_{n=0}^{\nu}f_n\,(1-x)^n+O((1-x)^{\nu+1})
\end{equation}
around $x=1$. In our concrete calculations we will set $\nu=8000$. Then replacing $x$ by the right-hand side of~\eqref{eq:x2s} we want to compute the first $\nu+1$ coefficients $\hat{h}_n$ of
\begin{equation}\label{Equ:SExpansion}
	\hat{F}(\hat{s})=f\Big(\tfrac{\sqrt{4-\hs}-\sqrt{-\hs}}{\sqrt{4-\hs}+\sqrt{-\hs}}\Big)=\sum_{n=0}^{\nu}\hat{h}_n\hs^n+O(\hs^{\nu+1}).
\end{equation}
We proceed in two steps. First, expand the right-hand side of~\eqref{eq:x2s} in $\hs$, 
and second plug this expansion into~\eqref{Equ:XExpansion} and extract the expansion coefficients 
$\hat{h}_n$ on the right-hand side of~\eqref{Equ:SExpansion}. 
Given these coefficients, we are interested in a third step to derive a linear differential equation for $\hat{F}(\hs)$ and a linear recurrence for $\hat{h}_n$. 
This latter equation is particularly important to produce further coefficients $\hat{h}_n$ with $100.000\geq n\geq\nu=8000$ that are needed in the calculations of Section~\ref{Sec:s4-expansions}.   

\medskip

\textit{Step 1.} As it turns out, the expansion of the right-hand side in~\eqref{eq:x2s} yields contributions in $\sqrt{\hs}$ where the root contributions cancel in the final expansion given in~\eqref{Equ:SExpansion}. As a consequence, one actually needs $2\,\nu+1$ coefficients in order to obtain $\nu+1$ expansion coefficients in the final result. To avoid these roots and to work purely with polynomial expressions, we replace $\hs$ by $\sigma^2$ and compute the expansion
$$g(\sigma)=f\Big(\tfrac{\sqrt{4-\sigma^2}-\sqrt{-\sigma^2}}{\sqrt{4-\sigma^2}+\sqrt{-\sigma^2}}\Big)=\sum_{n=0}^{2\nu}g_n\,\sigma^n+O(\sigma^{2\nu+1}).$$
More precisely, we compute the expansions of the numerator $p(\sigma)=\sqrt{4-\sigma^2}-\sqrt{-\sigma^2}=\sum_{n=0}^{2\nu}p_n \sigma^n+O(\sigma^{2\nu+1})$ and the denominator 
$q(\sigma)=\sqrt{4-\sigma^2}+\sqrt{-\sigma^2}=\sum_{n=0}^{2\nu} q_n \sigma^n+O(\sigma^{2\nu+1})$ with $q_0\neq0$, use afterwards the formula
$$q'_n(\sigma)=\begin{cases}
	\frac{1}{q_0}&\text{if $n=0$}\\
	\frac{-1}{q_0}\sum_{j=1}^nq_j\,q'_{n-j}&\text{if $n\geq1$}
\end{cases}$$
for $q^{-1}(\sigma) = \sum_{n=0}^{2\nu} q'_n \sigma^n + O(\sigma^{2\nu+1})$, 
and combine finally the two expansions of $p(\sigma)$ and $q^{-1}(\sigma)$ to obtain the coefficients $g_n$ in the expansion of $g(\sigma)$ using the Cauchy product. 
This task can be accomplished in about 2 hours.

\medskip

\textit{Step 2.} To accomplish the second step, we first compute the expansions of
\begin{equation}\label{Equ:PowerExpand}
	\tilde{H}_n(\sigma)=(1-g(\sigma))^n, \quad \quad n=0,\dots,2 \, \nu;
\end{equation}
note that the constant term of $g(\sigma)$ is $1$, thus the constant term of $1-g(\sigma)$ is zero and therefore the compositions~\eqref{Equ:PowerExpand} can be carried out purely formally. 
The following speed-ups for these calculations are relevant. 
First, one can use a divide-and-conquer paradigm to compute $h(\sigma)^n$ for a truncated power series $h(\sigma)$ by applying the formula
$$\texttt{FastPower}(h(\sigma),n)=\begin{cases}
	1&\text{if $n=0$}\\ 
	\texttt{FastPower}(h(\sigma),\tfrac{n}{2})^2&\text{if $n\geq2$ is even}\\
	\texttt{FastPower}(h(\sigma),\tfrac{n-1}{2})^2*h(\sigma)&\text{if $n\geq1$ is odd}
\end{cases}$$
recursively; note that on the right-hand sides one or two polynomial multiplications have to be executed. 
Here one may use the standard formula of the Cauchy-product $P(\sigma)\cdot Q(\sigma)$. 
Alternatively, one can use the {\tt Mathematica} command \texttt{Expand} that works extremely efficiently by parallelization mechanisms. 
However, starting with two truncated power series $P(\sigma)$ and $Q(\sigma)$, i.e., polynomials with degrees $2\nu$, 
such a naive expansion leads to a  polynomial of degree $4\nu=32000$ where the last $\nu=16000$ coefficients are not needed 
(and are actually incorrect if one compares them with the underlying power series expansion). 
These unnecessary calculations in terms of very large rational numbers can be avoided by the following refinement which leads to the method\footnote{The method \texttt{FastExpand} is also utilized within the package \texttt{SolveCoupledSystem}~\cite{Blumlein:2019oas,SolveCoupledSystem} that implements the large moments method of power series expansions~\cite{Blumlein:2019oas}.}
\texttt{FastExpand}. 
Here one truncates the polymomials $P(\sigma)$ and $Q(\sigma)$ with degrees $2\nu$ to polymomials $P'(\sigma)$ and $Q'(\sigma)$ up to the degree $\nu$, 
and denotes the tails by $\tilde{P}(\sigma)$ and $\tilde{Q}(\sigma)$, i.e.,
$P(\sigma)=P'(\sigma)+\tilde{P}(\sigma)$ and $Q(\sigma)=Q'(\sigma)+\tilde{Q}(\sigma)$ . Then 
\begin{equation}
P(\sigma)\cdot Q(\sigma)=P'(\sigma)\cdot Q'(\sigma)+P'(\sigma)\cdot\tilde{Q}(\sigma)+\tilde{P}(\sigma)\cdot Q'(\sigma)+\tilde{P}(\sigma)\cdot\tilde{Q}(\sigma),
\end{equation}
and the essential observation is that one can drop the calculation of $\tilde{P}(\sigma)\cdot\tilde{Q}(\sigma)$ to get the coefficients correct up to degree $2\,\nu$. 
The expansion $P'(\sigma)\cdot Q'(\sigma)$ fully contributes to the desired output. 
However, the expansions in $P'(\sigma)\cdot\tilde{Q}(\sigma)$ and $\tilde{P}(\sigma)\cdot Q'(\sigma)$ can be refined further by applying the proposed method on these subterms recursively. 
In short, one only expands those expressions which actually contribute to the final result. 
If an intermediate expression turns out to be small enough (base case), one simply expands normally and truncates the unwanted terms having degrees larger than $2\nu$.

Using \texttt{FastPower} in combination with \texttt{FastExpand} we proceed as follows. 
We compute 80 supporting points equally distributed within the range $1\leq n\leq2\,\nu$, i.e., $n_i=200\,i$ with $0\leq i\leq 79$. 
E.g., computing the expansion of~\eqref{Equ:PowerExpand} for $n=1000, 8000,15000$ it takes 1131s,  1525s, 1297s, respectively. 
Note that for larger $n$, the lower bounds in the expansions ~\eqref{Equ:PowerExpand} increase stepwise (since one can pull out $x$ from $1-g(\sigma)$). 
Hence the execution of \texttt{FastExpand} gets simpler, but on the other side, the recursion steps in \texttt{FastPower} increase when $n$ gets larger. 
Computing these 80 points in parallel with 10 kernels required around 5 hours. 
Finally, we compute for each segment the missing points $n_i\leq n<n_{i+1}$ by starting with the 
already computed expansion $\tilde{H}_{n_i}=(1-g(\sigma))^{n_i}$ 
and incrementally computing the corresponding expansions 
$\tilde{H}_{n_i+j+1}=\tilde{H}_{n,i+j}\,(1-g(\sigma))$ by using our \texttt{FastExpand} method. 
Doing this calculation with 15 kernels in parallel we needed 2 days and 5 hours to get all expansions~\eqref{Equ:PowerExpand} for $0\leq n\leq 2\,\nu$. 
In principle 15 supporting points $n_i$ would suffice for 15 kernels. 
However, in order to obtain a fair distribution to all available kernels, the generation of more jobs turned out to be appropriate.

Given the expansions in~\eqref{Equ:PowerExpand}, one could carry out the calculation
\begin{equation}\label{Equ:FinalStepNaive}
	\hat{F}(\sigma)=
f\Big(\tfrac{\sqrt{4-\sigma^2}-\sqrt{-\sigma^2}}{\sqrt{4-\sigma^2}+\sqrt{-\sigma^2}}\Big)=\sum_{n=0}^{2\,\nu}f_n\,\tilde{H}_n(\sigma)+O(\sigma^{2\nu+1})=\sum_{k=0}^{2\nu}\hat{h}_n 
\sigma^n \end{equation}
for each of the 18 cases.  But also this last step slows down considerably due to non-trivial rational number arithmetic. In order to streamline and speed up these last calculations, 
one can rearrange the calculation in~\eqref{Equ:FinalStepNaive} and obtain a matrix $A\in\QQ^{(2\nu+1)\times(2\nu+1)}$ with $(2\nu+1)^2$ rational number entries such that 
\begin{equation}\label{Equ:BasisTransformX2S}
	\left(\begin{matrix}h_0\\ \vdots\\ h_{2\nu}\end{matrix}\right)=A\left(\begin{matrix}f_0\\ \vdots\\ f_{2\nu}\end{matrix}\right)
\end{equation}
holds. Notice that the matrix $A$ with $\nu=8000$ requires about 22 Gb of memory within {\tt Mathematica}. 
Given this compact transformation from the $x$-expansion to the $s$-expansion, 
we parallelized the matrix operation in~\eqref{Equ:BasisTransformX2S}
by carrying out the row multiplications of $A$ with $(f_0,\dots,
f_{2\nu})$ using 16 {\tt Mathematica} kernels. 
As mentioned already above, in the output $(h_0,\dots, h_{2\nu})$ all entries $h_{2n+1}$ with $0\leq n\leq \nu-1$ are zero, 
and one obtains the desired coefficients $\hat{h}_n=h_{2n}$ with $0\leq n\leq\nu$ for~\eqref{Equ:SExpansion}.
The calculation time of this matrix multiplication in its parallelized implementation ranged between 2000s and 5000s for the given 18 problems. 
In total one needs 12 hours to perform this last calculation step. 

Summarizing, using all these techniques for steps~1 and~2, we needed around 3 days to obtain $\nu=8000$ coefficients in~\eqref{Equ:SExpansion} for all 18 cases.

\medskip

\textit{Step 3.} Given this data, one can now proceed to guess the 18 linear differential equations for the $\hat{F}(\hs)$ 
and the corresponding linear recurrences for their coefficients $\hat{h}(n)$ using the package \texttt{ore\_algebra}~\cite{SageOre} in {\tt Sage}.  
E.g., for $f(x)=F_{s,1}^{(0)}(\hs)$ one obtains a linear differential equation of order $42$ and maximal degree $622$ for the coefficients in $40$ minutes,
and a linear recurrence for the power series coefficients of order $33$ and maximal degree $569$ in $17$ minutes. 
We can then utilize the obtained recurrences up to order $36$ to produce the required coefficients 
$\hat{h}_{n}$ with $0\leq n\leq 100000$.

\medskip

\textit{Remark.} An alternative approach is to compute directly from the given differential equation for~\eqref{Equ:XExpansion} 
a differential equation for~\eqref{Equ:SExpansion} using holonomic closure properties~\cite{HOLONOMIC}. 
E.g., consider the linear differential equation for $f(x)=F_{s,1}^{(0)}(x)$ of order $43$ where the maximal degree of the coefficients is $2315$. 
Then applying the package \texttt{HolonomicFunctions}~\cite{Koutschan:13} to this equation yields in $2$ weeks a linear differential equation for~\eqref{Equ:SExpansion}; 
its order is $44$ and the maximal degree of the coefficients is $645$, which is of similar size to the equation that we obtained with the method described above (see details in Step~3). 
However, we need in addition a linear recurrence to extend the number of coefficients in~\eqref{Equ:SExpansion} from $8.000$ to 100.000. 
Thus we extract a linear recurrence of~\eqref{Equ:SExpansion} for the coefficients $\hat{h}_n$ from the derived linear differential equation 
(using again the holonomic closure properties~\cite{HOLONOMIC}). 
Performing this calculation one gets a recurrence of order $611$ where the coefficients have maximal degree $44$ 
(as expected by bounds in terms of the given order and degree of the input differential equation). 
The challenge is now to produce the first 100.000 coefficients of $\hat{h}_n$ with this monster recurrence. 
It seems rather challenging to produce the required $611$ initial values using, e.g., the plain Mathematica command \texttt{Series}, 
and one may opt to reuse at least in parts some of the tools described above. 
Using then this recurrence and the necessary initial values, one may produce 8000 coefficients and then proceed with our original approach: 
guess the minimal recurrence and produce the desired coefficients using this optimal recurrence. 
Summarizing, this approach seems to be far more time consuming and somehow throws one back to our original approach. 
One should mention that this second approach has to be applied step-wise to all 18 cases. 
Conversely, having the basis transformation~\eqref{Equ:BasisTransformX2S}, one can deal with all the cases (and problems that may arise in future calculations) straightforwardly within hours.
\medskip

\subsection{The two-particle threshold}
\label{Sec:s4-expansions}

\vspace*{1mm}
\noindent
Let us consider the expansions around $\hs=4$ now. Here the relation
\begin{equation}
z^2 = \frac{(1+x)^2}{x}
\end{equation}
holds, with  $x$ defined in (\ref{eq:1}).
The indicial equation associated to the guessed differential equations allow us to determine the value of $k_{\rm min}$ in Eq. (\ref{eq:s4ansatz}). 
In most cases $k_{\rm min} = -6$, with a few exceptions where $k_{\rm min} = -4$. Also, as in the previous section, the differential
equations confirm that the highest power of the log is 3, as written in Eq. (\ref{eq:s4ansatz}), since the insertion of an ansatz with
higher powers of the log in (\ref{diffeqinz1}) leads to equations showing that the corresponding coefficients vanish.
Unsurprisingly, all differential equations guessed from the expansions around $\hs=0$ were of the same order as the ones guessed 
from the expansions around $x=1$. However, the degrees of the polynomials in the differential equations were slightly smaller in $\hs$ than in $x$. 
Interestingly, unlike what happened in the previous section,
in the cases of $F_{I,1}^{(0)}(z)$ and $F_{I,2}^{(0)}(z)$, 
it was not necessary to remove inhomogeneous parts  arising from the fact that the first few coefficients in the expansions 
in $\hs$ must be ignored when guessing the differential equations. 
This had the welcomed effect of producing differential equations in $z$ that were simpler than they would be otherwise.

As in the previous section, 
the insertion of the ansatz (\ref{eq:s4ansatz}) in the differential equations (\ref{diffeqinz1}) leads to relations among the coefficients $\tc_{I,i}(j,k)$,
leaving only a few of them (as many as the order of the differential equation) to be determined by initial conditions. 
More precisely, these base coefficients are numerically determined by matching truncated versions of the expansions 
around $\hs=0$ and $\hs=4$ as well as (possibly) their derivatives at some intermediate points. We used 100.000 terms for the expansions at $\hs=0$,
and 3000 coefficients per power of the log in Eq. (\ref{eq:s4ansatz}), and matched at a set of equally spaced points,
\begin{equation}
\hs_0 +(k-1) \delta, \quad k=1,\dots, o_{I,i},
\label{eq:mps0}
\end{equation}
without using derivatives and taking $\delta=1/20.000$. In a first attempt, we used the middle point $\hs_0=2$, 
which allowed us to determine the simplest coefficients in terms of known constants using
the {\tt PSLQ} algorithm. For example, in the pseudoscalar case we found
\begin{eqnarray}
\tc_{p,1}(2,-2) &=& \frac{2048}{3} \zeta_2, \label{eq:PSLQ1} 
\\
\tc_{p,2}(1,-6) &=& \frac{16384}{3}, \label{eq:PSLQ2}
\\ 
\tc_{p,3}(0,-6) &=& -\frac{28672}{3}, \label{eq:PSLQ3}
\end{eqnarray}
and similar for other form factors. Simple coefficients $\tc_{I,i}(j,k)$ like these typically appear for higher values of $j$ and/or lower values of $k$. 
In every case, such simple coefficients, determined numerically using the settings mentioned above, 
differ from their exact values by $10^{-300}$ to $10^{-400}$. Such a precision is, of course, enough to determine a simple rational number 
(or a simple rational number times a simple constant such as $\zeta_2$) through the {\tt PSLQ} algorithm, but 
it is insufficient for more complicated coefficients involving
many known constants multiplied by increasingly larger rational numbers.
Therefore, we had to increase the numerical precision of our coefficients, which we did in the same way as in the case of the expansions around $x=0$, that is, using the
coefficients we have already determined, such as the ones in Eqs. (\ref{eq:PSLQ1}-\ref{eq:PSLQ3}), as an anchor to search for different values of $\delta$ 
and $\hs_0$ such that the precision of the coefficients is as high as possible. 
We found that by maintaining the same value of $\delta$ and choosing $\hs_0 = 381/100$, we could determine the base coefficients with a precision of around 1850 to 1950
decimal places. Notice that this value of $\hs_0$ is very close $\hs=4$, which most certainly has to do with the fact that we are using many more coefficients for
the expansions around $\hs=0$ than for the ones around $\hs=4$, and we are therefore better able to approximate the function away from its expansion point in the former
case than in the latter. In principle, we could also increase the precision of our coefficients by increasing the number of expansion terms around $\hs=4$, which should
shift the value of $\hs_0$ back to the middle, but this can be computationally very expensive, and it is also unnecessary, since 1800 decimal places for the coefficients turn 
out to be sufficient for our purposes. This level of precision allowed us to determine many of the remaining 
and more complicated coefficients using the {\tt PSLQ} algorithm.
In particular, we found that all coefficients $\tc_{I,i}(j,k)$ with $j > 1$ or $j=1$ and $k<-1$, can be written in terms of known constants. For example,
\begin{eqnarray}
\tc_{v,1,3}(2,-1) &=& \frac{4445056}{81 \pi}-\frac{50176}{\pi} l_2, 
\\
\tc_{v,1,2}(1,-2) &=& -\frac{3627008}{27}-\frac{26368}{3} \pi^2+4096 l_2, 
\\
\tc_{a,1,1}(3,1) &=& \frac{22400}{9 \pi} \zeta_3-\frac{173824}{243} \pi, 
\\
\tc_{a,2,1}(2,-1) &=& \frac{603008}{81 \pi} \zeta_3-\frac{50176}{3 \pi} \zeta_3 l_2+\frac{17536}{9} \pi+\frac{42880}{81} \pi^3, 
\\
\tc_{s,1}(2,3) &=& -\frac{118790}{81 \pi} \zeta_3-\frac{1064}{\pi} \zeta_3 l_2
-\frac{21674}{243} \pi+\frac{1016}{27} \pi^3+256 \pi l_2, 
\\
\tc_{p,1}(1,-2) &=& -\frac{7168}{9} \zeta_3+\frac{833536}{243} \pi^2+\frac{896}{3} \pi^4.
\end{eqnarray}

However, for $j \leq 1$, we were not able to express all base coefficients in terms of known constants, 
as we did in the case of the expansion around $x=0$. In particular, for $F^{(0)}_{I,i}(x)$ with $i=1,3$ the 
first coefficients for which this happened were the $\tc_{I,i}(1,-1)$'s, while for $F^{(0)}_{I,2}(x)$ only the 
$\tc_{I,i}(0,-1)$'s were left undetermined. Of course, it is quite possible that we are missing some known 
constants in our {\tt PSLQ} search that would allow to reveal their relation with the so far undetermined coefficients, 
but we have tried many different constants, including different (inverse) powers of $\pi$, 
different multiple zeta values and their products, different (powers of) logarithms and polylogarithms evaluated at different values, square roots of different integers and their
products with the constants already mentioned, as well as constants such as Clausen functions \cite{CLAUSEN,LEWIN1}
evaluated at $r \pi$, where $r$ is a rational 
number.\footnote{These constants typically appear in calculations of Feynman graphs of elliptic nature, such as the banana and kite graphs, 
which are among the master integrals required for the calculation of three-loop form factors.} All of these searches were unsuccessful. 
However, if we include more than one coefficient in a {\tt PSLQ} search, we can find relations among them. For example,
\begin{eqnarray}
  \tc_{a,1,3}(1,-1) &=& -\frac{31}{8} \tc_{v,2,3}(1,-1)-\frac{23128}{9} \pi-\frac{3644704}{81 \pi}-\frac{7168}{3 \pi} l_2^2-\frac{519232}{81 \pi} l_2,
  \\
  \tc_{a,2,3}(1,-1) &=& \frac{1}{8} \tc_{v,2,3}(1,-1)-\frac{45464}{9} \pi+\frac{747488}{81 \pi}-\frac{50176}{3 \pi} l_2^2+\frac{1226176}{81 \pi} l_2,
  \\
  \tc_{p,3}(1,-1) &=& - \tc_{a,2,3}(1,-1),
  \\
  \tc_{s,3}(1,-1) &=& \frac{3}{2} \tc_{v,2,3}(1,-1)-\frac{11360}{9} \pi+\frac{2377088}{81 \pi}-\frac{7168}{\pi} l_2^2+\frac{34048}{27 \pi} l_2,
  \\
\tc_{v,1,3}(1,-1) &=& -\frac{59}{8} \tc_{v,2,3}(1,-1)+\frac{137032}{9} \pi-\frac{798112}{81 \pi}+\frac{50176}{\pi} l_2^2 \nonumber \\ &&
-\frac{10079552}{81 \pi} l_2.
\end{eqnarray}

We can also find relations involving more than two coefficients. For example,
\begin{eqnarray}
\tc_{v,2,3}(0,-1) &=& \frac{2}{3} \tc_{s,3}(0,-1)-\frac{17}{12} \tc_{v,2,3}(1,-1)
+\frac{10752}{\pi} a_4+\frac{98224}{27 \pi} \zeta_3 \nonumber \\ &&
-\frac{3602}{45} \pi^3-\frac{326464}{243} \pi-\frac{4332160}{243 \pi}+\frac{448}{\pi} l_2^4
-\frac{14336}{9 \pi} l_2^3 \nonumber \\ &&
                                           +1472 \pi l_2^2+\frac{34048}{81 \pi} l_2^2-\frac{128512}{27} \pi l_2+\frac{4068736}{243 \pi} l_2, 
  \\
\tc_{a,1,3}(0,-1) &=& -\frac{31}{12} \tc_{s,3}(0,-1)+\frac{25}{96} \tc_{v,2,3}(1,-1)
-\frac{10752}{\pi} a_4-\frac{150724}{9 \pi} \zeta_3 \nonumber \\ &&
+\frac{13529}{90} \pi^3+\frac{3272722}{243} \pi-\frac{35407064}{729 \pi}-\frac{448}{\pi} l_2^4+\frac{62720}{9 \pi} l_2^3 \nonumber \\ &&
                                                                                                                                         -1472 \pi l_2^2+\frac{42560}{27 \pi} l_2^2-\frac{56752}{27} \pi l_2-\frac{7362320}{243 \pi} l_2.
                                                                                                                                         \end{eqnarray}

This suggests the idea of choosing some of these coefficients to define new constants and express all other coefficients in terms of them. 
We tried to reduce the number of new constants needed by finding as many relations among the base coefficients as possible. 
We ended up with 14 new constants, $\tk_i$, with $i=1,\dots,14$, defined as
\begin{alignat}{4}
\tk_1 &= \tc_{v,2,3}(1,-1), \quad & \tk_2    &= \tc_{v,2,2}(0,-1), \quad & \tk_3   &= \tc_{s,3}(0,-1), \quad           & \tk_4 &= \tc_{a,2,3}(0,0), \nonumber \\ 
\tk_5 &= \tc_{v,1,3}(0,0), \quad & \tk_6     &= \tc_{a,1,3}(0,2), \quad & \tk_7    &= \tc_{v,1,3}(0,2), \quad    & \tk_8 &= \tc_{s,3}(0,2), \nonumber \\
\tk_9 &= \tc_{v,2,1}(0,-1), \quad & \tk_{10} &= \tc_{a,1,1}(0,0), \quad & \tk_{11} &= \tc_{v,1,1}(0,0), \quad & \tk_{12} &= \tc_{a,1,1}(0,2), \nonumber \\
\tk_{13} &= \tc_{v,1,1}(0,2), \quad &\tk_{14} &= \tc_{s,1}(0,2). &&&&
                                                                      \label{eq:14consts}
\end{alignat}
In Appendix~\ref{app:kappa}, we show the numerical values of these constants with 60 digits. In an ancillary file, we provide these constants with 1800 digits. Of 
course, other choices of coefficients are also possible. 
One may even choose linear combinations of coefficients or even linear combinations of coefficients together with other known constants, 
but for the sake of simplicity, we chose the new constants to be directly equal to some of the $\tc_{I,i}(j,k)$'s with values of $k$ as low as possible. 
Notice also that we chose coefficients from different types of form factors, that is, with different values of $I$, 
since this choice was simple enough and there is no particular reason to stick to a single value of\footnote{We have not checked whether such choice is even possible.} $I$.

After making the choice given in Eqs. (\ref{eq:14consts}) we were able to express all coefficients as linear combinations of the following constants over rational numbers,
\begin{eqnarray}
&& \biggl\{
\tk_i, \pi^k, \pi^k l_2, \pi^j l_2^2, \pi^l l_2^3, \pi^n l_2^4, \pi l_2^5, \pi^n a_4, \pi l_2 a_4,
\pi^j \zeta_3, \pi^n l_2 \zeta_3, \nonumber \\ && \phantom{\biggl\{}
\frac{l_2^2 \zeta_3}{\pi}, \pi l_2^2 \zeta_3,
                                                  \frac{l_2^3 \zeta_3}{\pi}, \frac{\zeta_3^2}{\pi}, \pi \zeta_5, \tk_1 \zeta_3, \pi^2 \tk_2, \tk_3 \zeta_3\biggr\},
\label{eq:s4consts}
\end{eqnarray}
where $i=1,\ldots,14$, $j=-1,\ldots,3$, $k=-1,\ldots,5$, $l=-1,1,3$ and $n=-1,0,1$. Notice the presence in the list above of some products of the new constants with known constants,
such as $\tk_1 \zeta_3$ and $\pi^2 \tk_2$. 
Notice also the presence of odd powers of $\pi$ in some of these constants. These are expected since they appear even in the expansions of solvable parts. 
Here are a few other examples of coefficients given in terms of the constants in (\ref{eq:s4consts}),
\begin{eqnarray} 
\tc_{a,2,1}(0,-5) &=& \frac{8}{3} \tk_1 \zeta_3
-\frac{381440}{243} \pi \zeta_3
+\frac{13432832}{729 \pi} \zeta_3
+\frac{1347584}{81 \pi} \zeta_3 l_2 \nonumber \\ &&
+\frac{4096}{243} \pi^4
-\frac{75136}{243} \pi^3
                                                    +\frac{14336}{81} \pi^3 l_2,  \\
\label{eq:s4rel1}
\tc_{p,2}(0,1) &=& \frac{\tk_2}{64}
-15 \pi \zeta_3
-\frac{92327}{162} \pi^3
+\frac{128}{9} \pi^2
-\frac{35974507}{1944} \pi
+\frac{8}{3} \pi l_2^3 \nonumber \\ &&
-\frac{4222}{27} \pi l_2^2
+\frac{22}{3} \pi^3 l_2
                                       +\frac{776783}{81} \pi l_2,  \\
\label{eq:s4rel2}
\tc_{s,1}(0,-1) &=& -\frac{\pi^2}{4} \tk_2-\tk_3 \zeta_3
+\frac{1024}{3} \pi a_4
+80 \pi \zeta_5
-64 \pi^3 \zeta_3
+\frac{9472}{9} \pi \zeta_3 \nonumber \\ &&
-\frac{276928}{1215} \pi^5
-\frac{1088}{81} \pi^4
-\frac{312302}{729} \pi^3
-\frac{52640}{27} \pi
+\frac{128}{9} \pi l_2^4 \nonumber \\ &&
-\frac{128}{3} \pi^3 l_2^3
-\frac{12064}{27} \pi^3 l_2^2
-\frac{352}{3} \pi^5 l_2
+\frac{1194896}{243} \pi^3 l_2,\\
\label{eq:s4rel3}
\tc_{v,1,1}(0,-1) &=& \frac{185}{144} \pi^2 \tk_2 
+ \frac{\tk_9}{3} 
+ \frac{50176}{81} \pi a_4
-\frac{4096}{3} \pi l_2  a_4 
+ \frac{905}{288} \tk_1 \zeta_3 \nonumber \\ &&
+\frac{185}{36} \tk_3 \zeta_3 
+ 240 \pi \zeta_5 
- \frac{4025756}{81 \pi} \zeta_3^2 
+ \frac{33097}{54} \pi^3 \zeta_3 
+ \frac{7207298}{729} \pi  \zeta_3 \nonumber \\ &&
-\frac{72875656}{729 \pi } \zeta_3
+\frac{120064}{27 \pi } \zeta_3 l_2^3
-1152 \pi \zeta_3 l_2^2 
-\frac{722992}{81} \pi \zeta_3 l_2    \nonumber \\ &&
-\frac{14331968}{243 \pi } \zeta_3 l_2^2
+\frac{107318512}{729 \pi } \zeta_3 l_2 
+\frac{2668418}{3645} \pi^5
+\frac{800}{9} \pi^4   \nonumber \\ &&
-\frac{11892677}{1458} \pi^3
+\frac{420704}{405} \pi
-\frac{512}{9} \pi  l_2^5 
+\frac{6272}{243} \pi  l_2^4
+\frac{800}{3} \pi^3 l_2^3   \nonumber \\ &&
+\frac{294280}{243} \pi^3 l_2^2
+\frac{16408}{27} \pi^5 l_2
-\frac{1607788}{81} \pi^3 l_2, \\
\label{eq:s4rel4}
\tc_{v,2,3}(0,4) &=& -\frac{15531726824501}{297574847332017} \tk_4
+\frac{2425509036937}{793532926218712} \tk_5 \nonumber \\ &&
+\frac{50909627017694}{495958078886695} \tk_6
-\frac{37467581053953}{1983832315546780} \tk_7 \nonumber \\ &&
-\frac{8420376537572}{99191615777339} \tk_8
+\frac{343394686678188704}{40172604389822295} \pi^2 \nonumber \\ &&
-\frac{17742395132986623488}{66954340649703825}
-\frac{180289931782198336}{4463622709980255} l_2.
\label{eq:s4rel5}
\end{eqnarray}

The last relation, Eq. (\ref{eq:s4rel5}), is particularly curious. The rational numbers multiplying the constants have a similarly large number of digits in the numerator and denominator, 
which might make one suspect that the relation is spurious. 
However, this particular relation can be obtained using the {\tt PSLQ} algorithm with just 300 decimal 
places out of the more than 1800 available for the $\tk_i$'s.
The fact that the relation holds also when the full available precision is used gives us confidence that the relation is correct. Moreover, this relation is part of a
pattern that is followed by the $\tc_{I,i}(0,k)$'s starting from $k \geq -1$, namely, 
for $k$ even and $k \geq 2$, the base coefficients $\tc_{I,3}(0,k)$'s can be expressed as a linear combination of the constants
\begin{eqnarray}
\bigl\{\tk_4, \tk_5, \tk_6, \tk_7, \tk_8, \pi^2, l_2\bigr\},
\end{eqnarray}
for all values of $I$, while in the case of the $\tc_{I,1}(0,k)$'s, the following constants appear
\begin{eqnarray}
\bigl\{\tk_{10}, \tk_{11}, \tk_{12}, \tk_{13}, \tk_{14}, \pi^2, \pi^4, \pi^2 l_2, \pi^4 l_2, \pi^2 l_2^2, l_2^4, 
a_4, \zeta_3, \pi^2 \zeta_3, l_2 \zeta_3 \bigr\}.
\end{eqnarray}
There are also patterns when $k$ is odd. We find that the $\tc_{I,3}(0,k)$'s can be written in terms of
\begin{eqnarray}
\biggl\{\tk_1, \tk_3, \frac{1}{\pi}, \pi, \pi^3, \frac{l_2}{\pi}, \pi l_2,  \frac{l_2^2}{\pi}, \frac{l_2^3}{\pi}, \frac{\zeta_3}{\pi} \biggr\},
\end{eqnarray}
with $k \geq -1$ and $I \in \left\{(a,2), s, p\right\}$, while for $I \in \{(v,1), (v,2), (a,1)\}$ we have to include $\pi l_2^2$, $\frac{l_2^4}{\pi}$ and
$\frac{a_4}{\pi}$ in the previous list. In the case of the $\tc_{I,1}(0,k)$'s (with $k$ odd and $k \geq -1$) the list is
\begin{eqnarray}
&& 
\biggl\{\pi^2 \tk_2, \tk_1 \zeta_3, \tk_3 \zeta_3, \pi, \pi^3, \pi^4, \pi^5, \pi l_2,  \pi^3 l_2, \pi^5 l_2, \pi l_2^2, \pi^3 l_2^2, \pi l_2^3, \pi^3 l_2^3, \pi l_2^4, \pi a_4,
\nonumber \\ && \phantom{\biggl\{}
\frac{\zeta_3}{\pi}, \pi \zeta_3, \pi^3 \zeta_3, \frac{l_2 \zeta_3}{\pi}, \pi l_2 \zeta_3, \frac{l_2^2 \zeta_3}{\pi}, \frac{l_2^3 \zeta_3}{\pi}, \frac{\zeta_3^2}{\pi}, \pi \zeta_5 \biggr\},
\end{eqnarray}
for $I \in \left\{(a,2), s, p \right\}$, while for $I = (a,1)$ we have to add $\tk_9$ to the previous list, and for $I \in \left\{(v,1), (v,2)\right\}$, 
we also have to add $\pi l_2^5$, $\pi l_2 a_4$ and $\pi l_2^2 \zeta_3$. In all cases, the numerators and denominators of the rational numbers in the corresponding relations keep
increasing as we increase $k$. For example, the last two base coefficients in the scalar case $F^{(0)}_{s,1}$ are
\begin{eqnarray}
\tc_{s,1}(0,19) &=&
-\frac{5027881 \pi^2 \tk_2}{281474976710656}
+\frac{100060070143 \tk_1 \zeta_3}{354658470655426560} 
-\frac{5027881 \tk_3 \zeta_3}{70368744177664} \nonumber \\ &&
-\frac{27599 \pi a_4}{7247757312}
+\frac{230945 \pi \zeta_5}{17179869184}
+\frac{111888465185 \zeta_3^2}{118747255799808 \pi } \nonumber \\ &&
+\frac{67923409 \pi^3 \zeta_3}{35184372088832}
-\frac{59887782087324315547 \pi \zeta_3}{11163712136208753623040} \nonumber \\ &&
-\frac{12860526400251092894224460634578393 \zeta_3}{1828693524944337544832635203747840000 \pi}
-\frac{286229923 \zeta_3 l_2^3}{618475290624 \pi} \nonumber \\ &&
-\frac{173097268545048371 \zeta_3 l_2^2}{107982334650828718080 \pi}
+\frac{14447921549 \pi \zeta_3 l_2}{29686813949952} \nonumber \\ &&
+\frac{1663467490977652027449322769 \zeta_3 l_2}{62843710295357808508403712000 \pi}
+\frac{125362825561 \pi^5}{10687253021982720} \nonumber \\ &&
+\frac{46211464633 \pi^4}{56108078365409280}
-\frac{8372674776503657147978188451 \pi^3}{21115486659240223658823647232000} \nonumber \\ &&
+\frac{241809871326977039832281811355045074601 \pi}{70307069191793805111236142681607372800000}
-\frac{27599 \pi l_2^4}{173946175488} \nonumber \\ &&
-\frac{5027881 \pi^3 l_2^3}{1649267441664}
-\frac{185075295793 \pi l_2^3}{1098453111275520}
+\frac{4205606557 \pi^3 l_2^2}{178120883699712} \nonumber \\ &&
+\frac{5077431449402204323 \pi l_2^2}{19178378386034487459840}
-\frac{1867527173617354849 \pi^3 l_2}{36282064442678449274880} \nonumber \\ &&
-\frac{55306691 \pi^5 l_2}{6597069766656}
-\frac{236960916947338664853480017 \pi l_2}{139518243785426143252001587200}
\label{eq:cS1019}
\end{eqnarray}
and
\begin{eqnarray}
\tc_{s,1}(0,20) &=&
\frac{14952584516495142844794211005121159757399663}{29587498603927418202835305700260447979465848913920} \tk_{10} \nonumber \\ &&
-\frac{878453210567550928560288633987539397958571}{29587498603927418202835305700260447979465848913920} \tk_{11} \nonumber \\ &&
-\frac{11826878664137011934199273995883446541836167}{8453571029693548057952944485788699422704528261120} \tk_{12} \nonumber \\ &&
+\frac{1833853168547684958366205884106413005446099}{2817857009897849352650981495262899807568176087040} \tk_{13} \nonumber \\ &&
+\frac{39203399123485661004032269394602606971865}{44829543339283966973992887424637042393130074112} \tk_{14} \nonumber \\ &&
+\frac{2654371105794647318748259634400769208778218233}{4680834740074454832870429222111516184251433128960} a_4 \nonumber \\ &&
+\frac{377843701667024690186361039372089880848543401}{480085614366610752089274792011437557359121346560} \pi^2 \zeta_3 \nonumber \\ &&
-\frac{49806683454592960628621396049647345829942814397941}{1283886100134707611301603443779158724823250229657600} \zeta_3 \nonumber \\ &&
+\frac{96199190609477140229616469434841211096617849}{68583659195230107441324970287348222479874478080} \zeta_3 l_2 \nonumber \\ &&
+\frac{10408902722784322486445824075528059119015945087}{73265239409861032166667587824354166362196344627200} \pi^4 \nonumber \\ &&
+\frac{17366604873939319118314280454799960975921805411478493}{7549250268792080754453428249421453301960711350386688000} \pi^2 \nonumber \\ &&
+\frac{50974594014764148007869743942673992785127104647414623}{11795703544987626178833481639721020784313611484979200000} \nonumber \\ &&
+\frac{2654371105794647318748259634400769208778218233}{112340033761786915988890301330676388422034395095040} l_2^4 \nonumber \\ &&
-\frac{3725170028034232118514379268734358925597721567}{28085008440446728997222575332669097105508598773760} \pi^2 l_2^2 \nonumber \\ &&
+\frac{11426312723869372738005308114576710177621031}{174576587042403909850645378913250020857862307840} \pi^4 l_2 \nonumber \\ &&
+\frac{61643923761558393014266634711489371261054656913}{8009984581945591157853140914843236251125447065600} \pi^2 l_2.
\label{eq:cS1020}
\end{eqnarray}
In Eq. (\ref{eq:cS1019}), the largest numerator or denominator has 41 digits, while the smallest one has only seven digits. 
Eq. (\ref{eq:cS1020}) looks a bit more cumbersome, and it is also a bit more uniform, with the largest numerator or denominator having 56 digits, and the smallest having 41.
To get these relations, one needs a minimum of 1300 decimal places of precision in the case Eq. (\ref{eq:cS1019}), and at least
1100 decimal places in the case of Eq. (\ref{eq:cS1020}). These are the precisions required once we know exactly which constants
are involved and then limit ourselves to such constants when applying the {\tt PSLQ} algorithm, but of 
course, one does not usually know a priori which constants
are going to be present, and therefore one needs much higher precision in order to be able to include more constants in the search. We did searches
including all 1800 decimal places available for our base coefficients, and performed various checks of our results, including the introduction
of spurious constants, like we did in the case of the expansion at $x=0$, to make sure that the {\tt PSLQ} 
algorithm does in fact rule
out such constants from the result, which is unlikely to happen unless
the relation is not spurious. 

As we did in the case of the expansions at $x=0$, we can now obtain the form factors using Eq. (\ref{FFdecomp}), 
together with Eqs. (\ref{gidecomp2a}--\ref{gidecomp2d}) and Eqs. (\ref{Sdecomp}-\ref{Pdecomp}) 
rewritten in terms of $z$ instead of $x$ (using $z^2=(1+x)^2/x$). Since the expansions contain the unknown
constants $\tk_i$ we refrain from listing them here. The first 200 terms of the threshold
expansion of all form factors are given in the ancillary files. 
After this we can find recursions obeyed by the sequences of rational numbers associated to
the each combination of a given constant (including the known constants and the new ones, $\tk_i$) and a power of the log. For example, the sequence 
of rational number associated to the term $\pi \ln^2(z)$ in the non-solvable parts (at the expansion point $x=1$) of $F_{V,1}^{(0)}(z)$ are given by 
\begin{eqnarray}
&& \biggl\{0,0,0,\frac{27008}{27},0,\frac{2752}{81},0,-\frac{339913}{1215},0,-\frac{27836749}{170100},0,-\frac{2883493363}{76204800},0, \nonumber
\\ && \phantom{\biggl\{}
-\frac{33749896783}{3772137600},0, -\frac{32996303383481}{15343379251200},0,-\frac{749728845867049}{1436140297912320},0,\ldots\biggr\}.
\end{eqnarray}
In other words, the $n$th term in the sequence above is the rational number multiplying $x^{-7+n} \pi \ln^2(z)$ in the expansion, starting from $n=1$. 
As usual, the first few terms must be dropped in order to find a recursion. Notice that all terms with $n$ even are equal to zero. This is a
consequence of the patterns described above and requires us to find a sequence by considering the terms with $n$ odd only. We did so,
and found a recurrence of order five,
\begin{equation}
\sum_{i=0}^5 \tilde{p}_i(n) f_3(n+i) = 0,
\end{equation}
where $f_3(n)$ is the $n$th term of the sequence given above after dropping the first seven terms and all the zeroes, and the $\tilde{p}_i(n)$'s are polynomials in $n$. 
We found the following solution
\begin{eqnarray}
\lefteqn{f_3(n) =} \nonumber\\ && 
- \frac{2^{1-4 n}  T_3}{27 (1+n) (2+n) (3+n) (1+2 n)^2 (3+2 n)^2 (5+2 n)} \binom{2 
n}{n}
-\frac{\big(
        349+1164 n+612 n^2\big) 2^{5-2 n}}{27 (1+2 n) (3+2 n)}
\nonumber\\ &&
-\frac{\big(
        41967+66774 n+32884 n^2+5768 n^3\big) 2^{1-4 n} \binom{2 n}{n}}{27 (1+n) (2+n) (3+n)}
S_{\{2,1,2\}}(n)
-\frac{1}{27} 2^{8-2 n} 
\sum_{i_1=1}^n \frac{2^{-2 i_1} \binom{2 i_1}{i_1}}{\big(
        1+2 i_1\big)^2}
\nonumber\\ &&
+\Biggl[
        -\frac{5}{27} 2^{6-2 n}
        -\frac{3 \big(
                -7+4 n+4 n^2\big) 2^{5-4 n} \binom{2 n}{n}}{(1+2 n) (3+2 n)}
        -\frac{5}{27} 2^{6-2 n} 
        \sum_{i_1=1}^n \frac{2^{-2 i_1} \binom{2 i_1}{i_1}}{1+2 i_1}
\Biggr] S_{\{2,1,1\}}(n)
\nonumber\\ &&
-\frac{5 \big(
        65+220 n+116 n^2\big) 2^{5-2 n} 
\sum_{i_1=1}^n \frac{2^{-2 i_1} \binom{2 i_1}{i_1}}{1+2 i_1}}{27 (1+2 n) (3+2 n)}
-\frac{71}{27} 2^{5-2 n} 
\sum_{i_1=1}^n \frac{2^{-2 i_1} \binom{2 i_1}{i_1} 
S_{\{2,1,1\}}(i_1)}{1+2 i_1},
\end{eqnarray}
with
\begin{eqnarray}
T_3 &=& 434624 n^8+5219648 n^7+26709296 n^6+75033264 n^5+124288324 n^4+121056044 n^3
\nonumber\\ &&
+64493993 
n^2+15127189 n+517179 \,.
\end{eqnarray}
Here $S_{\{2,1,1\}}(n)$ and $S_{\{2,1,2\}}(n)$ are cyclotomic harmonic sums~\cite{Ablinger:2011te}, which are defined as
\begin{equation}
S_{\{a,b,c\}}(n) = \sum_{k=1}^n \frac{({\rm sign}(c))^k}{(a k+b)^{|c|}}.
\end{equation}
Resumming these expressions with the help of the  \texttt{HarmonicSums} command {\tt Compute
GeneratingFunction} one obtains expressions depending on iterative integrals of the kind
\begin{equation}
G\left(\{f_1(\tau), f_2(\tau), \ldots, f_n(\tau)\}, z\right) 
= \int_0^z d\tau_1 \,\, f_1(\tau_1) G\left(\{f_2(\tau), \ldots, f_n(\tau)\}, \tau_1\right),
\end{equation}
with 
\begin{eqnarray}
F_3(x) &=& \sum_{i=0}^\infty f_3(i) x^i =
-\frac{64 (1266+43 x)}{81 x^2}
-\frac{16 \text{G}\big(
        \sqrt{4-\tau } \sqrt{\tau };x\big) R_6}{27 x^{5/2} (4-x)}
+\frac{4 \sqrt{4-x} R_7}{27 (4-x) x^2}
\nonumber\\ &&
-48 (-2+x) \sqrt{4-x} \text{G}\left(
        \frac{1}{4-\tau };x\right)
-\frac{320 (2-x)^2}{27 x^{3/2}(4-x)} \text{G}\left(
        \frac{1}{4-\tau },\sqrt{4-\tau } \sqrt{\tau };x\right)
\nonumber\\ &&
-\frac{2848 (2-x)^2}{27 x^{3/2} (4-x)} \text{G}\left(
        \frac{1}{\tau },\sqrt{4-\tau } \sqrt{\tau };x\right)
-\frac{96 (2-x)^2}{x^{3/2}(4-x)} \text{G}\left(
        \sqrt{4-\tau } \sqrt{\tau },\frac{1}{4-\tau };x\right)
\nonumber\\ &&
-\frac{32 \big(
        4836-1636 x+19 x^2+67 x^3\big) \sqrt{4-x}}{27 (4-x) x^3} \text{G}\left(
        \sqrt{4-\tau } \sqrt{\tau },\sqrt{4-\tau } \sqrt{\tau };x\right),
\end{eqnarray}
and
\begin{eqnarray}
R_6 &=& 67 x^5-383 x^4-1202 x^3+13712 x^2-26888 x+1184,
\\
R_7 &=& 67 x^6-517 x^5-104 x^4+12936 x^3-39828 x^2+1928 x+13504.
\end{eqnarray}
Note that here the sum starts at $i = 0$, unlike the case for $f_{1(2)}(i)$ in Secton \ref{sec:3}. The
corresponding expression is 
\begin{eqnarray}
  F_{\pi \ln^2(z),V,1}^{(0)} = \frac{27008}{27 z^3} + \frac{2752}{81 z} + z F_3(z^2)
\end{eqnarray}
The $G$--functions are given as polynomials of rational terms and harmonic polylogarithms in the 
variable 
\begin{eqnarray} 
u = \frac{1}{2} \left(2 + i z \sqrt{4 - z^2} - z^2 \right) = - \frac{1}{x}. 
\end{eqnarray} 
One obtains 
\begin{eqnarray} 
\label{eq:G1func} 
G\left(\left\{\sqrt{4-\tau } \sqrt{\tau}\right\},z^2\right) 
&=& i\left[ - \frac{1}{2 u^2} + \frac{u^2}{2} - 2 \HA_0(u)\right], 
\\
G\left(\left\{\frac{1}{\tau },\sqrt{4-\tau } \sqrt{\tau }\right\},z^2\right), &=& i \Biggl[
        -\frac{(1 - u^2) \big(
                1+4 u+u^2\big)}{4 u^2}
        +4 \HA_0(u) \HA_1(u)
        -4 \HA_{0,1}(u)
\nonumber\\ && 
        +4 \zeta_2
        +\HA_0(u)
        +\HA_0^2(u)
\Biggr],
\\
\label{eq:G3func} 
G\left(\left\{\frac{1}{4-\tau },\sqrt{4-\tau } \sqrt{\tau }\right\},z^2\right)
&=& i \Biggl[
        \frac{(1-u^2)\big(
                1-4 u+u^2\big)}{4 u^2}
        -\HA_0^2(u)
        -4 \HA_{0,-1}(u)
\nonumber\\ &&
        +2 \zeta_2
        +\HA_0(u) \big(
                -1+4 \HA_{-1}(u)\big)
\Biggr]
\label{eq:G3func} 
\\
\label{eq:G4func}
G\left(\left\{
\sqrt{4-\tau } \sqrt{\tau},\frac{1}{4-\tau }\right\},z^2\right) 
&=&
G\left(\left\{\frac{1}{4-\tau} \right\},z^2\right)
G\left(\left\{\sqrt{4-\tau } \sqrt{\tau }\right\},z^2\right) 
\nonumber\\ &&
- 
G\left(\left\{\frac{1}{4-\tau },\sqrt{4-\tau } \sqrt{\tau }\right\},z^2\right),
\end{eqnarray}
by shuffling \cite{ALG}, where
\begin{eqnarray}
G\left(\left\{\frac{1}{4-\tau} \right\},z^2\right) &=& - \ln \left[1 - \frac{z^2}{4}\right],
\end{eqnarray}
Iterative 
$G$-functions over root--valued alphabets were studied in \cite{Ablinger:2014bra,TWOmass}.

After adding up all the terms for which the recurrences are solvable at the expansion point $\hs=4$, we find that the previously non-solvable parts
at the expansions point $x=1$ in Eq. (\ref{FFdecomp}) can be resummed as follows, 
\begin{eqnarray}
\lefteqn{F^{(0)}_{V,1,1}(z) + \zeta_2 F^{(0)}_{V,1,2}(z) + \zeta_3 F^{(0)}_{V,1,3}(z) =} \nonumber\\ &&
-\frac{32 \ln(2) \pi ^3 \frac{1}{\sqrt{4-z^2}} \ln\big(1-\frac{z^2}{4}\big) Q_{13}}{27 z^5}
+\frac{64 \ln(2) \pi ^3 \frac{1}{\sqrt{4-z^2}} \ln(z) Q_{24}}{27 z^5}
-\frac{4 \ln^2(2) \pi  \sqrt{4-z^2} Q_{42}}{27 z}
\nonumber\\ && 
-\frac{8 \ln(2) \pi  \sqrt{4-z^2} \ln(z) Q_{42}}{9 z}
-\frac{4 \pi ^3 \frac{1}{\sqrt{4-z^2}} \ln(z) Q_{48}}{27 z^3}
-\frac{32 \pi  \frac{1}{\sqrt{4-z^2}} \ln^3(z) Q_{51}}{81 z^3}
\nonumber\\ &&
-\frac{64 \pi ^2 \ln^2(z) Q_{55}}{27 (z-2) z^4 (z+2)}
-\frac{16 \pi  \frac{\zeta_3}{\sqrt{4-z^2}} Q_{64}}{729 z^5}
-\frac{2 \ln(2) \pi ^3 \frac{1}{\sqrt{4-z^2}} Q_{65}}{243 z^5}
+\frac{4 \pi  \frac{1}{\sqrt{4-z^2}} \ln^2(z) Q_{73}}{27 z^3}
\nonumber\\ && 
-\frac{\pi ^2 \ln(z) Q_{80}}{486 (z-2) z^2 (z+2)}
+\ln^2(2) \pi ^3 \Biggl[
        -\frac{32 \big(5981-3570 z^2+300 z^4\big) \frac{1}{\sqrt{4-z^2}}}{243 z^3}
\nonumber\\ &&  
      -\frac{64 \big(
                -12+5 z^2\big) \frac{1}{\sqrt{4-z^2}} \ln\big(
                1-\frac{z^2}{4}\big)}{3 z^5}
\Biggr]
+\ln^4(2) \pi  \Biggl[
        -\frac{128 \big(
                85+24 z^2+6 z^4\big) \frac{1}{\sqrt{4-z^2}}}{243 z^3}
\nonumber\\ &&   
     +
        \frac{64 \big(
                -12+5 z^2\big) \frac{1}{\sqrt{4-z^2}} \ln\big(
                1-\frac{z^2}{4}\big)}{3 z^5}
\Biggr]
+\Li_4\left(\frac{1}{2}\right) \pi  \Biggl[
        -\frac{1024 \big(
                85+24 z^2+6 z^4\big) \frac{1}{\sqrt{4-z^2}}}{81 z^3}
\nonumber\\ &&    
     +\frac{512 \big(
                -12+5 z^2\big) \frac{1}{\sqrt{4-z^2}} \ln\big(
                1-\frac{z^2}{4}\big)}{z^5}
\Biggr]
+\pi ^5 \Biggl[
        -\frac{4 \big(
                20915-14010 z^2+4836 z^4\big) 
\frac{1}{\sqrt{4-z^2}}}{3645 z^3}
\nonumber\\ &&     
    -\frac{8 \big(
                12+43 z^2\big) \frac{1}{\sqrt{4-z^2}} \ln\big(
                1-\frac{z^2}{4}\big)}{45 z^5}
\Biggr]
+\Biggl[
        -\frac{64 \ln^3(2) \pi  \big(
                z^2-2\big)^2}{9 (z-2) z^2 (z+2)}
        +\frac{128 \pi ^2 \frac{1}{\sqrt{4-z^2}} \ln^2(z) Q_8}{27 
z^5}
\nonumber\\ && 
        -\frac{224 \pi  \big(
                z^2-2\big) \zeta_3 Q_{16}}{27 (z-2) z^6 (z+2)}
        -\frac{8 \pi ^3 \big(
                z^2-2\big) \ln(z) Q_{29}}{27 (z-2) z^6 (z+2)}
        -\frac{16 \ln(2) \pi ^3 Q_{52}}{27 (z-2) z^6 (z+2)}
\nonumber\\ &&  
      +\frac{16 \ln^2(2) \pi  Q_{56}}{27 (z-2) z^4 (z+2)}
        +\frac{32 \ln(2) \pi  \ln(z) Q_{56}}{9 (z-2) z^4 
(z+2)}
        +\frac{16 \pi  \ln^2(z) Q_{57}}{27 (z-2) z^4 (z+2)}
        +\frac{\pi ^2 \frac{1}{\sqrt{4-z^2}} \ln(z) Q_{62}}{81 z^5}
\nonumber\\ &&        
 -
        \frac{64 \ln^2(2) \pi  \big(
                z^2-2\big)^2 \ln(z)}{(z-2) z^2 (z+2)}
        +\frac{64 \ln(2) \pi ^2 \big(
                -30+8 z^2-3 z^4+z^6\big) \frac{1}{\sqrt{4-z^2}} \ln
(z)}{9 z^5}
\nonumber\\ && 
        -\frac{192 \ln(2) \pi  \big(
                z^2-2\big)^2 \ln^2(z)}{(z-2) z^2 (z+2)}
        -\frac{10432 \pi  \big(
                z^2-2\big)^2 \ln^3(z)}{81 (z-2) z^2 (z+2)}
\Biggr] \GA\left(
        \sqrt{4-\tau } \sqrt{\tau };z^2\right)
\nonumber\\ && 
+\Biggl[
        -\frac{16 \ln^2(2) \pi  \frac{1}{\sqrt{4-z^2}} Q_{11}}{27 
z^5}
        -\frac{32 \ln(2) \pi  \frac{1}{\sqrt{4-z^2}} \ln(z) 
Q_{11}}{9 z^5}
        -\frac{16 \pi  \frac{1}{\sqrt{4-z^2}} \ln^2(z) Q_{22}}{27 
z^5}
\nonumber\\ && 
        +\frac{2 \pi ^2 \ln(z) Q_{43}}{9 (z-2) z^6 (z+2)}
        +\frac{8 \ln(2) \pi ^3 \big(
                -36+7 z^2\big) \frac{1}{\sqrt{4-z^2}}}{z^5}
\Biggr] \GA\big(
        \sqrt{4-\tau } \sqrt{\tau };z^2\big)^2
\nonumber\\ && 
+\Biggl[
        \frac{64 \ln^2(2) \pi  \big(
                z^2-2\big)^2}{27 (z-2) z^2 (z+2)}
        +\frac{128 \ln(2) \pi  \big(
                z^2-2\big)^2 \ln(z)}{9 (z-2) z^2 (z+2)}
        +
        \frac{320 \pi  \big(
                z^2-2\big)^2 \ln^2(z)}{27 (z-2) z^2 (z+2)}
\nonumber\\ &&  
       +\frac{64 \pi ^2 \big(
                3+z^2
        \big)
\big(6-4 z^2+z^4\big) \frac{1}{\sqrt{4-z^2}} \ln(z)}{27 z^5}
\Biggr] \GA\left(
        \frac{1}{4-\tau },\sqrt{4-\tau } \sqrt{\tau };z^2\right)
\nonumber\\ && 
+\Biggl[
        \frac{32 \ln^2(2) \pi  \big(
                z^2-2\big)^2}{3 (z-2) z^2 (z+2)}
        -\frac{16 \pi ^2 \frac{1}{\sqrt{4-z^2}} \ln(z) Q_{18}}{9 
z^5}
        +\frac{64 \ln(2) \pi  \big(
                z^2-2\big)^2 \ln(z)}{(z-2) z^2 (z+2)}
\nonumber\\ &&  
       +\frac{2848 \pi  \big(
                z^2-2\big)^2 \ln^2(z)}{27 (z-2) z^2 (z+2)}
\Biggr] \GA\left(
        \frac{1}{\tau },\sqrt{4-\tau } \sqrt{\tau };z^2\right)
+\Biggl[
        \frac{32 \ln^2(2) \pi  \big(
                z^2-2\big)^2}{3 (z-2) z^2 (z+2)}
\nonumber\\ &&     
    +\frac{64 \ln(2) \pi  \big(
                z^2-2\big)^2 \ln(z)}{(z-2) z^2 (z+2)}
        +\frac{96 \pi  \big(
                z^2-2\big)^2 \ln^2(z)}{(z-2) z^2 (z+2)}
\Biggr] \GA\left(
        \sqrt{4-\tau } \sqrt{\tau },\frac{1}{4-\tau };z^2\right)
\nonumber\\ && 
+\frac{4096 \Li_4\left(\frac{1}{2}\right) \ln(2) \pi  \frac{1}{\sqrt{4-z^2}}}{3 
z^3}
+\frac{512 \ln(2)^5 \pi  \frac{1}{\sqrt{4-z^2}}}{9 z^3}
-\frac{512 \ln^3(2) \pi ^3 \frac{1}{\sqrt{4-z^2}}}{9 z^3}
-\frac{1600 \ln(2) \pi ^5 \frac{1}{\sqrt{4-z^2}}}{27 z^3}
\nonumber\\ && 
-
\frac{\pi ^2 \big(
        -36+11 z^2\big) \kappa_2 \frac{1}{\sqrt{4-z^2}}}{36 
z^5}
-\frac{\big(
        -36+11 z^2\big) \kappa_9 \frac{1}{\sqrt{4-z^2}}}{6 
z^5}
+\frac{32}{9} \ln^3(2) \pi  z \big(
        z^2-2\big) \sqrt{4-z^2}
\nonumber\\ && 
+\frac{248}{27} \ln^2(2) \pi  z \big(
        4-z^2\big)^{3/2}
+\frac{2166080 \big(
        -36+11 z^2\big) \frac{\zeta_3}{\sqrt{4-z^2}}}{729 \pi  z^5}
\nonumber\\ &&
-\frac{1792 \Li_4\left(\frac{1}{2}\right) \big(
        -36+11 z^2\big) \frac{\zeta_3}{\sqrt{4-z^2}}}{\pi  z^5}
-\frac{2034368 \ln(2) \big(
        -36+11 z^2\big) \frac{\zeta_3}{\sqrt{4-z^2}}}{729 \pi  z^5}
\nonumber\\ && 
-\frac{17024 \ln^2(2) \big(
        -36+11 z^2\big) \frac{\zeta_3}{\sqrt{4-z^2}}}{243 \pi  z^5}
+\frac{7168 \ln^3(2) \big(
        -36+11 z^2\big) \frac{\zeta_3}{\sqrt{4-z^2}}}{27 \pi  z^5}
\nonumber\\ && 
-\frac{224 \ln^4(2) \big(
        -36+11 z^2\big) \frac{\zeta_3}{\sqrt{4-z^2}}}{3 \pi  z^5}
\nonumber\\ && 
+\frac{64256 \ln(2) \pi  \big(
        -36+11 z^2\big) \frac{\zeta_3}{\sqrt{4-z^2}}}{81 z^5}
-\frac{736 \ln^2(2) \pi  \big(
        -36+11 z^2\big) \frac{\zeta_3}{\sqrt{4-z^2}}}{3 z^5}
\nonumber\\ && 
+\frac{\pi ^3 \big(
        -64836+25451 z^2\big) \frac{\zeta_3}{\sqrt{4-z^2}}}{135 z^5}
+
\frac{17 \big(
        -36+11 z^2\big) \kappa_1 
\frac{\zeta_3}{\sqrt{4-z^2}}}{72 z^5}
-\frac{\big(
        -36+11 z^2\big) \kappa_3 
\frac{\zeta_3}{\sqrt{4-z^2}}}{9 z^5}
\nonumber\\ && 
-\frac{49112 \big(
        -36+11 z^2\big) \frac{\zeta_3^2}{\sqrt{4-z^2}}}{81 \pi  z^5}
-\frac{640 \pi ^4 \big(
        z^2-2\big)^2 \ln(z)}{27 (z-2) z^2 (z+2)}
-\frac{360 \pi  \frac{\zeta_5}{\sqrt{4-z^2}}}{z^3}
\nonumber\\ && 
-\frac{32 \ln(2) \pi ^2 \big(
        -16+10 z^2-5 z^4+z^6\big) \ln(z)}{9 z^2}
+32 \ln^2(2) \pi  z \big(
        z^2-2\big) \sqrt{4-z^2} \ln(z)
\nonumber\\ && 
+\frac{496}{9} \ln(2) \pi  z \big(
        4-z^2\big)^{3/2} \ln(z)
+\frac{4 \pi ^2 \big(
        -12+5 z^2\big) \frac{1}{\sqrt{4-z^2}} \GA\big(
        \sqrt{4-\tau } \sqrt{\tau };z^2\big)^3 \ln(z)}{3 z^5}
\nonumber\\ && 
+96 \ln(2) \pi  z \big(
        z^2-2\big) \sqrt{4-z^2} \ln^2(z)
+\frac{488}{9} \pi  z \big(
        4-z^2\big)^{3/2} \ln^2(z)
-\frac{4096 \pi ^2 \big(
        z^2-2\big)^2 \ln^3(z)}{81 (z-2) z^2 (z+2)}
\nonumber\\ && 
+\frac{16}{3} \ln^2(2) \pi  z \big(
        z^2-2\big) \sqrt{4-z^2} \ln\big(
        1-\frac{z^2}{4}\big)
+32 \ln(2) \pi  z \big(
        z^2-2\big) \sqrt{4-z^2}
 \ln(z) \ln\big(
        1-
        \frac{z^2}{4}\big)
\nonumber\\ && 
+48 \pi  z \big(
        z^2-2\big) \sqrt{4-z^2} \ln^2(z) \ln\big(
        1-\frac{z^2}{4}\big) 
+ \tilde{F}^{(0), \rm rest}_{V,1}(z) \,.
\end{eqnarray}


\noindent
The polynomials $Q_i$ are given in Appendix~\ref{app:z-resum}, where also the resummed expressions in $z$  
for the remaining form factors are presented. 
\subsection{The four-particle threshold}
\label{Sec:s16-expansions}

\vspace*{1mm}
\noindent
The differential equations in the case of the four-particle threshold were derived from the ones guessed using the expansions at $\hs=0$
by doing the change of variables $\tz$, after which we matched with the expansions at $\hs \rightarrow +\infty$ ($x = 0$).
In all cases the matching was done at a set of points given by Eq. (\ref{eq:mps0}) using $\delta=1/2000$. No derivatives were used.
Since the expansions at $\hs=16$ converge in the region $4<\hs<28$, and the high energy expansion converges for $\hs>16$,
we had to choose a value of $\hs_0$ in Eq. (\ref{eq:mps0}) in the region $16<\hs_0<28$. We used $\hs_0=20$ as the starting point.

We found agreement with two of the statements made in Ref. \cite{Fael:2022miw,Fael:2022rgm}, namely,
that these expansions contain no logarithms in the variable $\tilde{z}$, 
and that the first non-analytic terms (in other words, odd powers of $\tilde{z}$) start at order 7 in the case of the axial-vector and scalar form factors, 
and at order 9 in the case of the vector and pseudo-scalar form factors. 
The authors of \cite{Fael:2022miw,Fael:2022rgm} did find powers of logarithms as high as $\ln^5(\tilde{z})$,
together with powers of $\tilde{z}$ as low as $\tilde{z}^3$ at the level of the master integrals, which then disappear from the final results.
Since we do not deal with master integrals in our method, at no point in our calculations
do we find such terms, which makes matching with the high energy expansion easier in our case. 
We can even pinpoint from which of the form factors $F_{I,k}^{(0)}(\tilde{z})$ do the first non-analytic terms come from.
The non-solvable parts $F_{I,2}^{(0)}(\tilde{z})$ turn out to be analytic for all values of $I$. In all other cases, the first non-analytic terms
start at $\tilde{z}^9$, with the exception of $F_{a,1,1}^{(0)}(\tilde{z})$, $F_{a,1,3}^{(0)}(\tilde{z})$, $F_{s,1}^{(0)}(\tilde{z})$ and $F_{s,3}^{(0)}(\tilde{z})$,
where they start at order $\tilde{z}^7$. As examples, the non-solvable (at $x=1$) parts of the vector form factor, $F_{v,1,1}^{(0)}(\tz)$ and
$F_{v,1,3}^{(0)}(\tz)$ have the following numerical expansions at $\tz=0$,
\begin{eqnarray}
F_{v,1,1}^{(0)}(\tz) &=& 
708902.27306400652676-1166.85387707523430 i \nonumber \\ &&
-(61456.423805594065279-15405.248917971629065 i) \tz^2 \nonumber \\ &&
+(1716.6533220464368589-30.2703684431181752 i) \tz^4 \nonumber \\ &&
+(11.2780421205283601446+4.3654180757157056768 i) \tz^6 \nonumber \\ &&
+(0.74166620892322374519+0.87946094505024891071 i) \tz^8 \nonumber \\ &&
+(0.00029237832702742120027-0.00015868965691879171216 i) \tz^9 \nonumber \\ &&
+(0.063605792722744201987+0.086239707160403484962 i) \tz^{10} \nonumber \\ &&
+(0.000083577138055319044650-0.000045361868983569336325 i) \tz^{11} \nonumber \\ &&
+(0.0055226467252660130699+0.0075199442142108863165 i) \tz^{12} \nonumber \\ &&
+(0.000016701125786592152314-9.0646114169171 \times 10^{-6} i) \tz^{13} \nonumber \\ &&
+(0.00047286013902545145990+0.00063297686865989569848 i) \tz^{14} \nonumber \\ &&
+O\left(\tz^{15}\right) \,,
\end{eqnarray}
and
\begin{eqnarray}
F_{v,1,3}^{(0)}(\tz) &=& 
-(525309.50058104203958+59965.62896057212016 i) \nonumber \\ &&
+(38742.883772468588201+5783.453693303291125 i) \tz^2 \nonumber \\ &&
-(960.03909739184249814+38.76539648430706459 i) \tz^4 \nonumber \\ &&
-(7.5313243403421680570+0.2607012992726984795 i) \tz^6 \nonumber \\ &&
-(0.66891050212132794057-0.02251916817914504430 i) \tz^8 \nonumber \\ &&
+(5.98 \times 10^{-192}+0.000132015095543045899654 i) \tz^9 \nonumber \\ &&
-(0.056139396989125489861-0.002911370516926165654 i) \tz^{10} \nonumber \\ &&
+(1.71 \times 10^{-192}+0.000037736873241471453874 i) \tz^{11} \nonumber \\ &&
-(0.0046441851980261141902-0.0002618233295226892918 i) \tz^{12} \nonumber \\ &&
+(3.42 \times 10^{-193}+7.5409170673126 \times 10^{-6} i) \tz^{13} \nonumber \\ &&
+(-0.00038214879532394872786+0.00002100556849783558601 i) \tz^{14} \nonumber \\ &&
+O\left(\tz^{15}\right) \,.
\end{eqnarray}
Notice that in the case of $F_{v,1,3}^{(0)}(\tz)$ the real parts of the coefficients of the odd powers in $\tz$ are very small, namely, of the order of $10^{-192}$ or lower.
Clearly, the real parts of these coefficients are actually equal to zero, and the values obtained indicate that we are computing the coefficients with
a precision of around 200 digits. To get the final form factors, we also need to combine these results using Eq. (\ref{FFdecomp}). 
It can be shown that the imaginary parts of the coefficients of the odd powers in $\tz$ of the combination $F_{v,1,1}^{(0)}(\tz)+\zeta_3 F_{v,1,3}^{(0)}(\tz)$ are
of the order of $10^{-166}$ or lower. Again, this indicates that these imaginary parts actually cancel in the combination, and it is also a strong indication
that the calculation is correct, since such cancellations are unlikely
to happen otherwise.

At this point, we could proceed in a similar way as we did with previous expansions and use this type of  information about the coefficients to
improve the precision by matching at better points (that is, choosing better values of $\hs_0$ and $\delta$). 
The main reason for doing this would be to try to determine the base coefficients in terms of known constants, as we have done in the case of the expansions at the other points. However, this turns out to be impractical in this case.
Even the simplest coefficients of solvable parts are difficult to determine using the {\tt PSLQ} algorithm,
due to the large number of constants that would need to be included in a search. 
These constants involve all sorts of harmonic polylogarithms of weight up to 5 
evaluated at 
\begin{eqnarray}  
\xi=7-4\sqrt{3},
\end{eqnarray}  
on top
of the constants that appeared in previous expansions, such as $\zeta_k$, $a_4$, $l_2^k$ and their products.
For example, the coefficient of $\tz^0$ of the term proportional to $n_h^2$ in the vector form factor $F_{v,1}^{(0)}(\tz)$ (which is much simpler
than any of the terms proportional to $n_h$) contains the following constants and their products,
\begin{eqnarray}
&& \biggl\{\HA_{0,1}(\xi),
l_2, \, \pi, \, \zeta_2, \, \zeta_3, \, \frac{1}{\sqrt{3}}, \, \HA_0(\xi), \,  \HA_1(\xi), \, 
\HA_{-1}(\xi), \,  \HA_{0,1}(\xi), \,  \HA_{0,-1}(\xi), \, \HA_{0,0,1}(\xi), \,
\nonumber \\ && \phantom{\biggl\{}
\HA_{0,0,-1}(\xi), \, \HA_{0,1,1}(\xi), \,  \HA_{0,0,0,1}(\xi), \, \HA_{0,0,0,-1}(\xi), \, 
\HA_{0,0,1,1}(\xi), \, \HA_{0,1,1,1}(\xi)
\biggr\}.
\end{eqnarray}  
The non-solvable parts of the form factors will most likely be much more complicated, containing these and new constants, 
and we would need to determine the coefficients numerically with an unattainable precision in order to be able to do a {\tt PSLQ} search.
Fortunately, the expansions at $\hs=16$, which are very important since they are the only ones that cover the region $8<\hs<16$, 
are also very well behaved and rather smooth at the expansion point, and the numerical results we have obtained are enough for most applications.

With all the expansion thus far calculated, we cover the entire range of $\hs$.
In an ancillary file to this paper we provide the representations for the expansion around the four singular 
points for phenomenological use.
\section{Numerical Results} 
\label{sec:5}
\def\figh{5cm}

\vspace*{1mm}
\noindent
In the following we illustrate non--solvable constant parts, 
cf.~Ref.~\cite{Blumlein:2019oas}, of the real part to the different form factors 
${ F_{V,1}, F_{V,2}, F_{A,1}, F_{A,2}, F_{S}}$ and ${F_{P}}$ as a function of $\hat{s} 
= q^2/m^2$ for the constant 
\begin{center}
\begin{figure}[H]
\begin{center}
  \includegraphics[height=\figh]{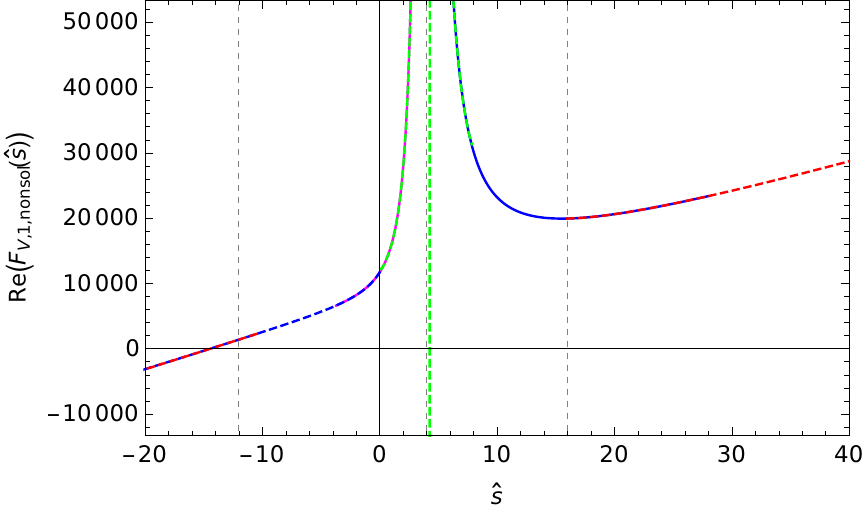}
     \includegraphics[height=\figh]{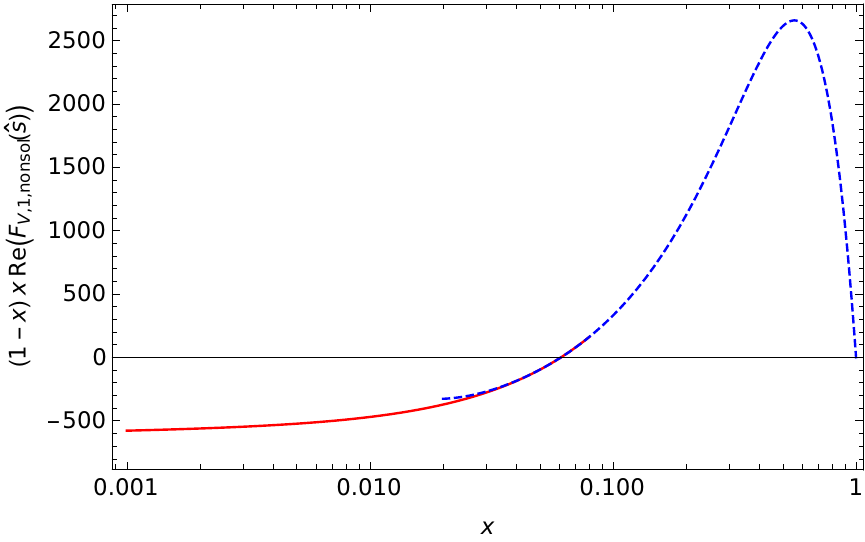}
\end{center}
\caption{\small\sf Real part of the non-solvable contribution for the
  vector form factor $F_{V,1}$ in dependence on $\hat s $ and $x$. The
  range of $x\in[0,1]$ corresponds to $\hat s \in [-\infty,0]$. The
  different line styles correspond to the various expansions. The
  vertical dashed lines indicate the (pseudo-)thresholds.
Left panel: the overlapping expansions are performed around $\hat{s} = -\infty, 0, 4, 16$ and $\hat{s} = \infty$.
Right panel: full line: expansion around $x=0$; dashed line: expansion around $x=1$.
}
\label{fig:NUMFV1}
\end{figure}
\end{center}
contribution  at $O(a_s^3)$ numerically. Here we use the 
first 200 expansion terms around $x=1$, $x=0$, and the (pseudo) thresholds $\hat{s} = 16$ 
and $4$. For this approach we estimated our numerical accuracy to be better than $10^{-40}$.
The numerical work is done by using {\tt Mathematica}. For harmonic, generalized harmonic  
\cite{HPL, Ablinger:2018sat}  and  cyclotomic harmonic polylogarithms \cite{Ablinger:2018sat,CPOLYF}  there 
are also different other numerical implementations.
\begin{center}
\begin{figure}[H]
\begin{center}
  \includegraphics[height=\figh]{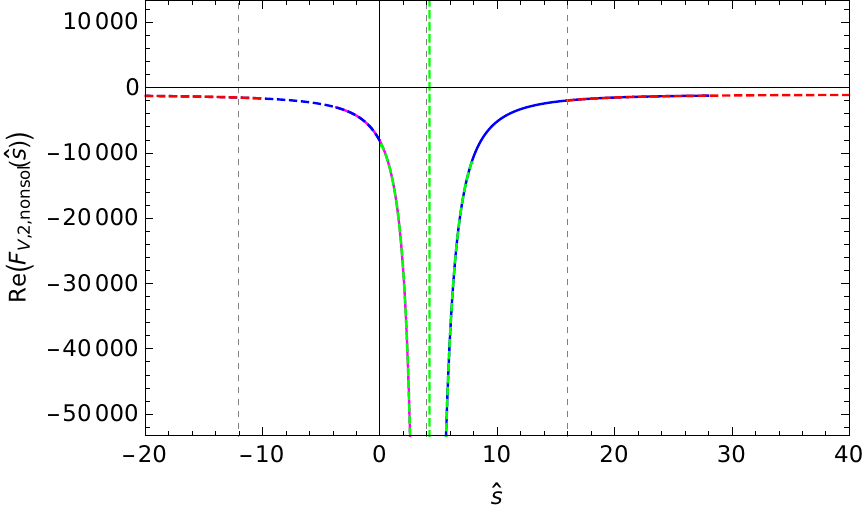}
  \includegraphics[height=\figh]{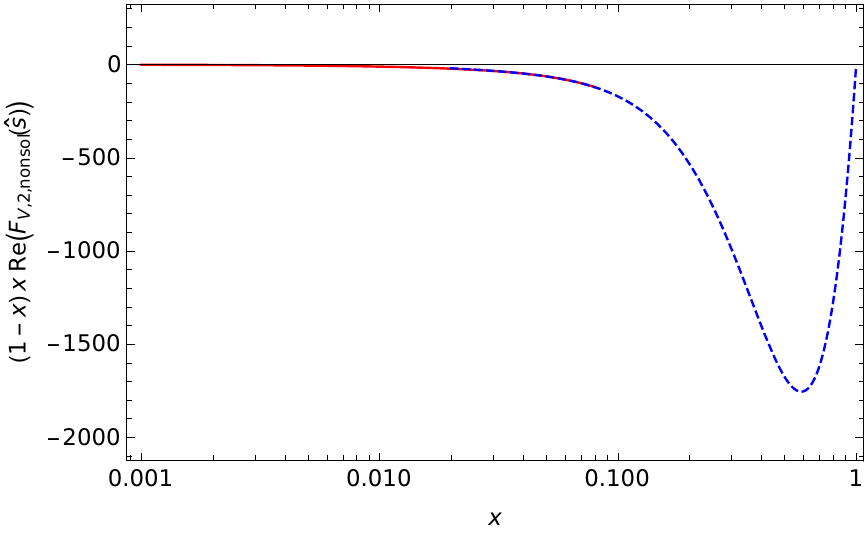}
\end{center}
\caption{\small\sf The same as Figure~\ref{fig:NUMFV1} for ${F_{V,2}}$.}
\label{fig:NUMFV2}
\end{figure}
\end{center}
\begin{center}
\begin{figure}[H]
\begin{center}
  \includegraphics[height=\figh]{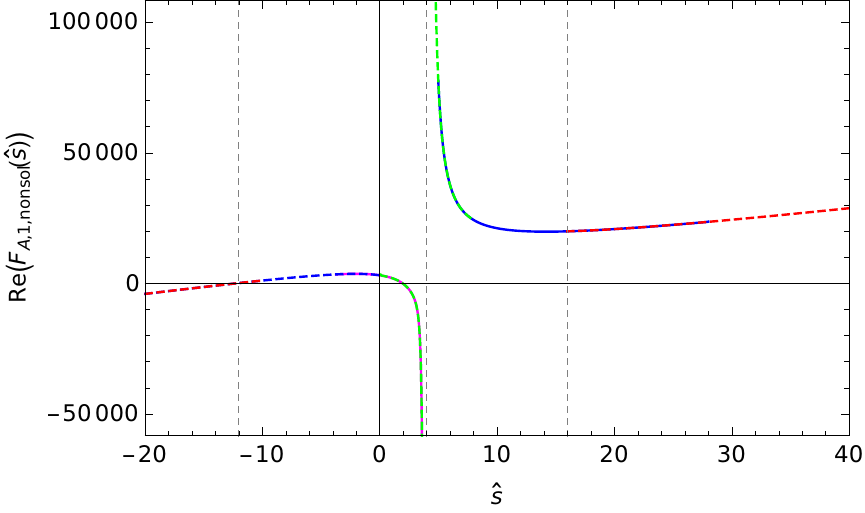}
  \includegraphics[height=\figh]{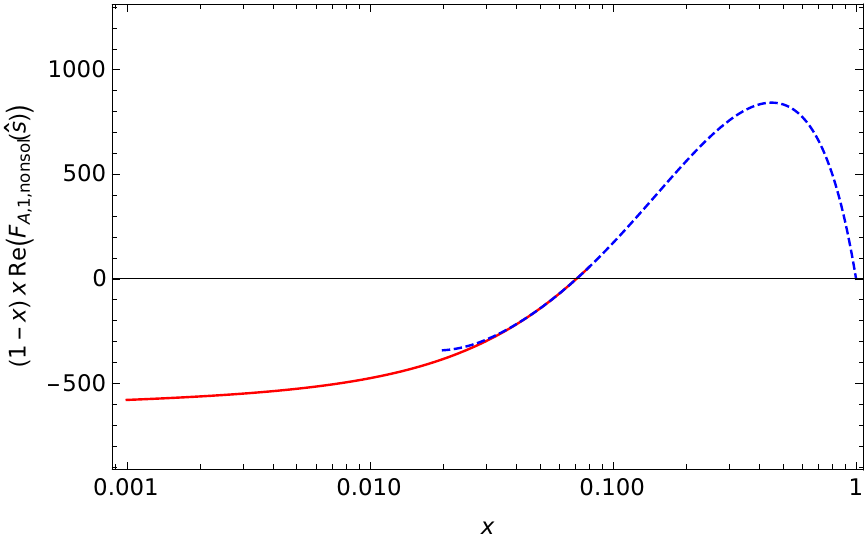}
\end{center}
\caption{\small\sf The same as Figure~\ref{fig:NUMFV1} for ${F_{A,1}}$.}
\label{fig:NUMFA1}
\end{figure}
\end{center}
\begin{center}
\begin{figure}[H]
\begin{center}
  \includegraphics[height=\figh]{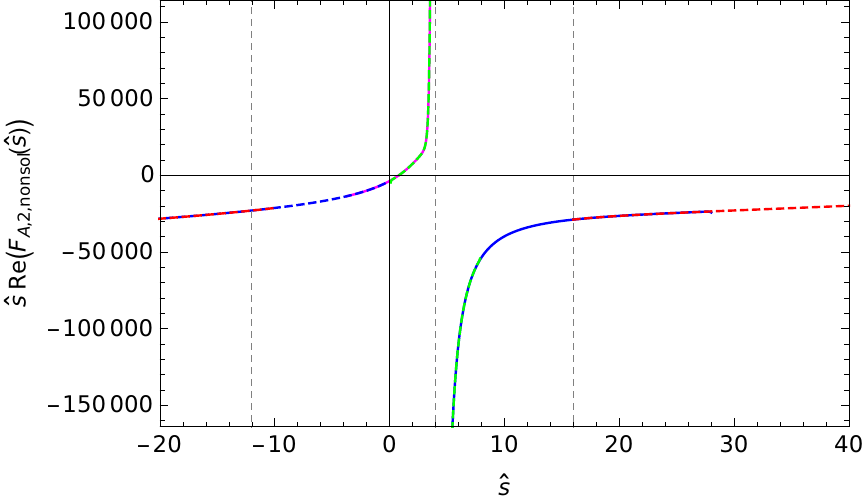}
  \includegraphics[height=\figh]{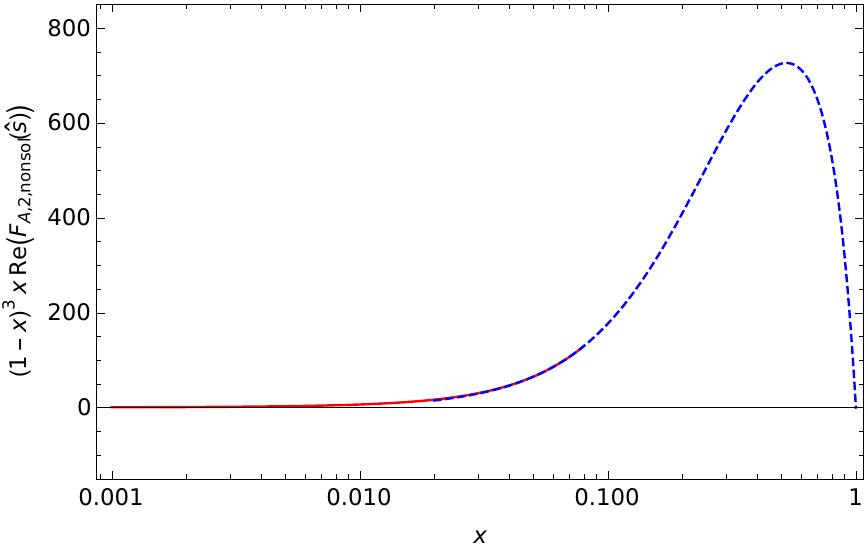}
\end{center}
\caption{\small\sf The same as Figure~\ref{fig:NUMFV1} for ${F_{A,2}}$.}
\label{fig:NUMFA2}
\end{figure}
\end{center}

\begin{center}
\begin{figure}[H]
\begin{center}
  \includegraphics[height=\figh]{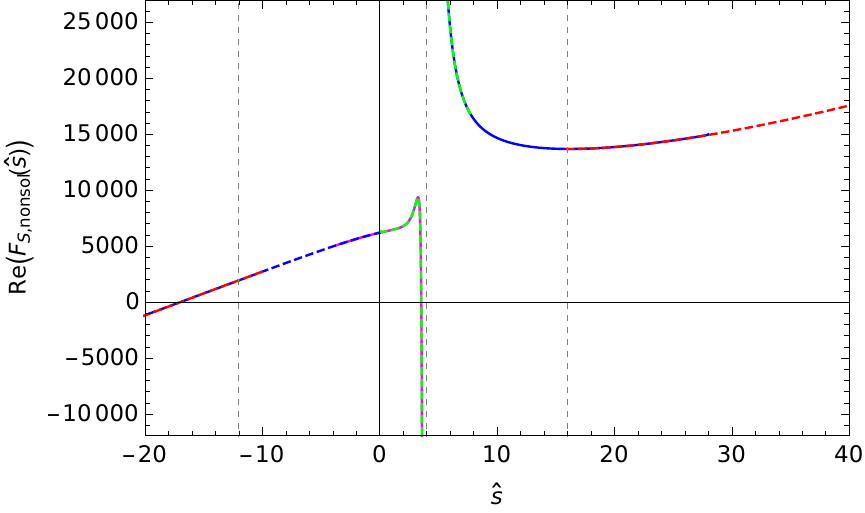}
  \includegraphics[height=\figh]{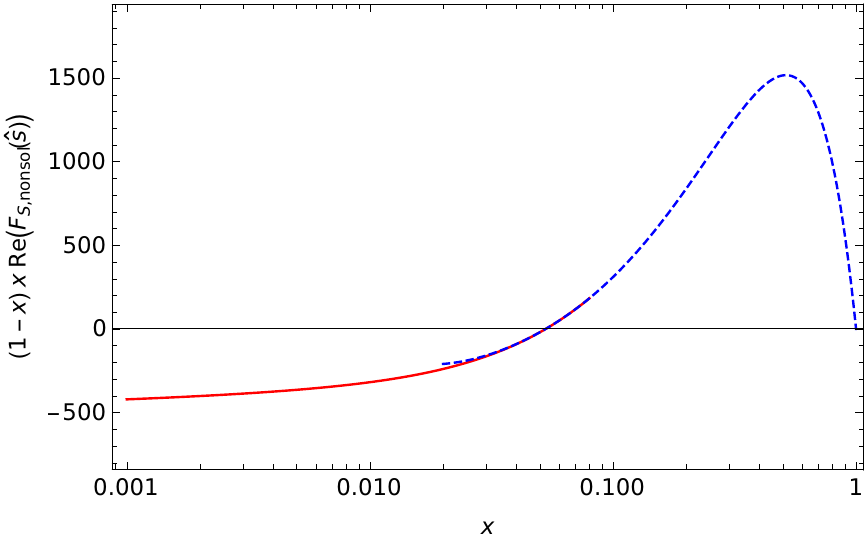}
\end{center}
\caption{\small\sf The same as Figure~\ref{fig:NUMFV1} for ${F_{S}}$.}
\label{fig:NUMFS}
\end{figure}
\end{center}

\begin{center}
\begin{figure}[H]
\begin{center}
  \includegraphics[height=\figh]{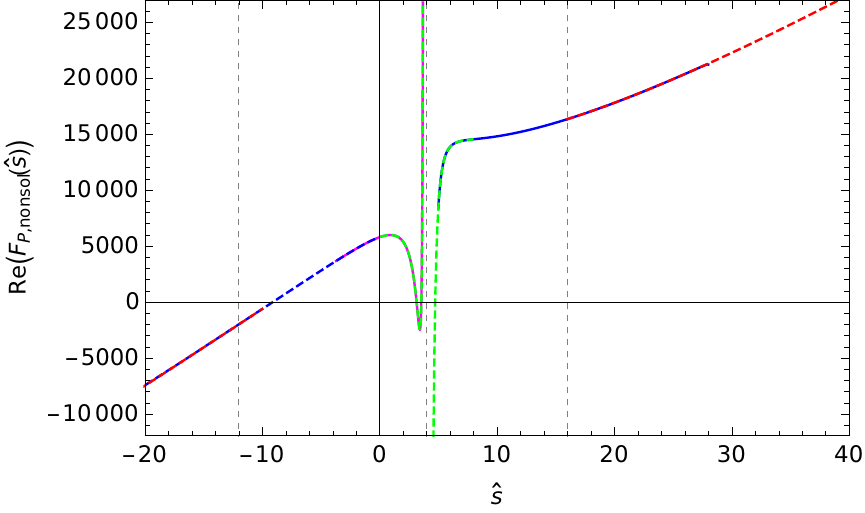}
  \includegraphics[height=\figh]{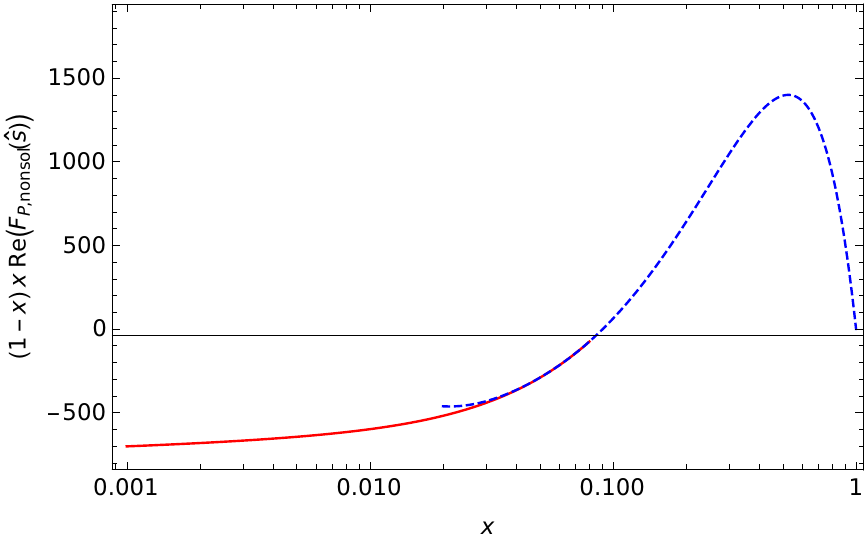}
\end{center}
\caption{\small\sf The same as Figure~\ref{fig:NUMFV1} for ${F_{P}}$.}
\label{fig:NUMFP}
\end{figure}
\end{center}

In Figures~\ref{fig:NUMFV1}--\ref{fig:NUMFP} we use different rescalings in $\hat{s}$
in part to allow for better visibility. Figures~\ref{fig:NUMFV1}, \ref{fig:NUMFV2} show 
the results for the vector form factors, Figures~\ref{fig:NUMFA1}, \ref{fig:NUMFA2} 
for the axial--vector form factors, and Figures~\ref{fig:NUMFS}  and 
\ref{fig:NUMFP} for the scalar and pseudo--scalar form factor, respectively. The left panels 
show the contributions as a function of $\hat{s}$, while in the 
right panel we illustrate the form factors in the region $x \in [0,1]$.
Similar numerical results are obtained for the corresponding imaginary parts in the 
non--solvable case, which we are not illustrating.
The singularities of the form factors in the variable $\hat{s}$ are indicated.
In the above figures it is shown, that the series expansions used do significantly 
overlap, which allows for a continuous representation. Concerning the expansion around 
$x = 0$, the singularity left to this value at $x = 4 \cdot \sqrt{3} - 7 \approx 
-0.0718$ requires that the expansion around $x=1$ has to be used even below $7 - 4 
\cdot \sqrt{3}$. We can extend our series representations to higher orders, to 
meet even higher accuracy.

\begin{center}
\begin{figure}[H]
\begin{center}
  \includegraphics[height=\figh]{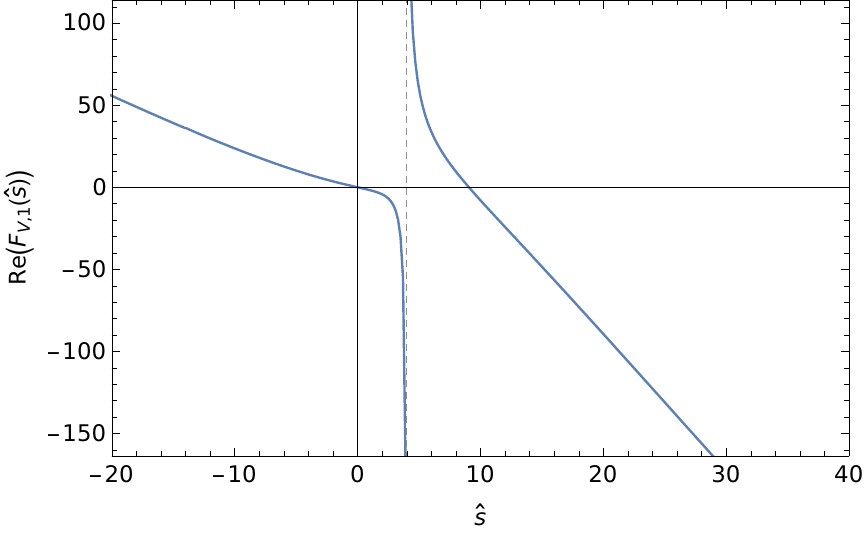}
     \includegraphics[height=\figh]{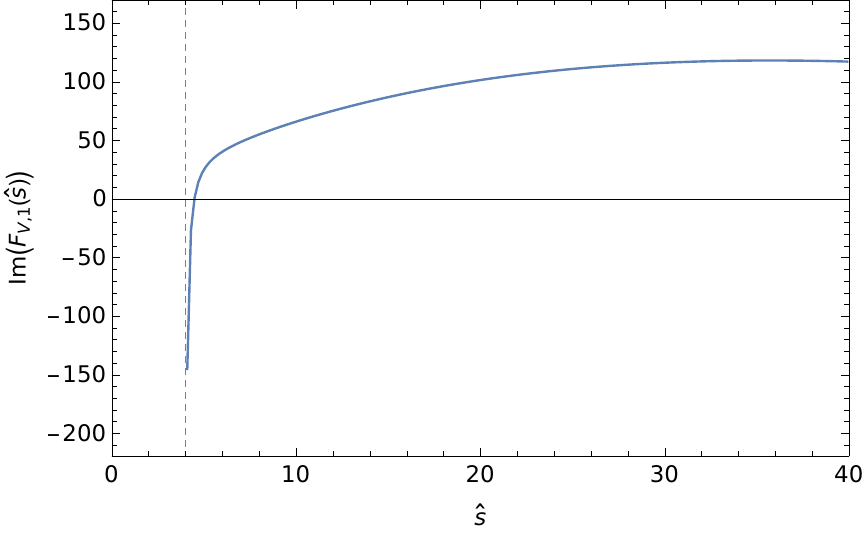}
\end{center}
\caption{\small\sf Real and imaginary part of the complete heavy-fermionic contributions to the
  vector form factor $F_{V,1}$ in dependence on $\hat s $.}
\label{fig:NUMFV1full}
\end{figure}
\end{center}

\begin{center}
\begin{figure}[H]
\begin{center}
  \includegraphics[height=\figh]{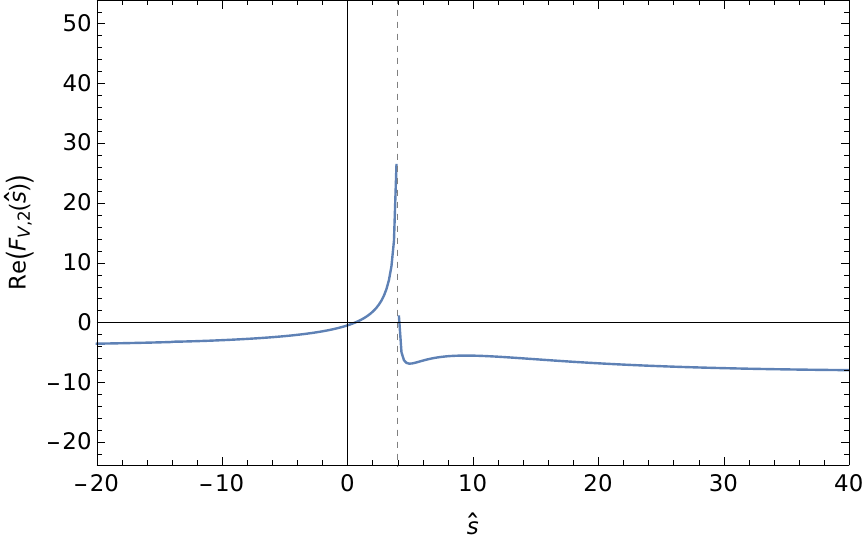}
     \includegraphics[height=\figh]{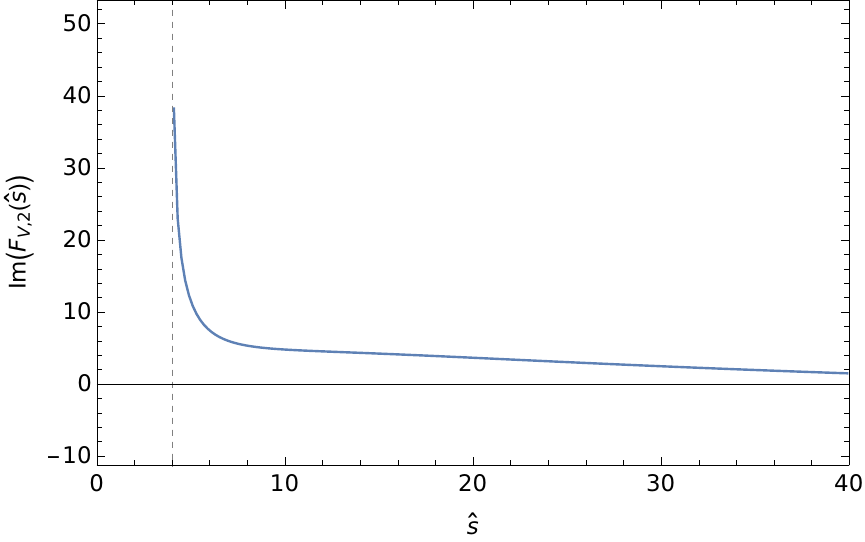}
\end{center}
\caption{\small\sf Real and imaginary part of the complete heavy-fermonic contributions to the
  vector form factor $F_{V,2}$ in dependence on $\hat s $.}
\label{fig:NUMFV2full}
\end{figure}
\end{center}

\begin{center}
\begin{figure}[H]
\begin{center}
  \includegraphics[height=\figh]{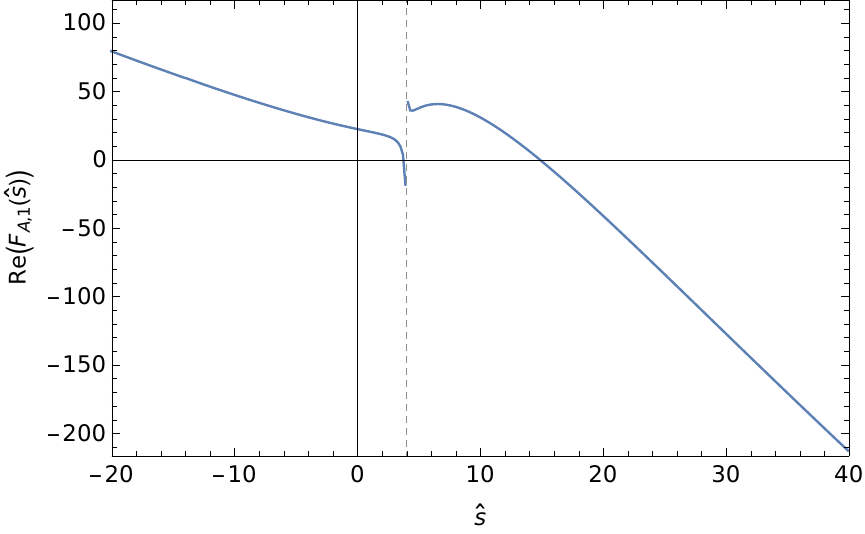}
     \includegraphics[height=\figh]{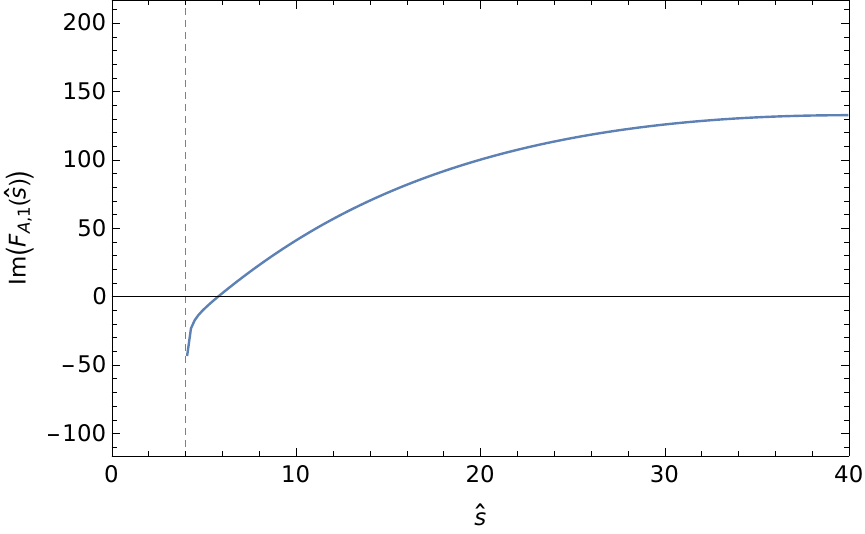}
\end{center}
\caption{\small\sf Real and imaginary part of the complete heavy-fermonic contributions to the
  axial-vector form factor $F_{A,1}$ in dependence on $\hat s $.}
\label{fig:NUMFA1full}
\end{figure}
\end{center}

\begin{center}
\begin{figure}[H]
\begin{center}
  \includegraphics[height=\figh]{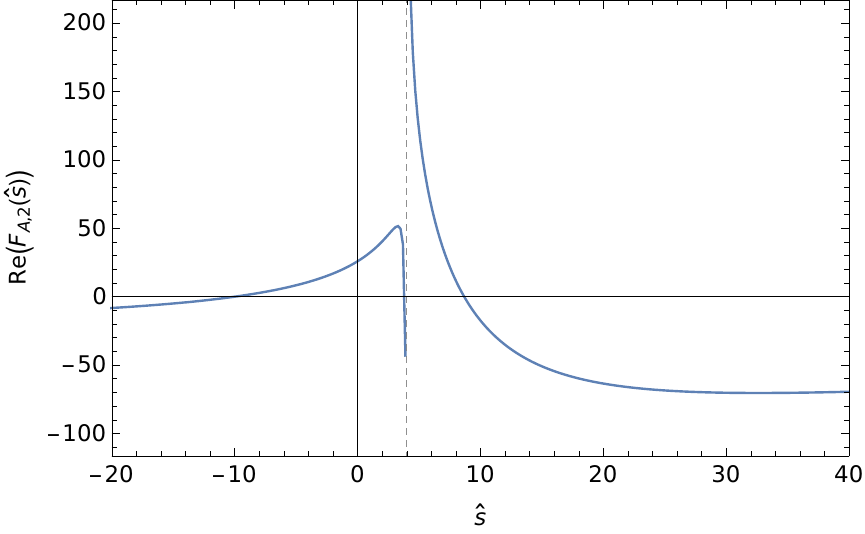}
     \includegraphics[height=\figh]{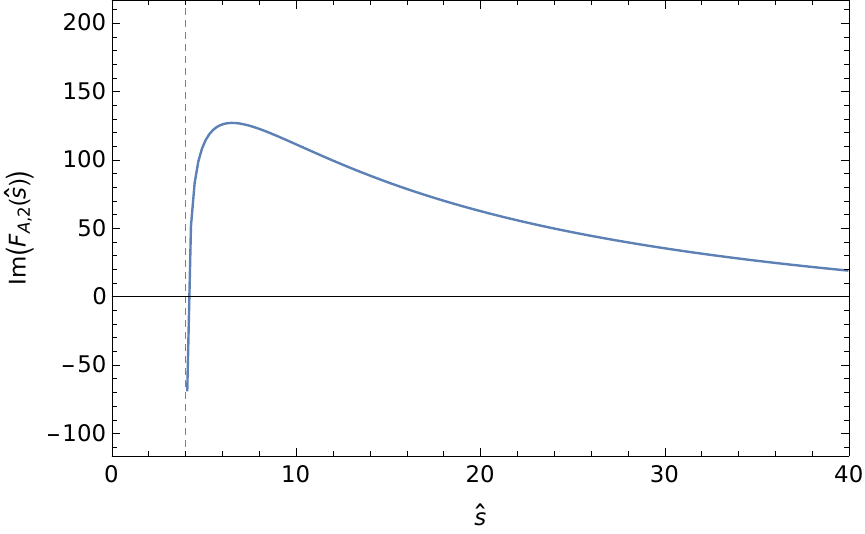}
\end{center}
\caption{\small\sf Real and imaginary part of the complete heavy-fermonic contributions to the
  axial-vector form factor $F_{A,2}$ in dependence on $\hat s $.}
\label{fig:NUMFA2full}
\end{figure}
\end{center}

\begin{center}
\begin{figure}[H]
\begin{center}
  \includegraphics[height=\figh]{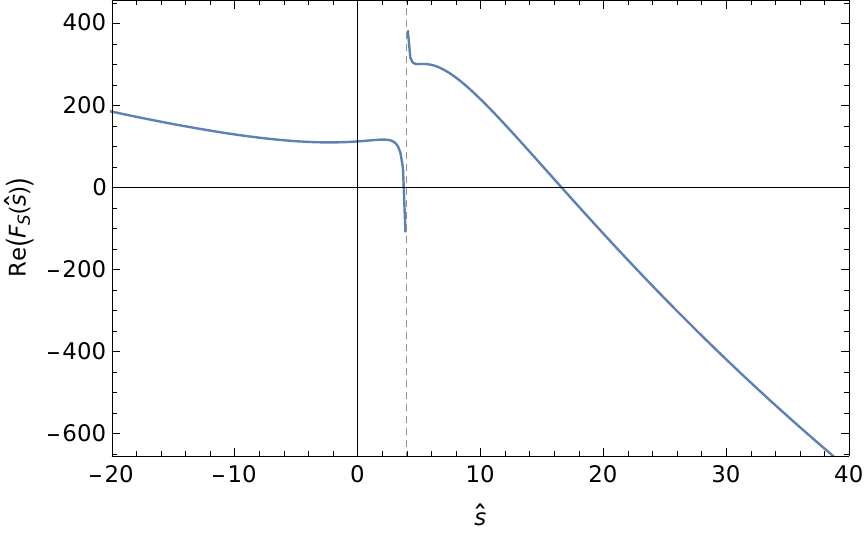}
     \includegraphics[height=\figh]{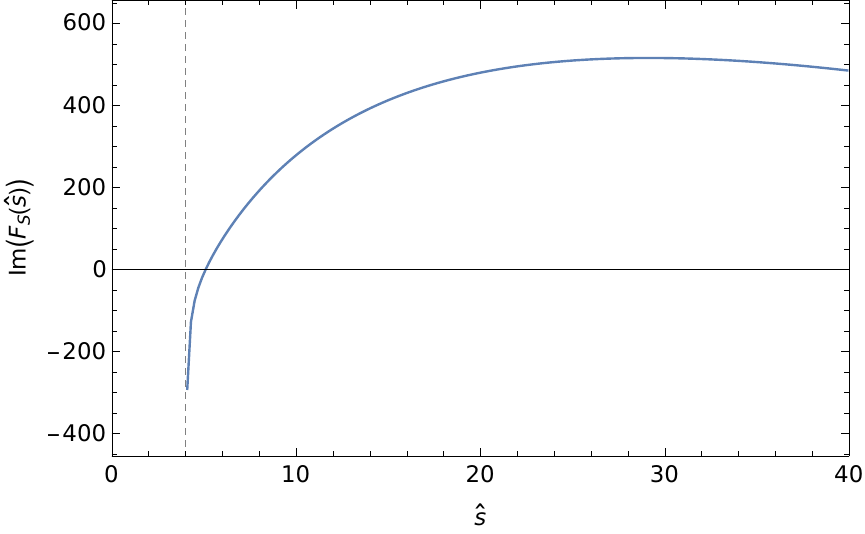}
\end{center}
\caption{\small\sf Real and imaginary part of the complete heavy-fermonic contributions to the
  scalar form factor $F_{S}$ in dependence on $\hat s $.}
\label{fig:NUMFSfull}
\end{figure}
\end{center}

\begin{center}
\begin{figure}[H]
\begin{center}
  \includegraphics[height=\figh]{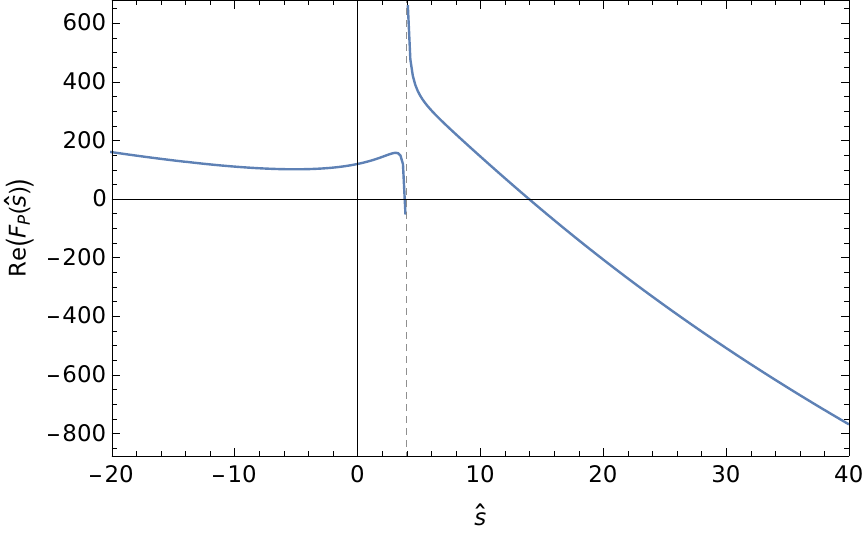}
     \includegraphics[height=\figh]{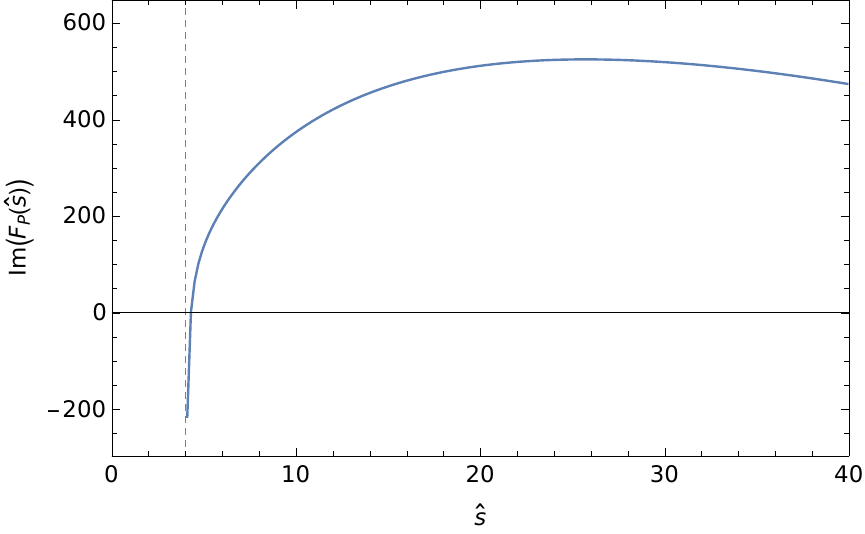}
\end{center}
\caption{\small\sf Real and imaginary part of the complete heavy-fermonic contributions to the
  pseudo-scalar form factor $F_{P}$ in dependence on $\hat s $.}
\label{fig:NUMFPfull}
\end{figure}
\end{center}

In Figures~\ref{fig:NUMFV1full}--\ref{fig:NUMFPfull} we illustrate the real and 
imaginary parts of the constant $O(a_s^3)$ contributions to the vector 
(Figures~\ref{fig:NUMFV1full}, \ref{fig:NUMFV2full}), axial--vector 
(Figures~\ref{fig:NUMFA1full}, \ref{fig:NUMFA2full}), scalar 
(Figure~\ref{fig:NUMFSfull}), and pseudo--scalar form factor 
(Figure~\ref{fig:NUMFPfull}) as functions of $\hat{s}$.

Finally, we would like to discuss the numerical comparison of the present 
results with those 
of Ref.~\cite{Fael:2022miw}. We compared the results at $\hat s = \{-300, -20, -10, -5, 
-0.1, -0.001,$ $-0.0001, 0.00001, 0.1, 3.9, 3.9999, 10, 15.9999, 16.0001, 
16.1, 20, 39.9999, 40\}$ using their {\tt Mathematica} files. In general we found a 
relative agreement of $O(10^{-10})$ or better, except for $\hat{s} = 39.9999$, where the 
deviation was in the range $[6.2 \cdot 10^{-6}, 1.6 \cdot 10^{-9}]$.

Furthermore, we compared the representations of \cite{Fael:2022miw}, Eq.~(35--40), in 
the high energy limit $s \rightarrow - \infty$ to our results for $n_h = 1$. Setting 
the color factors to those of QCD we obtain the following relative deviations
\begin{eqnarray}
\left|\frac{{ F_{I}^\mathrm{KIT}} - {F_{I}}} 
{{F_{I}}}\right| \in 
[2.27 \cdot 10^{-7},
 1.13 \cdot 10^{-4}], 
\end{eqnarray}
for $I = \{ {( V,1), (V,2), (A,1), (A,2), S, P} \}$.

We provide our numerical representation in {\tt Mathematica} format at \cite{AFILE}.
\section{Conclusions}
\label{sec:6}

\vspace*{1mm}
\noindent
We have calculated analytic representations for the quarkonic contributions to the massive three--loop 
form factors for different currents based on non--trivial computer
algebra methods, including guessing algorithms
to 
derive recursion relations and differential equation and to determine 
representations in terms of known constants.
Up to two--loop order, the massive form factors can be represented in terms of harmonic polylogarithms. 
From three--loop order onward there are two principal contributions \cite{Blumlein:2019oas}. One part results from 
first--order factorizable recurrences which in $x$--space can be represented in terms of harmonic and 
cyclotomic harmonic polylogarithms. The second part is related to the remaining non--first order factorizing 
recurrences. We first consider their representation in terms of a series expansion around $x=1$, which we then 
match with the symbolic series expansion around $x=0$. The latter also contains a finite number of 
series--modulated logarithmic contributions. This series representation has a convergence radius of 
$x=7-4\sqrt{3}$, due to the pseudo--threshold at $q^2/m^2 = 16$. Therefore, the matching of both series has to 
be performed for small positive values of $x$. We use a rather large number of expansion coefficients for 
this, at very high numerical precision. Thanks to this, the unknown expansion coefficients at $x=0$ can
be determined by using the {\tt PSLQ} algorithm. In this way, we determine the series representations for all 
form factors around $x=0$. The coefficients turn out to depend only on multiple zeta values linearly combined over $\mathbb{Q}$.
Moreover, the $x$--series can be resummed for a series of constants into harmonic polylogarithms in $x$ using the packages \texttt{Sigma} and \texttt{HarmonicSums}. 
This applies to cases in which the corresponding recurrence (guessed with the 
package {\tt  ore\_algebra}) factorizes to first order. We could determine 
all recurrences, which allows an even faster evaluation of the corresponding series, whenever needed.

This procedure was repeated for the other two singularities (thresholds) around $q^2/m^2 = 4$ and  $q^2/m^2 = 16$.
Here, however, unlike in the high energy limit $x=0$, 
our analysis showed that multiple zeta values are not sufficient for the corresponding series representations, 
which was expected due to elliptic (or even higher order) master integrals. 
We determined the new constants $\tilde{\kappa}_i$ at high numerical precision for further special analyses. 
Our series expansions can be used in phenomenological and experimental applications.
Still we provide analytic results in terms of series representations and in some cases also resumming the respective series 
into special functions. Our representations are given in terms of expansions around the four singular points to very high precision. 
We presented detailed numerical illustrations for the real and imaginary parts of the different form factors.

The present work resulted in a deeper analytic understanding of the massive three--loop form factors, beyond
the numerical solutions given in Ref.~\cite{Fael:2022miw,Fael:2022rgm}. We have compared to the numerical 
results given in \cite{Fael:2022miw,Fael:2022rgm} and found agreement, cf.~Section~\ref{sec:5} for details.  
Ref.~\cite{Fael:2022miw,Fael:2022rgm} estimated the numerical accuracy of the $O(\ep^0)$ contribution based on that 
obtained for the pole terms, which are known analytically. This method is not always reliable.
Ancillary files are provided for analytic and numerical
comparisons. The ancillary files to the present paper are given at \cite{AFILE}.

\vspace*{5mm}\noindent
{\bf Acknowledgments.}\\
\noindent
We would like to thank A.~Behring and K.~Sch\"onwald for discussions. This work  was supported  by the 
Austrian Science Fund (FWF) grant  P33530. N.R. has very essentially contributed to the calculation of the
physics amplitude. The Feynman diagrams have been drawn using {\tt Axodraw} 
\cite{Vermaseren:1994je}.

\appendix
\section{Vector coefficients at \boldmath $x=0$ in terms of known constants}
\label{app:coefficientresults}

\vspace*{1mm}
\noindent
In this appendix we show the first few coefficients of the expansions at $x=0$ of the vector form factors
$F_{v,1,i}^{(0)}(x)$ and $F_{v,2,i}^{(0)}(x)$, with $i=1,2,3$, as given in Eq.~(\ref{x=0exactexpansion}).
The corresponding constants are given in Eqs.~(\ref{zeta2consts1}), (\ref{zeta3consts1}) and (\ref{zeta0consts1}).
In other words, we provide the first few coefficients $c_{I,i}(j,k)$ in Eq.~(\ref{Fpowlog}) for
$I \in \left\{(v,1), \, (v,2)\right\}$. The coefficients in the case of other form factors look similar.
The coefficients of 
$F_{v,1,2}^{(0)}(x)$ are given by 
\begin{eqnarray}
c_{v,1,2}(4,-2) &=& 0, \\
c_{v,1,2}(4,-1) &=& 0, \\
c_{v,1,2}(4,0) &=& 0, \\
c_{v,1,2}(4,1) &=& 0, \\
c_{v,1,2}(4,2) &=& 0, \\
c_{v,1,2}(4,3) &=& 0, \\
c_{v,1,2}(3,-2) &=& 0,  \\
c_{v,1,2}(3,-1) &=& -128, \\
c_{v,1,2}(3,0) &=& -\frac{832}{9}, \\
 c_{v,1,2}(3,1) &=& 256l_2+\frac{92416}{81}, \\
 c_{v,1,2}(3,2) &=& \frac{41344}{27}, \\
c_{v,1,2}(3,3) &=& 256l_2+\frac{17920}{81}, \\
 c_{v,1,2}(2,-2) &=& 0, \\
 c_{v,1,2}(2,-1) &=& \frac{512}{9}l_2+\frac{18464}{81}, \\ 
 c_{v,1,2}(2,0) &=& -\frac{15616}{9}l_2-\frac{97792}{81}, \\
 c_{v,1,2}(2,1) &=& -256\zeta_2+2048l_2-\frac{995296}{81}, \\
 c_{v,1,2}(2,2) &=& -\frac{1024}{3}l_2+\frac{579520}{27}, \\
c_{v,1,2}(2,3) &=& -256\zeta_2-\frac{7168}{9}l_2-\frac{2448448}{81}, \\
 c_{v,1,2}(1,-2) &=& 0, \\
 c_{v,1,2}(1,-1) &=& \frac{14656}{27}\zeta_2+\frac{62080}{27}l_2-\frac{512}{3}l_2^2+\frac{125968}{81}, \\
 c_{v,1,2}(1,0) &=& \frac{256}{9}\zeta_2-\frac{432128}{27}l_2+\frac{2560}{3}l_2^2-\frac{870272}{81}, \\
 c_{v,1,2}(1,1) &=& -\frac{71360}{27}\zeta_2+\frac{290944}{27}l_2-1536l_2^2-512\zeta_3+\frac{9540560}{243}, \\
 c_{v,1,2}(1,2) &=& -\frac{207872}{27}\zeta_2-\frac{182272}{27}l_2+\frac{5120}{3}l_2^2-\frac{1943552}{27}, \\
c_{v,1,2}(1,3) &=& \frac{101888}{27}\zeta_2-\frac{131584}{3}l_2-\frac{5120}{3}l_2^2-512\zeta_3+\frac{18393728}{243}, \\
 c_{v,1,2}(0,-2) &=& -\frac{177152}{81}l_2, \\
 c_{v,1,2}(0,-1) &=& \frac{156256}{81}\zeta_2+\frac{1429312}{81}l_2-\frac{37120}{9}\zeta_2 l_2 \nonumber \\ &&
 -\frac{512}{3}l_2^2-\frac{224}{3}\zeta_3+\frac{2731388}{81}, \\
 c_{v,1,2}(0,0) &=& \frac{166528}{27}\zeta_2-\frac{6199040}{81}l_2+\frac{149504}{9}\zeta_2 l_2 \nonumber \\ &&
 +\frac{2048}{3}l_2^2+\frac{9856}{9}\zeta_3-\frac{43085848}{243}, \\
 c_{v,1,2}(0,1) &=& -\frac{1033504}{27}\zeta_2-128\zeta_2^2+\frac{1066304}{9}l_2-31488\zeta_2 l_2 \nonumber \\ &&
 -\frac{6656}{3}l_2^2+\frac{3296}{3}\zeta_3+\frac{46115828}{243}, \\
 c_{v,1,2}(0,2) &=& \frac{486400}{27}\zeta_2-\frac{14123776}{81}l_2+\frac{299008}{9}\zeta_2 l_2 \nonumber \\ &&
 +4096l_2^2-\frac{14080}{27}\zeta_3-\frac{159872992}{729}, \\
c_{v,1,2}(0,3) &=& -\frac{873344}{27}\zeta_2-128\zeta_2^2+\frac{5127808}{27}l_2-\frac{320512}{9}\zeta_2 l_2 \nonumber \\ &&
-6144l_2^2 -1024\zeta_3+\frac{706205288}{3645}\,. 
\end{eqnarray}
The coefficients of $F_{v,2,2}^{(0)}(x)$ read 
\begin{eqnarray}
c_{v,2,2}(4,-2) &=& 0,\\
c_{v,2,2}(4,-1) &=& 0,\\
c_{v,2,2}(4,0)  &=& 8,\\
c_{v,2,2}(4,1)  &=& 0,\\
c_{v,2,2}(4,2)  &=& 16,\\
c_{v,2,2}(4,3)  &=& 0,\\
c_{v,2,2}(3,-2) &=& 0, \\
c_{v,2,2}(3,-1) &=& -64, \\
c_{v,2,2}(3,0) &=& 128l_2+\frac{6592}{27}, \\ 
c_{v,2,2}(3,1) &=& -512l_2-\frac{47968}{27}, \\
c_{v,2,2}(3,2) &=& 256l_2+\frac{1888}{27}, \\ 
c_{v,2,2}(3,3) &=& -512l_2-\frac{79552}{27}, \\
c_{v,2,2}(2,-2) &=& 0, \\
c_{v,2,2}(2,-1) &=& \frac{256}{9}l_2-\frac{320}{81}, \\
c_{v,2,2}(2,0) &=& 64\zeta_2+\frac{10880}{9}l_2+\frac{113536}{81}, \\
c_{v,2,2}(2,1) &=& -64\zeta_2-\frac{8576}{9}l_2+\frac{183632}{27}, \\
c_{v,2,2}(2,2) &=& 128\zeta_2+\frac{896}{9}l_2-\frac{176864}{27}, \\
c_{v,2,2}(2,3) &=& -64\zeta_2-\frac{22528}{9}l_2+\frac{174184}{27}, \\
c_{v,2,2}(1,-2) &=&  0,\\
c_{v,2,2}(1,-1) &=& \frac{7328}{27}\zeta_2+\frac{27968}{27}l_2-\frac{256}{3}l_2^2+\frac{1240}{81}, \\
c_{v,2,2}(1,0) &=& -\frac{5056}{9}\zeta_2+\frac{8960}{3}l_2-\frac{512}{3}l_2^2+\frac{33008}{27}, \\
c_{v,2,2}(1,1) &=& \frac{161696}{27}\zeta_2+\frac{242368}{27}l_2+\frac{512}{3}l_2^2+128\zeta_3-\frac{63064}{27}, \\
c_{v,2,2}(1,2) &=& \frac{18560}{27}\zeta_2+\frac{63872}{9}l_2-\frac{1024}{3}l_2^2+\frac{861728}{81}, \\
c_{v,2,2}(1,3) &=& \frac{203392}{27}\zeta_2+\frac{319424}{27}l_2+\frac{256}{3}l_2^2+128\zeta_3-\frac{2598124}{243}, \\
c_{v,2,2}(0,-2) &=& -\frac{15872}{27}l_2, \\
c_{v,2,2}(0,-1) &=& \frac{82256}{81}\zeta_2+\frac{367808}{81}l_2-\frac{4736}{9}\zeta_2 l_2+\frac{64}{3}l_2^2 \nonumber \\ &&
-\frac{592}{3}\zeta_3+\frac{3186380}{243}, \\
c_{v,2,2}(0,0) &=& -6560\zeta_2+\frac{2112}{5}\zeta_2^2+\frac{103168}{9}l_2+\frac{4352}{9}\zeta_2 l_2 \nonumber \\ &&
-\frac{10496}{9}\zeta_3+\frac{10523224}{243}, \\
c_{v,2,2}(0,1) &=& \frac{3050672}{81}\zeta_2-\frac{3008}{5}\zeta_2^2+\frac{140480}{27}l_2+\frac{16000}{9}\zeta_2 l_2 \nonumber \\ &&
+\frac{64}{3}l_2^2+\frac{7664}{9}\zeta_3+\frac{59828}{243}, \\
c_{v,2,2}(0,2) &=& -\frac{529600}{27}\zeta_2+\frac{4224}{5}\zeta_2^2-\frac{475072}{81}l_2+\frac{8704}{9}\zeta_2 l_2 \nonumber \\ &&
-\frac{23936}{27}\zeta_3-\frac{18467960}{729}, \\
c_{v,2,2}(0,3) &=& \frac{3651488}{81}\zeta_2-\frac{3008}{5}\zeta_2^2-\frac{151184}{27}l_2+\frac{11264}{9}\zeta_2 l_2 \nonumber \\ &&
+1408\zeta_3-\frac{29436937}{1215}\,. 
\end{eqnarray}
The first coefficients of $F_{v,1,3}^{(0)}(x)$ are 
\begin{eqnarray}
c_{v,1,3}(5,-2) &=& 0,\\ 
c_{v,1,3}(5,-1) &=& 0,\\
c_{v,1,3}(5,0)  &=& 0,\\
c_{v,1,3}(5,1)  &=& \frac{224}{5 \pi^2},\\
c_{v,1,3}(5,2)  &=& 0,\\
c_{v,1,3}(5,3)  &=& \frac{224}{5 \pi^2},\\
c_{v,1,3}(4,-2) &=& 0,\\
c_{v,1,3}(4,-1) &=& -\frac{1904}{81 \pi ^2}, \\  
c_{v,1,3}(4,0)  &=& -\frac{3248}{81 \pi ^2}, \\
c_{v,1,3}(4,1)  &=& \frac{3080}{27 \pi ^2}, \\
c_{v,1,3}(4,2)  &=& \frac{52640}{27 \pi ^2}, \\ 
c_{v,1,3}(4,3)  &=& \frac{44128}{81 \pi ^2}, \\ 
c_{v,1,3}(3,-2) &=& 0, \\ 
c_{v,1,3}(3,-1) &=& -\frac{70336}{243 \pi ^2}, \\ 
c_{v,1,3}(3,0)  &=& \frac{250880}{243 \pi ^2}, \\
c_{v,1,3}(3,1)  &=& -\frac{512}{9}-\frac{117712}{81 \pi ^2}, \\
c_{v,1,3}(3,2)  &=& \frac{5944064}{243 \pi ^2}, \\
c_{v,1,3}(3,3)  &=& -\frac{512}{9}-\frac{5339824}{243 \pi ^2}, \\
c_{v,1,3}(2,-2) &=& 0, \\
c_{v,1,3}(2,-1) &=& -\frac{17696}{81}-\frac{447104}{243 \pi ^2}, \\ 
c_{v,1,3}(2,0)  &=& \frac{2964416}{243 \pi ^2}-\frac{58240}{81}, \\
c_{v,1,3}(2,1)  &=& 896l_2+\frac{896}{\pi ^2}\zeta_3+\frac{185888}{81}-\frac{11501560}{243 \pi ^2}, \\
c_{v,1,3}(2,2)  &=& \frac{143104}{81}+\frac{13919248}{81 \pi ^2}, \\
c_{v,1,3}(2,3)  &=& 896l_2+\frac{896}{\pi ^2}\zeta_3+\frac{69952}{27}-\frac{28909132}{243 \pi ^2}, \\
c_{v,1,3}(1,-2) &=& 0,\\
c_{v,1,3}(1,-1) &=& -\frac{1792}{9}l_2+\frac{60032}{27 \pi ^2}\zeta_3+\frac{87056}{243}-\frac{1974784}{243 \pi ^2}, \\
c_{v,1,3}(1,0)  &=& -\frac{7168}{9}l_2-\frac{12544}{\pi ^2}\zeta_3+\frac{11072}{9}+\frac{662144}{27 \pi 
^2}, \\
c_{v,1,3}(1,1)  &=& \frac{9856}{5}\zeta_2+\frac{490112}{27 \pi 
^2}\zeta_3-\frac{4687888}{243}-\frac{2968168}{243 \pi ^2}, \\
c_{v,1,3}(1,2)  &=& \frac{17920}{9}l_2-\frac{272384}{9 \pi 
^2}\zeta_3+\frac{7113728}{243}+\frac{145353152}{729 \pi ^2}, \\
c_{v,1,3}(1,3)  &=& -\frac{10150048}{243} + \frac{14336 l_2}{9} - \frac{115553998}{729 \pi^2} + \frac{9856 \zeta_2}{5} - \frac{1250816 \zeta_3}{27 \pi^2},\\
c_{v,1,3}(0,-2) &=& -\frac{221440}{243}, \\
c_{v,1,3}(0,-1) &=& -\frac{14560}{81}\zeta_2+\frac{26432}{81}l_2+\frac{794752}{81 \pi ^2}\zeta_3+\frac{11712400}{729}-\frac{4659760}{243 \pi ^2}, \\
c_{v,1,3}(0,0) &=& -\frac{590848}{135}\zeta_2-\frac{105728}{27}l_2-\frac{160384}{3 \pi ^2}\zeta_3+\frac{17780672}{243 \pi ^2}-\frac{71864320}{729}, \\
c_{v,1,3}(0,1) &=& \frac{2887936}{405}\zeta_2-\frac{688576}{81}l_2+3584\zeta_2 l_2+1344 \zeta_3
+\frac{10114048}{81 \pi ^2}\zeta_3 \nonumber \\ &&
 +\frac{448}{\pi ^2}\zeta_5+\frac{120607088}{729}-\frac{3788428}{27 \pi ^2}, \\
 c_{v,1,3}(0,2) &=& -\frac{1532672}{405}\zeta_2-\frac{476672}{81}l_2-\frac{8866816}{81 \pi ^2}\zeta_3+\frac{653277352}{2187 \pi ^2} \nonumber \\ &&
 -\frac{192621760}{729}, \\
c_{v,1,3}(0,3) &=& -\frac{2014528}{405}\zeta_2+\frac{68992}{3}l_2+3584\zeta_2 l_2+1344 \zeta_3 -\frac{905408}{81 \pi ^2}\zeta_3 \nonumber \\ &&
+\frac{448}{\pi ^2}\zeta_5+\frac{194264248}{729}-\frac{1214278177}{4374 \pi ^2}, 
\end{eqnarray}
and likewise for $F_{v,2,3}^{(0)}(x)$ 
\begin{eqnarray}
c_{v,2,3}(5,-2) &=& 0,\\
c_{v,2,3}(5,-1) &=& 0,\\
c_{v,2,3}(5,0) &=& -\frac{112}{15 \pi ^2},\\
c_{v,2,3}(5,1) &=& \frac{476}{15 \pi ^2},\\
c_{v,2,3}(5,2) &=& -\frac{224}{15 \pi ^2},\\
c_{v,2,3}(5,3) &=& \frac{476}{15 \pi ^2},\\ 
c_{v,2,3}(4,-2)&=& 0,\\
c_{v,2,3}(4,-1)&=& -\frac{952}{81 \pi^2},\\
c_{v,2,3}(4,0) &=& \frac{6944}{81 \pi ^2},\\
c_{v,2,3}(4,1) &=& -\frac{14056}{27 \pi^2},\\
c_{v,2,3}(4,2) &=& -\frac{728}{3 \pi ^2},\\
c_{v,2,3}(4,3) &=& -\frac{162400}{81 \pi^2},\\
c_{v,2,3}(3,-2)&=& 0,\\
c_{v,2,3}(3,-1)&=& -\frac{39200}{243 \pi ^2},\\
c_{v,2,3}(3,0)&=& \frac{320}{9}+\frac{7616}{81 \pi ^2},\\
c_{v,2,3}(3,1)&=& -\frac{1312}{9}-\frac{941248}{243 \pi^2},\\
c_{v,2,3}(3,2)&=& \frac{640}{9}-\frac{790216}{243 \pi ^2},\\
c_{v,2,3}(3,3)&=& -\frac{1312}{9}-\frac{176512}{27 \pi ^2},\\
c_{v,2,3}(2,-2)&=& 0,\\
c_{v,2,3}(2,-1)&=& -\frac{8848}{81}-\frac{322336}{243 \pi ^2},\\
c_{v,2,3}(2,0)&=& \frac{32560}{81}-448 l_2+\frac{677152}{243 \pi ^2}+\frac{896 \zeta_3}{\pi ^2},\\
c_{v,2,3}(2,1)&=& -\frac{273584}{81}-\frac{3312008}{243 \pi ^2}-\frac{2688 \zeta_3}{\pi ^2}\\
c_{v,2,3}(2,2)&=& -\frac{70624}{81}-896 l_2-\frac{2675876}{81 \pi ^2}
+
\frac{1792 \zeta_3}{\pi ^2},\\
c_{v,2,3}(2,3)&=& -\frac{183808}{27}
-\frac{1477532}{27 \pi ^2}
-\frac{2688 \zeta_3}{\pi ^2},\\
c_{v,2,3}(1,-2)&=& 0,\\
c_{v,2,3}(1,-1)&=& -\frac{126008}{243}
-\frac{896 l_2}{9}
-\frac{1073632}{243 \pi ^2}
+\frac{30016 \zeta_3}{27 \pi ^2},\\
c_{v,2,3}(1,0)&=& \frac{130192}{243}
-\frac{5824 l_2}{9}
-\frac{801472}{81 \pi ^2}
+64 \zeta_2
+\frac{25088 \zeta_3}{3 \pi ^2},\\
c_{v,2,3}(1,1)&=& \frac{1937720}{243}
-\frac{30464 l_2}{9}
-\frac{13104616}{243 \pi ^2}
-\frac{27232 \zeta_2}{15}
+\frac{169792 \zeta_3}{27 \pi ^2},\\
c_{v,2,3}(1,2)&=& -\frac{423584}{81}
-\frac{19712 l_2}{9}
-\frac{36706208}{729 \pi ^2}
+128 \zeta_2
+\frac{186368 \zeta_3}{9 \pi ^2},\\
c_{v,2,3}(1,3)&=& \frac{2046752}{243}
-\frac{52864 l_2}{9}
-\frac{54828256}{729 \pi ^2}
-\frac{27232 \zeta_2}{15}
+\frac{982016 \zeta_3}{27 \pi ^2},\\
c_{v,2,3}(0,-2)&=& -\frac{19840}{81},\\
c_{v,2,3}(0,-1)&=& \frac{4465256}{729}
+\frac{38752 l_2}{81}
-\frac{1908592}{243 \pi ^2}
+\frac{124304 \zeta_2}{405}
+\frac{85568 \zeta_3}{81 \pi ^2},\\
c_{v,2,3}(0,0)&=& \frac{13510576}{729}
+\frac{72128 l_2}{81}
-\frac{2291072}{243 \pi ^2}
+\frac{549968 \zeta_2}{135}
-1792 l_2 \zeta_2
-\frac{2432 \zeta_3}{3}
\nonumber\\ && 
+\frac{14336 \zeta_3}{81 \pi ^2}
+\frac{10752 \zeta_5}{\pi ^2},\\
c_{v,2,3}(0,1)&=& \frac{3356632}{729}
+7392 l_2
+\frac{1120700}{243 \pi ^2}
-\frac{2580928 \zeta_2}{405}
-\frac{1184 \zeta_3}{3}
-\frac{2202368 \zeta_3}{81 \pi ^2}
\nonumber\\ &&
-\frac{6944 \zeta_5}{\pi ^2},\\
c_{v,2,3}(0,2)&=& \frac{5003536}{729}
-\frac{26432 l_2}{9}
-\frac{110330920}{2187 \pi ^2}
+\frac{2573792 \zeta_2}{405}
-3584 l_2 \zeta_2
-\frac{4864 \zeta_3}{3}
\nonumber\\ &&
+\frac{674240 \zeta_3}{81 \pi ^2}
+\frac{21504 \zeta_5}{\pi ^2},\\
c_{v,2,3}(0,3)&=& -\frac{4741880}{729}
+\frac{363776 l_2}{81}
+\frac{102560675}{4374 \pi ^2}
-\frac{1971968 \zeta_2}{405}
-\frac{1184 \zeta_3}{3}
\nonumber\\ &&
+\frac{151648 \zeta_3}{81 \pi ^2}
-\frac{6944 \zeta_5}{\pi ^2}.
\end{eqnarray}
The coefficients of $F_{v,1,1}^{(0)}(x)$ are 
\begin{eqnarray}
c_{v,1,1}(6,-2) &=& 0,\\
c_{v,1,1}(6,-1) &=& 0,\\
c_{v,1,1}(6,0)  &=& 0,\\
c_{v,1,1}(6,1)  &=& 0,\\
c_{v,1,1}(6,2)  &=& 0,\\
c_{v,1,1}(6,3)  &=& 0,\\
c_{v,1,1}(5,-2) &=& 0,\\
c_{v,1,1}(5,-1) &=& -\frac{704}{81}, \\ 
c_{v,1,1}(5,0)  &=& \frac{13904}{405}, \\ 
c_{v,1,1}(5,1)  &=& -\frac{88}{45} -\frac{224}{5} \frac{\zeta_3}{\pi ^2}, \\ 
c_{v,1,1}(5,2)  &=& \frac{6944}{81}, \\ 
c_{v,1,1}(5,3)  &=& -\frac{22384}{405} - \frac{224 \zeta_3}{5 \pi^2},  \\ 
c_{v,1,1}(4,-2) &=& 0,\\
c_{v,1,1}(4,-1) &=& -\frac{688}{243} +\frac{1904}{81} \frac{\zeta_3}{\pi ^2}, \\ 
c_{v,1,1}(4,0) &=& \frac{6944}{81} +\frac{3248}{81} \frac{\zeta_3}{\pi ^2}, \\ 
c_{v,1,1}(4,1) &=& -\frac{193676}{243} +16 \zeta_2 -\frac{3080}{27} \frac{\zeta_3}{\pi ^2}, \\ 
c_{v,1,1}(4,2) &=& \frac{63224}{81} -\frac{52640}{27} \frac{\zeta_3}{\pi ^2}, \\ 
c_{v,1,1}(4,3) &=& -\frac{190484}{81} +16 \zeta_2 -\frac{44128}{81} \frac{\zeta_3}{\pi ^2}, \\ 
c_{v,1,1}(3,-2) &=& 0,\\
c_{v,1,1}(3,-1) &=& 240 + \frac{256}{3} \zeta_2 +\frac{70336}{243} \frac{\zeta_3}{\pi ^2}, \\ 
c_{v,1,1}(3,0) &=& -\frac{443584}{243} -\frac{7936}{81} \zeta_2 -\frac{250880}{243} \frac{\zeta_3}{\pi ^2}, \\ 
c_{v,1,1}(3,1) &=& \frac{4446616}{729} -\frac{16832}{81} \zeta_2 +\frac{512}{9} \zeta_3 +\frac{117712}{81} \frac{\zeta_3}{\pi ^2}, \\ 
c_{v,1,1}(3,2) &=& -\frac{4453616}{243} -\frac{8576}{9} \zeta_2 -\frac{5944064}{243} \frac{\zeta_3}{\pi ^2}, \\ 
c_{v,1,1}(3,3) &=& \frac{15976564}{729} +\frac{5632}{81} \zeta_2 +\frac{512}{9} \zeta_3 +\frac{5339824}{243} \frac{\zeta_3}{\pi ^2}, \\ 
c_{v,1,1}(2,-2) &=& 0,\\
c_{v,1,1}(2,-1) &=& \frac{73096}{27} -\frac{3136}{27} \zeta_2 -\frac{736}{81} \zeta_3 +\frac{447104}{243} \frac{\zeta_3}{\pi ^2}, \\ 
c_{v,1,1}(2,0) &=& -\frac{3005056}{243} +\frac{13952}{27} \zeta_2 +\frac{194752}{81} \zeta_3 -\frac{2964416}{243} \frac{\zeta_3}{\pi ^2}, \\ 
c_{v,1,1}(2,1) &=& \frac{6205580}{729} +\frac{157280}{27} \zeta_2 +\frac{1184}{5} \zeta_2^2 +256 \zeta_2 l_2^2-\frac{128}{3} l_2^4          \nonumber \\ &&
-1024 a_4-\frac{533408}{81} \zeta_3 +\frac{11501560}{243} \frac{\zeta_3}{\pi ^2} -896 \zeta_3 l_2 -896 \frac{\zeta_3^2}{\pi ^2}, \\ 
c_{v,1,1}(2,2) &=& -\frac{25654520}{729} -\frac{22144}{3} \zeta_2 +\frac{353408}{81} \zeta_3 -\frac{13919248}{81} \frac{\zeta_3}{\pi ^2}, \\ 
c_{v,1,1}(2,3) &=& \frac{10452319}{1215} +\frac{312752}{27} \zeta_2 +\frac{1184}{5} \zeta_2^2 +256 \zeta_2 l_2^2-\frac{128}{3} l_2^4      \nonumber \\ &&
-1024 a_4-\frac{151616}{27} \zeta_3 +\frac{28909132}{243} \frac{\zeta_3}{\pi ^2} -896 \zeta_3 l_2 -896 \frac{\zeta_3^2}{\pi ^2}, \\ 
c_{v,1,1}(1,-2) &=& 0,\\
c_{v,1,1}(1,-1) &=& \frac{2681840}{243} -\frac{239264}{243} \zeta_2 +\frac{4544}{45} \zeta_2^2 -\frac{512}{9} \zeta_2 l_2^2+\frac{256}{27} l_2^4     \nonumber \\ &&
+\frac{2048}{9} a_4 -\frac{755168}{243} \zeta_3 +\frac{1974784}{243} \frac{\zeta_3}{\pi ^2}
+\frac{1792}{9} \zeta_3 l_2 -\frac{60032}{27} \frac{\zeta_3^2}{\pi ^2}, \\ 
c_{v,1,1}(1,0) &=& -\frac{11847904}{243} +\frac{1762496}{243} \zeta_2 -\frac{13952}{135} \zeta_2^2 -\frac{2048}{9} \zeta_2 l_2^2+\frac{1024}{27} l_2^4     \nonumber \\ &&
+\frac{8192}{9} a_4 +\frac{960512}{81} \zeta_3 -\frac{662144}{27} \frac{\zeta_3}{\pi ^2} 
+\frac{7168}{9} \zeta_3 l_2 +12544 \frac{\zeta_3^2}{\pi ^2}, \\ 
c_{v,1,1}(1,1) &=& \frac{16988900}{2187} -\frac{4210432}{243} \zeta_2 -\frac{6016}{3} \zeta_2^2 +\frac{6957088}{243} \zeta_3            \nonumber \\ &&
+\frac{2968168}{243} \frac{\zeta_3}{\pi ^2} -\frac{9856}{5} \zeta_2 \zeta_3 -\frac{490112}{27} \frac{\zeta_3^2}{\pi ^2} +1280 \zeta_5, \\ 
c_{v,1,1}(1,2) &=& -\frac{23135480}{243} +\frac{9932032}{243} \zeta_2 -\frac{447872}{135} \zeta_2^2 +\frac{5120}{9} \zeta_2 l_2^2    \nonumber \\ &&
-\frac{2560}{27} l_2^4 -\frac{20480}{9} a_4 -\frac{3832448}{243} \zeta_3-\frac{145353152}{729} \frac{\zeta_3}{\pi ^2}          \nonumber \\ &&
-\frac{17920}{9} \zeta_3 l_2 +\frac{272384}{9} \frac{\zeta_3^2}{\pi ^2}, \\ 
c_{v,1,1}(1,3) &=& \frac{2962613437}{36450} -\frac{4402552}{81} \zeta_2 +\frac{51584}{45} \zeta_2^2 +\frac{4096}{9} \zeta_2 l_2^2    \nonumber \\ &&
-\frac{2048}{27} l_2^4 -\frac{16384}{9} a_4 +\frac{15899680}{243} \zeta_3 +\frac{115553998}{729} \frac{\zeta_3}{\pi ^2}   \nonumber \\ &&
-\frac{9856}{5} \zeta_2 \zeta_3 -\frac{14336}{9} \zeta_3 l_2 +\frac{1250816}{27} \frac{\zeta_3^2}{\pi ^2} +1280 \zeta_5, \\ 
c_{v,1,1}(0,-2) &=& \frac{221440}{243} \zeta_3, \\ 
c_{v,1,1}(0,-1) &=& -\frac{28904}{9} -\frac{743888}{243} \zeta_2 -\frac{95440}{81} \zeta_2^2 +\frac{256}{9} \zeta_2^2 l_2 +\frac{9344}{81} \zeta_2 l_2^2     \nonumber \\ &&
-\frac{4672}{243} l_2^4 -\frac{37376}{81} a_4 -\frac{10044736}{729} \zeta_3 +\frac{4659760}{243} \frac{\zeta_3}{\pi ^2}   \nonumber \\ &&
-\frac{38432}{81} \zeta_2 \zeta_3 -\frac{26432}{81} \zeta_3 l_2 -\frac{794752}{81} \frac{\zeta_3^2}{\pi ^2} +\frac{15856}{9} \zeta_5, \\ 
c_{v,1,1}(0,0) &=& \frac{10612672}{243} +\frac{3300608}{243} \zeta_2 +\frac{349888}{135} \zeta_2^2 -\frac{1024}{9} \zeta_2^2 l_2 -\frac{268480}{9} \zeta_5      \nonumber \\ &&
-\frac{25088}{27} \zeta_2 l_2^2 +\frac{12544}{81} l_2^4 +\frac{100352}{27} a_4 +\frac{46482880}{729} \zeta_3     \nonumber \\ &&
-\frac{17780672}{243} \frac{\zeta_3}{\pi ^2} +\frac{1078528}{135} \zeta_2 \zeta_3 +\frac{105728}{27} \zeta_3 l_2 +\frac{160384}{3} \frac{\zeta_3^2}{\pi ^2}, \\ 
c_{v,1,1}(0,1) &=& -\frac{641594606}{6561} +\frac{295424}{81} \zeta_2 +\frac{5199536}{405} \zeta_2^2 
+\frac{63296}{35} \zeta_2^3            \nonumber \\ &&
-\frac{256}{9} \zeta_2^2 l_2 -\frac{343936}{81} \zeta_2 l_2^2 +1024 \zeta_2^2 l_2^2 
-448 \frac{\zeta_3 \zeta_5}{\pi ^2}               \nonumber \\ &&
+\frac{171968}{243} l_2^4 -\frac{512}{3} \zeta_2 l_2^4 +\frac{1375744}{81} a_4 
+\frac{1642288}{27} \zeta_5               \nonumber \\ &&
-4096 \zeta_2 a_4 -\frac{82803488}{729} \zeta_3 +\frac{3788428}{27} \frac{\zeta_3}{\pi ^2} 
-\frac{5779456}{405} \zeta_2 \zeta_3     \nonumber \\ &&
+\frac{688576}{81} \zeta_3 l_2-3584 \zeta_2 \zeta_3 l_2 -1024 \zeta_3^2 
-\frac{10114048}{81} \frac{\zeta_3^2}{\pi ^2}, \\
c_{v,1,1}(0,2) &=& 
\frac{285175364}{6561}
+\frac{1200128 a_4}{81}
+\frac{150016 l_2^4}{243}
+\frac{5818816 \zeta_2}{243}
-\frac{300032}{81} l_2^2 \zeta_2
\nonumber\\ && 
-\frac{7099136 \zeta_2^2}{405}
-\frac{2048}{9} l_2 \zeta_2^2
+\frac{186528256 \zeta_3}{729}
+\frac{476672 l_2 \zeta_3}{81}
-\frac{653277352 \zeta_3}{2187 \pi ^2}
\nonumber\\ && 
+\frac{4719872 \zeta_2 \zeta_3}{405}
+\frac{8866816 \zeta_3^2}{81 \pi ^2}
-\frac{653696 \zeta_5}{9},\\
c_{v,1,1}(0,3) &=&  -\frac{1906863860207}{6561000}
-\frac{185344 a_4}{3}
-\frac{23168 l_2^4}{9}
-\frac{15147034 \zeta_2}{729}
-4096 a_4 \zeta_2
\nonumber\\ && 
+\frac{46336 l_2^2 \zeta_2}{3}
-\frac{512}{3} l_2^4 \zeta_2
+\frac{8162192 \zeta_2^2}{135}
+1024 l_2^2 \zeta_2^2
+\frac{63296 \zeta_2^3}{35}
\nonumber\\ && 
-\frac{169868872 \zeta_3}{729}
-\frac{68992 l_2 \zeta_3}{3}
+\frac{1214278177 \zeta_3}{4374 \pi ^2}
-\frac{1337792}{405} \zeta_2 \zeta_3
\nonumber\\ &&
-3584 l_2 \zeta_2 \zeta_3
-1024 \zeta_3^2
+\frac{905408 \zeta_3^2}{81 \pi ^2}
+\frac{1681792 \zeta_5}{27}
-\frac{448 \zeta_3 \zeta_5}{\pi ^2},
\end{eqnarray}
and those of $F_{v,2,1}^{(0)}(x)$ 
\begin{eqnarray}
c_{v,2,1}(6,-2) &=& 0,\\
c_{v,2,1}(6,-1) &=& 0,\\
c_{v,2,1}(6,0)  &=& 0,\\
c_{v,2,1}(6,1)  &=& -\frac{34}{45},\\ 
c_{v,2,1}(6,2)  &=& 0,\\
c_{v,2,1}(6,3)  &=& -\frac{34}{45},\\ 
c_{v,2,1}(5,-2) &=& 0,\\
c_{v,2,1}(5,-1) &=& -\frac{352}{81}, \\ 
c_{v,2,1}(5,0)  &=& -\frac{704}{81} +\frac{112}{15} \frac{\zeta_3}{\pi ^2}, \\ 
c_{v,2,1}(5,1)  &=& -\frac{17576}{405} -\frac{476}{15} \frac{\zeta_3}{\pi ^2}, \\ 
c_{v,2,1}(5,2)  &=& -\frac{6824}{405} +\frac{224}{15} \frac{\zeta_3}{\pi ^2}, \\ 
c_{v,2,1}(5,3)  &=& -\frac{31456}{405} -\frac{476}{15} \frac{\zeta_3}{\pi ^2}, \\ 
c_{v,2,1}(4,-2) &=& 0,\\
c_{v,2,1}(4,-1) &=& \frac{976}{243} +\frac{952}{81} \frac{\zeta_3}{\pi ^2}, \\ 
c_{v,2,1}(4,0)  &=& \frac{4784}{243} -8 \zeta_2 -\frac{6944}{81} \frac{\zeta_3}{\pi ^2}, \\ 
c_{v,2,1}(4,1)  &=& \frac{29620}{81} +\frac{14056}{27} \frac{\zeta_3}{\pi ^2}, \\ 
c_{v,2,1}(4,2)  &=& \frac{29512}{243} -16 \zeta_2 +\frac{728}{3} \frac{\zeta_3}{\pi ^2}, \\ 
c_{v,2,1}(4,3)  &=& \frac{113356}{243} +\frac{162400}{81} \frac{\zeta_3}{\pi ^2}, \\ 
c_{v,2,1}(3,-2) &=& 0,\\
c_{v,2,1}(3,-1) &=& \frac{5576}{27} +\frac{128}{3} \zeta_2 +\frac{39200}{243} \frac{\zeta_3}{\pi ^2}, \\ 
c_{v,2,1}(3,0)  &=& -\frac{31984}{81} -\frac{1888}{27} \zeta_2 -\frac{224}{9} \zeta_3 -\frac{7616}{81} 
\frac{\zeta_3}{\pi ^2}, \\ 
c_{v,2,1}(3,1)  &=& \frac{199544}{81} +\frac{58528}{81} \zeta_2 +\frac{928}{9} \zeta_3 +\frac{941248}{243} 
\frac{\zeta_3}{\pi ^2}, \\ 
c_{v,2,1}(3,2)  &=& -\frac{26168}{27} +\frac{9184}{81} \zeta_2 -\frac{448}{9} \zeta_3 +\frac{790216}{243} 
\frac{\zeta_3}{\pi ^2}, \\ 
c_{v,2,1}(3,3)  &=& \frac{1605320}{729} +\frac{101152}{81} \zeta_2 +\frac{928}{9} \zeta_3 +\frac{176512}{27} 
\frac{\zeta_3}{\pi ^2}, \\ 
c_{v,2,1}(2,-2) &=& 0,\\
c_{v,2,1}(2,-1) &=& \frac{43232}{27} -\frac{944}{27} \zeta_2 -\frac{368}{81} \zeta_3 +\frac{322336}{243} \frac{\zeta_3}{\pi ^2}, \\ 
c_{v,2,1}(2,0)  &=& \frac{130480}{27} -\frac{17872}{27} \zeta_2 -\frac{816}{5} \zeta_2^2 -128 \zeta_2 l_2^2 
+\frac{64}{3} l_2^4     \nonumber \\ &&
+512 a_4 -\frac{92944}{81} \zeta_3 -\frac{677152}{243} \frac{\zeta_3}{\pi ^2} +448 \zeta_3 l_2 -896 \frac{\zeta_3^2}{\pi ^2}, \\ 
c_{v,2,1}(2,1)  &=& \frac{875360}{243} -\frac{86704}{27} \zeta_2 +\frac{544}{5} \zeta_2^2 +\frac{375056}{81} 
\zeta_3 +\frac{3312008}{243} \frac{\zeta_3}{\pi ^2}     \nonumber \\ &&
+2688 \frac{\zeta_3^2}{\pi ^2}, \\ 
c_{v,2,1}(2,2)  &=& \frac{12501296}{729} +\frac{69008}{27} \zeta_2 -\frac{1632}{5} \zeta_2^2 -256 \zeta_2 
l_2^2 +\frac{128}{3} l_2^4      \nonumber \\ &&
+1024 a_4 -\frac{160448}{81} \zeta_3 +\frac{2675876}{81} \frac{\zeta_3}{\pi ^2} +896 \zeta_3 l_2      \nonumber \\ &&
-1792 \frac{\zeta_3^2}{\pi ^2}, \\ 
c_{v,2,1}(2,3)  &=& -\frac{9839219}{3645} -\frac{106528}{27} \zeta_2 +\frac{544}{5} \zeta_2^2 
+\frac{51712}{9} \zeta_3 +\frac{1477532}{27} \frac{\zeta_3}{\pi ^2}       \nonumber \\ &&
+2688 \frac{\zeta_3^2}{\pi ^2}, \\ 
c_{v,2,1}(1,-2) &=& 0,\\
c_{v,2,1}(1,-1) &=& \frac{1469552}{243} -\frac{128368}{243} \zeta_2 +\frac{2272}{45} \zeta_2^2 -\frac{256}{9} \zeta_2 l_2^2 +\frac{128}{27} l_2^4         \nonumber \\ &&
+\frac{1024}{9} a_4 -\frac{315184}{243} \zeta_3 +\frac{1073632}{243} \frac{\zeta_3}{\pi ^2} +\frac{896}{9} \zeta_3 l_2       \nonumber \\ &&
-\frac{30016}{27} \frac{\zeta_3^2}{\pi ^2}, \\ 
c_{v,2,1}(1,0)  &=& \frac{4850800}{243} -\frac{242336}{243} \zeta_2 -\frac{65056}{135} \zeta_2^2 
-\frac{1664}{9} \zeta_2 l_2^2 +\frac{832}{27} l_2^4       \nonumber \\ &&
+\frac{6656}{9} a_4 -\frac{804832}{243} \zeta_3 +\frac{801472}{81} \frac{\zeta_3}{\pi ^2} -64 \zeta_2 \zeta_3             \nonumber \\ &&
+\frac{5824}{9} \zeta_3 l_2 -\frac{25088}{3} \frac{\zeta_3^2}{\pi ^2} +256 \zeta_5, \\ 
c_{v,2,1}(1,1)  &=& \frac{32303216}{729} +\frac{24592}{243} \zeta_2 +\frac{261568}{135} \zeta_2^2 
-\frac{8704}{9} \zeta_2 l_2^2           \nonumber \\ &&
+\frac{4352}{27} l_2^4 +\frac{34816}{9} a_4 -\frac{7996304}{243} \zeta_3 +\frac{13104616}{243} \frac{\zeta_3}{\pi ^2}        \nonumber \\ &&
+\frac{27232}{15} \zeta_2 \zeta_3 +\frac{30464}{9} \zeta_3 l_2 -\frac{169792}{27} \frac{\zeta_3^2}{\pi ^2} -2880 \zeta_5, \\ 
c_{v,2,1}(1,2)  &=& -\frac{3166864}{243} -\frac{1317808}{243} \zeta_2 -\frac{21664}{135} \zeta_2^2 
-\frac{5632}{9} \zeta_2 l_2^2          \nonumber \\ &&
+\frac{2816}{27} l_2^4 +\frac{22528}{9} a_4 +\frac{1241824}{81} \zeta_3 +\frac{36706208}{729} \frac{\zeta_3}{\pi ^2}         \nonumber \\ &&
-128 \zeta_2 \zeta_3 +\frac{19712}{9} \zeta_3 l_2 -\frac{186368}{9} \frac{\zeta_3^2}{\pi ^2} +512 \zeta_5, \\ 
c_{v,2,1}(1,3)  &=& \frac{2238816019}{54675} +\frac{397808}{81} \zeta_2 +\frac{607072}{135} \zeta_2^2 
-\frac{15104}{9} \zeta_2 l_2^2          \nonumber \\ &&
+\frac{7552}{27} l_2^4 +\frac{60416}{9} a_4 -\frac{11472656}{243} \zeta_3 +\frac{54828256}{729} \frac{\zeta_3}{\pi ^2}       \nonumber \\ &&
+\frac{27232}{15} \zeta_2 \zeta_3 +\frac{52864}{9} \zeta_3 l_2 -\frac{982016}{27} \frac{\zeta_3^2}{\pi ^2} -2880 \zeta_5, \\ 
c_{v,2,1}(0,-2) &=& \frac{19840}{81} \zeta_3, \\ 
c_{v,2,1}(0,-1) &=& -\frac{936028}{243} -\frac{379792}{243} \zeta_2 -\frac{300856}{405} \zeta_2^2 +\frac{128}{9} \zeta_2^2 l_2  +\frac{10952}{9} \zeta_5      \nonumber \\ &&
+\frac{11200}{81} \zeta_2 l_2^2 -\frac{5600}{243} l_2^4 -\frac{44800}{81} a_4 -\frac{2901776}{729} \zeta_3           \nonumber \\ &&
+\frac{1908592}{243} \frac{\zeta_3}{\pi ^2} -\frac{179024}{405} \zeta_2 \zeta_3 -\frac{38752}{81} \zeta_3 l_2 -\frac{85568}{81} \frac{\zeta_3^2}{\pi ^2}, \\
c_{v,2,1}(0,0)  &=& \frac{102440}{27} -\frac{1816256}{243} \zeta_2 -\frac{47008}{405} \zeta_2^2 
-\frac{56288}{105} \zeta_2^3 +\frac{256}{9} \zeta_2^2 l_2      \nonumber \\ &&
+\frac{10112}{81} \zeta_2 l_2^2 -512 \zeta_2^2 l_2^2 -\frac{5056}{243} l_2^4 +\frac{256}{3} \zeta_2 l_2^4       \nonumber \\ &&
-\frac{40448}{81} a_4 +2048 a_4 \zeta_2 
-\frac{6183424}{729} \zeta_3 +\frac{2291072}{243} \frac{\zeta_3}{\pi ^2}       \nonumber \\ &&
-\frac{627088}{135} \zeta_2 \zeta_3 -\frac{72128}{81} \zeta_3 l_2 +1792 \zeta_2 \zeta_3 l_2 +\frac{2048}{3} \zeta_3^2       \nonumber \\ &&
-\frac{14336}{81} \frac{\zeta_3^2}{\pi ^2} +\frac{375584}{27} \zeta_5 -10752 \frac{\zeta_3 \zeta_5}{\pi ^2}, \\ 
c_{v,2,1}(0,1)  &=& \frac{19646248}{2187} -\frac{826384}{81} \zeta_2 -\frac{6101096}{405} \zeta_2^2 
-\frac{102272}{105} \zeta_2^3       \nonumber \\ &&
+\frac{1408}{9} \zeta_2^2 l_2 +\frac{32192}{9} \zeta_2 l_2^2 -\frac{16096}{27} l_2^4 + 6944 \frac{\zeta_3 \zeta_5}{\pi ^2}        \nonumber \\ &&
-\frac{128768}{9} a_4 -\frac{6590224}{729} \zeta_3 -\frac{1120700}{243} \frac{\zeta_3}{\pi ^2} -\frac{786136}{27} \zeta_5       \nonumber \\ &&
+\frac{3236608}{405} \zeta_2 \zeta_3 -7392 \zeta_3 l_2 -\frac{640}{3} \zeta_3^2 +\frac{2202368}{81} \frac{\zeta_3^2}{\pi ^2}, \\ 
c_{v,2,1}(0,2)  &=& 
\frac{777783418}{6561}
+\frac{65024 a_4}{9}
+\frac{8128 l_2^4}{27}
-\frac{1764008 \zeta_2}{243}
+4096 a_4 \zeta_2
-\frac{16256}{9} l_2^2 \zeta_2
\nonumber\\ && 
+\frac{512 l_2^4 \zeta_2}{3}
+\frac{163568 \zeta_2^2}{27}
+\frac{512 l_2 \zeta_2^2}{9}
-1024 l_2^2 \zeta_2^2
-\frac{112576 \zeta_2^3}{105}
-\frac{29704720 \zeta_3}{729}
\nonumber\\ && 
+\frac{26432 l_2 \zeta_3}{9}
+\frac{110330920 \zeta_3}{2187 \pi ^2}
-\frac{3101792}{405} \zeta_2 \zeta_3
+3584 l_2 \zeta_2 \zeta_3
+\frac{4096 \zeta_3^2}{3}
\nonumber\\ &&
-\frac{674240 \zeta_3^2}{81 \pi ^2}
+\frac{762688 \zeta_5}{27}
-\frac{21504 \zeta_3 \zeta_5}{\pi ^2}\\
c_{v,2,1}(0,3) &=& \frac{8943651091}{729000}
-\frac{481280 a_4}{81}
-\frac{60160 l_2^4}{243}
-\frac{8824004 \zeta_2}{729}
+\frac{120320 l_2^2 \zeta_2}{81}
\nonumber\\ && 
-\frac{3384272 \zeta_2^2}{135}
+\frac{512 l_2 \zeta_2^2}{3}
-\frac{102272 \zeta_2^3}{105}
-\frac{9461092 \zeta_3}{729}
-\frac{363776 l_2 \zeta_3}{81}
\nonumber\\ && 
-\frac{102560675 \zeta_3}{4374 \pi ^2}
+\frac{2475008 \zeta_2 \zeta_3}{405}
-\frac{640 \zeta_3^2}{3}
-\frac{151648 \zeta_3^2}{81 \pi ^2}
-\frac{479104 \zeta_5}{27}
\nonumber\\ &&
+\frac{6944 \zeta_3 \zeta_5}{\pi ^2}.
\end{eqnarray}

\section{\boldmath Expansions at $x=0$ ($\hs \rightarrow -\infty$)}
\label{app:x=0expansions}

\vspace*{1mm}
\noindent
In Eq. (\ref{eq:FV1x-expansion}), we showed the expansion up to $O(x^3)$ of the non-solvable parts of $F^{(0)}_{V,1}(x)$. 
In this appendix, we show the expansions up to $O(x^3)$ of the
remaining form factors,
\begin{eqnarray}
F^{(0)}_{V,2}(x) &=& F^{(0), \rm sol}_{V,2}(x)+n_h \biggl\{
-\frac{81920}{81} l_2 \zeta_2
+\biggl[
-3072 l_2 \zeta_2^2 - 64 \zeta_2 \zeta_3 - 672 \zeta_5 
\nonumber \\ &&
+ \frac{2176}{81} l_2^4 + \frac{9248}{45} \zeta_2^2 
- \frac{10112}{27} l_2^2 \zeta_2 + \frac{17408}{27} a_4 - \frac{18016}{9} \zeta_3 
\nonumber \\ &&
+ \frac{40768}{27} l_2 \zeta_2 + \frac{1091648}{243} 
+ \frac{1837100}{243} \zeta_2 
+ \biggl(
\frac{2048}{9} l_2 \zeta_2 + \frac{7936}{9} \zeta_3 
\nonumber \\ &&
+ \frac{14368}{9} \zeta_2 - \frac{257264}{243} 
\biggr) \ln(x)
+ \biggl(
-\frac{4456}{9} + \frac{5120}{27} \zeta_2
\biggr) \ln^2(x)  
\nonumber \\ &&
-\frac{4672}{27} \ln^3(x)
-\frac{880}{81} \ln^4(x) 
\biggr] x
+\biggl[
-4096 a_4 \zeta_2 + 1024 l_2^2 \zeta_2^2 
\nonumber \\ &&
+ 256 \zeta_3^2 - \frac{512}{3} l_2^4 \zeta_2 
- \frac{19456}{243} l_2^4 + \frac{23872}{105} \zeta_2^3 
+ \frac{83968}{9} l_2 \zeta_2^2 
\nonumber \\ &&
+ \frac{113216}{27} \zeta_2 \zeta_3 + \frac{114944}{81} l_2^2 \zeta_2 - \frac{155648}{81} a_4 
- \frac{295936}{9} l_2 \zeta_2 
\nonumber \\ &&
- \frac{1162336}{27} \zeta_5 - \frac{2363776}{81} \zeta_3 
- \frac{2463584}{243} - \frac{7840040}{81} \zeta_2 
\nonumber \\ &&
+ \frac{10280288}{405} \zeta_2^2 
+ \biggl(-512 \zeta_5 - \frac{256}{9} l_2^2 \zeta_2 + \frac{896}{27} l_2^4 + \frac{7168}{9} a_4 
\nonumber \\ &&
+ \frac{29376}{5} \zeta_2^2 - \frac{254848}{81} \zeta_3 
- \frac{257792}{27} l_2 \zeta_2 - \frac{817856}{81} \zeta_2 
\nonumber \\ &&
- \frac{1304960}{81} 
\biggr) \ln(x)
+ \biggl(
-1024 a_4 
+ 256 l_2^2 \zeta_2 
- \frac{128}{3} l_2^4 + \frac{992}{5} \zeta_2^2 
\nonumber \\ &&
- \frac{34304}{9} l_2 \zeta_2 + \frac{48896}{27} \zeta_3 - \frac{71200}{27} \zeta_2 
- \frac{684640}{243} 
\biggr) \ln^2(x) 
\nonumber \\ &&
+ \biggl(
-256 l_2 \zeta_2 - \frac{64}{3} \zeta_3 - \frac{64384}{81} \zeta_2 
+ \frac{350528}{243} 
\biggr) \ln^3(x)  
+ \frac{22976}{243} \ln^4(x)
\nonumber \\ &&
-\frac{176}{405} \ln^5(x)
\biggr] x^2
+\biggl[
20480 a_4 \zeta_2 - 5120 l_2^2 \zeta_2^2 + \frac{2560}{3} l_2^4 \zeta_2 
\nonumber \\ &&
- \frac{207872}{9} l_2 \zeta_2^2 + \frac{355520}{3} l_2 \zeta_2 
+ \frac{364096}{105} \zeta_2^3 + \frac{548480}{243} l_2^4 
+ \frac{766304}{3} \zeta_5 
\nonumber \\ &&
- \frac{647488}{27} \zeta_2 \zeta_3
- \frac{1356160}{81} l_2^2 \zeta_2 
+ \frac{4387840}{81} a_4 
+ \frac{10194944}{81} \zeta_3 
\nonumber \\ &&
- \frac{12707392}{81} \zeta_2^2 
+ \frac{92640452}{243} \zeta_2 
+ \frac{204738082}{6561} 
+ \biggl(
-768 \zeta_2 \zeta_3 
\nonumber \\ &&
+ 10112 \zeta_5 - \frac{1792}{27} l_2^4 - \frac{13312}{9} l_2^2 \zeta_2 
- \frac{14336}{9} a_4 + \frac{114944}{27} l_2 \zeta_2 
\nonumber \\ &&
- \frac{5501312}{135} \zeta_2^2 
+ \frac{5769280}{81} \zeta_3 + \frac{10840352}{243} \zeta_2 
+ \frac{25002020}{2187} 
\biggr) \ln(x)
\nonumber \\ &&
+ \biggl(
5120 a_4 - 1280 l_2^2 \zeta_2 + \frac{640}{3} l_2^4 - \frac{6496}{5} \zeta_2^2 
+ \frac{160000}{9} l_2 \zeta_2 
\nonumber \\ &&
- \frac{279616}{81} \zeta_2 - \frac{351296}{27} \zeta_3 + \frac{14492732}{729} 
\biggr) \ln^2(x) 
\nonumber \\ &&
+ \biggl(
2816 l_2 \zeta_2 
+ \frac{640}{3} \zeta_3 + \frac{152384}{27} \zeta_2 - \frac{6464984}{729} 
\biggr) \ln^3(x) 
\nonumber \\ &&
+ \biggl(
16 \zeta_2 - \frac{342692}{243} 
\biggr) \ln^4(x) 
+ \frac{6872}{81} \ln^5(x)
+ \frac{68}{45} \ln^6(x) 
\biggr] x^3
+O(x^4) \biggr\},
\end{eqnarray}
\begin{eqnarray}
F^{(0)}_{A,1}(x) &=& F^{(0), \rm sol}_{A,1}(x)+n_h \biggl\{
-\frac{44288}{81 x} l_2 \zeta_2
-1024 l_2 \zeta_2^2 - \frac{1120}{81} l_2^2 \zeta_2 - \frac{1168}{243} l_2^4 
\nonumber \\ &&
- \frac{1640}{9} \zeta_2 \zeta_3 
+ \frac{3964}{9} \zeta_5 
+ \frac{5068}{27} \zeta_2^2 
- \frac{7226}{9} - \frac{9344}{81} a_4 + \frac{46324}{81} \zeta_3 
\nonumber \\ &&
+ \frac{255952}{81} l_2 \zeta_2 
+ \frac{1862569}{243} \zeta_2 
+ \biggl(
\frac{64}{27} l_2^4 + \frac{512}{9} a_4 
- \frac{512}{9} l_2^2 \zeta_2 
\nonumber \\ &&
+ \frac{15520}{27} l_2 \zeta_2 + \frac{21728}{135} \zeta_2^2 
+ \frac{34660}{243} \zeta_2 
- \frac{55676}{81} \zeta_3 + \frac{670460}{243} 
\biggr) \ln(x) 
\nonumber \\ &&
+ \biggl(
\frac{128}{9} l_2 \zeta_2 - \frac{512}{9} \zeta_3
+ \frac{2264}{81} \zeta_2 
+ \frac{18274}{27} \biggr) \ln^2(x) 
\nonumber \\ &&
+ \biggl(
60 - \frac{32}{3} \zeta_2
\biggr) \ln^3(x) 
-\frac{172}{243} \ln^4(x) 
-\frac{176}{81} \ln^5(x)
\nonumber \\ &&
+\biggl[
\frac{1600}{243} l_2^4 + \frac{12352}{81} l_2^2 \zeta_2 + \frac{12800}{81} a_4 
+ \frac{16000}{27} \zeta_2 \zeta_3 
+ \frac{27392}{9} l_2 \zeta_2^2 
\nonumber \\ &&
- \frac{132320}{27} \zeta_5 
+ \frac{194336}{81} \zeta_2^2 - \frac{442960}{81} \zeta_3 - \frac{850912}{81} l_2 \zeta_2 
+ \frac{1637576}{243} 
\nonumber \\ &&
- \frac{2210888}{81} \zeta_2 
+ \biggl(
\frac{640}{27} l_2^4 - \frac{1664}{9} l_2^2 \zeta_2 + \frac{2912}{5} \zeta_2^2 
+ \frac{5120}{9} a_4 
\nonumber \\ &&
- \frac{13664}{81} \zeta_3 
- \frac{53248}{27} l_2 \zeta_2 - \frac{167312}{81} \zeta_2 
- \frac{1355456}{243} 
\biggr) \ln(x) 
\nonumber \\ &&
+ \biggl(
-\frac{1904}{9} \zeta_2 - \frac{3520}{9} l_2 \zeta_2 
+ \frac{4432}{27} \zeta_3 - \frac{86036}{81} 
\biggr) \ln^2(x)
\nonumber \\ &&
+ \biggl(
-\frac{7664}{81} \zeta_2 
- \frac{13088}{81} 
\biggr) \ln^3(x) 
+ \frac{17200}{243} \ln^4(x)
+ \frac{44}{405} \ln^5(x) 
\biggr] x
\nonumber \\ &&
+\biggl[
1024 a_4 \zeta_2 + 26304 l_2 \zeta_2 - 256 l_2^2 \zeta_2^2 - 48 \zeta_3^2 + \frac{128}{3} l_2^4 \zeta_2 
- \frac{24320}{3} l_2 \zeta_2^2 
\nonumber \\ &&
+ \frac{32176}{105} \zeta_2^3 + \frac{36608}{243} l_2^4 - \frac{87664}{27} \zeta_2 \zeta_3 
- \frac{100864}{81} l_2^2 \zeta_2 
- \frac{129970051}{13122} 
\nonumber \\ &&
+ \frac{292864}{81} a_4 + \frac{1359904}{81} \zeta_3 
+ \frac{4874924}{81} \zeta_2 - \frac{5137136}{405} \zeta_2^2 + \frac{249592}{9} \zeta_5
\nonumber \\ &&
+ \biggl(
6320 \zeta_2 - 128 \zeta_2 \zeta_3 + 576 \zeta_5 - \frac{1280}{27} l_2^4 + \frac{3328}{9} l_2^2 \zeta_2 - \frac{10240}{9} a_4 
\nonumber \\ &&
- \frac{36608}{27} l_2 \zeta_2 - \frac{138448}{45} \zeta_2^2 + \frac{520232}{81} \zeta_3 + \frac{22789421}{2187} 
\biggr) \ln(x) 
\nonumber \\ &&
+ \biggl(
256 a_4 - 64 l_2^2 \zeta_2 - 104 \zeta_2^2 + \frac{32}{3} l_2^4 + \frac{12032}{9} l_2 \zeta_2 
- \frac{31048}{27} \zeta_2 
\nonumber \\ &&
- \frac{43904}{27} \zeta_3 
+ \frac{4494101}{729} 
\biggr) \ln^2(x) 
+ \biggl(
192 l_2 \zeta_2 + \frac{32}{3} \zeta_3 + \frac{43696}{81} \zeta_2 
\nonumber \\ &&
+ \frac{91018}{729} 
\biggr) \ln^3(x) 
+ \biggl(
4 \zeta_2 - \frac{52199}{243} 
\biggr) \ln^4(x)
+ \frac{242}{405} \ln^5(x)
\biggr] x^2
\nonumber \\ &&
+\biggl[
-2048 a_4 \zeta_2 + 512 l_2^2 \zeta_2^2 
- 512 \zeta_3^2 - \frac{256}{3} l_2^4 \zeta_2 - \frac{18656}{15} \zeta_2^3
\nonumber \\ &&
- \frac{125888}{243} l_2^4 
+ \frac{136960}{9} l_2 \zeta_2^2 + \frac{162944}{27} \zeta_2 \zeta_3 
+ \frac{298432}{81} l_2^2 \zeta_2 
\nonumber \\ &&
- \frac{468128}{9} l_2 \zeta_2 
- \frac{1007104}{81} a_4 - \frac{1501472}{27} \zeta_5 - \frac{2243680}{81} \zeta_3 
\nonumber \\ &&
+ \frac{2364736}{81} \zeta_2^2 - \frac{47478772}{6561} 
- \frac{80144464}{729} \zeta_2 
+ \biggl(-640 \zeta_2 \zeta_3 
\nonumber \\ &&
- 1728 \zeta_5 
+ \frac{5632}{27} l_2^4 - \frac{13952}{9} l_2^2 \zeta_2 + \frac{45056}{9} a_4 
+ \frac{206204}{243} 
\nonumber \\ &&
+ \frac{213248}{27} l_2 \zeta_2 
+ \frac{1104608}{135} \zeta_2^2 
- \frac{2051936}{81} \zeta_3 
- \frac{5104624}{243} \zeta_2
\biggr) \ln(x) 
\nonumber \\ &&
+ \biggl(-512 a_4 + 128 l_2^2 \zeta_2 - \frac{64}{3} l_2^4 + \frac{176}{5} \zeta_2^2 
- \frac{21824}{9} l_2 \zeta_2 + \frac{75472}{27} \zeta_3 
\nonumber \\ &&
+ \frac{493280}{81} \zeta_2 - \frac{9056684}{729} 
\biggr) \ln^2(x) 
+ \biggl(
 \frac{64}{3} \zeta_3
-384 l_2 \zeta_2 
 - \frac{35312}{81} 
\nonumber \\ &&
- \frac{111536}{81} \zeta_2
\biggr) \ln^3(x)
+ \biggl(
 \frac{5900}{9} 
-8 \zeta_2 
\biggr) \ln^4(x) 
-\frac{3712}{405} \ln^5(x)
\nonumber \\ &&
-\frac{34}{45} \ln^6(x) 
\biggr] x^3
+O(x^4) \biggr\},
\end{eqnarray}
\begin{eqnarray}
F^{(0)}_{A,2}(x) &=& F^{(0), \rm sol}_{A,2}(x)+n_h \biggl\{
\frac{13312}{27} l_2 \zeta_2
+\biggl[
-\frac{128}{3} l_2^2 \zeta_2 - \frac{128}{9} l_2^4 - \frac{22528}{135} \zeta_2^2 
\nonumber \\ &&
- \frac{1024}{3} a_4 + \frac{96032}{27} \zeta_3 
+ \frac{97984}{27} l_2 \zeta_2 + \frac{273376}{81} + \frac{1417540}{81} \zeta_2 
\nonumber \\ &&
+ \biggl
(\frac{2816}{9} l_2 \zeta_2 - \frac{5888}{27} \zeta_3 + \frac{18016}{27} \zeta_2 + \frac{1883216}{243} 
\biggr) \ln(x) 
\nonumber \\ &&
+ \biggl(
\frac{992}{27} \zeta_2 
+ \frac{347672}{243} 
\biggr) \ln^2(x) 
-\frac{2368}{81} \ln^3(x)
-\frac{1360}{27} \ln^4(x) 
\biggr] x
\nonumber \\ &&
+\biggl[
-\frac{5888}{81} l_2^2 \zeta_2 + \frac{13312}{243} l_2^4 + \frac{14144}{27} \zeta_2 \zeta_3 + \frac{81920}{9} l_2 \zeta_2^2 
+ \frac{106496}{81} a_4 
\nonumber \\ &&
- \frac{397408}{27} \zeta_5 - \frac{579328}{27} l_2 \zeta_2 - \frac{920960}{81} \zeta_3 
+ \frac{1591712}{81} + \frac{3029984}{405} \zeta_2^2 
\nonumber \\ &&
- \frac{13386440}{243} \zeta_2 
+ \biggl(
\frac{2048}{3} \zeta_2^2 + \frac{3712}{27} l_2^4 - \frac{8960}{9} l_2^2 \zeta_2 + \frac{29696}{9} a_4 
\nonumber \\ &&
- \frac{41216}{9} l_2 \zeta_2 
+ \frac{45824}{81} \zeta_3 - \frac{49856}{81} - \frac{387008}{81} \zeta_2
\biggr) \ln(x)
\nonumber \\ &&
+ \biggl(
\frac{1280}{27} \zeta_3 - \frac{14080}{9} l_2 \zeta_2 - \frac{79136}{81} \zeta_2 - \frac{351616}{243} 
\biggr) \ln^2(x)
\nonumber \\ &&
+ \biggl(
\frac{5248}{27} 
- \frac{24896}{81} \zeta_2
\biggr) \ln^3(x) 
+ \frac{22624}{243} \ln^4(x)
+ \frac{176}{405} \ln^5(x) 
\biggr] x^2
\nonumber \\ &&
+\biggl[
4096 a_4 \zeta_2 + 53824 l_2 \zeta_2 - 1024 l_2^2 \zeta_2^2 - 13696 \zeta_2 \zeta_3 - 1408 \zeta_3^2 
\nonumber \\ &&
+ \frac{512}{3} l_2^4 \zeta_2 
- \frac{3776}{105} \zeta_2^3 - \frac{53248}{9} l_2 \zeta_2^2 + \frac{108416}{243} l_2^4 - \frac{206464}{81} l_2^2 \zeta_2 
\nonumber \\ &&
+ \frac{867328}{81} a_4 + \frac{1905088}{27} \zeta_5 + \frac{4172704}{81} \zeta_3 - \frac{5531966}{2187} + \frac{7887212}{81} \zeta_2 
\nonumber \\ &&
- \frac{12842176}{405} \zeta_2^2 
+ \biggl(
-2304 \zeta_2 \zeta_3 + 1152 \zeta_5 - \frac{1024}{3} l_2^2 \zeta_2 + \frac{1280}{9} l_2^4 
\nonumber \\ &&
+ \frac{10240}{3} a_4 
- \frac{345856}{27} l_2 \zeta_2 - \frac{2236736}{135} \zeta_2^2 + \frac{2694272}{81} \zeta_3 + \frac{4301152}{243} \zeta_2 
\nonumber \\ &&
+ \frac{24011348}{729} 
\biggr) \ln(x)
+ \biggl(
1024 a_4 - 256 l_2^2 \zeta_2 
+ \frac{128}{3} l_2^4 - \frac{3808}{5} \zeta_2^2 
\nonumber \\ &&
+ \frac{15872}{9} l_2 \zeta_2 
- \frac{48256}{9} \zeta_3 - \frac{100384}{27} \zeta_2 
+ \frac{126532}{9} 
\biggr) \ln^2(x) 
\nonumber \\ &&
+ \biggl(768 l_2 \zeta_2 
+ 128 \zeta_3 - \frac{76184}{243} + \frac{144704}{81} \zeta_2
\biggr) \ln^3(x) 
\nonumber \\ &&
+ \biggl(
16 \zeta_2 - \frac{66932}{81} 
\biggr) \ln^4(x) 
+ \frac{184}{15} \ln^5(x)
-\frac{68}{45} \ln^6(x) 
\biggr] x^3
+O(x^4) \biggr\},
\end{eqnarray}
\begin{eqnarray}
F^{(0)}_{S}(x) &=& F^{(0), \rm sol}_{S}(x)+n_h \biggl\{
-\frac{34304}{81 x} l_2 \zeta_2
-1024 l_2 \zeta_2^2 - \frac{256}{81} l_2^2 \zeta_2 - \frac{304}{243} l_2^4 
\nonumber \\ &&
- \frac{1640}{9} \zeta_2 \zeta_3 
- \frac{2432}{81} a_4 + \frac{3964}{9} \zeta_5 + \frac{10324}{45} \zeta_2^2 + \frac{107668}{81} \zeta_3 
\nonumber \\ &&
+ \frac{209600}{81} l_2 \zeta_2 
- \frac{1676170}{243} + \frac{1828108}{243} \zeta_2 
+ \biggl(
\frac{64}{27} l_2^4 + \frac{512}{9} a_4 
\nonumber \\ &&
- \frac{512}{9} l_2^2 \zeta_2 - \frac{5876}{243} \zeta_2 
+ \frac{13408}{27} l_2 \zeta_2 
+ \frac{21728}{135} \zeta_2^2 
- \frac{51260}{81} \zeta_3
\nonumber \\ &&
+ \frac{22184}{27} 
\biggr) \ln(x) 
+ \biggl(
\frac{128}{9} l_2 \zeta_2 
- \frac{512}{9} \zeta_3 
+ \frac{1520}{81} \zeta_2 
+ \frac{77548}{243} 
\biggr) \ln^2(x) 
\nonumber \\ &&
+ \biggl(
-\frac{32}{3} \zeta_2 + \frac{5452}{81} 
\biggr) \ln^3(x) 
+ \frac{2888}{243} \ln^4(x)
-\frac{176}{81} \ln^5(x) 
\nonumber \\ &&
+\biggl[
2304 l_2 \zeta_2^2 
- \frac{1472}{81} l_2^4 + \frac{8128}{27} l_2^2 \zeta_2 - \frac{11776}{27} a_4 + \frac{13328}{27} \zeta_2 \zeta_3 
\nonumber \\ &&
- \frac{23896}{27} \zeta_5 + \frac{62168}{405} \zeta_2^2 - \frac{99776}{81} \zeta_3 - \frac{534016}{81} l_2 \zeta_2 
+ \frac{1087216}{243} 
\nonumber \\ &&
- \frac{4558496}{243} \zeta_2 
+ \biggl(
64 l_2^2 \zeta_2 - 256 a_4 - \frac{32}{3} l_2^4 + \frac{6176}{15} \zeta_2^2 
- \frac{40912}{27} \zeta_2 
\nonumber \\ &&
- \frac{25408}{27} l_2 \zeta_2 
- \frac{44704}{81} \zeta_3 - \frac{2297624}{243} 
\biggr) \ln(x) 
- \biggl(
\frac{2344}{81} \zeta_2 
+ \frac{138080}{81} 
\nonumber \\ &&
- \frac{4112}{27} \zeta_3 
\biggr) \ln^2(x) 
- \biggl(
\frac{160}{9} \zeta_2 + \frac{9088}{27} 
\biggr) \ln^3(x) 
+ \frac{5456}{81} \ln^4(x)
\biggr] x
\nonumber \\ &&
+\biggl[
-272 \zeta_3^2 - \frac{2144}{35} \zeta_2^3 - \frac{24320}{3} l_2 \zeta_2^2 + \frac{34816}{243} l_2^4 - \frac{41776}{27} \zeta_2 \zeta_3 
\nonumber \\ &&
- \frac{100736}{81} l_2^2 \zeta_2 + \frac{259208}{27} \zeta_5 + \frac{278528}{81} a_4 + \frac{510944}{27} l_2 \zeta_2 
+ \frac{1004896}{81} \zeta_3 
\nonumber \\ &&
- \frac{1411504}{405} \zeta_2^2 + \frac{11143964}{243} \zeta_2 - \frac{62619035}{6561} 
+ \biggl(
-192 \zeta_2 \zeta_3 - 544 \zeta_5 
\nonumber \\ &&
+ \frac{512}{27} l_2^4 
- \frac{640}{9} l_2^2 \zeta_2 + \frac{4096}{9} a_4 + \frac{12032}{9} l_2 \zeta_2 - \frac{19904}{27} \zeta_2^2 
+ \frac{68696}{27} \zeta_3 
\nonumber \\ &&
+ \frac{963968}{243} \zeta_2 + \frac{15324602}{2187} 
\biggr) \ln(x) 
+ \biggl(
\frac{4736}{9} l_2 \zeta_2 
-\frac{432}{5} \zeta_2^2
- \frac{20480}{27} \zeta_3 
\nonumber \\ &&
- \frac{29176}{81} \zeta_2 
+ \frac{4544258}{729} 
\biggr) \ln^2(x) 
- \biggl(
\frac{32}{3} \zeta_3 
+ \frac{2324}{729} 
- \frac{4544}{27} \zeta_2
\biggr) \ln^3(x) 
\nonumber \\ &&
+ \biggl(8 \zeta_2 - \frac{9028}{243} 
\biggr) \ln^4(x) 
-\frac{3032}{405} \ln^5(x)
-\frac{17}{45} \ln^6(x) 
\biggr] x^2
\nonumber \\ &&
+\biggl[
544 \zeta_3^2 + \frac{4288}{35} \zeta_2^3 + \frac{43264}{3} l_2 \zeta_2^2 + \frac{82448}{27} \zeta_2 \zeta_3 
- \frac{2633152}{81} \zeta_3 
\nonumber \\ &&
+ \frac{218560}{81} l_2^2 \zeta_2 - \frac{424736}{9} l_2 \zeta_2 - \frac{431096}{27} \zeta_5 + \frac{652576}{81} \zeta_2^2 
- \frac{687616}{81} a_4
\nonumber \\ &&
- \frac{85952}{243} l_2^4 
+ \frac{29263327}{2187} - \frac{72816460}{729} \zeta_2 
+ \biggl(
384 \zeta_2 \zeta_3 + 1088 \zeta_5 
\nonumber \\ &&
- \frac{64}{3} l_2^2 \zeta_2 - \frac{352}{9} l_2^4 - \frac{2176}{3} l_2 \zeta_2 
- \frac{2816}{3} a_4 + \frac{530096}{135} \zeta_2^2 - \frac{873344}{81} \zeta_3 
\nonumber \\ &&
- \frac{1344104}{81} \zeta_2 - \frac{50224357}{2187} 
\biggr) \ln(x) 
+ \biggl(
\frac{864}{5} \zeta_2^2 - \frac{3008}{3} l_2 \zeta_2 
+ \frac{9584}{9} \zeta_3 
\nonumber \\ &&
+ \frac{142976}{81} \zeta_2 - \frac{2236508}{243} 
\biggr) \ln^2(x)
+ \biggl(
\frac{64}{3} \zeta_3 - \frac{41008}{81} \zeta_2 
\nonumber \\ &&
- \frac{1157750}{729} 
\biggr) \ln^3(x)
- \biggl(
16 \zeta_2 - \frac{115906}{243} 
\biggr) \ln^4(x) 
+ \frac{722}{135} \ln^5(x)
\nonumber \\ &&
+ \frac{34}{45} \ln^6(x) 
\biggr] x^3
+O(x^4) \biggr\},
\end{eqnarray}
\begin{eqnarray}
F^{(0)}_{P}(x) &=& F^{(0), \rm sol}_{P}(x)+n_h \biggl\{
-\frac{54272}{81 x} l_2 \zeta_2
-1024 l_2 \zeta_2^2 - \frac{256}{81} l_2^2 \zeta_2 - \frac{304}{243} l_2^4 
\nonumber \\ &&
- \frac{1640}{9} \zeta_2 \zeta_3 
- \frac{2432}{81} a_4 
+ \frac{3964}{9} \zeta_5 + \frac{10324}{45} \zeta_2^2 
+ \frac{202432}{81} l_2 \zeta_2 
\nonumber \\ &&
+ \frac{107668}{81} \zeta_3 
- \frac{1676170}{243} 
+ \frac{1828108}{243} \zeta_2 
+ \biggl(
\frac{64}{27} l_2^4 + \frac{512}{9} a_4 - \frac{512}{9} l_2^2 \zeta_2 
\nonumber \\ &&
- \frac{5876}{243} \zeta_2 
+ \frac{13408}{27} l_2 \zeta_2 
+ \frac{21728}{135} \zeta_2^2 + \frac{22184}{27} - \frac{51260}{81} \zeta_3
\biggr) \ln(x) 
\nonumber \\ &&
+ \biggl(
\frac{128}{9} l_2 \zeta_2 
- \frac{512}{9} \zeta_3 + \frac{1520}{81} \zeta_2 + \frac{77548}{243} 
\biggr) \ln^2(x) 
\nonumber \\ &&
+ \biggl(
-\frac{32}{3} \zeta_2 + \frac{5452}{81} 
\biggr) \ln^3(x) 
+ \frac{2888}{243} \ln^4(x)
-\frac{176}{81} \ln^5(x) 
\nonumber \\ &&
+\biggl[
768 l_2 \zeta_2^2 - \frac{128}{9} l_2^4 + \frac{448}{3} l_2^2 \zeta_2 - \frac{1024}{3} a_4 + \frac{12464}{27} \zeta_2 \zeta_3 - \frac{32968}{27} \zeta_5 
\nonumber \\ &&
- \frac{68672}{81} \zeta_3 + \frac{180392}{405} \zeta_2^2 - \frac{279424}{81} l_2 \zeta_2 + \frac{853856}{243} - \frac{1159744}{243} \zeta_2 
\nonumber \\ &&
+ \biggl(
-256 a_4 + 64 l_2^2 \zeta_2 - \frac{32}{3} l_2^4 + \frac{6176}{15} \zeta_2^2 
- \frac{14512}{27} \zeta_2 - \frac{18112}{27} l_2 \zeta_2 
\nonumber \\ &&
- \frac{33952}{81} \zeta_3 - \frac{376456}{243} 
\biggr) \ln(x) 
+ \biggl(
\frac{3632}{243} + \frac{4112}{27} \zeta_3 + \frac{4136}{81} \zeta_2
\biggr) \ln^2(x) 
\nonumber \\ &&
+ \biggl(
-\frac{160}{9} \zeta_2 - \frac{18208}{81} 
\biggr) \ln^3(x) 
+ \frac{1808}{81} \ln^4(x)
\biggr] x
\nonumber \\ &&
+\biggl[
304 \zeta_3^2 + \frac{2208}{7} \zeta_2^3 + \frac{3952}{9} \zeta_2 \zeta_3 + \frac{8200}{3} \zeta_5 + \frac{10912}{9} l_2 \zeta_2 + \frac{17024}{243} l_2^4 
\nonumber \\ &&
- \frac{18688}{9} l_2 \zeta_2^2 - \frac{43856}{45} \zeta_2^2 - \frac{51328}{81} l_2^2 \zeta_2 + \frac{136192}{81} a_4 - \frac{215776}{81} \zeta_3 
\nonumber \\ &&
+ \frac{952988}{243} \zeta_2 - \frac{1907579}{6561} 
+ \biggl(448 \zeta_2 \zeta_3 + 288 \zeta_5 - \frac{128}{3} l_2^2 \zeta_2 - \frac{128}{9} l_2^4 
\nonumber \\ &&
- \frac{1024}{3} a_4 - \frac{14080}{27} l_2 \zeta_2 + \frac{37984}{27} \zeta_2^2 - \frac{126008}{81} \zeta_3 - \frac{129382}{2187} 
\nonumber \\ &&
- \frac{160576}{243} \zeta_2
\biggr) \ln(x) 
+ \biggl(
\frac{432}{5} \zeta_2^2 + \frac{1024}{9} l_2 \zeta_2 - \frac{7072}{27} \zeta_3 - \frac{58168}{81} \zeta_2 
\nonumber \\ &&
+ \frac{1143650}{729} 
\biggr) \ln^2(x)
+ \biggl(
-\frac{64}{3} \zeta_3 - \frac{4928}{81} \zeta_2 + \frac{224332}{729} 
\biggr) \ln^3(x) 
\nonumber \\ &&
 + \frac{4124}{81} \ln^4(x)
 -\frac{304}{135} \ln^5(x) 
+ \frac{17}{45} \ln^6(x) 
\biggr] x^2
\nonumber \\ &&
+\biggl[
608 \zeta_3^2 + \frac{4416}{7} \zeta_2^3 + \frac{6400}{9} l_2 \zeta_2^2 - \frac{6560}{3} l_2 \zeta_2 + \frac{8320}{81} l_2^4 - \frac{11456}{27} l_2^2 \zeta_2 
\nonumber \\ &&
- \frac{19048}{27} \zeta_5 + \frac{33328}{27} \zeta_2 \zeta_3 + \frac{66560}{27} a_4 - \frac{120800}{81} \zeta_2^2 + \frac{275728}{81} \zeta_3 
\nonumber \\ &&
+ \frac{7061252}{729} \zeta_2 + \frac{57246137}{6561} 
+ \biggl(896 \zeta_2 \zeta_3 + 576 \zeta_5 - \frac{1312}{27} l_2^4 
\nonumber \\ &&
+ \frac{1856}{9} l_2^2 \zeta_2 - \frac{10496}{9} a_4 - \frac{25472}{9} l_2 \zeta_2 
+ \frac{113296}{135} \zeta_2^2 - \frac{187424}{81} \zeta_3 
\nonumber \\ &&
+ \frac{336520}{243} \zeta_2 + \frac{529369}{729} 
\biggr) \ln(x) 
+ \biggl(
\frac{864}{5} \zeta_2^2 
- \frac{1328}{3} \zeta_2 + \frac{1600}{9} l_2 \zeta_2 
\nonumber \\ &&
+ \frac{2384}{9} \zeta_3 + \frac{579628}{729} 
\biggr) \ln^2(x) 
+ \biggl(
\frac{400}{9} \zeta_2-\frac{128}{3} \zeta_3 - \frac{42146}{243} 
\biggr) \ln^3(x) 
\nonumber \\ &&
+ \frac{15370}{243} \ln^4(x)
+ \frac{238}{405} \ln^5(x)
+ \frac{34}{45} \ln^6(x) 
\biggr] x^3
+O(x^4) \biggr\}.
\end{eqnarray}
As we did in the case of Eq. (\ref{eq:FV1sinf}) for $F^{n_h}_{V,1}(x)$, we can also perform the corresponding expansions in $\hs$ up to $O(\hs^{-3})$ 
for the other form factors. These expansions include the solvable parts, as well as the renormalization and infrared subtraction terms. One obtains

\begin{eqnarray}
F^{n_h}_{V,2}(\hs) &=&
\biggl[
832 \zeta_5 + \frac{256}{27} l_2^4 - \frac{512}{9} l_2^2 \zeta_2 + \frac{2048}{9} a_4 - \frac{32896}{27} \zeta_2^2 + \frac{47776}{27} \zeta_3 
\nonumber \\ &&
+ \frac{114688}{9} l_2 \zeta_2 - \frac{535360}{243} - \frac{2382464}{243} \zeta_2 
+ \biggl(\frac{8192}{27} \zeta_3 - \frac{54592}{81} \zeta_2 
\nonumber \\ &&
- \frac{176576}{243} \biggr) l_s 
+ \biggl(-\frac{1696}{27} \zeta_2 - \frac{1808}{243} \biggr) l_s^2 
+ \frac{1664}{81} l_s^3
+ \frac{320}{81} l_s^4 
\biggr] \frac{1}{\hs}
\nonumber \\ &&
+\biggl[
- \frac{35264}{105} \zeta_2^3 - \frac{36352}{243} l_2^4 
+ \frac{72704}{81} l_2^2 \zeta_2 
+ \frac{1091648}{81} \zeta_3 
- \frac{116192}{81} \zeta_2^2 
\nonumber \\ &&
- \frac{6016}{27} \zeta_5 + \frac{7808}{9} \zeta_2 \zeta_3 
- \frac{290816}{81} a_4 - \frac{481984}{81} 
+ \frac{2732032}{27} l_2 \zeta_2 
\nonumber \\ &&
+ 256 \zeta_3^2
- \frac{23493952}{243} \zeta_2 
+ \biggl(
 \frac{724352}{81} \zeta_3 
- \frac{508864}{81} \zeta_2 
- \frac{1299712}{81} 
\nonumber \\ &&
+ \frac{768}{5} \zeta_2^2
- 512 \zeta_5 
\biggr) l_s 
+ \biggl(\frac{288}{5} \zeta_2^2 + \frac{896}{3} \zeta_3 - \frac{37792}{27} \zeta_2 
- \frac{389248}{243} \biggr) l_s^2 
\nonumber \\ &&
+ \biggl(
-\frac{64}{3} \zeta_3 - \frac{8320}{27} \zeta_2 + \frac{159104}{243} 
\biggr) l_s^3 
+ \frac{11648}{243} l_s^4
-\frac{16}{81} l_s^5 
\biggr] \frac{1}{\hs^2}
\nonumber \\ &&
+\biggl[
1024 l_2 \zeta_2^2 + 1024 \zeta_3^2 + \frac{65920}{9} \zeta_2 \zeta_3 - \frac{88768}{27} \zeta_5 - \frac{202432}{35} \zeta_2^3 
\nonumber \\ &&
- \frac{226304}{81} l_2^4 
+ \frac{452608}{27} l_2^2 \zeta_2 - \frac{1810432}{27} a_4 + \frac{4540000}{81} \zeta_3 
\nonumber \\ &&
- \frac{5931872}{9} \zeta_2 - \frac{7876978}{6561} + \frac{15570016}{405} \zeta_2^2 
+ \frac{16488448}{27} l_2 \zeta_2 
\nonumber \\ &&
+ \biggl(2304 \zeta_2 \zeta_3 - 13440 \zeta_5 - \frac{1024}{27} l_2^4 + \frac{2048}{9} l_2^2 \zeta_2 - \frac{8192}{9} a_4 
\nonumber \\ &&
+ \frac{57472}{45} \zeta_2^2 
+ \frac{212992}{27} l_2 \zeta_2 + \frac{6275776}{81} \zeta_3 - \frac{10957184}{243} \zeta_2 
\nonumber \\ &&
- \frac{262638740}{2187} \biggr) l_s
+ \biggl(\frac{6432}{5} \zeta_2^2 + \frac{104896}{27} \zeta_3 - \frac{772160}{81} \zeta_2 
\nonumber \\ &&
- \frac{5223476}{729} \biggr) l_s^2 
+ \biggl(-\frac{896}{3} \zeta_3 - \frac{291136}{81} \zeta_2 + \frac{5233208}{729} \biggr) l_s^3 
\nonumber \\ &&
+ \biggl(-16 \zeta_2 + \frac{161668}{243} \biggr) l_s^4 
-\frac{16984}{405} l_s^5
-\frac{68}{45} l_s^6 
\biggr] \frac{1}{\hs^3}
+O\left(\frac{1}{\hs^4}\right),
\end{eqnarray}
\begin{eqnarray}
F^{n_h}_{A,1}(\hs) &=&
\frac{64}{3} l_2 \zeta_2^2 - \frac{464}{3} \zeta_2 \zeta_3 - \frac{3808}{243} l_2^4 - \frac{4856}{9} \zeta_5 + \frac{5828}{27} \zeta_2^2 + \frac{7616}{81} l_2^2 \zeta_2 
\nonumber \\ &&
- \frac{30464}{81} a_4 - \frac{52708}{81} \zeta_3 - \frac{107008}{81} l_2 \zeta_2 + \frac{281176}{243} \zeta_2 + \frac{461144}{243} 
\nonumber \\ &&
+ \biggl(
-\frac{64}{81} l_2^4 + \frac{128}{27} l_2^2 \zeta_2 + \frac{368}{27} \zeta_2^2 
- \frac{512}{27} a_4 - \frac{1804}{9} \zeta_3 + \frac{47936}{243} \zeta_2 
\nonumber \\ &&
+ \frac{171704}{243} 
\biggr) l_s 
+ \biggl(
-\frac{160}{9} \zeta_3 + \frac{896}{81} \zeta_2 + \frac{46064}{243} 
\biggr) l_s^2 
\nonumber \\ &&
+ \biggl(
-\frac{416}{81} \zeta_2 + \frac{4616}{243} 
\biggr) l_s^3 
-\frac{136}{81} l_s^4
-\frac{32}{81} l_s^5 
+\biggl[
-\frac{1600}{27} \zeta_5 + \frac{3584}{243} l_2^4
\nonumber \\ &&
- \frac{2240}{9} \zeta_2 \zeta_3 - \frac{7168}{81} l_2^2 \zeta_2 + \frac{14224}{81} \zeta_2^2 
+ \frac{28672}{81} a_4 
- \frac{77576}{81} \zeta_3 
\nonumber \\ &&
- \frac{207760}{243} - \frac{625408}{27} l_2 \zeta_2 + \frac{5245840}{243} \zeta_2 
+ \biggl(
\frac{1696}{45} \zeta_2^2 + \frac{86320}{81} \zeta_2 
\nonumber \\ &&
- \frac{118048}{81} \zeta_3 + \frac{296048}{81} 
\biggr) l_s
+ \biggl(
\frac{32}{9} \zeta_2 + \frac{112}{9} \zeta_3 + \frac{54928}{81} \biggr) l_s^2 
\nonumber \\ &&
+ \biggl(\frac{848}{27} \zeta_2 + \frac{1088}{81} 
\biggr) l_s^3 
-\frac{5896}{243} l_s^4
-\frac{4}{81} l_s^5 
\biggr] \frac{1}{\hs}
\nonumber \\ &&
+\biggl[
-48 \zeta_3^2 - \frac{128}{3} l_2 \zeta_2^2 + \frac{16832}{81} l_2^4 + \frac{30832}{105} \zeta_2^3 - \frac{33664}{27} l_2^2 \zeta_2 
\nonumber \\ &&
- \frac{36560}{27} \zeta_2 \zeta_3 
- \frac{58952}{27} \zeta_5 - \frac{70276}{81} \zeta_3 + \frac{134656}{27} a_4 - \frac{498592}{405} \zeta_2^2 
\nonumber \\ &&
- \frac{8287744}{81} l_2 \zeta_2 + \frac{23853932}{243} \zeta_2 
- \frac{78457723}{13122} 
\nonumber \\ &&
+ \biggl(
-288 \zeta_2 \zeta_3 
+ 656 \zeta_5 + \frac{128}{81} l_2^4 
- \frac{256}{27} l_2^2 \zeta_2 + \frac{1024}{27} a_4 + \frac{7024}{45} \zeta_2^2 
\nonumber \\ &&
- \frac{19456}{27} l_2 \zeta_2 
- \frac{658360}{81} \zeta_3 + \frac{1181320}{243} \zeta_2 
+ \frac{40251689}{2187} 
\biggr) l_s 
\nonumber \\ &&
+ \biggl(
-\frac{536}{5} \zeta_2^2 + \frac{3272}{27} \zeta_2 - \frac{9632}{27} \zeta_3 + \frac{2847341}{729} 
\biggr) l_s^2 
\nonumber \\ &&
+ \biggl(
\frac{32}{3} \zeta_3 + \frac{21296}{81} \zeta_2 - \frac{66110}{729} 
\biggr) l_s^3 
+ \biggl(
4 \zeta_2 - \frac{9493}{81} 
\biggr) l_s^4 
+ \frac{986}{405} l_s^5
\biggr] \frac{1}{\hs^2}
\nonumber \\ &&
+\biggl[
320 \zeta_3^2 - \frac{1024}{3} l_2 \zeta_2^2 + \frac{83744}{35} \zeta_2^3 - \frac{105920}{9} \zeta_5 + \frac{108032}{81} l_2^4 
\nonumber \\ &&
- \frac{151808}{27} \zeta_2 \zeta_3 
- \frac{216064}{27} l_2^2 \zeta_2 + \frac{591248}{243} \zeta_3 + \frac{864256}{27} a_4 
\nonumber \\ &&
- \frac{5388448}{405} \zeta_2^2 - \frac{36299776}{81} l_2 \zeta_2 - \frac{180817228}{6561} 
+ \frac{320235272}{729} \zeta_2
\nonumber \\ &&
+ \biggl(
-1280 \zeta_2 \zeta_3 + 4992 \zeta_5 + \frac{1024}{81} l_2^4 
- \frac{2048}{27} l_2^2 \zeta_2 + \frac{2336}{5} \zeta_2^2 
\nonumber \\ &&
+ \frac{8192}{27} a_4 - \frac{54272}{9} l_2 \zeta_2 - \frac{3323392}{81} \zeta_3 + \frac{5432992}{243} \zeta_2 + \frac{7229564}{81} 
\biggr) l_s 
\nonumber \\ &&
+ \biggl(
-\frac{416}{3} \zeta_2 - \frac{2352}{5} \zeta_2^2 - \frac{53728}{27} \zeta_3 + \frac{4294144}{243} 
\biggr) l_s^2 
\nonumber \\ &&
+ \biggl(
\frac{126080}{81} \zeta_2
+ \frac{64}{3} \zeta_3  
- \frac{296848}{243} 
\biggr) l_s^3 
+ \biggl(
24 \zeta_2 - \frac{141088}{243} 
\biggr) l_s^4 
\nonumber \\ &&
+ \frac{5324}{405} l_s^5
+ \frac{34}{45} l_s^6 
\biggr] \frac{1}{\hs^3}
+O\left(\frac{1}{\hs^4}\right),
\end{eqnarray}
\begin{eqnarray}
F^{n_h}_{A,2}(\hs) &=&
\biggl[
384 \zeta_2 \zeta_3 - 320 \zeta_5 - \frac{256}{81} l_2^4 + \frac{512}{27} l_2^2 \zeta_2 - \frac{2048}{27} a_4 - \frac{4256}{27} \zeta_3 
\nonumber \\ &&
- \frac{67936}{135} \zeta_2^2 + \frac{111296}{243} 
+ \frac{327680}{27} l_2 \zeta_2 - \frac{329024}{27} \zeta_2 
\nonumber \\ &&
+ \biggl(
\frac{512}{3} \zeta_3 
- \frac{13312}{27} \zeta_2 - \frac{637504}{243} 
\biggr) l_s 
\nonumber \\ &&
+ \biggl(
\frac{256}{3} \zeta_2 - \frac{49168}{81} 
\biggr) l_s^2 
+ \frac{832}{81} l_s^3
+ \frac{320}{27} l_s^4 
\biggr] \frac{1}{\hs}
\nonumber \\ &&
+\biggl[
\frac{5504}{27} \zeta_2 \zeta_3 - \frac{9280}{3} - \frac{15872}{243} l_2^4 + \frac{31744}{81} l_2^2 \zeta_2 - \frac{46720}{27} \zeta_5 
\nonumber \\ &&
- \frac{119584}{405} \zeta_2^2 
- \frac{126976}{81} a_4 + \frac{506560}{81} \zeta_3 + \frac{2017280}{27} l_2 \zeta_2 
\nonumber \\ &&
- \frac{17172416}{243} \zeta_2 
+ \biggl(
\frac{306176}{81} \zeta_3
-\frac{70592}{135} \zeta_2^2 - \frac{110272}{81} \zeta_2 
\nonumber \\ &&
- \frac{2675584}{243} 
\biggr) l_s 
+ \biggl(
-\frac{7904}{81} \zeta_2 - \frac{11648}{27} \zeta_3 - \frac{576256}{243} 
\biggr) l_s^2 
\nonumber \\ &&
+ \biggl(
-\frac{3520}{27} \zeta_2 + \frac{6848}{27} 
\biggr) l_s^3 
+ \frac{15904}{243} l_s^4
+ \frac{16}{81} l_s^5 
\biggr] \frac{1}{\hs^2}
\nonumber \\ &&
+\biggl[
1408 \zeta_3^2 + \frac{1024}{3} l_2 \zeta_2^2 + \frac{9152}{105} \zeta_2^3 + \frac{93952}{27} \zeta_2 \zeta_3 - \frac{126976}{243} l_2^4 
\nonumber \\ &&
+ \frac{212224}{27} \zeta_5 
+ \frac{251008}{135} \zeta_2^2 + \frac{253952}{81} l_2^2 \zeta_2 + \frac{614656}{81} \zeta_3 
\nonumber \\ &&
- \frac{1015808}{81} a_4 + \frac{8947712}{27} l_2 \zeta_2 + \frac{17539406}{2187} 
- \frac{79538816}{243} \zeta_2 
\nonumber \\ &&
+ \biggl(
2048 \zeta_2 \zeta_3 - 512 \zeta_5 - \frac{1024}{81} l_2^4 + \frac{2048}{27} l_2^2 \zeta_2 + \frac{7616}{135} \zeta_2^2 
\nonumber \\ &&
- \frac{8192}{27} a_4 
+ \frac{623488}{27} \zeta_3 - \frac{3769600}{243} \zeta_2 - \frac{48018980}{729} 
\biggr) l_s 
\nonumber \\ &&
+ \biggl(
\frac{3872}{5} \zeta_2^2 + \frac{13952}{27} \zeta_3 - \frac{174656}{81} \zeta_2 - \frac{2940964}{243} 
\biggr) l_s^2 
\nonumber \\ &&
+ \biggl(
-128 \zeta_3 - \frac{99008}{81} \zeta_2 + \frac{324088}{243} \biggr) l_s^3 
- \biggl(
16 \zeta_2 - \frac{115612}{243} 
\biggr) l_s^4 
\nonumber \\ &&
+ \frac{824}{405} l_s^5
+ \frac{68}{45} l_s^6 
\biggr] \frac{1}{\hs^3}
+O\left(\frac{1}{\hs^4}\right),
\end{eqnarray}
\begin{eqnarray}
F^{n_h}_{S}(\hs) &=&
\frac{64}{3} \zeta_2^2 l_2-\frac{107008}{81} \zeta_2 l_2+\frac{7616}{81} \zeta_2 l_2^2
-\frac{3808}{243} l_2^4+\frac{29116}{135} \zeta_2^2-\frac{464}{3} \zeta_2 \zeta_3
\nonumber \\ &&
+\frac{177760}{243} \zeta_2-\frac{37876}{81} \zeta_3-\frac{4856}{9} \zeta_5
+\frac{200698}{243}-\frac{30464}{81} a_4
\nonumber \\ &&
+\biggl(
-\frac{512}{27} a_4+\frac{128}{27} \zeta_2 l_2^2-\frac{64}{81} l_2^4
+\frac{368}{27} \zeta_2^2+\frac{20288}{243} \zeta_2-\frac{1420}{9} \zeta_3
\nonumber \\ &&
+\frac{19496}{243}
\biggr) l_s
+\biggl(
\frac{1472}{81} \zeta_2-\frac{160}{9} \zeta_3+\frac{16256}{243}
\biggr) l_s^2
\nonumber \\ &&
+\biggl(-\frac{416}{81} \zeta_2+\frac{5624}{243}
\biggr) l_s^3
+\frac{104}{81} l_s^4
-\frac{32}{81} l_s^5
+\biggl[
\frac{256}{9} \zeta_2 l_2^2+\frac{78008}{81} \zeta_2^2
\nonumber \\ &&
-\frac{128}{27} l_2^4-\frac{152320}{9} \zeta_2 l_2-\frac{10528}{27} \zeta_2 \zeta_3
+\frac{131744}{9} \zeta_2-\frac{5080}{81} \zeta_3
\nonumber \\ &&
-\frac{20192}{27} \zeta_5-\frac{4432}{3}-\frac{1024}{9} a_4
+\biggl(
\frac{108416}{81} \zeta_2-\frac{2512}{27} \zeta_2^2-\frac{59552}{81} \zeta_3
\nonumber \\ &&
+\frac{203600}{81}
\biggr) l_s
+\biggl(
-\frac{2600}{81} \zeta_2-\frac{2576}{27} \zeta_3+\frac{46400}{81}
\biggr) l_s^2
\nonumber \\ &&
+\biggl(-\frac{32}{27} \zeta_2+\frac{8464}{81}\biggr) l_s^3
-\frac{1280}{81} l_s^4
\biggr] \frac{1}{\hs}
+\biggl[
-\frac{22226279}{6561}+\frac{128}{3} \zeta_2^2 l_2
\nonumber \\ &&
-\frac{6406144}{81} \zeta_2 l_2-\frac{44672}{81} \zeta_2 l_2^2+\frac{22336}{243} l_2^4
-\frac{2144}{35} \zeta_2^3+\frac{1023424}{405} \zeta_2^2
\nonumber \\ &&
-\frac{43408}{27} \zeta_2 \zeta_3+\frac{17358044}{243} \zeta_2+\frac{203516}{81} \zeta_3
-\frac{105608}{27} \zeta_5+\frac{178688}{81} a_4
\nonumber \\ &&
-272 \zeta_3^2
+\biggl(
\frac{256}{27} \zeta_2 l_2^2-\frac{1024}{27} a_4-\frac{19456}{27} \zeta_2 l_2-\frac{128}{81} l_2^4-\frac{61216}{135} \zeta_2^2
\nonumber \\ &&
+\frac{475480}{81} \zeta_2-\frac{328664}{81} \zeta_3-288 \zeta_2 \zeta_3-624 \zeta_5+\frac{25638926}{2187}
\biggr) l_s
\nonumber \\ &&
+\biggl(
-\frac{432}{5} \zeta_2^2+\frac{11944}{81} \zeta_2-\frac{3616}{9} \zeta_3+\frac{2984282}{729}
\biggr) l_s^2
+\biggl(
-\frac{32}{3} \zeta_3
\nonumber \\ &&
+\frac{1664}{27} \zeta_2+\frac{236572}{729}
\biggr) l_s^3
+\biggl(
8 \zeta_2
-\frac{1424}{27}
\biggr) l_s^4
-\frac{832}{405} l_s^5
-\frac{17}{45} l_s^6
\biggr] \frac{1}{\hs^2}
\nonumber \\ &&
+\biggl[
-\frac{95621945}{6561}+\frac{512}{3} \zeta_2^2 l_2-\frac{28773376}{81} \zeta_2 l_2-\frac{324608}{81} \zeta_2 l_2^2
\nonumber \\ &&
+\frac{162304}{243} l_2^4-\frac{12864}{35} \zeta_2^3+\frac{565208}{81} \zeta_2^2+\frac{241177964}{729} \zeta_2
\nonumber \\ &&
-\frac{179552}{27} \zeta_2 \zeta_3+\frac{4438880}{243} \zeta_3-\frac{59792}{3} \zeta_5-1632 \zeta_3^2+\frac{1298432}{81} a_4
\nonumber \\ &&
+\biggl(
-\frac{4096}{27} a_4-\frac{152576}{27} \zeta_2 l_2+\frac{1024}{27} \zeta_2 l_2^2
-\frac{512}{81} l_2^4-\frac{120976}{45} \zeta_2^2
\nonumber \\ &&
+\frac{7481048}{243} \zeta_2-\frac{1568096}{81} \zeta_3-1728 \zeta_2 \zeta_3-3744 \zeta_5+\frac{138016885}{2187}
\biggr) l_s
\nonumber \\ &&
+\biggl(
-\frac{2592}{5} \zeta_2^2+\frac{68104}{81} \zeta_2-\frac{15776}{9} \zeta_3+\frac{14330044}{729}
\biggr) l_s^2
\nonumber \\ &&
+\biggl(
\frac{43600}{81} \zeta_2-64 \zeta_3+\frac{870614}{729}
\biggr) l_s^3
+\biggl(
48 \zeta_2-\frac{83306}{243}
\biggr) l_s^4
\nonumber \\ &&
-\frac{2342}{135} l_s^5
-\frac{34}{15} l_s^6
\biggr] \frac{1}{\hs^3}
+O\left(\frac{1}{\hs^4}\right),
\end{eqnarray}
\begin{eqnarray}
F^{n_h}_{P}(\hs) &=&
\frac{64}{3} \zeta_2^2 l_2-\frac{107008}{81} \zeta_2 l_2+\frac{7616}{81} \zeta_2 l_2^2
-\frac{3808}{243} l_2^4+\frac{29116}{135} \zeta_2^2-\frac{464}{3} \zeta_2 \zeta_3
\nonumber \\ &&
+\frac{177760}{243} \zeta_2-\frac{37876}{81} \zeta_3-\frac{4856}{9} \zeta_5+\frac{200698}{243}-\frac{30464}{81} a_4
\nonumber \\ &&
+\biggl(
-\frac{512}{27} a_4+\frac{128}{27} \zeta_2 l_2^2-\frac{64}{81} l_2^4+\frac{368}{27} \zeta_2^2
+\frac{20288}{243} \zeta_2-\frac{1420}{9} \zeta_3
\nonumber \\ &&
+\frac{19496}{243}
\biggr) l_s
+\biggl(
\frac{1472}{81} \zeta_2-\frac{160}{9} \zeta_3+\frac{16256}{243}
\biggr) l_s^2
\nonumber \\ &&
+\biggl(
-\frac{416}{81} \zeta_2+\frac{5624}{243}
\biggr) l_s^3
+\frac{104}{81} l_s^4
-\frac{32}{81} l_s^5
+\biggl[
\frac{256}{27} \zeta_2 l_2^2-\frac{121088}{27} \zeta_2 l_2
\nonumber \\ &&
-\frac{405008}{243}-\frac{128}{81} l_2^4+\frac{41224}{405} \zeta_2^2-\frac{5344}{27} \zeta_2 \zeta_3
+\frac{984992}{243} \zeta_2+\frac{49064}{81} \zeta_3
\nonumber \\ &&
-\frac{13280}{27} \zeta_5-\frac{1024}{27} a_4
+\biggl(
-\frac{2512}{27} \zeta_2^2+\frac{5504}{9} \zeta_2-\frac{41504}{81} \zeta_3
\nonumber \\ &&
+\frac{261968}{243}
\biggr) l_s
+\biggl(
-\frac{1688}{81} \zeta_2-\frac{2576}{27} \zeta_3+\frac{25328}{243}
\biggr) l_s^2
\nonumber \\ &&
+\biggl(
-\frac{32}{27} \zeta_2+\frac{6224}{81}
\biggr) l_s^3
-\frac{640}{81} l_s^4
\biggr] \frac{1}{\hs}
+\biggl[
-\frac{25532903}{6561}+\frac{128}{3} \zeta_2^2 l_2
\nonumber \\ &&
-\frac{1576960}{81} \zeta_2 l_2-\frac{37504}{81} \zeta_2 l_2^2+\frac{18752}{243} l_2^4
+\frac{2208}{7} \zeta_2^3-\frac{287296}{405} \zeta_2^2
\nonumber \\ &&
-\frac{13072}{27} \zeta_2 \zeta_3+\frac{1327348}{81} \zeta_2+\frac{27796}{27} \zeta_3
-\frac{5896}{27} \zeta_5+\frac{150016}{81} a_4
\nonumber \\ &&
+304 \zeta_3^2
+\biggl(
-\frac{1024}{27} a_4-\frac{19456}{27} \zeta_2 l_2+\frac{256}{27} \zeta_2 l_2^2
-\frac{128}{81} l_2^4+\frac{22976}{135} \zeta_2^2
\nonumber \\ &&
+\frac{62744}{81} \zeta_2-\frac{190744}{81} \zeta_3+224 \zeta_2 \zeta_3+528 \zeta_5
+\frac{5114702}{2187}
\biggr) l_s
\nonumber \\ &&
+\biggl(
\frac{432}{5} \zeta_2^2-\frac{36152}{81} \zeta_2-\frac{2048}{9} \zeta_3+\frac{701306}{729}
\biggr) l_s^2
\nonumber \\ &&
+\biggl(
-\frac{128}{3} \zeta_2-\frac{64}{3} \zeta_3+\frac{179260}{729}
\biggr) l_s^3
+\frac{424}{243} l_s^4
+\frac{1192}{405} l_s^5
+\frac{17}{45} l_s^6
\biggr] \frac{1}{\hs^2}
\nonumber \\ &&
+\biggl[
-\frac{85198093}{6561}+\frac{512}{3} \zeta_2^2 l_2-\frac{5825536}{81} \zeta_2 l_2
-\frac{26624}{27} \zeta_2 l_2^2+\frac{13312}{81} l_2^4
\nonumber \\ &&
+\frac{4416}{7} \zeta_2^3+\frac{101672}{405} \zeta_2^2-\frac{44192}{27} \zeta_2 \zeta_3
+\frac{38327036}{729} \zeta_2+\frac{364016}{243} \zeta_3
\nonumber \\ &&
+\frac{4400}{9} \zeta_5+608 \zeta_3^2
+\frac{106496}{27} a_4
+\biggl(
1056 \zeta_5-\frac{4096}{27} a_4-\frac{5120}{3} \zeta_2 l_2
\nonumber \\ &&
+\frac{1024}{27} \zeta_2 l_2^2-\frac{512}{81} l_2^4+\frac{9232}{9} \zeta_2^2
+\frac{2008}{243} \zeta_2-\frac{532288}{81} \zeta_3+448 \zeta_2 \zeta_3
\nonumber \\ &&
+\frac{5075447}{729}
\biggr) l_s
+\biggl(
\frac{864}{5} \zeta_2^2-\frac{36776}{27} \zeta_2-\frac{25792}{27} \zeta_3+\frac{965428}{243}
\biggr) l_s^2
\nonumber \\ &&
+\biggl(
-\frac{17584}{81} \zeta_2-\frac{128}{3} \zeta_3+\frac{235522}{243}
\biggr) l_s^3
+\frac{10862}{243} l_s^4
+\frac{6254}{405} l_s^5
\nonumber \\ &&
+\frac{34}{45} l_s^6
\biggr] \frac{1}{\hs^3}
+O\left(\frac{1}{\hs^4}\right).
\end{eqnarray}

\section{\boldmath Partially resummed expansions at $x=0$}
\label{app:resum}

\vspace*{1mm}
\noindent
In Eq. (\ref{eq:FV1resummed2}) we showed the result for the partially resummed expansion of $F^{(0)}_{V,1,1}(x) + \zeta_2 F^{(0)}_{V,1,2}(x) + \zeta_3 F^{(0)}_{V,1,3}(x)$. 
In this appendix, we show the corresponding results for the other form factors,
\begin{eqnarray}
\lefteqn{F^{(0)}_{V,2,1}(x) + \zeta_2 F^{(0)}_{V,2,2}(x) + \zeta_3 F^{(0)}_{V,2,3}(x) =}  
\nonumber\\ &&
\ln(2) \Biggl\{
        \frac{1024 x \zeta_2^2 P_{48}}{9 (x-1)^3 (1+x)^5}
        +\Biggl[
                 \frac{256 x \ln(1+x) P_{18}}{9 (x-1)^2 (1+x)^4}
                +\frac{512 x \Li_2(-x) P_{46}}{9 (x-1)^3 (1+x)^5}
\nonumber\\ &&
                -\frac{256 x^2 \ln^2(x) P_{49}}{9 (x-1)^3 (1+x)^6}
    -\frac{64 P_{66}}{81 (x-1)^2 (1+x)^6}
\nonumber\\ &&
                -\ln(1-x) \Biggl(
                        \frac{1024 x^2 \ln(x) P_1}{9 (x-1)^3 (1+x)^5}
                        -\frac{1024 x P_4}{9 (x-1)^2 (1+x)^4}
                \Biggr)
\nonumber\\ &&                
 +\ln(x) \Biggl(
                        \frac{512 x \ln(1+x) P_{47}}{9 (x-1)^3 (1+x)^5}
                        -\frac{256 x P_{55}}{27 (x-1)^3 (1+x)^5}
                \Biggr)
                +\frac{256 x^2 \big( 1-7 x+x^2\big) \ln^3(x)}{(x-1) (1+x)^5}
\nonumber\\ &&                
 -\frac{1024 x^2 \big(1-12 x+x^2\big)\big(1-x+x^2\big) \Li_2(x)}{9 (x-1)^3 (1+x)^5}
        \Biggr] \zeta_2
\Biggr\}
+\ln^2(2) \Biggl\{
        \Biggr[
                \frac{256 x^2 \ln(x) P_5}{9 (x-1)^3 (1+x)^5}
\nonumber\\ &&                
 -\frac{128 x P_{60}}{81 (x-1)^3 (1+x)^5}
                -\ln(1-x) \Biggl(
                        -\frac{1024 x \big(1-x+x^2\big)}{(x-1) (1+x)^3}
                        -\frac{1024 x^2 \big(1-x+x^2\big) \ln(x)}{(x-1)(1+x)^5}
                \Biggr)
\nonumber\\ &&                
 -\frac{256 x^2 \big(1-x+x^2\big) \ln^2(x)}{(x-1) (1+x)^5}
                -\frac{1024 x \big(1-x+x^2\big) \ln(1+x)}{(x-1) (1+x)^3}
                +\frac{1024 x^2 \big(1-x+x^2\big) \Li_2(x)}{(x-1) (1+x)^5}
        \Biggr] \zeta_2
\nonumber\\ &&       
  -\frac{1024 x^2 \big(1-x+x^2\big) \zeta_2^2}{(x-1) (1+x)^5}
\Biggr\}
+\Li_4\left(\frac{1}{2}\right) \Biggl\{
        -\frac{1024 x^2 \ln(x) P_{11}}{9 (x-1)^3 (1+x)^5}
        +\frac{1024 x P_{56}}{81 (x-1)^3 (1+x)^5}
\nonumber\\ &&    
     -\ln(1-x) \Biggl[
                 \frac{4096 x \big(1-x+x^2\big)}{(x-1) (1+x)^3}
                +\frac{4096 x^2 \big(1-x+x^2\big) \ln(x)}{(x-1) (1+x)^5}
        \Biggr]
\nonumber\\ && 
        +\frac{1024 x^2 \big(1-x+x^2\big) \ln^2(x)}{(x-1) (1+x)^5}
        +\frac{4096 x \big(1-x+x^2\big) \ln(1+x)}{(x-1) (1+x)^3}
        -\frac{4096 x^2 \big(1-x+x^2\big) \Li_2(x)}{(x-1) (1+x)^5}
\Biggr\}
\nonumber\\ &&
+\ln^4(2) \Biggl\{
        -\frac{128 x^2 \ln(x) P_{11}}{27 (x-1)^3 (1+x)^5}
        +\frac{128 x P_{56}}{243 (x-1)^3 (1+x)^5}
        -\ln(1-x) \Biggl[
                \frac{512 x \big(1-x+x^2\big)}{3 (x-1) (1+x)^3}
\nonumber\\ &&               
  +\frac{512 x^2 \big(1-x+x^2\big) \ln(x)}{3 (x-1) (1+x)^5}
        \Biggr]
        +\frac{512 x^2 \big(1-x+x^2\big) \zeta_2}{3 (x-1) (1+x)^5}
        +\frac{128 x^2 \big(1-x+x^2\big) \ln^2(x)}{3 (x-1) (1+x)^5}
\nonumber\\ &&  
      +\frac{512 x \big(1-x+x^2\big) \ln(1+x)}{3 (x-1) (1+x)^3}
        -\frac{512 x^2 \big(1-x+x^2\big) \Li_2(x)}{3 (x-1) (1+x)^5}
\Biggr\}
+\Biggl\{
        \frac{4096 \Li_4\left(\frac{1}{2}\right) x^2 \big(1-x+x^2\big)}{(x-1) (1+x)^5}
\nonumber\\ &&   
      +\frac{768 x^3 \big(1+x^2\big) \zeta_3 \ln(x)}{(x-1)^3 (1+x)^5}
        -\frac{16 x^3 \ln^4(x)}{(x-1)^3 (1+x)^3}
\Biggr\} \zeta_2
+\frac{64 x^2 \big(1-4 x+x^2\big)^2 \zeta_3 \ln^3(x)}{3 (x-1)^3 (1+x)^5}
\nonumber\\ && 
-\frac{68 x^3 \big(1-4 x+x^2\big) \ln^6(x)}{45 (x-1)^3 (1+x)^5}
-\frac{128 x^2 \zeta_3^2 P_8}{(x-1)^3 (1+x)^5}
+\frac{128 x^2 \zeta_5 \ln(x) P_{10}}{(x-1)^3 (1+x)^5}
-\frac{32 x^2 \zeta_2^2 \ln^2(x) P_{24}}{5 (x-1)^3 (1+x)^5}
\nonumber\\ &&
-\frac{64 x^2 \zeta_2^3 P_{38}}{105 (x-1)^3 (1+x)^5}   +\tilde{F}^{(0), \rm rest}_{V,2}(x),
\end{eqnarray}
\begin{eqnarray}
\lefteqn{F^{(0)}_{A,1,1}(x) + \zeta_2 F^{(0)}_{A,1,2}(x) + \zeta_3 F^{(0)}_{A,1,3}(x) =}  
\nonumber\\ &&
\ln(2) \Biggl\{
        \frac{256 \big(1+x^2\big) \zeta_2^2 P_{27}}{9 (x-1)^3 (1+x)^3}
        +\Biggl[
                \frac{128 \ln(1+x) P_{15}}{9 (x-1)^2 (1+x)^2}
                +\frac{64 \big(1+x^2\big) \Li_2(-x) P_{36}}{9 (x-1)^3 (1+x)^3}
\nonumber\\ &&              
                +\frac{64 \Li_2(x) P_{51}}{9 (x-1)^3 (1+x)^3}
   -\frac{64 \ln^2(x) P_{61}}{9 (x-1)^3 (1+x)^4}
\nonumber\\ &&              
                -\frac{16 P_{83}}{81 (x-1)^4 x (1+x)^6}
                -\ln(1-x) \Biggl(
                        \frac{128 \big(3-2 x^2+3 x^4\big)}{9 (x-1)^2 (1+x)^2}
\nonumber\\ &&                    
     -\frac{64 \ln(x) P_{52}}{9 (x-1)^3 (1+x)^3}
                \Biggr)
                +\ln(x) \Biggl(
                        \frac{64 \ln(1+x) P_{57}}{9 (x-1)^3 (1+x)^3}
                        -\frac{32 P_{79}}{27 (x-1)^5 (1+x)^5}
                \Biggr)
\nonumber\\ &&                 
-\frac{192 x^2 \ln^3(x)}{(x-1) (1+x)^3}
        \Biggr] \zeta_2
\Biggr\}
+\ln^2(2) \Biggl\{
        \Biggl[
                \frac{128 \ln(x) P_{44}}{9 (x-1)^3 (1+x)^3}
                -\frac{32 P_{74}}{81 (x-1)^5 (1+x)^5}
\nonumber\\ &&                
 -\ln(1-x) \Biggl(
                        -\frac{128 \big(1+x^2\big)\big(11-2 x+11 x^2\big)}{3 (x-1) (1+x)^3}
                        +\frac{256 x^2 \ln(x)}{(x-1) (1+x)^3}\Biggr)
                +\frac{64 x^2 \ln^2(x)}{(x-1) (1+x)^3}
\nonumber\\ &&                
 -\frac{128 \big(1+x^2\big)\big(11-2 x+11 x^2\big) \ln(1+x)}{3 (x-1) (1+x)^3}
                -\frac{256 x^2 \Li_2(x)}{(x-1) (1+x)^3}
        \Biggr] \zeta_2
        +\frac{256 x^2 \zeta_2^2}{(x-1) (1+x)^3}
\Biggr\}
\nonumber\\ &&
+\ln^4(2) \Biggl\{
        -\frac{64 \ln(x) P_{42}}{27 (x-1)^3 (1+x)^3}
        -\frac{16 P_{67}}{243 (x-1)^5 (1+x)^5}
        -\ln(1-x) 
\nonumber\\ &&
\times 
\Biggl[
                 \frac{64 \big(1+x^2\big)\big(11-2 x+11 x^2\big)}{9 (x-1) (1+x)^3}
                -\frac{128 x^2 \ln(x)}{3 (x-1) (1+x)^3}
        \Biggr]
        -\frac{128 x^2 \zeta_2}{3 (x-1) (1+x)^3}
\nonumber\\ &&        
 -\frac{32 x^2 \ln^2(x)}{3 (x-1) (1+x)^3}
        +\frac{64 \big(1+x^2\big)\big(11-2 x+11 x^2\big) \ln(1+x)}{9 (x-1) (1+x)^3}
        +\frac{128 x^2 \Li_2(x)}{3 (x-1) (1+x)^3}
\Biggr\}
\nonumber\\ && 
+\Li_4\left(\frac{1}{2}\right) \Biggl\{
        -\frac{512 \ln(x) P_{42}}{9 (x-1)^3 (1+x)^3}
        -\frac{128 P_{67}}{81 (x-1)^5 (1+x)^5}
        -\ln(1-x) \nonumber\\ &&
\times 
\Biggl[
                \frac{512 \big(1+x^2\big)\big(11-2 x+11 x^2\big)}{3 (x-1) (1+x)^3}
                -\frac{1024 x^2 \ln(x)}{(x-1) (1+x)^3}
        \Biggr]
        -\frac{256 x^2 \ln^2(x)}{(x-1) (1+x)^3}
\nonumber\\ &&        
 +\frac{512 \big(1+x^2\big)\big(11-2 x+11 x^2\big) \ln(1+x)}{3 (x-1) (1+x)^3}
        +\frac{1024 x^2 \Li_2(x)}{(x-1) (1+x)^3}
\Biggr\}
+\Biggl\{
        -\frac{1024 \Li_4\left(\frac{1}{2}\right) x^2}{(x-1) (1+x)^3}
\nonumber\\ &&    
     +\frac{128 x^2 \big(1+5 x+x^2\big) \zeta_3 \ln(x)}{(x-1)^3 (1+x)^3}
        -\frac{4 x^2 \ln^4(x)}{(x-1) (1+x)^3}
\Biggr\} \zeta_2
+\frac{16 x^2 \big(3+32 x+3 x^2\big) \zeta_3^2}{(x-1)^3 (1+x)^3}
\nonumber\\ && 
-\frac{16 x^2 \big(2011-8162 x+2011 x^2\big) \zeta_2^3}{105 (x-1)^3 (1+x)^3}
-\frac{576 x^2 \big(1-3 x+x^2\big) \zeta_5 \ln(x)}{(x-1)^3 (1+x)^3}
\nonumber\\ &&
+\frac{8 x^2 \big(65-22 x+65 x^2\big) \zeta_2^2 \ln^2(x)}{5 (x-1)^3 (1+x)^3}
-\frac{32 x^2 \zeta_3 \ln^3(x)}{3 (x-1)^3 (1+x)}
+\frac{34 x^3 \ln^6(x)}{45 (x-1)^3 (1+x)^3}    \nonumber\\ &&           +\tilde{F}^{(0), \rm rest}_{A,1}(x),
\end{eqnarray}
\begin{eqnarray}
\lefteqn{F^{(0)}_{A,2,1}(x) + \zeta_2 F^{(0)}_{A,2,2}(x) + \zeta_3 F^{(0)}_{A,2,3}(x) =}  
\nonumber\\ &&
\ln(2) \Biggl\{
        -\frac{2048 x^2 \zeta_2^2 P_{29}}{9 (x-1)^5 (1+x)^3}
        +\Biggl[
                \frac{1024 x^2 \Li_2(x) P_2}{9 (x-1)^5 
(1+x)^3}
                +\frac{256 x \ln(1+x) P_{14}}{9 (x-1)^4 (1+x)^2}
\nonumber\\ && 
                -\frac{512 x^2 \Li_2(-x) P_{32}}{9 (x-1)^5 
(1+x)^3}
                +\frac{256 x^2 \ln^2(x) P_{39}}{9 (x-1)^5 (1+x)^4}
                +\frac{64 P_{75}
                }{27 (x-1)^4 (1+x)^6}
\nonumber\\ &&               
 -\ln(1-x) \Biggl(
                        \frac{1024 x^2 \big(
                                1+4 x+x^2\big)}{9 (x-1)^4 (1+x)^2}
                        -\frac{1024 x^2 \ln(x) P_3}{9 (x-1)^5 
(1+x)^3}
                \Biggr)
                +\ln(x) 
\nonumber\\ &&
\times \Biggl(
                        -\frac{512 x^2 \ln(1+x) P_{33}}{9 (x-1)^5 
(1+x)^3}
                        -\frac{256 x P_{64}}{27 (x-1)^5 (1+x)^5}
                \Biggr)
                -\frac{768 x^3 \ln^3(x)}{(x-1)^3 (1+x)^3}
        \Biggr] \zeta_2
\Biggr\}
\nonumber\\ && 
+\ln^2(2) \Biggl\{
                \frac{256 x^2 \ln(x) P_{26}}{9 (x-1)^5 (1+x)^3}
                -\frac{128 x P_{63}}{81 (x-1)^5 (1+x)^5}
                -\ln(1-x)  
\nonumber\\ && \times       \Biggl[
                        \frac{1024 x^2 \big(
                                3-2 x+3 x^2\big)}{(x-1)^3 (1+x)^3}
                        +\frac{1024 x^3 \ln(x)}{(x-1)^3 (1+x)^3}
                \Biggr]
                +\frac{256 x^3 \ln^2(x)}{(x-1)^3 (1+x)^3}
\nonumber\\ &&           
      +\frac{1024 x^2 \big(
                        3-2 x+3 x^2\big) \ln(1+x)}{(x-1)^3 
(1+x)^3}
                -\frac{1024 x^3 \Li_2(x)}{(x-1)^3 (1+x)^3}
        \Biggr] \zeta_2
        +\frac{1024 x^3 \zeta_2^2}{(x-1)^3 (1+x)^3}
\Biggr\}
\nonumber\\ && 
+\ln^4(2) \Biggl\{
        -\frac{128 x^2 \ln(x) P_{23}}{27 (x-1)^5 (1+x)^3}
        -\frac{128 x P_{62}}{243 (x-1)^5 (1+x)^5}
        -\ln(1-x)
\nonumber\\ &&  \times
 \Biggl[
                -\frac{512 x^2 \big(
                        3-2 x+3 x^2\big)}{3 (x-1)^3 (1+x)^3}
                -\frac{512 x^3 \ln(x)}{3 (x-1)^3 (1+x)^3}
        \Biggr]
\nonumber\\ && 
        -\frac{512 x^3 \zeta_2}{3 (x-1)^3 (1+x)^3}
        -
        \frac{128 x^3 \ln^2(x)}{3 (x-1)^3 (1+x)^3}
        -\frac{512 x^2 \big(
                3-2 x+3 x^2\big) \ln(1+x)}{3 (x-1)^3 (1+x)^3}
\nonumber\\ && 
        +\frac{512 x^3 \Li_2(x)}{3 (x-1)^3 (1+x)^3}
\Biggr\}
+\Li_4\left(\frac{1}{2}\right) \Biggl\{
        -\frac{1024 x^2 \ln(x) P_{23}}{9 (x-1)^5 (1+x)^3}
        -\frac{1024 x P_{62}}{81 (x-1)^5 (1+x)^5}
\nonumber\\ &&         
-\ln(1-x) \Biggl[
                -\frac{4096 x^2 \big(
                        3-2 x+3 x^2\big)}{(x-1)^3 (1+x)^3}
                -\frac{4096 x^3 \ln(x)}{(x-1)^3 (1+x)^3}
        \Biggr]
        -\frac{1024 x^3 \ln^2(x)}{(x-1)^3 (1+x)^3}
\nonumber\\ &&         
-\frac{4096 x^2 \big(
                3-2 x+3 x^2\big) \ln(1+x)}{(x-1)^3 (1+x)^3}
        +\frac{4096 x^3 \Li_2(x)}{(x-1)^3 (1+x)^3}
\Biggr\}
\nonumber\\ && 
+\Biggl\{
        -\frac{4096 \Li_4\left(\frac{1}{2}\right) x^3}{(x-1)^3 (1+x)^3}
        +\frac{768 x^3 \big(
                3+8 x+3 x^2\big) \zeta_3 \ln(x)}{(x-1)^5 (1+x)^3}
        -\frac{16 x^3 \ln^4(x)}{(x-1)^3 (1+x)^3}
\Biggr\} \zeta_2
\nonumber\\ && 
+\frac{64 x^3 \big(
        59+12302 x+59 x^2\big) \zeta_2^3}{105 (x-1)^5 (1+x)^3}
+\frac{128 x^3 \big(
        11+35 x+11 x^2\big) \zeta_3^2}{(x-1)^5 (1+x)^3}
\nonumber\\ && 
-\frac{1152 x^3 \big(
        1-8 x+x^2\big) \zeta_5 \ln(x)}{(x-1)^5 (1+x)^3}
+\frac{32 x^3 \big(
        119+86 x+119 x^2\big) \zeta_2^2 \ln^2(x)}{5 (x-1)^5 
(1+x)^3}
\nonumber\\ && 
-
\frac{128 x^3 \zeta_3 \ln^3(x)}{(x-1)^5 (1+x)}
+\frac{68 x^3 \big(
        1+4 x+x^2\big) \ln^6(x)}{45 (x-1)^5 (1+x)^3} +\tilde{F}^{(0), \rm rest}_{A,2}(x),
\end{eqnarray}
\begin{eqnarray}
\lefteqn{F^{(0)}_{S,1}(x) + \zeta_2 F^{(0)}_{S,2}(x) + \zeta_3 F^{(0)}_{S,3}(x) =}  
\nonumber\\ &&
\ln^4(2) \Biggl\{
        -\frac{32 \ln(x) P_7}{27 (x-1) (1+x)^3}
        -\frac{64 \ln(1-x) P_{19}}{9 (x-1) (1+x)^3}
        +\frac{64 \ln(1+x) P_{19}}{9 (x-1) (1+x)^3}
\nonumber\\ && 
        +\frac{16 P_{71}}{243 (x-1)^5 (1+x)^5}
\Biggr\}
+\Li_4\left(\frac{1}{2}\right) \Biggl\{
        -\frac{256 \ln(x) P_7}{9 (x-1) (1+x)^3}
        -\frac{512 \ln(1-x) P_{19}}{3 (x-1) (1+x)^3}
\nonumber\\ &&       
  +\frac{512 \ln(1+x) P_{19}}{3 (x-1) (1+x)^3}
        +\frac{128 P_{71}}{81 (x-1)^5 (1+x)^5}
\Biggr\}
+\ln(2) \Biggl\{
        \frac{256 \zeta_2^2 P_{20}}{3 (x-1) (1+x)^3}
\nonumber\\ &&     
    +\Biggl[
                \frac{64 \Li_2(x) P_{31}}{9 (x-1) (1+x)^3}
                +\frac{64 \Li_2(-x) P_{37}}{9 (x-1) (1+x)^3}
                -\frac{64 \ln^2(x) P_{41}}{9 (x-1)^2 (1+x)^3}
\nonumber\\ &&                
 -\frac{64 P_{80}}{81 (x-1)^4 x (1+x)^6}
                -\ln(1-x) \Biggl(
                        \frac{128 \big(
                                3+8 x+3 x^2\big)}{9 (1+x)^2}
                        -\frac{64 \ln(x) P_{22}}{3 (x-1) (1+x)^3}
                \Biggr)
\nonumber\\ &&                
 +\ln(x) \Biggl(
                        \frac{64 \ln(1+x) P_{25}}{3 (x-1) (1+x)^3}
                        -\frac{32 P_{76}}{27 (x-1)^5 (1+x)^5}
                \Biggr)
                +\frac{64 \big(
                        9-32 x+9 x^2\big) \ln(1+x)}{9 (1+x)^2}
        \Biggr] \zeta_2
\Biggr\}
\nonumber\\ && 
+\ln(2)^2 \Biggl\{
        \frac{64 \ln(x) P_{13}}{9 (x-1) (1+x)^3}
        +\frac{128 \ln(1-x) P_{19}}{3 (x-1) (1+x)^3}
        -\frac{128 \ln(1+x) P_{19}}{3 (x-1) (1+x)^3}
\nonumber\\ && 
        -\frac{64 P_{69}}{81 (x-1)^5 (1+x)^5}
\Biggr\} \zeta_2
+\Biggl\{
        \frac{192 x^2 \zeta_3 \ln(x)}{(x-1) (1+x)^3}
        -\frac{8 x^2 \ln^4(x)}{(x-1) (1+x)^3}
\Biggr\} \zeta_2
\nonumber\\ && 
+\frac{2144 x^2 \zeta_2^3}{35 (x-1) (1+x)^3}
+\frac{272 x^2 \zeta_3^2}{(x-1) (1+x)^3}
+\frac{544 x^2 \zeta_5 \ln(x)}{(x-1) (1+x)^3}
+\frac{432 x^2 \zeta_2^2 \ln^2(x)}{5 (x-1) (1+x)^3}
\nonumber\\ &&
+\frac{32 x^2 \zeta_3 \ln^3(x)}{3 (x-1) (1+x)^3}
+\frac{17 x^2 \ln^6(x)}{45 (x-1) (1+x)^3} +\tilde{F}^{(0), \rm rest}_{S}(x) ,
\end{eqnarray}
\begin{eqnarray}
\lefteqn{F^{(0)}_{P,1}(x) + \zeta_2 F^{(0)}_{P,2}(x) + \zeta_3 F^{(0)}_{P,3}(x) =}  
\nonumber\\ 
&& \ln(2) \Biggl\{
        \frac{256 \zeta_2^2 P_{28}}{9 (x-1)^3 (1+x)}
        +\Biggl[
                \frac{64 \ln(1+x) P_{16}}{9 (x-1)^2 (1+x)^2}
                +\frac{64 \Li_2(x) P_{21}}{3 (x-1)^3 (1+x)}
                +\frac{64 \Li_2(-x) P_{35}}{9 (x-1)^3 (1+x)}
\nonumber\\ && 
                -\frac{64 \ln^2(x) P_{40}}{9 (x-1)^3 (1+x)^2}
                -\frac{64 P_{81}}{81 (x-1)^4 x (1+x)^6}
                -\ln(1-x) \Biggl(
                        \frac{128 \big(
                                3-8 x+3 x^2\big)}{9 (x-1)^2}
\nonumber\\ &&               
         -\frac{64 \ln(x) P_{30}}{9 (x-1)^3 (1+x)}
                \Biggr)
                +\ln(x) \Biggl(
                        \frac{64 \ln(1+x) P_{34}
                        }{9 (x-1)^3 (1+x)}
                        -
                        \frac{32 P_{77}}{27 (x-1)^5 (1+x)^5}
                \Biggr)
        \Biggr] \zeta_2
\Biggr\}
\nonumber\\ && 
+\ln^4(2) \Biggl\{
        -\frac{32 \ln(x) P_6}{27 (x-1)^3 (1+x)}
        +\frac{16 P_{72}}{243 (x-1)^5 (1+x)^5}
        -\frac{64 \big(
                11-6 x+11 x^2\big) \ln(1-x)}{9 (x-1) (1+x)}
\nonumber\\ &&         
+\frac{64 \big(
                11-6 x+11 x^2\big) \ln(1+x)}{9 (x-1) (1+x)}
\Biggr\}
+\Li_4\left(\frac{1}{2}\right) \Biggl\{
        -\frac{256 \ln(x) P_6}{9 (x-1)^3 (1+x)}
        +\frac{128 P_{72}}{81 (x-1)^5 (1+x)^5}
\nonumber\\ &&        
 -\frac{512 \big(
                11-6 x+11 x^2\big) \ln(1-x)}{3 (x-1) (1+x)}
        +\frac{512 \big(
                11-6 x+11 x^2\big) \ln(1+x)}{3 (x-1) (1+x)}
\Biggr\}
\nonumber\\ &&
+\ln^2(2) \Biggl\{
        \frac{64 \ln(x) P_{12}}{9 (x-1)^3 (1+x)}
        -\frac{64 P_{70}}{81 (x-1)^5 (1+x)^5}
        +\frac{128 \big(
                11-6 x+11 x^2\big) \ln(1-x)}{3 (x-1) (1+x)}
\nonumber\\ &&   
     -\frac{128 \big(
                11-6 x+11 x^2\big) \ln(1+x)}{3 (x-1) (1+x)}
\Biggr\} \zeta_2
-\frac{2208 x^2 \zeta_2^3}{7 (x-1)^3 (1+x)}
-\frac{304 x^2 \zeta_3^2}{(x-1)^3 (1+x)}
\nonumber\\ && 
-\frac{448 x^2 \zeta_2 \zeta_3 \ln(x)}{(x-1)^3 (1+x)}
-\frac{288 x^2 \zeta_5 \ln(x)}{(x-1)^3 (1+x)}
-\frac{432 x^2 \zeta_2^2 \ln^2(x)}{5 (x-1)^3 (1+x)}
+\frac{64 x^2 \zeta_3 \ln^3(x)}{3 (x-1)^3 (1+x)}
\nonumber\\ && 
-\frac{17 x^2 \ln^6(x)}{45 (x-1)^3 (1+x)} +\tilde{F}^{(0), \rm rest}_{P}(x).
\end{eqnarray}
The polynomials are given by
\begin{eqnarray}
P_1&=&x^4-21 x^3+34 x^2-21 x+1,\\
P_2&=&x^4-5 x^3-10 x^2-5 x+1,\\
P_3&=&x^4-x^3-6 x^2-x+1,\\
P_4&=&x^4+5 x^3-10 x^2+5 x+1,\\
P_5&=&x^4+90 x^3+46 x^2+90 x+1,\\
P_6&=&2 x^4-13 x^3+6 x^2-13 x+2,\\
P_7&=&2 x^4-5 x^3-2 x^2-5 x+2,\\
P_8&=&2 x^4+4 x^3-29 x^2+4 x+2,\\P_9&=&3 x^4+16 x^3+62 x^2+16 
x+3,\\P_{10}&=&4 x^4-71 x^3+168 x^2-71 x+4,\\P_{11}&=&7 x^4+36 x^3+106 x^2+36 
x+7,\\P_{12}&=&8 x^4-25 x^3+24 x^2-25 x+8,\\P_{13}&=&8 x^4+7 x^3-8 x^2+7 
x+8,\\P_{14}&=&9 x^4-32 x^3+34 x^2-32 x+9,\\P_{15}&=&9 x^4-27 x^3+34 x^2-27 
x+9,\\P_{16}&=&9 x^4-22 x^3+34 x^2-22 x+9,\\P_{17}&=&9 x^4-5 x^3+98 x^2-5 
x+9,\\P_{18}&=&11 x^4-56 x^3+118 x^2-56 x+11,\\P_{19}&=&11 x^4+4 x^3+34 x^2+4 
x+11,\\P_{20}&=&12 x^4-3 x^3+41 x^2-3 x+12,\\P_{21}&=&19 x^4-38 x^3+42 x^2-38 
x+19,\\P_{22}&=&19 x^4+38 x^3+34 x^2+38 x+19,\\P_{23}&=&29 x^4-28 x^3+94 
x^2-28 x+29,\\P_{24}&=&31 x^4-141 x^3+112 x^2-141 x+31,\\P_{25}&=&33 x^4+12 
x^3+98 x^2+12 x+33,\\P_{26}&=&35 x^4-58 x^3+106 x^2-58 x+35,\\P_{27}&=&36 
x^4-107 x^3+141 x^2-107 x+36,\\P_{28}&=&36 x^4-99 x^3+127 x^2-99 
x+36,\\P_{29}&=&40 x^4-106 x^3+135 x^2-106 x+40,\\P_{30}&=&57 x^4-114 x^3+118 
x^2-114 x+57,\\P_{31}&=&57 x^4+114 x^3+110 x^2+114 x+57,\\P_{32}&=&73 x^4-192 
x^3+262 x^2-192 x+73,\\P_{33}&=&79 x^4-214 x^3+282 x^2-214 x+79,\\P_{34}&=&99 
x^4-252 x^3+310 x^2-252 x+99,\\P_{35}&=&111 x^4-276 x^3+338 x^2-276 
x+111,\\P_{36}&=&111 x^4-200 x^3+170 x^2-200 x+111,\\P_{37}&=&111 x^4+60 
x^3+274 x^2+60 x+111,\\P_{38}&=&373 x^4+6435 x^3-14420 x^2+6435 
x+373,\\P_{39}&=&28 x^5-42 x^4+35 x^3+29 x^2-117 x+55,\\P_{40}&=&70 x^5-97 
x^4+13 x^3+12 x^2-2 x+2,\\P_{41}&=&70 x^5-11 x^4+71 x^3-70 x^2-2 
x-2,\\P_{42}&=&x^6+10 x^5-23 x^4+40 x^3-23 x^2+10 x+1,\\P_{43}&=&x^6+12 
x^5+47 x^4+96 x^3+47 x^2+12 x+1,\\P_{44}&=&4 x^6+13 x^5-38 x^4+52 x^3-38 
x^2+13 x+4,\\P_{45}&=&4 x^6+21 x^5+80 x^4+60 x^3+80 x^2+21 x+4,\\P_{46}&=&27 
x^6-35 x^5+85 x^4-114 x^3+85 x^2-35 x+27,\\P_{47}&=&27 x^6-29 x^5+15 
x^4-14 x^3+15 x^2-29 x+27,\\P_{48}&=&27 x^6-28 x^5-15 x^4+38 x^3-15 
x^2-28 x+27,\\P_{49}&=&27 x^6+136 x^5-245 x^4-221 x^3+202 x^2+223 
x-134,\\P_{50}&=&36 x^6-19 x^5-67 x^4-66 x^3-67 x^2-19 x+36,\\P_{51}&=&57 
x^6+4 x^5-65 x^4-16 x^3-65 x^2+4 x+57,\\P_{52}&=&57 x^6+4 x^5-57 x^4-16 
x^3-57 x^2+4 x+57,\\P_{53}&=&57 x^6+224 x^5+423 x^4+344 x^3+423 x^2+224 
x+57,\\P_{54}&=&57 x^6+224 x^5+431 x^4+216 x^3+431 x^2+224 x+57,\\P_{55}&=&69 
x^6-905 x^5-1667 x^4+6264 x^3-1856 x^2-797 x+24,\\P_{56}&=&75 x^6-58 
x^5-2637 x^4+5456 x^3-2907 x^2-598 x-51,\\P_{57}&=&99 x^6-212 x^5+333 
x^4-448 x^3+333 x^2-212 x+99,\\P_{58}&=&99 x^6+68 x^5+77 x^4-96 x^3+77 
x^2+68 x+99,\\P_{59}&=&111 x^6+104 x^5+217 x^4-368 x^3+217 x^2+104 
x+111,\\P_{60}&=&285 x^6-440 x^5-4707 x^4+10912 x^3-6381 x^2-872 
x-237,\\P_{61}&=&70 x^7+15 x^6-69 x^5-94 x^4+6 x^3+127 x^2-53 
x+2,\\P_{62}&=&27 x^8-328 x^7-1934 x^6-2456 x^5-7952 x^4-2024 x^3-962 
x^2+104 x-27,\\P_{63}&=&27 x^8+494 x^7+3382 x^6+4750 x^5+15904 x^4+4210 
x^3+2410 x^2-46 x-27,\\P_{64}&=&60 x^8-537 x^7-2164 x^6+5223 x^5+8561 
x^4+5277 x^3-2002 x^2-483 x+33,\\P_{65}&=&70 x^8+173 x^7-342 x^6+245 
x^5+860 x^4+241 x^3-458 x^2+45 x-2,\\
P_{66}&=&1280 x^8+3209 x^7+41288 x^6+2239 x^5-120320 x^4+2239 x^3+41288 x^2+3209 x
\nonumber\\ &&
+1280,\\
P_{67}&=&17 x^{10}+1828 x^9-2261 x^8-2392 x^7-7422 x^6+24304 x^5-6162 x^4-3688 x^3
\nonumber\\ &&
-2531 x^2+2476 x-73,\\P_{68}&=&17 x^{10}+2276 x^9-2885 x^8-24696 x^7+91794 x^6
\nonumber\\ &&
-126000 x^5+85278 x^4-20808 x^3-2867 x^2+2060 x-73,\\
P_{69}&=&32 x^{10}-1617 x^9+1830 x^8+1114 x^7+1234 x^6-9448 x^5+5734 x^4-506 x^3
\nonumber\\ &&
+390 x^2-807 x-4,\\
P_{70}&=&32 x^{10}-1377 x^9+82 x^8+2442 x^7-1938 x^6-728 x^5-1110 x^4+1686 x^3
\nonumber\\ &&
-134 x^2-999 x-4,\\
P_{71}&=&37 x^{10}-2748 x^9+2823 x^8+1256 x^7+5114 x^6-18896 x^5+8822 x^4-40 x^3
\nonumber\\ &&
+1617 x^2-2100 x+19,\\
P_{72}&=&37 x^{10}-2592 x^9-241 x^8+4560 x^7-2526 x^6-1456 x^5-3570 x^4+3696 x^3
\nonumber\\ &&
+137 x^2-2160 x+19,\\
P_{73}&=&91 x^{10}-3086 x^9+5585 x^8+19836 x^7-87474 x^6 +126000 x^5-89598 x^4
\nonumber\\ &&
+25668 x^3+167 x^2-1250 x-35,\\
P_{74}&=&91 x^{10}-2314 x^9+2585 x^8+3364 x^7+5910 x^6-24304 x^5+7674 x^4+2716 x^3
\nonumber\\ &&
+2207 x^2-1990 x-35,\\
P_{75}&=&208 x^{10}+1947 x^9-6722 x^8+104 x^7+33106 x^6+61370 x^5+33106 x^4+104 x^3
\nonumber\\ &&
-6722 x^2+1947 x+208,\\
P_{76}&=&437 x^{10}-866 x^9-1849 x^8+6412 x^7-7694 x^6+5232 x^5-5930 x^4+5764 x^3
\nonumber\\ &&
-2407 x^2-542 x+419,\\P_{77}&=&437 x^{10}-422 x^9-3057 x^8+148 x^7+3786 x^6-3312 x^5
\nonumber\\ &&
+4686 x^4-68 x^3-3327 x^2-314 x+419,\\
P_{78}&=&557 x^{10}-2012 x^9+4515 x^8+10256 x^7-65160 x^6+103636 x^5-68040 x^4
\nonumber\\ &&
+13496 x^3+2715 x^2-1472 x+485,\\P_{79}&=&557 x^{10}-1736 x^9-5117 x^8+18176 x^7
\nonumber\\ &&
-4312 x^6-16556 x^5-3304 x^4+17528 x^3-5333 x^2-1412 x+485,\\
P_{80}&=&536 x^{12}-2203 x^{11}+807 x^{10}-2747 x^9+13604 x^8-42 x^7-16838 x^6-42 x^5
\nonumber\\ &&
+13604 x^4-2747 x^3+807 x^2-2203 x+536,\\
P_{81}&=&848 x^{12}-1467 x^{11}-3883 x^{10}+10033 x^9+11148 x^8-4342 x^7-21602 x^6
\nonumber\\ &&
-4342 x^5+11148 x^4+10033 x^3-3883 x^2-1467 x+848,\\
P_{82}&=&2768 x^{12}-13405 x^{11}+1776 x^{10}-116585 x^9+235760 x^8+331206 x^7
\nonumber\\ &&
-870752 x^6+331206 x^5+235760 x^4-116585 x^3+1776 x^2-13405 x+2768,\\
P_{83}&=&2768 x^{12}-10461 x^{11}+15692 x^{10}-6353 x^9-33936 x^8-3154 x^7+83176 x^6
\nonumber\\ &&
-3154 x^5-33936 x^4-6353 x^3+15692 x^2-10461 x+2768.
\end{eqnarray}

\section{\boldmath Partially resummed expansions at $\hs=4$ }
\label{app:z-resum}

\vspace*{1mm}
\noindent
In the main text, we showed the partially resummed expansion of 
$F^{(0)}_{V,1,1}(z) + \zeta_2 F^{(0)}_{V,1,2}(z) + \zeta_3 F^{(0)}_{V,1,3}(z)$. 
We give now the corresponding results for the remaining factors. The $G$--functions appearing in 
these expressions are given in Eqs.~(\ref{eq:G1func}--\ref{eq:G4func}).

\begin{eqnarray}
\lefteqn{F^{(0)}_{V,2,1}(z) + \zeta_2 F^{(0)}_{V,2,2}(z) + \zeta_3 F^{(0)}_{V,2,3}(z) =} \nonumber\\ &&
\frac{64 \pi ^2 \frac{1}{\big(
        4-z^2\big)^{3/2}} \GA\left(
        \frac{1}{\tau },\sqrt{4-\tau } \sqrt{\tau };z^2\right) \ln(z) 
Q_5}{9 z^5}
+\frac{128 \ln(2) \pi ^3 \frac{1}{\big(
        4-z^2\big)^{3/2}} \ln\big(
        1-\frac{z^2}{4}\big) Q_{14}}{27 z^5}
\nonumber\\ && 
+\frac{512 \pi ^2 \ln^2(z) Q_{19}}{27 (z-2) z^4 (z+2)}
-\frac{16 \pi  \frac{1}{\big(
        4-z^2\big)^{3/2}} \ln^2(z) Q_{60}}{27 z^3}
+\frac{2 \ln(2) \pi ^3 \frac{1}{\sqrt{4-z^2}} Q_{63}}{243 z^5}
\nonumber\\ && 
-\frac{2 \pi ^2 \ln(z) Q_{68}}{81 (z-2) z^2 (z+2)}
+\ln(2) \pi ^3 \ln(z) \Biggl[
        -\frac{4 \frac{1}{\sqrt{4-z^2}} Q_3}{27 z^5}
        -\frac{4}{9} z^3 \frac{1}{\big(
                4-z^2\big)^{3/2}}
\Biggr]
\nonumber\\ && 
+\Li_4\left(\frac{1}{2}\right) \pi  \Biggl[
        \frac{256}{3} z^3 \frac{1}{\big(
                4-z^2\big)^{3/2}}
        +\frac{256 \big(
                340+108 z^2+27 z^4\big) \frac{1}{\sqrt{4-z^2}}}{81 
z^3}
\nonumber\\ && 
        -\frac{2048 \big(
                z^2-3\big) \frac{1}{\sqrt{4-z^2}} \ln\big(
                1-\frac{z^2}{4}\big)}{z^5}
\Biggr]
+\ln^4(2) \pi  \Biggl[
        \frac{32}{9}
         z^3 
        \frac{1}{\big(
                4-z^2\big)^{3/2}}
\nonumber\\ &&         
+\frac{32 \big(
                340+108 z^2+27 z^4\big) \frac{1}{\sqrt{4-z^2}}}{243 
z^3}
        -\frac{256 \big(
                z^2-3\big) \frac{1}{\sqrt{4-z^2}} \ln\big(
                1-\frac{z^2}{4}\big)}{3 z^5}
\Biggr]
\nonumber\\ && 
+\ln^2(2) \pi ^3 \Biggl[
        -\frac{14}{3} z^3 \frac{1}{\big(
                4-z^2\big)^{3/2}}
        -\frac{2 \big(
                -95696+33372 z^2+567 z^4\big) 
\frac{1}{\sqrt{4-z^2}}}{243 z^3}
\nonumber\\ &&       
  +\frac{256 \big(
                z^2-3\big) \frac{1}{\sqrt{4-z^2}} \ln\big(
                1-\frac{z^2}{4}\big)}{3 z^5}
\Biggr]
+\pi ^5 \Biggl[
        -\frac{16 \big(
                -20915+6761 z^2\big) \frac{1}{\big(
                4-z^2\big)^{3/2}}}{3645 z^3}
\nonumber\\ && 
        +\frac{32 \big(
                3+11 z^2\big) \frac{1}{\sqrt{4-z^2}} \ln\big(
                1-\frac{z^2}{4}\big)}{45 z^5}
\Biggr]
+\Biggl[
        \frac{128 \ln^2(2) \pi  \big(
                -36-4 z^2+5 z^4\big)}{27 (z-2) z^4 (z+2)}
\nonumber\\ && 
        +\frac{64 \pi  \ln^2(z) Q_{33}}{27 (z-2) z^4 (z+2)}
        +\frac{32 \ln(2) \pi ^3 Q_{35}}{3 (z-2) z^6 (z+2)}
        -\frac{4 \pi ^2 \frac{1}{\big(
                4-z^2\big)^{3/2}} \ln(z) Q_{69}}{27 z^5}
        -\frac{3584 \pi  \big(
                z^2-2\big) \zeta_3}{3 (z-2) z^6 (z+2)}
\nonumber\\ &&    
     +\frac{1024 \pi ^3 \big(
                z^2-2\big) \ln(z)}{9 (z-2) z^6 (z+2)}
        +\frac{256 \ln(2) \pi ^2 \big(
                -10+z^2
        \big)
\big(z^2-3\big) \frac{1}{\big(
                4-z^2\big)^{3/2}} \ln(z)}{9 z^5}
\nonumber\\ &&  
        +
        \frac{256 \ln(2) \pi  \big(
                -36-4 z^2+5 z^4\big) \ln(z)}{9 (z-2) z^4 (z+2)}
      +\frac{512 \pi ^2 \big(
                1758-855 z^2+104 z^4\big) \frac{1}{\big(
                4-z^2\big)^{3/2}} \ln^2(z)}{27 z^5}
\Biggr] 
\nonumber\\ && \times 
\GA\left(
        \sqrt{4-\tau } \sqrt{\tau };z^2\right) 
+\Biggl[
        \frac{8 \pi ^2 \ln(z) Q_{36}}{3 (z-2) z^6 (z+2)}
        -\frac{32 \ln(2) \pi ^3 (-3+z) (3+z) 
\frac{1}{\sqrt{4-z^2}}}{z^5}
\nonumber\\ && 
        -\frac{512 \ln(2) \pi  \frac{1}{\sqrt{4-z^2}} \ln(z)}{3 
z^5}
        -\frac{256 \ln^2(2) \pi  \frac{1}{\sqrt{4-z^2}}}{9 z^5}
        +\frac{64 \pi  \big(
                1612-814 z^2+103 z^4\big) \frac{1}{\big(
                4-z^2\big)^{3/2}} \ln^2(z)}{9 z^5}
\Biggr] 
\nonumber\\ &&  \times
\GA\big(
        \sqrt{4-\tau } \sqrt{\tau };z^2\big)^2
+\frac{778}{81} \ln(2) \pi ^3 z^3 \frac{1}{\big(
        4-z^2\big)^{3/2}}
-\frac{4096 \Li_4\left(\frac{1}{2}\right) \ln(2) \pi  \frac{1}{\sqrt{4-z^2}}}{3 
z^3}
\nonumber\\ && 
-\frac{512 \ln(2)^5 \pi  \frac{1}{\sqrt{4-z^2}}}{9 z^3}
+\frac{512 \ln^3(2) \pi ^3 \frac{1}{\sqrt{4-z^2}
}}{9 z^3}
+
\frac{1600 \ln(2) \pi ^5 \frac{1}{\sqrt{4-z^2}}}{27 z^3}
+\frac{\pi ^2 \big(
        -9+2 z^2\big) \kappa_2 \frac{1}{\sqrt{4-z^2}}}{9 
z^5}
\nonumber\\ && 
+\frac{2 \big(
        -9+2 z^2\big) \kappa_9 \frac{1}{\sqrt{4-z^2}}}{3 
z^5}
-\frac{128 \ln^2(2) \pi  \big(
        9+z^2\big) \sqrt{4-z^2}}{27 z}
+\pi ^3 \Biggl[
        \frac{\frac{1}{\sqrt{4-z^2}} Q_{27}}{270 z^5}
        +\frac{1}{2} z^3 \frac{1}{\big(
                4-z^2\big)^{3/2}}
\Biggr] \zeta_3
\nonumber\\ && 
-\frac{8664320 \big(-9+2 z^2\big) \frac{\zeta_3}{\sqrt{4-z^2}}}{729 \pi  z^5}
+\frac{7168 \Li_4\left(\frac{1}{2}\right) \big(-9+2 z^2\big) \frac{\zeta_3}{\sqrt{4-z^2}}}{\pi  z^5}
\nonumber\\ &&
+\frac{8137472 \ln(2) \big(-9+2 z^2\big) \frac{\zeta_3}{\sqrt{4-z^2}}}{729 \pi  z^5}
+\frac{68096 \ln^2(2) \big(-9+2 z^2\big) \frac{\zeta_3}{\sqrt{4-z^2}}}{243 \pi  z^5}
\nonumber\\ &&
-\frac{28672 \ln^3(2) \big(-9+2 z^2\big) \frac{\zeta_3}{\sqrt{4-z^2}}}{27 \pi  z^5}
+\frac{896 \ln^4(2) \big(-9+2 z^2\big) \frac{\zeta_3}{\sqrt{4-z^2}}}{3 \pi  z^5}
\nonumber\\ &&
-\frac{257024 \ln(2) \pi  \big(-9+2 z^2\big) \frac{\zeta_3}{\sqrt{4-z^2}}}{81 z^5}
+\frac{2944 \ln(2)^2 \pi  \big(-9+2 z^2\big) \frac{\zeta_3}{\sqrt{4-z^2}}}{3 z^5}
\nonumber\\ &&
-\frac{128 \pi  \big(-45909-4864 z^2+7533 z^4\big) \frac{\zeta_3}{\sqrt{4-z^2}}}{729 z^5}
-\frac{17 \big(-9+2 z^2\big) \kappa_1 
\frac{\zeta_3}{\sqrt{4-z^2}}}{18 z^5}
\nonumber\\ &&
+\frac{4 \big(-9+2 z^2\big) \kappa_3 \frac{\zeta_3}{\sqrt{4-z^2}}}{9 z^5}
+\frac{196448 \big(-9+2 z^2\big) \frac{\zeta_3^2}{\sqrt{4-z^2}}}{81 \pi  z^5}
+\pi  \Biggl[
        \frac{5}{2} z^3 \frac{1}{\big(
                4-z^2\big)^{3/2}}
\nonumber\\ && 
 +\frac{5 \big(
                144+4 z^2+z^4\big) \frac{1}{\sqrt{4-z^2}}}{2 z^3}
\Biggr] \zeta_5
+\frac{128 \ln(2) \pi ^2 \big(
        z^2-6\big) \ln(z)}{9 z^2} \nonumber\\ && +\frac{704 \pi ^3 \big(
        z^2-2\big) \frac{1}{\sqrt{4-z^2}} \ln(z)}{9 z^3}
-\frac{256 \ln(2) \pi  \big(
        9+z^2\big) \sqrt{4-z^2} \ln(z)}{9 z}
\nonumber\\ && 
-\frac{16 \pi ^2 \big(
        z^2-3\big) \frac{1}{\sqrt{4-z^2}} \GA\big(
        \sqrt{4-\tau } \sqrt{\tau };z^2\big)^3 \ln(z)}{3 z^5}
\nonumber\\ && 
-\frac{256 \pi ^2 \big(
        z^2-6
\big)\big(z^2-3\big) \frac{1}{\big(
        4-z^2\big)^{3/2}} \GA\left(
        \frac{1}{4-\tau },\sqrt{4-\tau } \sqrt{\tau };z^2\right) \ln
(z)}{27 z^5}
\nonumber\\ &&
+
\frac{32}{9} \pi  z^3 \frac{1}{\big(
        4-z^2\big)^{3/2}} \ln^2(z)
-\frac{512 \pi  \big(
               z^2-2\big) \frac{1}{\sqrt{4-z^2}} \ln^3(z)}{9 z^3}
               +\tilde{F}^{(0), \rm rest}_{V,2}(z),
\end{eqnarray}
\begin{eqnarray}
\lefteqn{F^{(0)}_{A,1,1}(z) + \zeta_2 F^{(0)}_{A,1,2}(z) + \zeta_3 F^{(0)}_{A,1,3}(z) =} \nonumber\\ &&
-\frac{8}{9} \ln(2) \pi  z \sqrt{4-z^2} \ln(z) Q_9
-\frac{4}{27} \ln^2(2) \pi  z \sqrt{4-z^2} Q_{10}
+\frac{32 \ln(2) \pi ^3 \frac{1}{\big(
        4-z^2\big)^{3/2}} \ln\big(
        1-\frac{z^2}{4}\big) Q_{12}}{27 z^3}
\nonumber\\ && 
-\frac{64 \ln(2) \pi ^3 \frac{1}{\big(
        4-z^2\big)^{3/2}} \ln(z) Q_{23}}{27 z^3}
-\frac{64 \pi ^2 \ln^2(z) Q_{37}}{27 (z-2) z^2 (z+2)}
-\frac{4 \pi ^3 \frac{1}{\sqrt{4-z^2}} \ln(z) Q_{47}}{27 z^3}
-\frac{16 \pi  \frac{\zeta_3}{\sqrt{4-z^2}} Q_{54}}{243 z^3}
\nonumber\\ && 
-\frac{32 \pi  \frac{1}{\sqrt{4-z^2}} \ln^3(z) Q_{51}}{81 z^3}
+\frac{4 \pi  \frac{1}{\sqrt{4-z^2}} \ln^2(z) Q_{74}}{27 z^3}
+\frac{2 \ln(2) \pi ^3 \frac{1}{\big(
        4-z^2\big)^{3/2}} Q_{81}}{81 z^5}
\nonumber\\ && 
-\frac{\pi ^2 \ln(z) Q_{82}}{486 (z-2) z^2 (z+2)}
+\ln^2(2) \pi ^3 \Biggl[
        \frac{32 \frac{1}{\big(
                4-z^2\big)^{3/2}} Q_{25}}{81 z^3}
        -\frac{64 \frac{1}{\sqrt{4-z^2}} \ln\big(
                1-\frac{z^2}{4}\big)}{3 z^3}
\Biggr]
\nonumber\\ && 
+\pi ^5 \Biggl[
        \frac{4 \frac{1}{\big(
                4-z^2\big)^{3/2}} Q_{34}}{1215 z^3}
        +\frac{8 \frac{1}{\sqrt{4-z^2}} \ln\big(
                1-\frac{z^2}{4}\big)}{45 z^3}
\Biggr]
+\ln^4(2) \pi  \Biggl[
        \frac{128 \frac{1}{\big(
                4-z^2\big)^{3/2}} Q_2}{81 z^3}
        +\frac{64 \frac{1}{\sqrt{4-z^2}} \ln\big(
                1-\frac{z^2}{4}\big)}{3 z^3}
\Biggr]
\nonumber\\ && 
+\Li_4\left(\frac{1}{2}\right) \pi  \Biggl[
        \frac{1024 \frac{1}{\big(
                4-z^2\big)^{3/2}} Q_2}{27 z^3}
        +\frac{512 \frac{1}{\sqrt{4-z^2}} \ln\big(
                1-\frac{z^2}{4}\big)}{z^3}
\Biggr]
+\Biggl[
        -\frac{64 \ln^3(2) \pi  \big(
                z^2-2\big)^2}{9 (z-2) z^2 (z+2)}
\nonumber\\ &&        
 -\frac{128 \pi ^2 \frac{1}{\big(
                4-z^2\big)^{3/2}} \ln^2(z) Q_6}{27 z^3}
        -\frac{16 \ln(2) \pi ^3 Q_{30}}{27 (z-2) z^4 (z+2)}
        +\frac{16 \ln^2(2) \pi  Q_{41}}{27 (z-2) z^2 (z+2)}
     +\frac{32 \ln(2) \pi  \ln(z) Q_{41}}{9 (z-2) z^2 
(z+2)}
\nonumber\\ &&   
        +\frac{16 \pi  \ln^2(z) Q_{46}}{27 (z-2) z^2 (z+2)}
        -\frac{\pi ^2 \frac{1}{\big(
                4-z^2\big)^{3/2}} \ln(z) Q_{79}}{81 z^5}
        -\frac{224 \pi  \big(
                z^2-2
        \big)
\big(12-62 z^2+31 z^4\big) \zeta_3}{27 (z-2) z^4 (z+2)}
\nonumber\\ &&        
 -\frac{64 \ln(2)
        ^2 \pi  \big(
                z^2-2\big)^2 \ln(z)}{(z-2) z^2 (z+2)}
        -
        \frac{8 \pi ^3 \big(
                z^2-2
        \big)
\big(-32-278 z^2+139 z^4\big) \ln(z)}{27 (z-2) z^4 (z+2)}
\nonumber\\ &&    
     -\frac{64 \ln(2) \pi ^2 \big(
                14-5 z^4+z^6\big) \frac{1}{\big(
                4-z^2\big)^{3/2}} \ln(z)}{9 z^3}
        -\frac{192 \ln(2) \pi  \big(
                z^2-2\big)^2 \ln^2(z)}{(z-2) z^2 (z+2)}
\nonumber\\ &&   
     -\frac{10432 \pi  \big(
                z^2-2\big)^2 \ln^3(z)}{81 (z-2) z^2 (z+2)}
\Biggr] \GA\left(
        \sqrt{4-\tau } \sqrt{\tau };z^2\right)
+\Biggl[
        \frac{2 \pi ^2 \ln(z) Q_{20}}{9 (z-2) z^4 (z+2)}
\nonumber\\ &&    
     +\frac{16 \pi  \frac{1}{\big(
                4-z^2\big)^{3/2}} \ln^2(z) Q_{21}}{27 z^3}
        +\frac{24 \ln(2) \pi ^3 \frac{1}{\sqrt{4-z^2}}}{z^3}
        -\frac{16 \ln^2(2) \pi  \big(
                4-22 z^2+11 z^4\big) \frac{1}{\sqrt{4-z^2}}}{27 z^3}
\nonumber\\ &&    
     -\frac{32 \ln(2) \pi  \big(
                4-22 z^2+11 z^4\big) \frac{1}{\sqrt{4-z^2}} \ln
(z)}{9 z^3}
\Biggr] \GA\left(
        \sqrt{4-\tau } \sqrt{\tau };z^2\right)^2
+\Biggl[
        \frac{64 \ln^2(2) \pi  \big(
                z^2-2\big)^2}{27 (z-2) z^2 (z+2)}
\nonumber\\ &&   
     +\frac{128 \ln(2) \pi  \big(
                z^2-2\big)^2 \ln(z)}{9 (z-2) z^2 (z+2)}
        -\frac{64 \pi ^2 \big(
                -26+18 z^2-7 z^4+z^6\big) \frac{1}{\big(
                4-z^2\big)^{3/2}} \ln(z)}{27 z^3}
\nonumber\\ &&       
  +
        \frac{320 \pi  \big(
                z^2-2\big)^2 \ln^2(z)}{27 (z-2) z^2 (z+2)}
\Biggr] \GA\left(
        \frac{1}{4-\tau },\sqrt{4-\tau } \sqrt{\tau };z^2\right)
+\Biggl[
        \frac{32 \ln^2(2) \pi  \big(
                z^2-2\big)^2}{3 (z-2) z^2 (z+2)}
\nonumber\\ &&    
     +\frac{16 \pi ^2 \frac{1}{\big(
                4-z^2\big)^{3/2}} \ln(z) Q_{17}}{9 z^3}
        +\frac{64 \ln(2) \pi  \big(
                z^2-2\big)^2 \ln(z)}{(z-2) z^2 (z+2)}
        +\frac{2848 \pi  \big(
                z^2-2\big)^2 \ln^2(z)}{27 (z-2) z^2 (z+2)}
\Biggr] 
\nonumber\\ && \times 
\GA\left(
        \frac{1}{\tau },\sqrt{4-\tau } \sqrt{\tau };z^2\right)
+\Biggl[
        \frac{32 \ln^2(2) \pi  \big(
                z^2-2\big)^2}{3 (z-2) z^2 (z+2)}
    +\frac{64 \ln(2) \pi  \big(
                z^2-2\big)^2 \ln(z)}{(z-2) z^2 (z+2)}
\nonumber\\ &&     
        +\frac{96 \pi  \big(
                z^2-2\big)^2 \ln^2(z)}{(z-2) z^2 (z+2)}
\Biggr] \GA\left(
        \sqrt{4-\tau } \sqrt{\tau },\frac{1}{4-\tau };z^2\right)
-\frac{\pi ^2 \kappa_2 \frac{1}{\sqrt{4-z^2}}}{12 z^3}
-\frac{\kappa_9 \frac{1}{\sqrt{4-z^2}}}{2 z^3}
\nonumber\\ && 
+\frac{32}{9} \ln^3(2) \pi  z \big(
        z^2-2\big) \sqrt{4-z^2}
-\frac{\pi ^3 \big(
        -2404+1441 z^2\big) \frac{\zeta_3}{\big(
        4-z^2\big)^{3/2}}}{45 z^3}
+
\frac{2166080 \frac{\zeta_3}{\sqrt{4-z^2}}}{243 \pi  z^3}
\nonumber\\ && 
-\frac{5376 \Li_4\left(\frac{1}{2}\right) \frac{\zeta_3}{\sqrt{4-z^2}}}{\pi  z^3}
-\frac{2034368 \ln(2) \frac{\zeta_3}{\sqrt{4-z^2}}}{243 \pi  z^3}
-\frac{17024 \ln^2(2) \frac{\zeta_3}{\sqrt{4-z^2}}}{81 \pi  z^3}
+\frac{7168 \ln^3(2) \frac{\zeta_3}{\sqrt{4-z^2}}}{9 \pi  z^3}
\nonumber\\ && 
-\frac{224 \ln^4(2) \frac{\zeta_3}{\sqrt{4-z^2}}}{\pi  z^3}
+\frac{64256 \ln(2) \pi  \frac{\zeta_3}{\sqrt{4-z^2}}}{27 z^3}
-\frac{736 \ln^2(2) \pi  \frac{\zeta_3}{\sqrt{4-z^2}}}{z^3}
+\frac{17 \kappa_1 \frac{\zeta_3}{\sqrt{4-z^2}}}{24 z^3}
-\frac{\kappa_3 \frac{\zeta_3}{\sqrt{4-z^2}}}{3 z^3}
\nonumber\\ && 
-\frac{49112 \frac{\zeta_3^2}{\sqrt{4-z^2}}}{27 \pi  z^3}
+\frac{40 \pi  \big(
        8+z^2\big) \frac{\zeta_5}{\big(
        4-z^2\big)^{3/2}}}{z^3}
-\frac{32 \ln(2) \pi ^2 \big(
        24-6 z^2-3 z^4+z^6\big) \ln(z)}{9 z^2}
\nonumber\\ && 
-\frac{640 \pi ^4 \big(
        z^2-2\big)^2 \ln(z)}{27 (z-2) z^2 (z+2)}
+32 \ln^2(2) \pi  z \big(
        z^2-2\big) \sqrt{4-z^2} \ln(z)
\nonumber\\ && 
+\frac{496}{9} \ln(2)
 \pi  z \big(
        4-z^2\big)^{3/2} \ln(z)
+
\frac{4 \pi ^2 \frac{1}{\sqrt{4-z^2}} \GA\left(
        \sqrt{4-\tau } \sqrt{\tau };z^2\right)^3 \ln(z)}{3 z^3}
\nonumber\\ && 
+96 \ln(2) \pi  z \big(
        z^2-2\big) \sqrt{4-z^2} \ln^2(z)
+\pi ^2 \Bigl[
        -\frac{1024}{27}
        +\frac{4096}{81 z^2}
        -\frac{1024 z^2}{81 (z-2) (z+2)}
\Biggr] \ln^3(z)
\nonumber\\ && 
+\frac{16}{3} \ln^2(2) \pi  z \big(
        z^2-2\big) \sqrt{4-z^2} \ln\big(
        1-\frac{z^2}{4}\big)
+32 \ln(2) \pi  z \big(
        z^2-2\big) \sqrt{4-z^2} \ln(z) 
\nonumber\\ && \times
\ln\big(
        1-\frac{z^2}{4}\big)
+48 \pi  z \big(
        z^2-2\big) \sqrt{4-z^2} \ln^2(z) 
\ln\big(
               1-\frac{z^2}{4}\big)
                              +\tilde{F}^{(0), \rm rest}_{A,1}(z) ,
\end{eqnarray}
\begin{eqnarray}
\lefteqn{F^{(0)}_{A,2,1}(z) + \zeta_2 F^{(0)}_{A,2,2}(z) + \zeta_3 F^{(0)}_{A,2,3}(z) =} \nonumber\\ &&
\frac{512 \pi ^2 \ln^2(z) Q_4}{9 (z-2)^2 z^2 (z+2)^2}
-\frac{16 \pi  \frac{1}{\big(
        4-z^2\big)^{3/2}} \ln^2(z) Q_{58}}{27 z^3}
+\frac{2 \pi ^2 \ln(z) Q_{70}}{81 (z-2)^2 z^2 (z+2)^2}
\nonumber\\ && 
-\frac{8 \ln(2) \pi ^3 \frac{1}{\big(
        4-z^2\big)^{5/2}} Q_{75}}{27 z^5}
+\Li_4\left(\frac{1}{2}\right) \pi  \Biggl[
        -\frac{1024 \big(
                192-140 z^2+29 z^4\big) \frac{1}{\big(
                4-z^2\big)^{5/2}}}{9 z^3}
\nonumber\\ &&   
      -\frac{2048 \frac{1}{\big(
                4-z^2\big)^{3/2}} \ln\big(
                1-\frac{z^2}{4}\big)}{z^3}
\Biggr]
+\ln^4(2) \pi  \Biggl[
        -
        \frac{128 \big(
                192-140 z^2+29 z^4\big) \frac{1}{\big(
                4-z^2\big)^{5/2}}}{27 z^3}
\nonumber\\ &&   
      -\frac{256 \frac{1}{\big(
                4-z^2\big)^{3/2}} \ln\big(
                1-\frac{z^2}{4}\big)}{3 z^3}
\Biggr]
+\pi ^5 \Biggl[
        \frac{16 \big(
                6288-4393 z^2+805 z^4\big) \frac{1}{\big(
                4-z^2\big)^{5/2}}}{405 z^3}
\nonumber\\ &&   
      -\frac{32 \frac{1}{\big(
                4-z^2\big)^{3/2}} \ln\big(
                1-\frac{z^2}{4}\big)}{45 z^3}
\Biggr]
+\ln^2(2) \pi ^3 \Biggl[
        \frac{128 \big(
                720-445 z^2+67 z^4\big) \frac{1}{\big(
                4-z^2\big)^{5/2}}}{9 z^3}
\nonumber\\ &&    
    +\frac{256 \frac{1}{\big(
                4-z^2\big)^{3/2}} \ln\big(
                1-\frac{z^2}{4}\big)}{3 z^3}
\Biggr]
+\Biggl[
        \frac{128 \ln^2(2) \pi  \big(
                92-84 z^2+17 z^4\big)}{27 (z-2)^2 z^2 (z+2)^2}
        -\frac{32 \ln(2) \pi ^3 Q_{15}}{9 (z-2)^2 z^4 (z+2)^2}
\nonumber\\ &&    
     +\frac{64 \pi  \ln^2(z) Q_{26}}{27 (z-2)^2 z^2 (z+2)^2}
        +\frac{4 \pi ^2 \frac{1}{\big(
                4-z^2\big)^{5/2}} \ln(z) Q_{67}}{27 z^5}
        -\frac{3584 \pi  \big(
                z^2-2\big) \zeta_3}{9 (z-2)^2 z^4 (z+2)^2}
\nonumber\\ &&   
        +\frac{256 \ln(2) \pi  \big(
                92-84 z^2+17 z^4\big) \ln(z)}{9 (z-2)^2 z^2 
(z+2)^2}
        +
        \frac{256 \ln(2) \pi ^2 \big(
                z^2-7
        \big)
\big(z^2-2\big) \frac{1}{\big(
                4-z^2\big)^{5/2}} \ln(z)}{9 z^3}
\nonumber\\ && 
      +\frac{1024 \pi ^3 \big(
                z^2-2\big) \ln(z)}{27 (z-2)^2 z^4 (z+2)^2}
        +\frac{512 \pi ^2 \big(
                1130-425 z^2+36 z^4\big) \sqrt{4-z^2} \ln^2(z)}{27 
(z-2)^3 z^3 (z+2)^3}
\Biggr] \GA\left(
        \sqrt{4-\tau } \sqrt{\tau };z^2\right)
\nonumber\\ && 
+\Biggl[
        \frac{256 \ln^2(2) \pi  \frac{1}{\big(
                4-z^2\big)^{3/2}}}{27 z^3}
        -\frac{96 \ln(2) \pi ^3 \frac{1}{\big(
                4-z^2\big)^{3/2}}}{z^3}
        -\frac{8 \pi ^2 \big(
                -256-16 z^2+9 z^6\big) \ln(z)}{9 (z-2)^2 z^4 
(z+2)^2}
\nonumber\\ && 
        +\frac{512 \ln(2) \pi  \frac{1}{\big(
                4-z^2\big)^{3/2}} \ln(z)}{9 z^3}
        +\frac{64 \pi  \big(
                3340-1270 z^2+111 z^4\big) \sqrt{4-z^2} \ln
^2(z)}{27 (z-2)^3 z^3 (z+2)^3}
\Biggr] 
\nonumber\\ &&  \times
\GA\left(
        \sqrt{4-\tau } \sqrt{\tau };z^2\right)^2
+\frac{128}{27} \ln^2(2) \pi  z \big(
        -23+8 z^2\big) \frac{1}{\sqrt{4-z^2}}
+\frac{\pi ^2 \kappa_2 \frac{1}{\big(
        4-z^2\big)^{3/2}}}{3 z^3}
\nonumber\\ && 
+\frac{2 \kappa_9 \frac{1}{\big(
        4-z^2\big)^{3/2}}}{z^3}
+\frac{1}{\pi } \Biggl[
        -\frac{135380}{243} z^3 \frac{1}{\big(
                4-z^2\big)^{3/2}}
        -\frac{135380 \big(
                4-2 z+z^2
        \big)
\big(4+2 z+z^2\big) \frac{1}{\sqrt{4-z^2}}}{243 z^3}
\Biggr] 
\nonumber\\ &&  \times \zeta_3
+
\frac{4 \pi ^3 \big(
        -2404+3121 z^2\big) \frac{\zeta_3}{\big(
        4-z^2\big)^{5/2}}}{45 z^3}
+\frac{21504 \Li_4\left(\frac{1}{2}\right) \frac{\zeta_3}{\big(
        4-z^2\big)^{3/2}}}{\pi  z^3}
+\frac{8137472 \ln(2) \frac{\zeta_3}{\big(
        4-z^2\big)^{3/2}}}{243 \pi  z^3}
\nonumber\\ && 
+\frac{68096 \ln^2(2) \frac{\zeta_3}{\big(
        4-z^2\big)^{3/2}}}{81 \pi  z^3}
-\frac{28672 \ln^3(2) \frac{\zeta_3}{\big(
        4-z^2\big)^{3/2}}}{9 \pi  z^3}
+\frac{896 \ln^4(2) \frac{\zeta_3}{\big(
        4-z^2\big)^{3/2}}}{\pi  z^3}
\nonumber\\ && 
-\frac{257024 \ln(2) \pi  \frac{\zeta_3}{\big(
        4-z^2\big)^{3/2}}}{27 z^3}
+\frac{128 \pi  \big(
        2891-5994 z^2+999 z^4\big) \frac{\zeta_3}{\big(
        4-z^2\big)^{3/2}}}{243 z^3}
+\frac{4 \kappa_3 \frac{\zeta_3}{\big(
        4-z^2\big)^{3/2}}}{3 z^3}
\nonumber\\ && 
+\frac{2944 \ln^2(2) \pi  \frac{\zeta_3}{\big(
        4-z^2\big)^{3/2}}}{z^3}
-\frac{17 \kappa_1 \frac{\zeta_3}{\big(
        4-z^2\big)^{3/2}}}{6 z^3}
+\frac{196448 \frac{\zeta_3^2}{\big(
        4-z^2\big)^{3/2}}}{27 \pi  z^3}
-\frac{160 \pi  \big(
        8+7 z^2\big) \frac{\zeta_5}{\big(
        4-z^2\big)^{5/2}}}{z^3}
\nonumber\\ && 
-
\frac{128 \ln(2) \pi ^2 \big(
        -8+z^2
\big)
\big(z^2-2\big) \ln(z)}{9 (z-2) z^2 (z+2)}
-\frac{512 \ln(2) \pi ^3 \big(
        866-535 z^2+80 z^4\big) \frac{\ln(z)}{\big(
        4-z^2\big)^{5/2}}}{27 z^3}
\nonumber\\ && 
+\frac{64 \pi ^3 \big(
        z^2-2\big) \frac{1}{\sqrt{4-z^2}} \ln(z)}{27 z^3}
+\frac{256}{9} \ln(2) \pi  z \big(
        -23+8 z^2\big) \frac{1}{\sqrt{4-z^2}} \ln(z)
\nonumber\\ && 
-\frac{16 \pi ^2 \frac{1}{\big(
        4-z^2\big)^{3/2}} \GA\left(
        \sqrt{4-\tau } \sqrt{\tau };z^2\right)^3 \ln(z)}{3 z^3}
-\frac{256 \pi ^2 \big(
        26-9 z^2+z^4\big) \frac{1}{\big(
        4-z^2\big)^{5/2}} 
\ln(z)}{27 z^3} 
\nonumber\\ &&  \times
\GA\left(
        \frac{1}{4-\tau },\sqrt{4-\tau } \sqrt{\tau };z^2\right) 
+\frac{64 \pi ^2 \big(
        4808-1880 z^2+171 z^4\big) \frac{1}{\big(
        4-z^2\big)^{5/2}} 
\ln
(z)}{9 z^3}
\nonumber\\ &&  \times
\GA\left(
        \frac{1}{\tau },\sqrt{4-\tau } \sqrt{\tau };z^2\right) 
+\frac{128 \ln(2) \pi ^3 \big(
        860-550 z^2+83 z^4\big) \frac{1}{\big(
        4-z^2\big)^{5/2}} \ln\big(
        1-\frac{z^2}{4}\big)}{27 z^3}
\nonumber\\ &&
-\frac{512 \pi  \big(
               z^2-2\big) \frac{1}{\sqrt{4-z^2}} \ln^3(z)}{9 z^3}
                              +\tilde{F}^{(0), \rm rest}_{A,2}(z) ,
\end{eqnarray}
\begin{eqnarray}
\lefteqn{F^{(0)}_{S,1}(z) + \zeta_2 F^{(0)}_{S,2}(z) + \zeta_3 F^{(0)}_{S,3}(z) =} \nonumber\\ &&
\frac{4}{27} \ln^2(2) \pi  z \sqrt{4-z^2} Q_1
+\frac{8}{9} \ln(2) \pi  z \sqrt{4-z^2} \ln(z) Q_1
-\frac{64 \pi ^2 \ln^2(z) Q_{38}}{27 (z-2) z^2 (z+2)} 
\nonumber \\ &&
-\frac{4 \pi ^3 \frac{1}{\sqrt{4-z^2}} \ln(z) Q_{49}}{27 z^3}
-\frac{32 \pi  \frac{1}{\sqrt{4-z^2}} \ln^3(z) Q_{50}}{81 z^3}
-\frac{16 \pi  \frac{\zeta_3}{\sqrt{4-z^2}} Q_{53}}{27 z^3}
-\frac{4 \ln(2) \pi ^3 \frac{1}{\sqrt{4-z^2}} Q_{61}}{81 z^5} 
\nonumber \\ &&
+\frac{4 \pi  \frac{1}{\sqrt{4-z^2}} \ln^2(z) Q_{71}}{27 z^3}
-\frac{\pi ^2 \ln(z) Q_{77}}{243 (z-2) z^2 (z+2)}
+\Biggl[-\frac{64 \ln^3(2) \pi  \big(z^2-2\big)^2}{9 (z-2) z^2 (z+2)} 
\nonumber \\ &&
-\frac{8 \ln(2) \pi ^3 \big(1172-1118 z^2+239 z^4\big)}{27 (z-2) z^2 (z+2)}
+\frac{16 \ln^2(2) \pi  Q_{39}}{27 (z-2) z^2 (z+2)}
+\frac{32 \ln(2) \pi  \ln(z) Q_{39}}{9 (z-2) z^2 (z+2)} 
\nonumber \\ &&
+\frac{16 \pi  \ln^2(z) Q_{44}}{27 (z-2) z^2 (z+2)}
+\frac{4 \pi ^2 \frac{1}{\sqrt{4-z^2}} \ln(z) Q_{59}}{81 z^5}
-\frac{6944 \pi  \big(z^2-2\big)^2 \zeta_3}{27 (z-2) z^2 (z+2)} 
\nonumber \\ &&
-\frac{64 \ln^2(2) \pi  \big(z^2-2\big)^2 \ln(z)}{(z-2) z^2 (z+2)}
-\frac{1112 \pi ^3 \big(z^2-2\big)^2 \ln(z)}{27 (z-2) z^2 (z+2)}
-\frac{192 \ln(2) \pi  \big(z^2-2\big)^2 \ln^2(z)}{(z-2) z^2 (z+2)} 
\nonumber \\ &&
+\frac{64 \ln(2) \pi ^2 \big(z^2-3\big)\big(1+z^2\big) \frac{1}{\sqrt{4-z^2}} \ln(z)}{9 z^3}
+\frac{128 \pi ^2 \big(101-34 z^2+9 z^4\big) \frac{1}{\sqrt{4-z^2}} \ln^2(z)}{27 z^3} 
\nonumber \\ &&
-\frac{10432 \pi  \big(z^2-2\big)^2 \ln^3(z)}{81 (z-2) z^2 (z+2)}\Biggr] \GA\left(\sqrt{4-\tau } \sqrt{\tau };z^2\right)
+\Biggl[-\frac{176 \ln^2(2) \pi  \big(z^2-2\big) \frac{1}{\sqrt{4-z^2}}}{27 z} 
\nonumber \\ &&
+\frac{4 \pi ^2 \big(68-53 z^2+11 z^4\big) \ln(z)}{3 (z-2) z^2 (z+2)}
-\frac{16 \pi  \big(-318-86 z^2+67 z^4\big) \frac{1}{\sqrt{4-z^2}} \ln^2(z)}{27 z^3} 
\nonumber \\ &&
-\frac{352 \ln(2) \pi  \big(z^2-2\big) \frac{1}{\sqrt{4-z^2}} \ln(z)}{9 z}\Biggr] \GA\left(\sqrt{4-\tau } \sqrt{\tau };z^2\right)^2
+\Biggl[\frac{64 \ln^2(2) \pi  \big(z^2-2\big)^2}{27 (z-2) z^2 (z+2)} 
\nonumber \\ &&
+\frac{128 \ln(2) \pi  \big(z^2-2\big)^2 \ln(z)}{9 (z-2) z^2 (z+2)}
+\frac{64 \pi ^2 \big(5-2 z^2+z^4\big) \frac{1}{\sqrt{4-z^2}} \ln(z)}{27 z^3}
+\frac{320 \pi  \big(z^2-2\big)^2 \ln^2(z)}{27 (z-2) z^2 (z+2)}\Biggr]   
\nonumber \\ &&
\times \GA\left(\frac{1}{4-\tau },\sqrt{4-\tau } \sqrt{\tau };z^2\right)
+\Biggl[\frac{32 \ln^2(2) \pi  \big(z^2-2\big)^2}{3 (z-2) z^2 (z+2)}
+\frac{64 \ln(2) \pi  \big(z^2-2\big)^2 \ln(z)}{(z-2) z^2 (z+2)} 
\nonumber \\ &&
-\frac{16 \pi ^2 \big(476-181 z^2+45 z^4\big) \frac{1}{\sqrt{4-z^2}} \ln(z)}{9 z^3}
+\frac{2848 \pi  \big(z^2-2\big)^2 \ln^2(z)}{27 (z-2) z^2 (z+2)}\Biggr] \GA\left(\frac{1}{\tau },\sqrt{4-\tau } \sqrt{\tau };z^2\right) 
\nonumber \\ &&
+\Biggl[\frac{32 \ln^2(2) \pi  \big(z^2-2\big)^2}{3 (z-2) z^2 (z+2)}
+\frac{64 \ln(2) \pi  \big(z^2-2\big)^2 \ln(z)}{(z-2) z^2 (z+2)}
+\frac{96 \pi  \big(z^2-2\big)^2 \ln^2(z)}{(z-2) z^2 (z+2)}\Biggr]   
\nonumber \\ &&
\times \GA\left(\sqrt{4-\tau } \sqrt{\tau },\frac{1}{4-\tau };z^2\right)
-\frac{256 \Li_4\left(\frac{1}{2}\right) \pi  \big(36-43 z^2+8 z^4\big) \frac{1}{\sqrt{4-z^2}}}{27 z^3} 
\nonumber \\ &&
-\frac{32 \ln^4(2) \pi  \big(36-43 z^2+8 z^4\big) \frac{1}{\sqrt{4-z^2}}}{81 z^3}
-\frac{32 \ln^2(2) \pi ^3 \big(1350-659 z^2+100 z^4\big) \frac{1}{\sqrt{4-z^2}}}{81 z^3} 
\nonumber \\ &&
-\frac{4 \pi ^5 \big(1818-2711 z^2+1612 z^4\big) \frac{1}{\sqrt{4-z^2}}}{1215 z^3}
+\frac{32}{9} \ln^3(2) \pi  z \big(z^2-2\big) \sqrt{4-z^2}
-\frac{16 \pi ^3 \frac{\zeta_3}{\sqrt{4-z^2}}}{z^3} 
\nonumber \\ &&
-\frac{80 \pi  \frac{\zeta_5}{\sqrt{4-z^2}}}{z^3}
-\frac{640 \pi ^4 \big(z^2-2\big)^2 \ln(z)}{27 (z-2) z^2 (z+2)}
-\frac{32 \ln(2) \pi ^2 \big(2+z^2\big)\big(12-6 z^2+z^4\big) \ln(z)}{9 z^2} 
\nonumber \\ &&
+\frac{64 \ln(2) \pi ^3 \big(142-120 z^2+33 z^4\big) \frac{1}{\sqrt{4-z^2}} \ln(z)}{9 z^3}
+32 \ln^2(2) \pi  z \big(z^2-2\big) \sqrt{4-z^2} \ln(z) 
\nonumber \\ &&
+96 \ln(2) \pi  z \big(z^2-2\big) \sqrt{4-z^2} \ln^2(z)
-\frac{32 \ln(2) \pi ^3 \big(74-41 z^2+7 z^4\big) \frac{1}{\sqrt{4-z^2}} \ln\big(1-\frac{z^2}{4}\big)}{9 z^3} 
\nonumber \\ &&
-\frac{4096 \pi ^2 \big(z^2-2\big)^2 \ln^3(z)}{81 (z-2) z^2 (z+2)}
+\frac{16}{3} \ln^2(2) \pi  z \big(z^2-2\big) \sqrt{4-z^2} \ln\big(1-\frac{z^2}{4}\big) 
\nonumber \\ &&
+32 \ln(2) \pi  z \big(z^2-2\big) \sqrt{4-z^2} \ln(z) \ln\left(1-\frac{z^2}{4}\right)+48 \pi z \big(z^2-2\big) \sqrt{4-z^2} \ln^2(z)  
\nonumber \\ &&
                \times \ln\left(1-\frac{z^2}{4}\right)
                               +\tilde{F}^{(0), \rm rest}_{S}(z),
\end{eqnarray}
\begin{eqnarray}
\lefteqn{F^{(0)}_{P,1}(z) + \zeta_2 F^{(0)}_{P,2}(z) + \zeta_3 F^{(0)}_{P,3}(z) =} \nonumber\\ &&
-\frac{64 \pi ^2 \ln^2(z) Q_7}{27 (z-2) (z+2)}
-\frac{4 \pi ^3 \frac{1}{\sqrt{4-z^2}} \ln(z) Q_{28}}{27 z}
-\frac{32 \pi  \frac{1}{\sqrt{4-z^2}} \ln^3(z) Q_{31}}{81 z}
-\frac{16 \pi  \frac{\zeta_3}{\sqrt{4-z^2}} Q_{32}}{27 z} 
\nonumber \\ &&
-\frac{4 \ln(2) \pi ^3 \frac{1}{\big(4-z^2\big)^{3/2}} Q_{66}}{81 z^5}
-\frac{4 \pi (z-2) (z+2) \frac{1}{\big(4-z^2\big)^{3/2}} \ln^2(z) Q_{72}}{27 z^3}
-\frac{\pi ^2 \ln(z) Q_{78}}{243 (z-2) z^2 (z+2)} 
\nonumber \\ &&
+\Li_4\left(\frac{1}{2}\right) \pi \Biggl[-\frac{256}{9} z^3 \frac{1}{\big(4-z^2\big)^{3/2}}
-\frac{256 \big(-31+11 z^2\big) \frac{1}{\sqrt{4-z^2}}}{27 z}\Biggr]
+\Biggl[-\frac{64 \ln^3(2) \pi  \big(z^2-2\big)^2}{9 (z-2) z^2 (z+2)}         
\nonumber \\ &&
-\frac{8 \ln(2) \pi ^3 \big(740-1010 z^2+239 z^4\big)}{27 (z-2) z^2 (z+2)}
+\frac{16 \ln^2(2) \pi  Q_{40}}{27 (z-2) z^2 (z+2)}     
+\frac{32 \ln(2) \pi  \ln(z) Q_{40}}{9 (z-2) z^2 (z+2)} 
\nonumber \\ &&
+\frac{16 \pi  \ln^2(z) Q_{45}}{27 (z-2) z^2 (z+2)}
-\frac{4 \pi ^2 \frac{1}{\big(4-z^2\big)^{3/2}} \ln(z) Q_{76}}{81 z^5}
-\frac{6944 \pi  \big(z^2-2\big)^2 \zeta_3}{27 (z-2) z^2 (z+2)} 
\nonumber \\ &&
-\frac{64 \ln^2(2) \pi  \big(z^2-2\big)^2 \ln(z)}{(z-2) z^2 (z+2)}
-\frac{64 \ln(2) \pi ^2 \big(z^2-3\big)^2 \frac{1}{\big(4-z^2\big)^{3/2}} \ln(z)}{9 z} 
-\frac{1112 \pi ^3 \big(z^2-2\big)^2 \ln(z)}{27 (z-2) z^2 (z+2)} 
\nonumber \\ &&
-\frac{192 \ln(2) \pi  \big(z^2-2\big)^2 \ln^2(z)}{(z-2) z^2 (z+2)}
-\frac{128 \pi ^2 \big(137-70 z^2+9 z^4\big) \frac{1}{\big(4-z^2\big)^{3/2}} \ln^2(z)}{27 z} 
\nonumber \\ &&
-\frac{10432 \pi \big(z^2-2\big)^2 \ln^3(z)}{81 (z-2) z^2 (z+2)}\Biggr] \GA\left(\sqrt{4-\tau } \sqrt{\tau };z^2\right)
+\Biggl[-\frac{176 \ln^2(2) \pi  \big(z^2-2\big) \frac{1}{\sqrt{4-z^2}}}{27 z}
\nonumber \\ &&
+\frac{4 \pi ^2 \big(44-47 z^2+11 z^4\big) \ln(z)}{3 (z-2) z^2 (z+2)}
+\frac{16 \pi \big(338-354 z^2+67 z^4\big) \frac{1}{\big(4-z^2\big)^{3/2}} \ln^2(z)}{27 z} 
\nonumber \\ &&
-\frac{352 \ln(2) \pi \big(z^2-2\big) \frac{1}{\sqrt{4-z^2}} \ln(z)}{9 z}\Biggr] \GA^2\left(\sqrt{4-\tau } \sqrt{\tau };z^2\right)
+\Biggl[\frac{64 \ln^2(2) \pi  \big(z^2-2\big)^2}{27 (z-2) z^2 (z+2)}     
\nonumber \\ &&
+\frac{128 \ln(2) \pi  \big(z^2-2\big)^2 \ln(z)}{9 (z-2) z^2 (z+2)}
-\frac{64 \pi ^2 \big(z^2-3\big)^2 \frac{1}{\big(4-z^2\big)^{3/2}} \ln(z)}{27 z}
+\frac{320 \pi  \big(z^2-2\big)^2 \ln^2(z)}{27 (z-2) z^2 (z+2)}\Biggr]   
\nonumber \\ &&
\times \GA\left(\frac{1}{4-\tau },\sqrt{4-\tau } \sqrt{\tau };z^2\right)
+\Biggl[\frac{32 \ln^2(2) \pi  \big(z^2-2\big)^2}{3 (z-2) z^2 (z+2)}
+\frac{64 \ln(2) \pi  \big(z^2-2\big)^2 \ln(z)}{(z-2) z^2 (z+2)}         
\nonumber \\ &&
+\frac{16 \pi ^2 \big(656-343 z^2+45 z^4\big) \frac{1}{\big(4-z^2\big)^{3/2}} \ln(z)}{9 z}
+\frac{2848 \pi  \big(z^2-2\big)^2 \ln^2(z)}{27 (z-2) z^2 (z+2)}\Biggl]   
\nonumber \\ &&
\times \GA\left(\frac{1}{\tau },\sqrt{4-\tau } \sqrt{\tau };z^2\right)
+\Biggl[\frac{32 \ln^2(2) \pi  \big(z^2-2\big)^2}{3 (z-2) z^2 (z+2)}
+\frac{64 \ln(2) \pi  \big(z^2-2\big)^2 \ln(z)}{(z-2) z^2 (z+2)}        
\nonumber \\ &&
+\frac{96 \pi  \big(z^2-2\big)^2 \ln^2(z)}{(z-2) z^2 (z+2)}\Biggr] \GA\left(\sqrt{4-\tau } \sqrt{\tau },\frac{1}{4-\tau };z^2\right) 
\nonumber \\ &&
-\frac{32 \ln^4(2) \pi  \big(-124+75 z^2-8 z^4\big) \frac{1}{\big(4-z^2\big)^{3/2}}}{81 z}
+\frac{32 \ln^2(2) \pi ^3 \big(1358-735 z^2+100 z^4\big) \frac{1}{\big(4-z^2\big)^{3/2}}}{81 z} 
\nonumber \\ &&
+\frac{4 \pi ^5 \big(11642-9159 z^2+1612 z^4\big) \frac{1}{\big(4-z^2\big)^{3/2}}}{1215 z}
+\frac{32}{9} \ln^3(2) \pi  z \big(z^2-2\big) \sqrt{4-z^2} 
\nonumber \\ &&
-\frac{4}{27} \ln^2(2) \pi  z \big(z^2-2\big)\big(28-44 z^2+11 z^4\big) \sqrt{4-z^2}
+\frac{112 \pi ^3 \frac{\zeta_3}{\big(4-z^2\big)^{3/2}}}{3 z}
-\frac{240 \pi  \frac{\zeta_5}{\big(4-z^2\big)^{3/2}}}{z} 
\nonumber \\ &&
-\frac{32 \ln(2) \pi ^2 \big(8+4 z^2-4 z^4+z^6\big) \ln(z)}{9 z^2}
-\frac{64 \ln(2) \pi ^3 \big(1010-648 z^2+99 z^4\big) \frac{1}{\big(4-z^2\big)^{3/2}} \ln(z)}{27 z}
\nonumber \\ &&
-\frac{640 \pi ^4 \big(z^2-2\big)^2 \ln(z)}{27 (z-2) z^2 (z+2)}
+32 \ln^2(2) \pi  z \big(z^2-2\big) \sqrt{4-z^2} \ln(z) 
\nonumber \\ &&
-\frac{8}{9} \ln(2) \pi  z \big(z^2-2\big)\big(28-44 z^2+11 z^4\big) \sqrt{4-z^2} \ln(z)
+96 \ln(2) \pi z \big(z^2-2\big) \sqrt{4-z^2} \ln^2(z) 
\nonumber \\ &&
-\frac{4096 \pi ^2 \big(z^2-2\big)^2 \ln^3(z)}{81 (z-2) z^2 (z+2)}
+\frac{32 \ln(2) \pi ^3 \big(274-153 z^2+21 z^4\big) \frac{1}{\big(4-z^2\big)^{3/2}} \ln\big(1-\frac{z^2}{4}\big)}{27 z} 
\nonumber \\ &&
+\frac{16}{3} \ln^2(2) \pi z \big(z^2-2\big) \sqrt{4-z^2} \ln\left(1-\frac{z^2}{4}\right)
+32 \ln(2) \pi z \big(z^2-2\big) \sqrt{4-z^2} \ln(z) \ln\left(1-\frac{z^2}{4}\right)
\nonumber \\ &&
+48 \pi z \big(z^2-2\big) \sqrt{4-z^2} \ln^2(z) \ln\left(1-\frac{z^2}{4}\right)                +\tilde{F}^{(0), \rm rest}_{P}(z) \,.
\end{eqnarray}
The polynomials $Q_i$ are given by
\begin{eqnarray}
Q_1&=&-11 z^6+66 z^4-116 z^2+312,\\
Q_2&=&2 z^6+3 z^4-74 z^2+144,\\
Q_3&=&3 z^6-3444 z^4+10544 z^2-2496,\\
Q_4&=&6 z^6-69 z^4+196 z^2+32,\\
Q_5&=&9 z^6-479 z^4+3528 z^2-7032,\\
Q_6&=&9 z^6-106 z^4+562 z^2-1130,\\
Q_7&=&9 z^6-88 z^4+292 z^2-272,\\
Q_8&=&9 z^6-70 z^4+562 z^2-1758,\\
Q_9&=&11 z^6-66 z^4+118 z^2+8,\\
Q_{10}&=&11 z^6-66 z^4+180 z^2-240,\\
Q_{11}&=&11 z^6-22 z^4+4 z^2-48,\\
Q_{12}&=&21 z^6-236 z^4+824 z^2-860,\\
Q_{13}&=&21 z^6-202 z^4+300 z^2+156,\\
Q_{14}&=&27 z^6-187 z^4+274 z^2+156,\\
Q_{15}&=&27 z^6-144 z^4+80 z^2+128,\\
Q_{16}&=&31 z^6-62 z^4+12 z^2-144,\\
Q_{17}&=&45 z^6-514 z^4+2536 z^2-4808,\\
Q_{18}&=&45 z^6-352 z^4+2356 z^2-7032,\\
Q_{19}&=&52 z^6-505 z^4+1188 z^2-32,\\
Q_{20}&=&57 z^6-282 z^4+280 z^2+256,\\
Q_{21}&=&67 z^6-243 z^4-932 z^2+3340,\\
Q_{22}&=&67 z^6+19 z^4-1636 z^2+4836,\\
Q_{23}&=&99 z^6-808 z^4+2080 z^2-1732,\\
Q_{24}&=&99 z^6-524 z^4+672 z^2-156,\\
Q_{25}&=&100 z^6-1338 z^4+5363 z^2-6480,\\
Q_{26}&=&111 z^6-1230 z^4+3448 z^2+536,\\
Q_{27}&=&135 z^6+540 z^4-40096 z^2+129672,\\
Q_{28}&=&139 z^6-834 z^4+200 z^2+1832,\\
Q_{29}&=&139 z^6-278 z^4-32 z^2+384,\\
Q_{30}&=&160 z^6-721 z^4+490 z^2+192,\\
Q_{31}&=&163 z^6-978 z^4+692 z^2+1152,\\
Q_{32}&=&217 z^6-1302 z^4+1145 z^2+738,\\
Q_{33}&=&309 z^6-2914 z^4+6744 z^2-296,\\
Q_{34}&=&1612 z^6-11574 z^4+24821 z^2-18864,\\
Q_{35}&=&3 z^8-51 z^6+156 z^4+64 z^2-128,\\
Q_{36}&=&3 z^8-21 z^6+36 z^4-128 z^2+256,\\
Q_{37}&=&9 z^8-124 z^6+706 z^4-1448 z^2-192,\\
Q_{38}&=&9 z^8-88 z^6+392 z^4-544 z^2-256,\\
Q_{39}&=&11 z^8-88 z^6+196 z^4-200 z^2+288,\\
Q_{40}&=&11 z^8-88 z^6+196 z^4-72 z^2+32,\\
Q_{41}&=&11 z^8-88 z^6+230 z^4-240 z^2+216,\\
Q_{42}&=&11 z^8-66 z^6+118 z^4+232 z^2-288,\\
Q_{43}&=&21 z^8+6 z^6-296 z^4+1792 z^2-3072,\\
Q_{44}&=&67 z^8-488 z^6+484 z^4+992 z^2+1216,\\
Q_{45}&=&67 z^8-488 z^6+796 z^4+208 z^2+384,\\
Q_{46}&=&67 z^8-377 z^6-434 z^4+3656 z^2+920,\\
Q_{47}&=&139 z^8-834 z^6+196 z^4+1856 z^2-32,\\
Q_{48}&=&139 z^8-834 z^6+196 z^4+2368 z^2-1056,\\
Q_{49}&=&139 z^8-834 z^6+200 z^4+1848 z^2-32,\\
Q_{50}&=&163 z^8-978 z^6+692 z^4+1008 z^2+288,\\
Q_{51}&=&163 z^8-978 z^6+728 z^4+936 z^2+288,\\
Q_{52}&=&214 z^8-1585 z^6+3082 z^4+1344 z^2-2304,\\
Q_{53}&=&217 z^8-1302 z^6+1145 z^4-150 z^2+1776,\\
Q_{54}&=&1953 z^8-11718 z^6+12303 z^4-5346 z^2+5782,\\
Q_{55}&=&9 z^{10}-124 z^8+986 z^6-4976 z^4+9568 z^2-256,\\
Q_{56}&=&11 z^{10}-88 z^8+230 z^6-152 z^4-8 z^2+288,\\
Q_{57}&=&67 z^{10}-383 z^8-1202 z^6+13712 z^4-26888 z^2+1184,\\
Q_{58}&=&111 z^{10}-1190 z^8+2952 z^6+2784 z^4+4912 z^2+3520,\\
Q_{59}&=&297 z^{10}-2025 z^8+2792 z^6-754 z^4+12112 z^2+10240,\\
Q_{60}&=&309 z^{10}-4622 z^8+22854 z^6-38000 z^4-816 z^2+15808,\\
Q_{61}&=&717 z^{10}-4788 z^8+10572 z^6-4640 z^4-13584 z^2+18496,\\
Q_{62}&=&783 z^{10}-4860 z^8+2132 z^6-832 z^4+315328 z^2-1149440,\\
Q_{63}&=&972 z^{10}-20412 z^8+78927 z^6+27852 z^4-206992 z^2+499680,\\
Q_{64}&=&5859 z^{10}-35154 z^8+36909 z^6-52326 z^4+8306 z^2+367272,\\
Q_{65}&=&6003 z^{10}-50112 z^8+123660 z^6+25152 z^4-329080 z^2+477024,\\
Q_{66}&=&-717 z^{12}+7332 z^{10}-26484 z^8+35984 z^6+2160 z^4-20160 z^2-15104,\\
Q_{67}&=&-27 z^{12}+876 z^8-4872 z^6+61696 z^4-180480 z^2-65536,\\
Q_{68}&=&27 z^{12}-297 z^{10}+756 z^8+3660 z^6-59760 z^4+331520 z^2-667648,\\
Q_{69}&=&27 z^{12}-297 z^{10}+972 z^8-724 z^6-20808 z^4+176384 z^2-380416,\\
Q_{70}&=&27 z^{12}+108 z^{10}-2052 z^8+4512 z^6-76608 z^4+277504 z^2+212992,\\
Q_{71}&=&67 z^{12}-622 z^{10}+1530 z^8-1768 z^6+3468 z^4+12696 z^2+2368,\\
Q_{72}&=&67 z^{12}-622 z^{10}+1842 z^8-2064 z^6-1396 z^4+3000 z^2-2304,\\
Q_{73}&=&67 z^{12}-517 z^{10}-470 z^8+15864 z^6-45684 z^4+1928 z^2+13504,\\
Q_{74}&=&67 z^{12}-511 z^{10}+652 z^8+888 z^6+1388 z^4+7912 z^2+1216,\\
Q_{75}&=&81 z^{12}-864 z^{10}+2804 z^8-3136 z^6-13848 z^4+87456 z^2-59392,\\
Q_{76}&=&297 z^{12}-3051 z^{10}+9272 z^8-6222 z^6-6088 z^4+5184 z^2+8192,\\
Q_{77}&=&891 z^{12}-11421 z^{10}+41232 z^8+10668 z^6-263024 z^4+321536 z^2-180224,\\
Q_{78}&=&891 z^{12}-10935 z^{10}+35400 z^8+37092 z^6-341072 z^4+447488 z^2+65536,\\
Q_{79}&=&1107 z^{12}-12204 z^{10}+39716 z^8-39504 z^6+160736 z^4-520704 z^2
\nonumber\\ &&
-163840,\\
Q_{80}&=&1377 z^{12}-17658 z^{10}+56220 z^8+64968 z^6+129440 z^4-3691520 z^2
\nonumber\\ &&
+8142848,\\
Q_{81}&=&1677 z^{12}-17256 z^{10}+61380 z^8-81376 z^6-45864 z^4+302688 z^2-147968,\\
Q_{82}&=&1701 z^{12}-22194 z^{10}+76956 z^8+60648 z^6-452320 z^4+62464 z^2-507904.
\end{eqnarray}

\section{Numerical values of the new constants }
\label{app:kappa}

\vspace*{1mm}
\noindent
Here we present the 14 new constants needed to express the coefficients of the expansion around $\hs=4$ with 60 digits,
\begin{eqnarray}
\tk_1 &=& -1264.94322242780923299577505233621720067624211086209986111296, \\
\tk_2 &=& -26176.4667608724683949216820111127329755051498931864672207674, \\
\tk_3 &=& -2729.29921775058112000342259069251066915697435521878829616461, \\
\tk_4 &=& 55185.6670430603029362317458218280389428429637759659305766923, \\
\tk_5 &=& -231417.543320624197335029133354832277513762956642168670916934, \\
\tk_6 &=& 27058.0674155939392402733850737674942036176399269710732266681, \\
\tk_7 &=& 37228.1393096283192321319569484136035748028723926780936227023, \\
\tk_8 &=& -13339.4468993806410955294285663003095854470302119871183700172, \\
\tk_9 &=& 36376.0677825693690120778060832493123585788086425881920483389, \\
\tk_{10} &=& 44168.3670154020748917804528969924640054915510728808520969412, \\
\tk_{11} &=& 216837.119105601604298423515472074350527268068308535384925274, \\
\tk_{12} &=& -5730.87155843894481719264039344225664636604380605461996009706, \\
\tk_{13} &=& -135665.066806256268480389800559366792285769481271731824723568, \\
\tk_{14} &=& 25026.2194317039528218591802514512389969169209802143192666245.
\end{eqnarray}
In the ancillary files, we provide these constants with 1800 digits of precision.
{\footnotesize

}


\begin{thebibliography}{99}
%
\bibitem{Bernreuther:2004ih}
W.~Bernreuther, R.~Bonciani, T.~Gehrmann, R.~Heinesch, T.~Leineweber,
P.~Mastrolia, and E.~Remiddi,
{\it {Two-loop QCD corrections to the heavy quark form-factors: The Vector contributions}},
Nucl. Phys. B
  {\bfseries 706} (2005) 245--324 [hep-ph/0406046].
%
\bibitem{Bernreuther:2004th}
W.~Bernreuther, R.~Bonciani, T.~Gehrmann, R.~Heinesch, T.~Leineweber, P.~Mastrolia, and E.~Remiddi,
{\it Two-loop QCD corrections to the heavy quark form-factors: Axial vector contributions},
Nucl. Phys. B
  {\bfseries 712} (2005) 229--286
[hep-ph/0412259].
%
\bibitem{Bernreuther:2005rw}
W.~Bernreuther, R.~Bonciani, T.~Gehrmann, R.~Heinesch, T.~Leineweber, and E.~Remiddi,
{\it Two-loop QCD corrections to the heavy quark form-factors:Anomaly contributions},
Nucl. Phys. B
  {\bfseries 723} (2005) 91--116 [hep-ph/0504190].
%
\bibitem{Bernreuther:2005gw}
W.~Bernreuther, R.~Bonciani, T.~Gehrmann, R.~Heinesch, P.~Mastrolia, and E.~Remiddi,
{\it Decays of scalar and pseudoscalar Higgs bosons into fermions: Two-loop QCD 
corrections to the Higgs-quark-antiquark amplitude},
Phys. Rev. D
  {\bfseries 72} (2005) 096002 [hep-ph/0508254].
%
\bibitem{Gluza:2009yy}
J.~Gluza, A.~Mitov, S.~Moch, and T.~Riemann,
{\it The QCD form factor of heavy quarks at NNLO},
JHEP {\bfseries
  07} (2009) 001 [arXiv:0905.1137 [hep-ph]].
%
\bibitem{Ablinger:2017hst}
  J.~Ablinger, A.~Behring, J.~Bl\"umlein, G.~Falcioni, A.~De Freitas, P.~Marquard, N.~Rana and C.~Schneider,
  {\it Heavy quark form factors at two loops},
  Phys.\ Rev.\ D {\bf 97} (2018) no.9,  094022
  [arXiv:1712.09889 [hep-ph]].
%
\bibitem{Henn:2016kjz}
J.M.~Henn, A.V.~Smirnov, and V.A.~Smirnov,
{\it {Analytic results for planar three-loop integrals for massive form factors},
 JHEP} {\bfseries 12}
  (2016) 144 [arXiv:1611.06523 [hep-ph]].
%
\bibitem{Henn:2016tyf}
  J.~Henn, A.V.~Smirnov, V.A.~Smirnov and M.~Steinhauser,
  {\it Massive three-loop form factor in the planar limit},
  JHEP {\bf 1701} (2017) 074
  [arXiv:1611.07535 [hep-ph]].
%
\bibitem{Ahmed:2017gyt}
  T.~Ahmed, J.M.~Henn and M.~Steinhauser,
  {\it High energy behaviour of form factors},
  JHEP {\bf 06} (2017) 125
  [arXiv:1704.07846 [hep-ph]].
%
\bibitem{Ablinger:2018yae}
  J.~Ablinger, J.~Bl\"umlein, P.~Marquard, N.~Rana and C.~Schneider,
  {\it Heavy Quark Form Factors at Three Loops in the Planar Limit}
  Phys.\ Lett.\ B {\bf 782} (2018) 528--532   
  [arXiv:1804.07313 [hep-ph]].
%
\bibitem{Ablinger:2018zwz}
  J.~Ablinger, J.~Bl\"umlein, P.~Marquard, N.~Rana and C.~Schneider,
  {\it Automated Solution of First Order Factorizable Systems of Differential Equations in One Variable},
  Nucl. Phys. B {\bf 939} (2019) 253--291.
  [arXiv:1810.12261 [hep-ph]].
%
\bibitem{Lee:2018nxa}
  R.N.~Lee, A.V.~Smirnov, V.A.~Smirnov and M.~Steinhauser,
 {\it {Three-loop massive form factors: complete light-fermion corrections for the vector current}},
  JHEP {\bf 03} (2018) 136
  [arXiv:1801.08151 [hep-ph]].
%
\bibitem{Lee:2018rgs} 
  R.N.~Lee, A.V.~Smirnov, V.A.~Smirnov and M.~Steinhauser,
  {\it Three-loop massive form factors: complete light-fermion and large-N$_{c}$ corrections for vector, 
  axial-vector, scalar and pseudo-scalar currents},
  JHEP {\bf 05} (2018) 187
  [arXiv:1804.07310 [hep-ph]].  
%
\bibitem{Blumlein:2018tmz}
  J.~Bl\"umlein, P.~Marquard and N.~Rana,
  {\it Asymptotic behavior of the heavy quark form factors at higher order}
  Phys.\ Rev.\ D {\bf 99} (2019) no.1,  016013
  [arXiv:1810.08943 [hep-ph]].
%
\bibitem{Blumlein:2019oas}
J.~Bl\"umlein, P.~Marquard, N.~Rana and C.~Schneider,
{\it The Heavy Fermion Contributions to the Massive Three Loop Form Factors},
Nucl. Phys. B \textbf{949} (2019) 114751
[arXiv:1908.00357 [hep-ph]].
%
\bibitem{Fael:2022miw}
M.~Fael, F.~Lange, K.~Sch\"onwald and M.~Steinhauser,
{\it Singlet and nonsinglet three-loop massive form factors},
Phys. Rev. D \textbf{106} (2022) 034029
[arXiv:2207.00027 [hep-ph]].
%
\bibitem{Fael:2022rgm}
M.~Fael, F.~Lange, K.~Sch\"onwald and M.~Steinhauser,
{\it Massive Vector Form Factors to Three Loops},
Phys. Rev. Lett. \textbf{128} (2022) 172003
[arXiv:2202.05276 [hep-ph]];\\
%
K.~Sch\"onwald,
{\it Massive form factors at $\mathcal{O}(\alpha_s^3)$},
PoS (LL2022)  006
[arXiv:2207.06705 [hep-ph]].
%
\bibitem{Fael:2023zqr}
M.~Fael, F.~Lange, K.~Sch\"onwald and M.~Steinhauser,
{\it Massive three-loop form factors: Anomaly contribution},
Phys. Rev. D \textbf{107} (2023) no.9, 094017
[arXiv:2302.00693 [hep-ph]].
%
\bibitem{Blumlein:2017dxp}
  J.~Bl\"umlein and C.~Schneider,
  {\it The Method of Arbitrarily Large Moments to Calculate Single Scale Processes in Quantum Field 
  Theory},
  Phys.\ Lett.\ B {\bf 771} (2017) 31--36,
  [arXiv:1701.04614 [hep-ph]].
%
\bibitem{SolveCoupledSystem}
J. Ablinger, C. Schneider, A. Behring, J. Bl\"umlein, A. {De Freitas}, {\it Algorithms to solve coupled 
systems of differential equations in terms of power series}, PoS(LL2016)005, [arXiv:1608.05376 [cs.SC]];
\\
J. Bl\"umlein, P. Marquard, C. Schneider, \textit{Heavy quark form factors at three loops}, 
PoS(RADCOR2019)013, [arXiv:1912.04390 [cs.SC]].
%
\bibitem{Blumlein:2009cf}
J.~Bl\"umlein, D.J.~Broadhurst and J.A.M.~Vermaseren,
{\it The Multiple Zeta Value Data Mine},
Comput. Phys. Commun. \textbf{181} (2010) 582--625
[arXiv:0907.2557 [math-ph]].
%
\bibitem{LEWIN1}
A.~Devoto and D.W.~Duke,
{\it Table of Integrals and Formulae for Feynman Diagram Calculations},
Riv. Nuovo Cim. \textbf{7N6} (1984) 1--39;\\
L. Lewin, {\sf Dilogarithms and associated functions}, (Macdonald, London, 1958);
{\sf Polylogarithms and associated functions}, (North Holland, New York,
1981).
%
\bibitem{SageOre}
M.~Kauers, {\it Guessing Handbook}, JKU Linz, Technical Report RISC 09--07;\\
  J.~Bl\"umlein, M.~Kauers, S.~Klein and C.~Schneider,
  {\it Determining the closed forms of the $O(a_s^3)$ anomalous
dimensions and Wilson coefficients from Mellin
moments by means of computer algebra},
  Comput.\ Phys.\ Commun.\  {\bf 180} (2009) 2143--2165
  [arXiv:0902.4091 [hep-ph]];\\
M.~Kauers, M.~Jaroschek, and F.~Johansson, {\it Ore Polynomials in
Sage}, in:
{\sf Computer Algebra and Polynomials},
Editors: J.~Gutierrez, J.~Schicho, Josef, M.~Weimann, Eds..
Lecture Notes in Computer Science {\bf 8942} (Springer, Berlin, 2015) 105--125
[arXiv:1306.4263 [cs.SC]].
%
\bibitem{HOLONOMIC}
B. Salvy and P. Zimmermann, 
{\it GFUN: a Maple package for the manipulation of generating and holonomic functions in one variable},
ACM Trans. Math. Software {\bf 20} (1994) 163--177;\\
C. Mallinger, {\it Algorithmic Manipulations and Transformations of
	Univariate Holonomic Functions and Sequences} Master's thesis RISC, J.
Kepler University Linz, 1996;\\
M. Kauers and P. Paule, {\sf The Concrete Tetrahedron}, Text and
Monographs in Symbolic Computation (Springer Wien), 2011.
%
\bibitem{DRAlgorithms}
M.~Karr, 
{\it Summation in finite terms},
{J.~ACM} {\bf 28} (1981) 305--350;\\
C.~Schneider,
{\sf Symbolic summation in difference fields} Ph.D. Thesis
RISC, Johannes Kepler University, Linz technical report 01-17 (2001);
{\it A symbolic summation approach to find optimal nested sum representations} in:
{\sf Motives, Quantum Field Theory, and Pseudodifferential
	Operators} ({\sf Clay Mathematics Proceedings Vol.}~{\bf{12}}) ed. A.~Carey,
D.~Ellwood, S.~Paycha and S.~Rosenberg,(Amer. Math. Soc) (2010), 285--308
[arXiv:0904.2323];
\textit{Fast algorithms for refined parameterized telescoping in difference fields},
in: Computer Algebra and Polynomials, Applications of Algebra and Number Theory, 
J.~Gutierrez, J.~Schicho, M.~Weimann (ed.), Lecture Notes in Computer Science 
(LNCS) 8942 (2015), 157--191
[arXiv:13077887 [cs.SC]];
{\it A refined difference field theory for symbolic summation},
{J. Symbolic Comput.} {\bf 43} (2008) 611--644
[arXiv:0808.2543]; {\it A difference ring theory for symbolic summation},
J. Symb. Comput. {\bf 72} (2016) 82--127  
[arXiv:1408.2776 [cs.SC]];
{\it Summation Theory II: Characterizations of ${R\Pi\Sigma^*}$-extensions and algorithmic aspects},
J. Symb. Comput. {\bf 80} (2017) 616--664  
[arXiv:1603.04285 [cs.SC]];
\\
E.D. Ocansey and C. Schneider, {\it Representation of hypergeometric products of
	higher nesting depths in difference rings}, J. Symbol. Comput, in press, 2023. \url{doi.org/10.1016/j.jsc.2023.03.002}\\
C. Schneider, {\it Refined telescoping algorithms in $R\Pi\Sigma$-extensions to reduce the degrees of the denominators}, 2023. arXiv:2302.03563 [cs.SC]. 
%
\bibitem{Blumlein:2018cms}
J.~Bl\"umlein and C.~Schneider,
{\it Analytic computing methods for precision calculations in quantum field theory},
Int. J. Mod. Phys. A \textbf{33} (2018) no.17, 1830015
[arXiv:1809.02889 [hep-ph]].
%
\bibitem{Nogueira:1991ex}
P.~Nogueira,
{\it {Automatic Feynman graph generation}},
 J. Comput. Phys.
  {\bfseries 105} (1993) 279--289.
%
\bibitem{Vermaseren:2000nd}
J.A.M.~Vermaseren, {\it {New features of FORM}},
math-ph/0010025.
%
\bibitem{Tentyukov:2007mu}
M.~Tentyukov and J.A.M. Vermaseren,
{\it The Multithreaded version of FORM},
Comput. Phys. Commun. {\bfseries 181} (2010) 1419--1427 [hep-ph/0702279].
%
\bibitem{vanRitbergen:1998pn}
T.~van Ritbergen, A.~N.~Schellekens and J.~A.~M.~Vermaseren,
{\it Group theory factors for Feynman diagrams},
Int. J. Mod. Phys. A \textbf{14} (1999), 41-96
[hep-ph/9802376].
%
\bibitem{IBP}
J. Lagrange, {\sf Nouvelles recherches sur la nature et la propagation
du son}, Miscellanea Taurinensis, t. II, 1760-61; Oeuvres t. I, p. 263;\\
C.F. Gau\ss{}, {Theoria attractionis corporum sphaeroidicorum ellipticorum
homogeneorum methodo novo tractate}, Commentationes societas scientiarum
Gottingensis recentiores, Vol III, 1813, Werke Bd. {\bf V} pp. 5--7;\\
G. Green, {\sf Essay on the Mathematical Theory of Electricity and
Magnetism}, Nottingham, 1828 [Green Papers, pp. 1--115];\\
M. Ostrogradsky (presented: November 5, 1828 ; published: 1831) {\it Premi\`ere note sur la th\'eorie de la
chaleur}, M\'emoires de l'Acad\'emie imp\'eriale des sciences de St. P\'etersbourg, series 6,
1: 129--133;\\
  K.G.~Chetyrkin and F.V.~Tkachov,
  {\it Integration by Parts: The Algorithm to Calculate $\beta$ Functions in 4 Loops},
  Nucl.\ Phys.\ B {\bf 192} (1981) 159--204;\\
  S.~Laporta,
  {\it High precision calculation of multiloop Feynman integrals by difference equations},
  Int.\ J.\ Mod.\ Phys.\ A {\bf 15} (2000) 5087--5159
  [hep-ph/0102033].
%
\bibitem{CRUSHER}
 P.~Marquard and D.~Seidel, {\it The {\tt Crusher} algorithm}, unpublished.
%
\bibitem{LinearSolver}
M. Bronstein, {\it On solutions of linear ordinary difference equations in their coefficient field}, J.~Symbolic Comput. 
{\bf 29}, pp. 841--877;\\
S.A. Abramov, M. Petkov{\v s}ek, {\it D'{A}lembertian solutions of linear
	differential and difference equations} in Proc. ISSAC'94, ed von~zur
Gathen J (ACM Press) pp. 169--174;\\
C. Schneider, 
{\it A collection of denominator bounds to solve parameterized linear difference 
	equations in {${\Pi}{\Sigma}$}-extensions},
An. Univ. Timisoara Ser. Mat.-Inform. {\bf 42} (2004) 163;
{\it Solving parameterized linear difference equations in terms of indefinite nested 
	sums and products},
J. Differ. Equations Appl.  {\bf 11} (2005) 799--821;
{\it Degree bounds to find polynomial solutions of parameterized linear difference 
	equations in {$\Pi\Sigma$}-Fields},
Appl. Algebra Engrg. Comm. Comput. {\bf 16}(2005) 1--32;\\
S.A. Abramov, M. Bronstein, M. Petkov\v{s}ek, C. Schneider, 
{\it On rational and hypergeometric solutions of linear ordinary difference equations 
	in {$\Pi \Sigma^*$}-field extensions},
{J. Symbolic Comput.} {\bf 107} (2021) 23--66
[arXiv:2005.04944].
%
\bibitem{SIG1}
C.~Schneider, {\it Symbolic Summation Assists Combinatorics}, 
{S\'em.~Lothar. Combin.\/} {\bf 56} (2007) 1--36
 article B56b.
%
\bibitem{SIG2}
C.~Schneider, {\it Simplifying Multiple Sums in Difference Fields}, in:~{{\sf Computer
Algebra in Quantum Field Theory: Integration,
  Summation and Special Functions}\/} Texts and Monographs in Symbolic
  Computation eds. C.~Schneider and J.~Bl\"umlein  (Springer, Wien, 2013) 325--360
  [arXiv:1304.4134 [cs.SC]].
%
\bibitem{Schneider:2013zna}
  C.~Schneider,
  {\it Modern Summation Methods for Loop Integrals in Quantum Field Theory: The Packages Sigma, 
EvaluateMultiSums and SumProduction},
  J.\ Phys.\ Conf.\ Ser.\  {\bf 523} (2014) 012037
  [arXiv:1310.0160 [cs.SC]].
%
\bibitem{Ablinger:2011te}
J.~Ablinger, J.~Bl\"umlein and C.~Schneider,
{\it Harmonic Sums and Polylogarithms Generated by Cyclotomic Polynomials},
J. Math. Phys. \textbf{52} (2011) 102301
[arXiv:1105.6063 [math-ph]].
%
\bibitem{Ablinger:2014bra}
J.~Ablinger, J.~Bl\"umlein, C.G.~Raab and C.~Schneider,
{\it Iterated Binomial Sums and their Associated Iterated Integrals},
J. Math. Phys. \textbf{55} (2014) 112301
[arXiv:1407.1822 [hep-th]].
%
\bibitem{Vermaseren:1998uu}
  J.A.M.~Vermaseren,
  {\it Harmonic sums, Mellin transforms and integrals},
  Int.\ J.\ Mod.\ Phys.\ A {\bf 14} (1999) 2037--2076
  [hep-ph/9806280].   
%
\bibitem{Blumlein:1998if} 
  J.~Bl\"umlein and S.~Kurth,
  {\it Harmonic sums and Mellin transforms up to two loop order},
  Phys.\ Rev.\ D {\bf 60} (1999) 014018 
  [hep-ph/9810241].
%
\bibitem{ALG}
M.E.~Hoffman,
{\it The Algebra of Multiple Harmonic Series} 
J. Algebra  {\bf 194} (1997) 477--495;\\
J.~Bl\"umlein,
{\it Algebraic relations between harmonic sums and associated quantities},
Comput. Phys. Commun. \textbf{159} (2004) 19--54
[hep-ph/0311046].
%
\bibitem{Remiddi:1999ew}
E.~Remiddi and J.A.M.~Vermaseren,
{\it Harmonic polylogarithms},
Int. J. Mod. Phys. A \textbf{15} (2000) 725--754
[hep-ph/9905237].
%
\bibitem{HARMSU}
J.~Ablinger, J.~Bl\"umlein and C.~Schneider,
{\it Generalized Harmonic, Cyclotomic, and Binomial Sums, their Polylogarithms and Special Numbers},
J. Phys. Conf. Ser. \textbf{523} (2014) 012060
[arXiv:1310.5645 [math-ph]];\\
J.~Ablinger,
{\it The package HarmonicSums: Computer Algebra and Analytic aspects of
Nested Sums},
PoS (LL2014) 019 [arXiv:1407.6180[cs.SC]];
  {\it A Computer Algebra Toolbox for Harmonic Sums Related to Particle Physics},
  Diploma Thesis, JKU Linz, 2009,
  arXiv:1011.1176[math-ph];
{\it Computer Algebra Algorithms for Special Functions in
  Particle Physics}, Ph.D. Thesis, Linz U. (2012) arXiv:1305.0687[math-ph];
{\it Inverse Mellin Transform of Holonomic Sequences},
PoS (LL2016) 067;
{\it Discovering and Proving Infinite Binomial Sums Identities},
Experimental Mathematics 26 (2017) [arXiv:1507.01703 [math.CO]];
{\it Computing the Inverse Mellin Transform of Holonomic Sequences
using Kovacic's Algorithm},
PoS (RADCOR2017) 001 [arXiv:1801.01039 [cs.SC]];
{\it Discovering and Proving Infinite Pochhammer Sum Identities},
arXiv:1902.11001 [math.CO];
{\it An Improved Method to Compute the Inverse Mellin Transform of Holonomic Sequences},
PoS (LL2018) 063.
%
\bibitem{Ablinger:2021fnc}
J.~Ablinger, J.~Bl\"umlein and C.~Schneider,
{\it Iterated integrals over letters induced by quadratic forms},
Phys. Rev. D \textbf{103} (2021) no.9, 096025
[arXiv:2103.08330 [hep-th]].
%
\bibitem{Blumlein:2009ta}
J.~Bl\"umlein,
{\it Structural Relations of Harmonic Sums and Mellin Transforms up to Weight w = 5},
Comput. Phys. Commun. \textbf{180} (2009) 2218-2249
[arXiv:0901.3106 [hep-ph]].
%
\bibitem{Ablinger:2013cf}
J.~Ablinger, J.~Bl\"umlein and C.~Schneider,
{\it Analytic and Algorithmic Aspects of Generalized Harmonic Sums and Polylogarithms},
J. Math. Phys. \textbf{54} (2013) 082301
[arXiv:1302.0378 [math-ph]].
%
\bibitem{Tarasov:1980au}
O.V.~Tarasov, A.A.~Vladimirov and A.Y.~Zharkov,
{\it The Gell-Mann-Low Function of QCD in the Three Loop Approximation},
Phys. Lett. B \textbf{93} (1980) 429--432

\bibitem{Larin:1993tp}
S.A.~Larin and J.A.M.~Vermaseren,
{\it The Three loop QCD Beta function and anomalous dimensions},
Phys. Lett. B \textbf{303} (1993) 334--336
[arXiv:hep-ph/9302208 [hep-ph]].

\bibitem{vanRitbergen:1997va}
T.~van Ritbergen, J.A.M.~Vermaseren and S.A.~Larin,
{\it The Four loop beta function in quantum chromodynamics}
Phys. Lett. B \textbf{400} (1997) 379--384
[arXiv:hep-ph/9701390 [hep-ph]].

\bibitem{Czakon:2004bu}
M.~Czakon,
{\it The Four-loop QCD beta-function and anomalous dimensions},
Nucl. Phys. B \textbf{710} (2005) 485--498
[arXiv:hep-ph/0411261 [hep-ph]].

\bibitem{Chetyrkin:2004mf}
K.G.~Chetyrkin,
{\it Four-loop renormalization of QCD: Full set of renormalization constants and anomalous dimensions},
Nucl. Phys. B \textbf{710} (2005) 499--510
[arXiv:hep-ph/0405193 [hep-ph]].

\bibitem{Baikov:2016tgj}
P.A.~Baikov, K.G.~Chetyrkin and J.H.~K\"uhn,
{\it Five-Loop Running of the QCD coupling constant},
Phys. Rev. Lett. \textbf{118} (2017) no.8, 082002
[arXiv:1606.08659 [hep-ph]].
\bibitem{Herzog:2017ohr}
F.~Herzog, B.~Ruijl, T.~Ueda, J.A.M.~Vermaseren and A.~Vogt,
{\it The five-loop beta function of Yang-Mills theory with fermions},
JHEP \textbf{02} (2017) 090
[arXiv:1701.01404 [hep-ph]].

\bibitem{Luthe:2017ttg}
T.~Luthe, A.~Maier, P.~Marquard and Y.~Schr\"oder,
{\it The five-loop Beta function for a general gauge group and anomalous dimensions beyond Feynman gauge}
JHEP \textbf{10} (2017) 166
[arXiv:1709.07718 [hep-ph]].
\bibitem{Luthe:2017ttc}
T.~Luthe, A.~Maier, P.~Marquard and Y.~Schr\"oder,
{\it Complete renormalization of QCD at five loops},
JHEP \textbf{03} (2017) 020
[arXiv:1701.07068 [hep-ph]].
\bibitem{Chetyrkin:2017bjc}
K.~G.~Chetyrkin, G.~Falcioni, F.~Herzog and J.~A.~M.~Vermaseren,
{\it Five-loop renormalisation of QCD in covariant gauges},
JHEP \textbf{10} (2017) 179
[arXiv:1709.08541 [hep-ph]].

\bibitem{Schroder:2005hy}
Y.~Schr\"oder and M.~Steinhauser,
{\it Four-loop decoupling relations for the strong coupling},
JHEP \textbf{01} (2006) 051
[arXiv:hep-ph/0512058 [hep-ph]].

\bibitem{Chetyrkin:2005ia}
K.G.~Chetyrkin, J.H.~K\"uhn and C.~Sturm,
{\it QCD decoupling at four loops},
Nucl. Phys. B \textbf{744} (2006) 121--135
[arXiv:hep-ph/0512060 [hep-ph]].



\bibitem{Chetyrkin:1999ys}
K.G.~Chetyrkin and M.~Steinhauser,
{\it Short distance mass of a heavy quark at order $\alpha_s^3$},
Phys. Rev. Lett. \textbf{83} (1999) 4001--4004
[arXiv:hep-ph/9907509 [hep-ph]].
\bibitem{Chetyrkin:1999qi}
K.G.~Chetyrkin and M.~Steinhauser,
{\it The Relation between the MS-bar and the on-shell quark mass at order $\alpha_s^3$},
Nucl. Phys. B \textbf{573} (2000) 617--651
[arXiv:hep-ph/9911434 [hep-ph]].
\bibitem{Melnikov:2000qh}
K.~Melnikov and T.~v.~Ritbergen,
{\it The Three loop relation between the MS-bar and the pole quark masses},
Phys. Lett. B \textbf{482} (2000) 99--108
[arXiv:hep-ph/9912391 [hep-ph]].
\bibitem{Broadhurst:1991fy}
D.J.~Broadhurst, N.~Gray and K.~Schilcher,
{\it Gauge invariant on-shell Z(2) in QED, QCD and the effective field theory of a static quark},
Z. Phys. C \textbf{52} (1991) 111--122
\bibitem{Marquard:2018rwx}
P.~Marquard, A.V.~Smirnov, V.A.~Smirnov and M.~Steinhauser,
{\it Four-loop wave function renormalization in QCD and QED},
Phys. Rev. D \textbf{97} (2018) no.5, 054032
[arXiv:1801.08292 [hep-ph]].
\bibitem{Marquard:2016dcn}
P.~Marquard, A.V.~Smirnov, V.A.~Smirnov, M.~Steinhauser and D.~Wellmann,
{\it $\overline{\rm MS}$-on-shell quark mass relation up to four loops in QCD and a general SU$(N)$ gauge group},
Phys. Rev. D \textbf{94} (2016) no.7, 074025
[arXiv:1606.06754 [hep-ph]].
\bibitem{Marquard:2015qpa}
P.~Marquard, A.V.~Smirnov, V.A.~Smirnov and M.~Steinhauser,
{\it Quark Mass Relations to Four-Loop Order in Perturbative QCD},
Phys. Rev. Lett. \textbf{114} (2015) no.14, 142002
[arXiv:1502.01030 [hep-ph]].

\bibitem{Becher:2009kw}
T.~Becher and M.~Neubert,
{\it Infrared singularities of QCD amplitudes with massive partons},
Phys. Rev. D \textbf{79} (2009) 125004
[erratum: Phys. Rev. D \textbf{80} (2009) 109901]
[arXiv:0904.1021 [hep-ph]].
\bibitem{Mitov:2006xs}
A.~Mitov and S.~Moch,
{\it The Singular behavior of massive QCD amplitudes},
JHEP \textbf{05} (2007) 001
[arXiv:hep-ph/0612149 [hep-ph]].

\bibitem{Korchemsky:1987wg}
G.P.~Korchemsky and A.~V.~Radyushkin,
{\it Renormalization of the Wilson Loops Beyond the Leading Order},
Nucl. Phys. B \textbf{283} (1987) 342--364
\bibitem{Kidonakis:2009ev}
N.~Kidonakis,
{\it Two-loop soft anomalous dimensions and NNLL resummation for heavy quark production},
Phys. Rev. Lett. \textbf{102} (2009) 232003
[arXiv:0903.2561 [hep-ph]].
\bibitem{Grozin:2014hna}
A.~Grozin, J.M.~Henn, G.P.~Korchemsky and P.~Marquard,
{\it Three Loop Cusp Anomalous Dimension in QCD},
Phys. Rev. Lett. \textbf{114} (2015) no.6, 062006
[arXiv:1409.0023 [hep-ph]].
\bibitem{Grozin:2015kna}
A.~Grozin, J.M.~Henn, G.P.~Korchemsky and P.~Marquard,
{\it The three-loop cusp anomalous dimension in QCD and its supersymmetric extensions},
JHEP \textbf{01} (2016) 140
[arXiv:1510.07803 [hep-ph]].

\bibitem{PSLQ}
H.~Ferguson and R.~Forcade, {\it Generalization of the Euclidean algorithm for real numbers to all 
dimensions higher than two},
Bull. Am. Math. Soc. \textbf{1} (1979) 912--914; \\
%
H.~Ferguson and D.~Bailey,
{\it A Polynomial Time, Numerically Stable Integer Relation Algorithm},
Technical Report RNR-91-032 (1991); \\
%
D.H.~Bailey and D.J.~Broadhurst,
{\it Parallel integer relation detection: Techniques and applications},
Math. Comput. \textbf{70} (2001) 1719--1736
[arXiv:math/9905048 [math.NA]].
%
\bibitem{Koutschan:13}
C. Koutschan, {\it Creative telescoping for holonomic functions} in Computer
Algebra in Quantum Field Theory, Texts and Monographs in Symbolic
Computation (Springer), ed C. Schneider and J. Bl\"umlein, pp. 171--194.
%
\bibitem{CLAUSEN}
T.~Clausen, {\it \"Uber die Function $\sin(\varphi) + (1/22) \sin(2\varphi) + (1/32) \sin(3 \varphi)$ + etc}, 
Journal f\"r die reine und angewandte Mathematik (Crelle)  {\bf 8} (1832) 298--300;\\
J.M.~Borwein, D.J.~Broadhurst and J.~Kamnitzer,
{\it Central binomial sums, multiple Clausen values and Zeta values},
Exper. Math. \textbf{10} (2001) 25--34
[arXiv:hep-th/0004153 [hep-th]];\\
A.I.~Davydychev and M.Y.~Kalmykov,
{\it Massive Feynman diagrams and inverse binomial sums},
Nucl. Phys. B \textbf{699} (2004) 3--64
[arXiv:hep-th/0303162 [hep-th]].
%
\bibitem{TWOmass}
J.~Ablinger, J.~Bl\"umlein, A.~De Freitas, M.~Saragnese, C.~Schneider and K.~Sch\"onwald,
{\it The three-loop polarized pure singlet operator matrix element with two different masses}
Nucl. Phys. B \textbf{952} (2020) 114916
[arXiv:1911.11630 [hep-ph]];\\
J.~Ablinger, J.~Bl\"umlein, A.~De Freitas, A.~Goedicke, M.~Saragnese, C.~Schneider and 
K.~Sch\"onwald,
{\it The two-mass contribution to the three-loop polarized gluonic operator matrix element 
$A_{gg,Q}^{(3)}$},
Nucl. Phys. B \textbf{955} (2020) 115059
[arXiv:2004.08916 [hep-ph]];\\
J.~Ablinger, J.~Bl\"umlein, A.~De Freitas, M.~Saragnese, C.~Schneider and K.~Sch\"onwald,
{\it The three-loop polarized pure singlet operator matrix element with two different masses},
Nucl. Phys. B \textbf{952} (2020) 114916
[arXiv:1911.11630 [hep-ph]];\\
J.~Ablinger, J.~Bl\"umlein, A.~De Freitas, A.~Goedicke, M.~Saragnese, C.~Schneider and 
K.~Sch\"onwald,
{\it The two-mass contribution to the three-loop polarized gluonic operator matrix element 
$A_{gg,Q}^{(3)}$}
Nucl. Phys. B \textbf{955} (2020) 115059
[arXiv:2004.08916 [hep-ph]].
%
\bibitem{HPL}
T.~Gehrmann and E.~Remiddi,
{\it Numerical evaluation of harmonic polylogarithms},
Comput. Phys. Commun. \textbf{141} (2001) 296--312
[hep-ph/0107173];\\
J.~Vollinga and S.~Weinzierl,
{\it Numerical evaluation of multiple polylogarithms},
Comput. Phys. Commun. \textbf{167} (2005) 177--194
[hep-ph/0410259];\\
D.~Maitre,
{\it {\tt HPL}, a mathematica implementation of the harmonic polylogarithms},
Comput. Phys. Commun. \textbf{174} (2006) 222--240
[hep-ph/0507152];
{\it Extension of {\tt HPL} to complex arguments},
Comput. Phys. Commun. \textbf{183} (2012) 846
[hep-ph/0703052];\\
S.~Buehler and C.~Duhr,
{\it {\tt CHAPLIN} - Complex Harmonic Polylogarithms in Fortran},
Comput. Phys. Commun. \textbf{185} (2014) 2703--2713
[arXiv:1106.5739 [hep-ph]];\\
L.~Naterop, A.~Signer and Y.~Ulrich,
{\it {\tt handyG} \textemdash{}Rapid numerical evaluation of generalised polylogarithms in Fortran},
Comput. Phys. Commun. \textbf{253} (2020) 107165
[arXiv:1909.01656 [hep-ph]].
%
\bibitem{Ablinger:2018sat}
J.~Ablinger, J.~Bl\"umlein, M.~Round and C.~Schneider,
{\it Numerical Implementation of Harmonic Polylogarithms to Weight w = 8},
Comput. Phys. Commun. \textbf{240} (2019) 189--201
[arXiv:1809.07084 [hep-ph]].
%
\bibitem{CPOLYF}
J.~Bl\"umlein, Code {\tt Cpoly.f}, {\tt https://www-zeuthen.desy.de/$\sim${}blumlein/}
%
\bibitem{AFILE}
\href{https://doi.org/10.5281/zenodo.8109907}{doi:10.5281/zenodo.8109907}
%
%
\bibitem{Vermaseren:1994je}
 J.~Vermaseren, {\it Axodraw}, Comput. Phys. Commun. {\bf 83} (1994) 45--58.
\end{thebibliography}
\end{document}